\documentclass[useAMS,usenatbib,referee]{mn2e}
\usepackage{graphicx,graphics,aas_macros,amsmath}
\usepackage[authoryear]{natbib}
\usepackage{pdflscape}
\usepackage{longtable}
\usepackage{booktabs}
\usepackage{color}
\usepackage{hyperref}
\usepackage{amssymb}
\usepackage{wrapfig}
\usepackage{picture}
\title[Comprehensive analysis]{Comprehensive broadband study of accreting neutron stars with \emph{Suzaku}: Is there a bi-modality in the X-ray spectrum?}
\author[P. Pradhan, B.Paul, E. Bozzo, C. Maitra, B.C. Paul]{Pragati Pradhan$^{1,2}$ \thanks{E-mail:pragati@mit.edu}
 Biswajit Paul$^{3}$ Enrico Bozzo$^{4}$ Chandreyee Maitra $^{5}$ B.C. Paul$^{6}$ \\
 $^{1}$ Massachusetts Institute of Technology, Kavli Institute for Astrophysics and Space Research, Cambridge, MA, 02139, USA \\
$^{2}$ St. Joseph's College, Singamari, Darjeeling-734104, West Bengal, India\\
$^{3}$ Raman Research Institute, Sadashivnagar, Bangalore-560080, India\\
$^{4}$ ISDC, University of Geneva, Chemin d'Ecogia 16, Versoix, 1290, Switzerland \\
$^{5}$ Max Planck Institute For Extraterrestrial Physics, 85748 Garching, Germany \\
$^{6}$ North Bengal University, Raja Rammohanpur,  District Darjeeling-734013, West Bengal, India \\
 }
 \newcommand{\kte}{\ensuremath{kT_{\rm e}}}

 \newcommand{\lumin}{\ensuremath{L_{\rm{X} }}}
\newcommand{\wgamma}{\ensuremath{\Gamma} }
\newcommand{\ecut}{\ensuremath{E_{\rm{cut} }}}
\newcommand{\efold}{\ensuremath{E_{\rm{fold} }}}
\newcommand{\suzaku}{\emph{Suzaku}}
\newcommand{\ecyc}[1]{\ensuremath{E_{\rm{C}}}}

 \newcommand{\plcuteq}{\ensuremath{
  \mathrm{HIGHECUT}(E) = A\ E^{-\Gamma}\times\
  \begin{cases}
     1                           & (E\leq\ecut) \\
     {\rm e}^{-(E-\ecut)/\efold} & (E>\ecut)    \\
  \end{cases}
}}

\newcommand{\eqfwhmpropto}{
   W_{\rm c}\ensuremath{ \propto \ecyc{} \sqrt{\kte} \left|\cos(\theta)\right|
}}

\newcommand{\crsfeq}[1]{\ensuremath{
   \ecyc{#1}= 11.6\frac{B}{10^{12}\,\rm{G}}(1 + z)^{-1}
}}

\begin{document}
\date{}
 \maketitle
 \label{firstpage}
 \begin{abstract}
 We present a broadband spectral analysis of accreting neutron stars using data from XIS and PIN onboard \emph{Suzaku}. From spectral fits of these sources with a single continuum model including a powerlaw and high energy cut-off, cyclotron lines (where required), we studied the correlation between various spectral parameters.  Among 39 sources we studied, 16 are those where the existence of a cyclotron line is known in literature, and 29 need a cutoff energy. Among these 29 sources, 18 have cutoff energy bunched in a range of 3-10 keV while for 11 sources, it spreads over 12-25 keV. This bi-modal behaviour is not based on the specific nature of the systems being a Be XRB or supergiant HMXB, nor on different beaming patterns characterizing their X-ray emission (as inferred from simultaneous study of their pulse profiles). The broadband coverage of \emph{Suzaku} also shows that the cutoff energies saturate for higher values of cyclotron line energies - consistent with previous works in literature - for both the groups and the width of the cyclotron line show a weak correlation with the cyclotron line energy. We also find an anticorrelation with luminosity for both spectral index and folding energy, respectively. Unlike previous works, we did not detect any anticorrelation between X-ray luminosity and EW of  K$\alpha$ lines. Finally, we show that the EW and flux of the iron  K$\alpha$ line are smaller in SFXTs than classical NS-HMXBs. We discuss these findings in terms of different properties of stellar winds and accretion mechanisms.
  \end{abstract}

\begin{keywords}
X-rays: binaries--pulsars: general
\end{keywords}

\section{Introduction}
High-mass X-ray binaries (HMXBs) are binary systems comprising a compact object and a massive star orbiting around a common centre of mass. 
The compact object can be either a neutron star or a black hole. In this work, we mainly deal with HMXBs 
that harbour a neutron star as the compact object accreting from the powerful wind of the massive companion \citep[]{white1983, nagase1989, bilsden1997}. 
These neutron stars in HMXBs have magnetic fields of the order of $\gtrsim$ 10$^{12}$ G. The accreted material is thus expected to be channelled along 
the magnetic field lines at relatively large distances from the compact object, leading to the formation of
extended accretion columns. The exact details of how the magnetic field affects the accretion flow is still a topic of investigation \citep{becker_wolf2007}.  
It is generally expected that the inflowing material is directed towards the the magnetic poles of the neutron star where two hot spots are formed.
The gravitational potential energy of the inflowing material is first converted into kinetic energy and then released as X-rays due to shocks and dissipations
into the accretion column and on the hot spots \citep{basko1975}. 
If the rotation axis and the magnetic axis of the neutron star are not completely aligned, the X-ray emission from the source appears pulsed to any distant 
observer whose line of sight to the object intersects the beam periodically. \\
The high magnetic fields of neutron stars can be measured from the cyclotron resonance scattering features (CRSFs). 
These features are known to be produced as a consequence of cyclotron resonant scattering of X-ray photons in the presence of an intense magnetic field. Their centroid 
energy is related to the NS magnetic field intensity by the equation: 
\begin{equation}\label{eq:crsfeq}
\crsfeq{}\rm{\,keV} ,
\end{equation}
where $z$ is the gravitational redshift. \\
Classical HMXBs can be classified as either Be/X-ray binaries (BeXBs; \citealt{reig2011}) or Supergiant X-ray binaries (SGXBs; \citealt{walter2015}). 
Relatively recently discovered HMXBs with supergiant companions, called Supergiant Fast X-ray Transients (SFXTs), are characterized by short outbursts with fast 
rise times ($\sim$ tens of minutes) and typical durations of a few hours \citep{sguera2006}. For this study, along with both classical HMXBs and SFXTs observed 
with \emph{Suzaku} \citep{M07}, we have also included a few sources which have a low mass companion star but are known to harbour neutron star with strong magnetic fields. These four sources are  Her X-1, 4U 1626-67, GX 1+4, and 4U 1822-37 and have been classified here as Low Mass X-ray Binaries, LMXBs (although strictly speaking, Her X-1 is an intermediate mass X-ray binary).

\section{Data Reduction}

\emph{Suzaku} \citep{M07} is a broad-band X-ray observatory covering the energy range 0.2-600 \rmfamily{keV}. 
There are two main instruments on-board \emph{Suzaku}: the X-ray Imaging Spectrometer {XIS} \citep{K07}, covering the 0.2-12
\rmfamily{keV} energy range, and the Hard X-ray Detector (HXD). 
The XIS consists of four CCD detectors of which three (XIS 0, 2 and 3) are front illuminated 
(FI) and one (XIS 1) is back illuminated (BI). The HXD comprises PIN diodes \citep{T07} that cover the 10-70 \rmfamily{keV} energy range 
and GSO crystal scintillator detectors that cover the 70-600 \rmfamily{keV} energy range. 
 \\ 
For the XIS and the HXD data, we used the filtered cleaned event files which are obtained using the pre-determined screening criteria as suggested in 
the Suzaku ABC guide\footnote{ \url{http://heasarc.gsfc.nasa.gov/docs/suzaku/analysis/abc/}}. The data reduction for both instruments was carried out following the 
reduction technique mentioned in the same Suzaku ABC guide. We applied the barycentric correction to all event files using \texttt{aepipeline}. 
For the XIS data reduction we applied the following procedure: for sources that showed jitters in the image, the event files were corrected by using the \texttt{aeattcorr} 
and \texttt{xiscoord} tools to update the attitude information; for those sources affected by pile-up, we discarded 
photons collected within the portion of the PSF where the estimated 
pile-up fraction was 
greater than 4 \%. This was done by using the FTOOLS task \texttt{pileest}. XIS spectra were then extracted by
choosing circular regions of $2^{'}$, $3^{'}$, or $4^{'}$ radius from the source position depending on whether the observation was made in 
1/8, 1/4, or 0 window mode, respectively. Background spectra for the XIS were extracted by selecting regions of
the same size as mentioned above in a portion of the CCD that was not significantly contaminated by the source X-ray emission. 
For relatively fainter objects observed in window `Off' mode (like IGR J16493-4348, IGR J16465-4507, IGR J16479-4514, and IGR J08408-4503) 
choosing the same radius for the source and background extraction region gave rise to a dip-like artefact in the spectra visible around 6 \rm{keV}. Hence, for these 
sources, a larger background region was chosen by adopting an annulus with inner (outer) radius of 5 $^{'}$ (7 $^{'}$) centered on the source best known position. \\
The PIN source spectra were additionally corrected for dead-time effects by using the \texttt{FTOOLS} task \texttt{hxddtcor}. 
For the HXD/PIN, simulated `tuned' non X-ray background event files (NXB) corresponding to the month and year of the respective observations 
were used to estimate the non X-ray background \footnote{\url{https://heasarc.gsfc.nasa.gov/docs/suzaku/analysis/pinbgd.html}}\citep{F09}. \\
The XIS spectra were extracted with 2048 channels and the PIN spectra with 255 channels. 
Response files for the XIS were created using the CALDB version `20150312'. For the HXD/PIN spectrum, response files corresponding to the epoch 
of the observation were obtained from the \emph{Suzaku} guest observer
facility\footnote{\url{https://heasarc.gsfc.nasa.gov/docs/heasarc/caldb/suzaku/}}. 
\\
The sources considered for the present study and all the OBSID corresponding to the observations used are listed in Table \ref{obslog}. 

\section{Spectral Analysis}
\label{sect:spectral analysis}
For the spectral analysis of the selected sources, we have used the spectra from all the available XIS units (0, 1, 2 and 3) and the PIN. 
In some cases we noticed systematic differences between the spectra obtained from the BI XIS 1 and the rest of the XIS units. In all these cases, we did not 
make use of the BI XIS 1 data in the spectral fitting as they gave no additional information and led to a poorer fit with larger $\chi^{2}_{\rm red}$. 
Spectral fitting was performed by using \texttt{XSPEC} v12.9.0. Artificial residuals are known to arise in the XIS spectra around the Si
edge and the Au edge. We have thus discarded the energy range $\sim$ 1.7-2.3 \rmfamily{keV} for spectral analysis. 
For most sources, we have limited our analysis to the energy range $\sim$ 0.8-10.0 \rm{keV} for XIS and 15.0-70.0 \rm{keV} for the HXD-PIN. 
For sources like OAO 1657-415, 4U 1909+07, IGR J16393-4643, GX 301-2, and GX 1+4 where the absorption at soft X-ray is larger, we did not use data below 
$\sim$ 3 \rm{keV} either due to the poor signal-to-noise ratio (S/N) or because the spectra below 3 \rm{keV} show a `low energy tail' which is not 
characteristic of the source but has an instrumental origin \citep{suchy2011}.
For other sources characterized by a PIN spectrum with limited statistics, like V 0332+53, 4U 1909+07, IGR J16393-4643, 4U 2206+54, SW J2000.6+3210, and 4U 1822-37, we have 
discarded the data points in the PIN spectrum, above the energy where the S/N is very low. 
For some sources like 4U 1907+09, 4U 1700-37, and IGR J17544-2619, we noticed an excess above the power-law continuum in the data around 10 \rm{keV}, 
with an amplitude of 10 \%, 10\% and 5 \% respectively. This `10 \rm{keV}' feature was also previously reported in the cases of 4U 1907+09 and 4U 1538-522 
\citep[]{4u1907_energyrange,coburn2002}. Since the aim of this work is to perform a homogeneous fit as much as possible, we discarded in all cases the data around this feature. \\
For IGR J16318-4848, we limited our analysis to the energy range 5.0-60.0 \rm{keV}, owing to very strong absorption below 5.0 \rm{keV} and the low S/N above 60 
\rm{keV} \citep{igrj16318-4848_energyrange}. 
For seven sources (IGR J16195-4945, IGR J16493-4348, IGR J16465-4507, IGR J16479-4514, IGR J17391-3021, IGR J08408-4503, and IGR J00370+6122), we could make use only of the XIS data as their PIN spectrum was contaminated either by nearby bright sources, or by the diffuse emission from 
Galactic Ridge, or due to the high CXB and NXB \citep[]{igrj16465-4507-energyrange, igrj16479-4514-time, igrj17391-energyrange, igrj08408-energyrange}. 
For eclipsing HMXBs like Her X-1, LMC X-4, Cen X-3, 4U 1700-37, 4U 1538-522, SMC X-1, Vela X-1, OAO 1657-415, 4U 1822-37, IGR J16479-4514 and 
IGR J16195-4945, we have checked whether 
the source was in eclipse or not during the considered observations by folding its lightcurve together with the corresponding RXTE-ASM orbital 
lightcurve\footnote{\url{https://heasarc.gsfc.nasa.gov/docs/xte/ASM/sources.html}} at the known orbital period of the source. 
Except for Cen X-3, IGR J16479-4514, and 4U 1822-37, all the other sources were not observed during an X-ray eclipse. For Cen X-3 and IGR J16479-4514, we have extracted the time 
filtered spectrum corresponding to the times when the source was not in eclipse. For Cen X-3 we extracted spectra only for `Segment E' of \citet{cenx3_time} while for 
IGR J16479-4514, we extracted the spectrum starting 143 \rm{ks} after the beginning of the observation (as is done in \citealt{igrj16479-4514-time}). 
For 4U 1822-37, the full \emph{Suzaku} observation spanned nearly four times the orbital period of the neutron star. 
Hence for 4U 1822-37, we extracted a spectrum only for that phase interval during which the source was outside the dips and eclipse (0.1-0.6 when Phase 0 is at MJD 54010.48). 
In the cases of Vela X-1 and 4U 1538-522, spectra were extracted only for that part of the observation during which the hardness ratio remained relatively stable 
as done in \citet{vela_time} and \citet{4u1538_time}. 
\\
We fitted the spectra of all sources by using all available instruments simultaneously. We kept all spectral parameters of the different instruments tied together 
during the fit. Only the inter-calibration constants were left free to vary. We have used the most commonly adopted models to describe the high energy emission of HMXBs 
and LMXBs consisting 
of a power law component with a high energy cutoff (HIGHECUT; \citealt{white1983, coburn2002}). This model gave formally acceptable fits ($\chi^{2}_{\rm red}$=0.89-1.45) 
to all data, the only exceptions being A0535+026, LMC X-4, and GX 1+4. In these three cases, the NPEX \citep[]{M95, makishima1999}, FDCUT \citep{T86} and 
NEWHCUT \citep{burderi2000} models respectively provided a better description of the data. We noticed that no formal acceptable fit could be obtained with the selected models 
to the data of GX 1+4 ($\chi^{2}_{\rm red}$= 1.66) and thus we decided not to include this source for the discussion on the spectral parameters' correlations. Such a high value of $\chi^{2}_{\rm red}$ for GX 1+4 has also been reported in a recent analysis of the same data set to which refer the readers for further details \citep{Yoshida_2017}. 

Furthermore, although we fit Her X-1 and 4U 0115+63 with the HIGHECUT model, we did not include their CRSF parameters for correlation studies. The reason being that the former is known to show a complicated cyclotron energy variation \citep{staubert2014} and for latter, the CRSF lie at $\sim$ 11\,keV, which is outside the usable PIN band \citep{kuhnel2019}.

The analytical form of the main model considered in this work: HIGHECUT is: 

\begin{equation}\label{eq:plcut}
\plcuteq
\end{equation} 

(where \wgamma\ is the photon index, \ecut\ the
cutoff energy, and \efold\ the folding energy), 

%\begin{equation}\label{eq:npex}
%\npexeq
%\end{equation}
%(where $\Gamma_{1}$ and $\Gamma_{2}$ are the photon indices with positive
%values - $\Gamma_{1}$ was frozen to 2.0, and \efold\ the folding energy),  

%\begin{equation}\label{eq:fdco}
%\fdcoeq
%\end{equation}
%(where $\Gamma$ is the photon index, \ecut\ the
%cutoff energy, and \efold\ the folding energy), and
%\\

%\begin{equation}\label{eq:newhcut}
%\newhcuteq
%\end{equation}
%(where \wgamma\ is the photon index, \ecut\ the
%cutoff energy, \efold\ the folding energy, and W is the width used to smoothen the abrupt break in HIGHECUT at \ecut\ ). 
For other sources where the cutoff energy was not required (IGR J16195-4945, IGR J16493-4348, IGR J16465-4507, 
IGR J16479-4514, IGR J17391-3021, IGR J08408-4503, and
IGR J00370+6122), we fitted the spectrum with a simple powerlaw corrected for photoelectric absorption. 
In many of the analysed sources, CRSFs were clearly detected as broad absorption features in the X-ray spectra. 
Where required, those features have been fit with 
pseudo Lorentzian optical depth profiles (CYCLABS in Xspec). We used additional Gaussian components to take into account 
the presence of emission lines, mostly due to the flourescence of neutral iron. 
In addition to the above components, in some cases, a partial covering model and/or 
blackbody component was used to account for fractional local absorption and for thermal emission components in the spectrum, respectively. 
The best fit continuum parameters for all the sources are reported in Table \ref{continuum}. The harmonics of the detected cyclotron 
line are given in Table \ref{harmonics}. We report in Table \ref{emission} all significantly measured emission lines.

It should be noted here that to have an overview and for the purpose of this analysis, we have mostly considered X-ray spectrum over the entire 
observation as the representative of that source - except eclipses and a few observations showing large variation in spectrum which were filtered likewise as mentioned 
earlier. The details on the X-ray variability of the sources that we have considered can be found in the already published papers on the corresponding data \citep[see e.g.,][]{pradhan2013_swj,pradhan2014_oao,pradhan2015_0114,vela_time,maitra2013_ao_xt_1907}. While such studies on short term variability study will help addressing the effect of the wind inhomogeneities on the overall 
accretion process on timescales as short as tens to hundreds of seconds, the average spectrum considered in this current work allows a clean and relatively straightforward comparison between the different classes and sub-classes of sources independent from the short term variation of the winds which are washed out on the long integration times we used in the analysis here.

We should also mention here that our choice of HIGHECUT model is motivated by finding a representative model to fit {\it all} our sources while at the same time having the least degeneracy between model parameters. Since HIGHECUT has shown to have the least degeneracy while describing the X-ray spectra of accreting neutron stars \citep{coburn2002}, we use this model throughout our analysis. To minimize degeneracy in fitting though, we have cross-checked the fitting parameters at each step to literature values (where available), and if it is physical. We are therefore confident that the correlation presented in this paper is real within the limits to what a phenomenological model can provide.

\section{Results}
The main purpose of this work is to carry out a correlation study among the various spectral parameters that are measured from 
known HMXBs and a few strong magnetic field LMXBs observed by \emph{Suzaku}. 
We have also investigated the variation of the spectral parameters with luminosity, \lumin.~The distances used 
for \lumin\ calculations\footnote{For sources where distance errors are not determined, we assumed an error of 1\,kpc.} are reported in Table \ref{obslog}. \lumin\ is calculated over the energy range for which the 
individual spectra were fitted. For most sources, it is 0.8-70.0 keV. 
In other cases, as mentioned in Section \ref{sect:spectral analysis}, we have discarded spectral data below 3 keV. Due to the relatively high absorption, we verified that in these cases the difference in the luminosity estimated in 
the 0.8-70 \rm{keV} and 3.0-70.0 \rm{keV} bands are negligible. We also verified that for those sources where the PIN spectrum is truncated prior 
to 70 keV due to the poor S/N (mentioned in Section \ref{sect:spectral analysis}), the X-ray luminosity evaluated in the energy 
range of the fit or in the full 0.8-70.0 \rm{keV} energy range would not change significantly. 
All the fainter sources for which only the XIS spectrum is used, are marked in grey in all relevant figures. 
We discuss in the following sub-sections all the correlations between the different parameters that we could find from 
the analysis of the \emph{Suzaku} data of the considered sources. 

\subsection[]{Correlation among the spectral parameters}

\subsubsection{\ecut\ versus \lumin}

\begin{figure} 
\hspace{-2cm}
\includegraphics[scale=0.65,angle=-90]{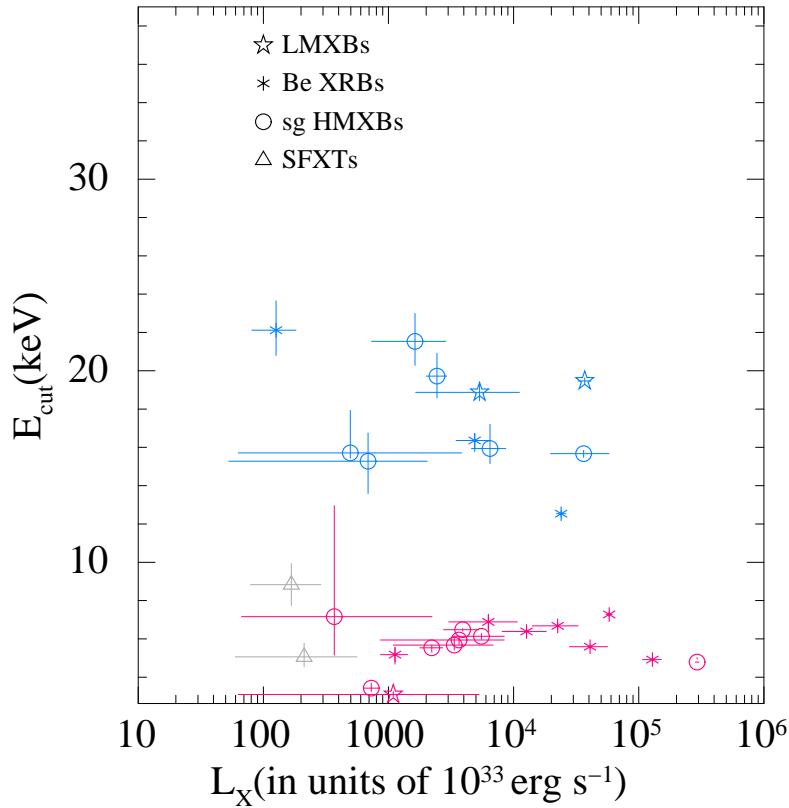}
\vspace{2.5cm}
\caption{Plot of cutoff energy versus the X-ray luminosity. To identify the apparent dichotomy, we mark the two groups in blue and magenta with SFXTs marked in grey. }
\label{lumin_ecut}
\end{figure} 

\begin{figure*}
\centering
\includegraphics[height=10cm,width=8cm,angle=-90]{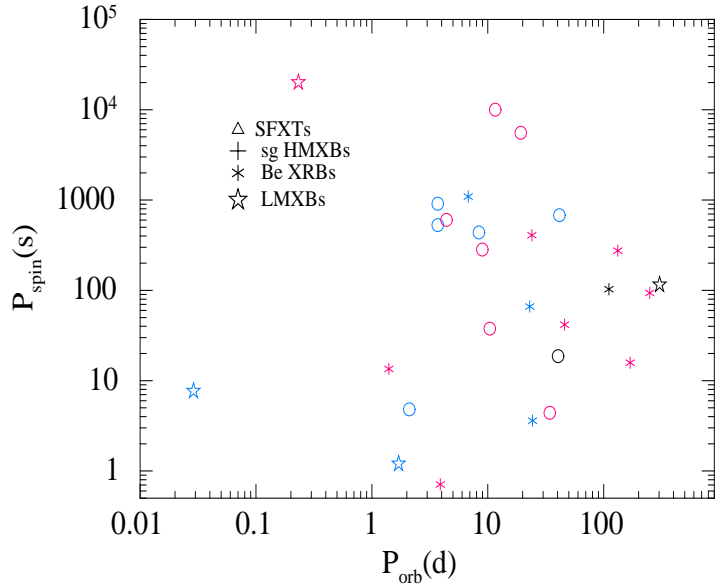}
\vspace{2.6cm}
\caption{Corbet diagram of the pulsars in our study. Sources marked in blue and magenta are for the sources with cutoff energy greater than and less than 10\,keV respectively. As seen in the plot, the distinction is not on the basis of the companion star being a Be X-ray binary or and OB star.}
\label{corbett}
\end{figure*}

When the measured values of the \ecut\ are plotted against \lumin, we note a bimodality in the \ecut~distribution (see Fig.~\ref{lumin_ecut}). 
For clarity, we have used different colour schemes to distinguish the two groups. 
The upper branch where the cutoff energy range from $\sim$ 12-25 keV (henceforth Branch 1) has been plotted in blue while the lower branch 
which have cutoff energy range of 3-10 \rm{keV} (henceforth Branch 2) has been plotted in magenta. Although there is not a clear correlation or anti-correlation among the sources in the two groups\footnote{Note that we did not use A 0535+026, LMC X-4, and GX 1+4 in this plot, as \ecut\ for these three sources is obtained 
from the NPEX, FDCUT, and NEWHCUT models, respectively. All these models are mathematically different from HIGHECUT and thus the measured 
value of \ecut\ has a different meaning.}, they appear clearly distinguished on the \ecut\--\lumin\ plane. The sources in Branch 1 are: Her X-1, 4U 0115+63, Cen X-3, 4U 1626-67, 4U 1907+09, 4U 1538-522, GX 301-2, Cep X-4, IGR J16393-4643, IGR J16318-4848, and V0332+54. 
The sources in Branch 2 are: Vela X-1, XTE J1946+274, 1A 1118-61, 4U 0114+65, GX 304-1, OAO 1657-415, 4U 1700-37, GRO J1008-57, 4U 1909+07, 4U 2206+54, SW J2000.6+3210, SMC X-1, EXO 2030+375, 4U 1822-37, KS 1947+300,  IGR J16207-5129, IGR J17544-2619, and IGR J18410-0535. Given that the nature of companion stars among the two lists is mixed \citep[see Table.~1 of][]{sidoli2019}, we see that this bi-modal behaviour cannot be distinguished on the basis of their 
companion (Be XRBs or supergiants)  as illustrated in the Corbet diagram\footnote{\url{https://www.sternwarte.uni-erlangen.de/wiki/index.php/List\_of\_accreting\_X-ray\_pulsars}} for the X-ray pulsars in our study in Fig.\ref{corbett}. 

Since  we find only three sources with \ecut\ within the range of 7-14\,keV, there appears to be a paucity of sources with \ecut\ around 10\,keV, which is the same energy range where there is a gap between the XIS and PIN energy band. We therefore checked if this distribution is an artifact of the energy ranges considered here by simulating several fake spectra with varying cutoff energies distribute from 4-20\,keV and fitting them. For all these fake spectra, we recovered the same spectral parameters as the input model thereby validating the robustness of our results. 

 Additionally, since the cutoff energies in the second group are spread around 6\,keV, and most of the sources exhibit an iron emission line around 6.4\,keV, there may be concerns about a possible mix-up between the iron line and cutoff energies for these sources. However, it should be noted that since the iron line emission in these sources are narrow features and prominently stand out in the X-ray emission, we can easily disentangle\footnote{If the equivalent width of the iron line is smaller, it only gives large uncertainty for the line flux but does not affect the continuum parameters} between iron line and continuum parameters for a CCD spectrum. We also cross-checked this by simulating several fake spectra with varying cutoff energies (3-7\,keV) and narrow iron lines of different equivalent width and were able to fully recover the input parameters, hence supporting our claim. Furthermore, there is no correlation between the cutoff energy and the iron line parameters as seen in the confidence plots for the iron line energy versus the cutoff energy, and the iron line normalization versus the cutoff energy for this group in Fig.~\ref{cutoff-ironenergy} and Fig \ref{cutoff-ironnorm} in the Appendix.

%\subsubsection{Statistical tests}

\begin{figure*}
\includegraphics[height=7cm,width=10cm,angle=0]{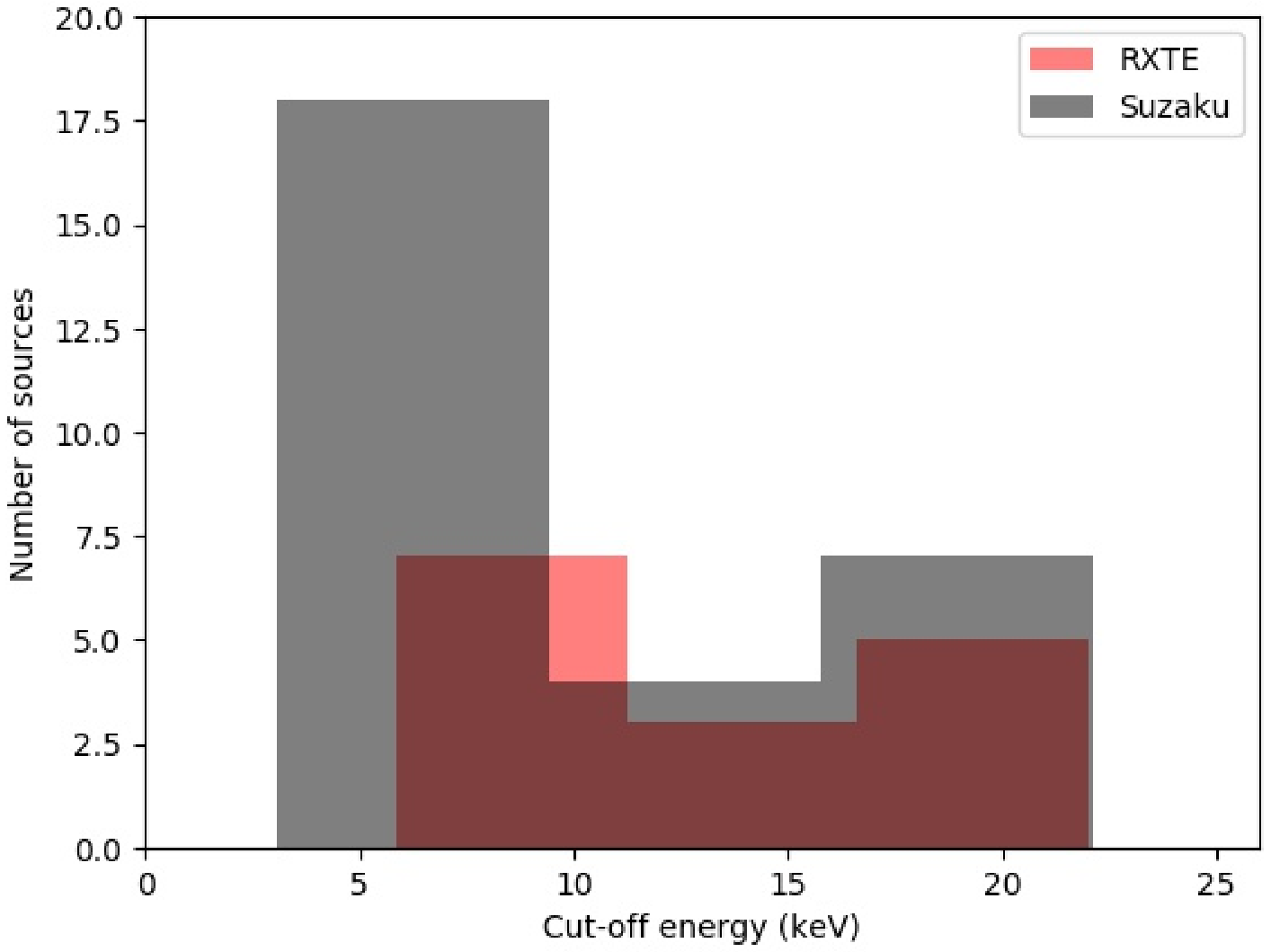}
\includegraphics[height=7cm,width=10cm,angle=0]{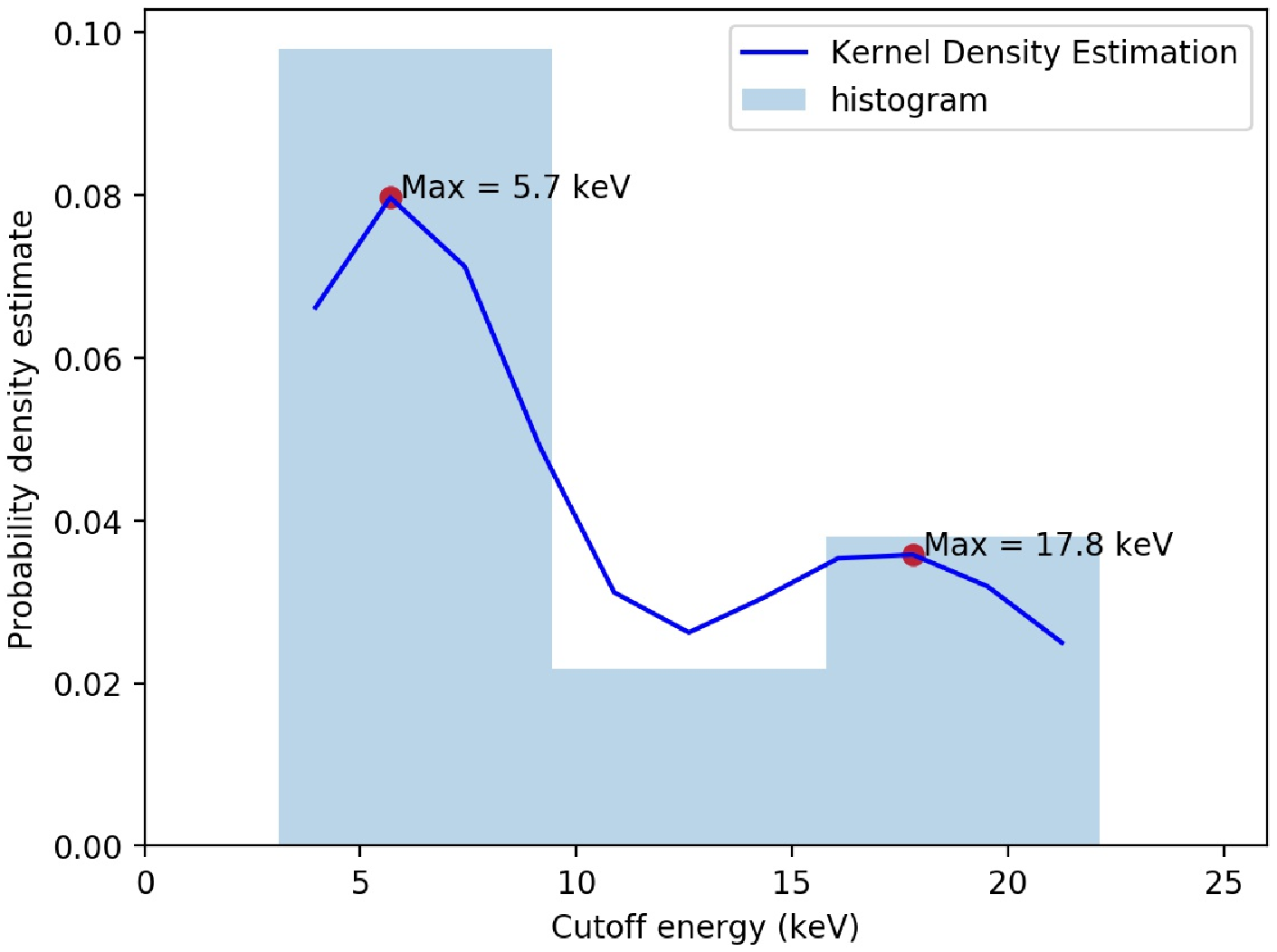}
\includegraphics[height=7cm,width=10cm,angle=0]{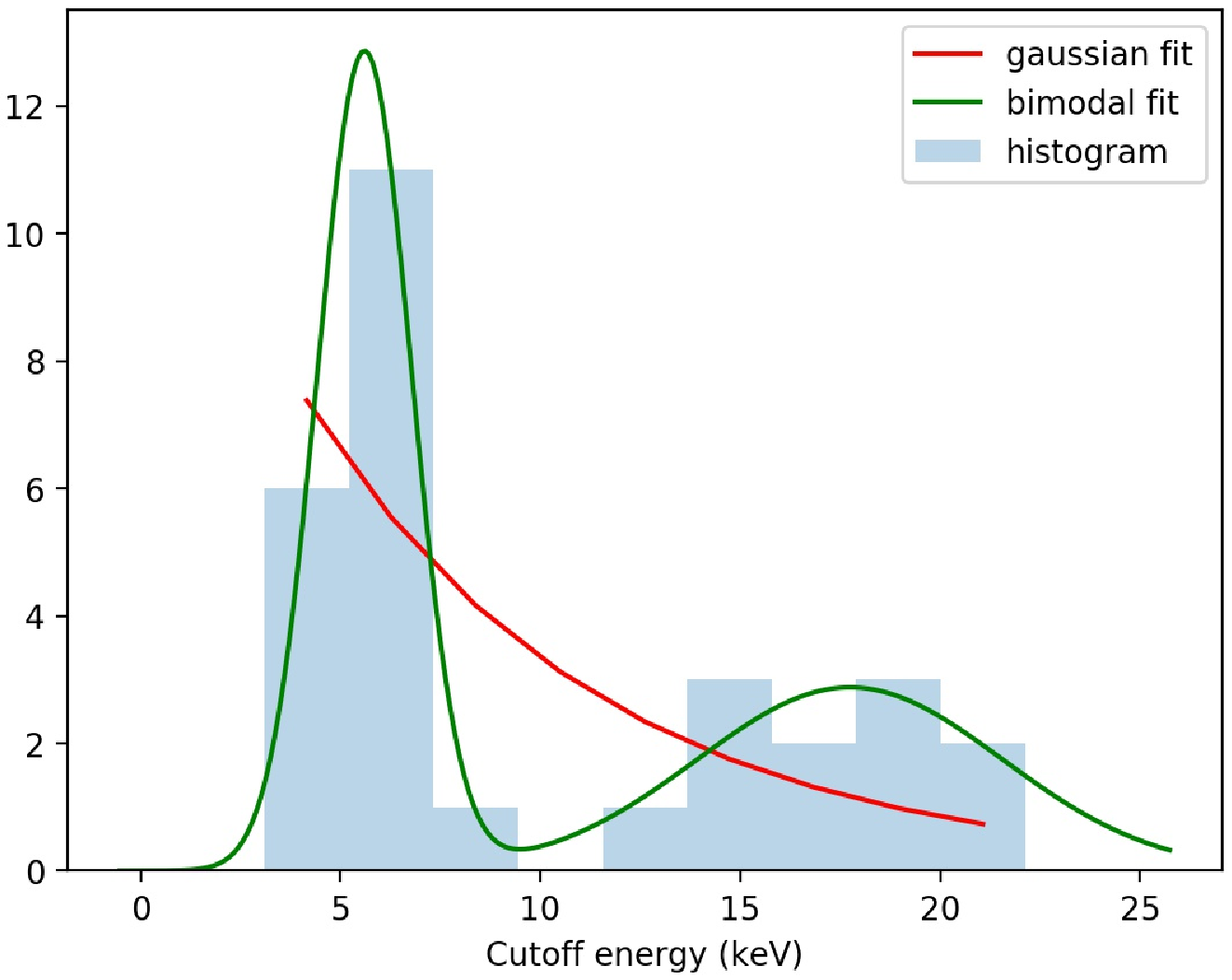}

\caption{Top: Histogram of cutoff energies from RXTE in red and the current work in gray. The RXTE measurements are indicative of the bi-modality as found in the \emph{Suzaku} measurements. Middle: The KDE plotted (with a chosen Silverman bandwidth of 3) demonstrate that there are two peaks in the data at $\sim$ 5.7\,keV and $\sim$ 17.8\,keV respectively.
Bottom: The shaded blue region is the histogram for \emph{Suzaku} data with finer binning compared to top and middle figure. The histogram fits well with a bi-modal function plotted in green as compared to single Gaussian fits shown by red line (the trailing edge of Gaussian). }
\label{histogram}
\end{figure*}

In order to test the bi-modality against reported values in literature, we also checked the cutoff energy in literature (with RXTE) obtained using the same model as in this work. These values are tabulated in Table \ref{literature-cutoff}. We make a histogram of these literature values and the ones we obtain from our work and find that although for some sources, the cut off energy is different in these two instruments, an apparent bi-modality is visible in both the cut-off energy distribution as seen in top panel of Fig.~\ref{histogram}. It should be pointed out that we do not expect a {\it one-to-one} correspondence on the values of cutoff energies from these two instruments since they cover different energy ranges and also because for some sources, the X-ray spectrum evolve with flux (see \citealt{bexrbs_reig}). The interesting finding here is the apparent bi-modal distribution of cut-off energy for sources even within the same class.

We also carried out a number of statistical tests to quantify the bi-modality in the distribution of cut-off energy so obtained. We first performed a 2D KS test\footnote{\url{https://github.com/syrte/ndtest/blob/master/ndtest.py}} for these two branches. The 2D test allow us to compare how different the two distributions are from each other. For the \lumin~ versus \ecut\ variation for red and green groups, the $p$ value is low ($\sim$ 1$\times$10$^{-5}$) indicating that these two distributions are different. 

However, since our main finding is the bi-modality in the distribution of cut-off energy, we next focus on this. We performed a Dip test\footnote{\url{https://github.com/BenjaminDoran/unidip}} \citep{hartigan1985} and obtained a $p$-value of 0.15. The rule of thumb for interpretation of this test is that $p$-values less than 0.05 indicate significant bimodality and values in the range of 0.05-0.10 suggest bimodality with marginal significance. Therefore, while the Dip test do not favor multimodality in the cut-off energy, it should be remarked that this Hartigan’s Dip test work best for small samples but with large bumps\footnote{\url{https://www.paspk.org/wp-content/uploads/2019/11/9-ES-609A-Comparison-of-Modality.pdf}}. The alternative approach is therefore to search for modes in the data distribution by generating the probability densities\footnote{\url{https://towardsdatascience.com/modality-tests-and-kernel-density-estimations-3f349bb9e595}} of the cut-off energies. In order to do this, we first calculate the best value of band-width based on Silverman's band-width test and then look for points of inflexion in the Kernel (probability) Density Estimates (KDE) using this band-width as the smoothing function. As seen in the middle of Fig.~\ref{histogram}, we find two points of inflexion at $\sim$ 5.7 and 17.8\,keV. In the same plot, we have also over-plotted the probability density of histogram using the bin size as the band-width to demonstrate bi-modality. 

Finally, we also fit the frequency estimate obtained from histogram with a single Gaussian and two Gaussian functions. As seen from the lower panel of Fig.~\ref{histogram}, the histogram is better described by a two Gaussian function ($\chi^{2}_{\rm red}$ $\sim$ 0.7) and a single Gaussian simply fail to fit and result in large ($>$30) values of $\chi^{2}_{\rm red}$.

\begin{figure}
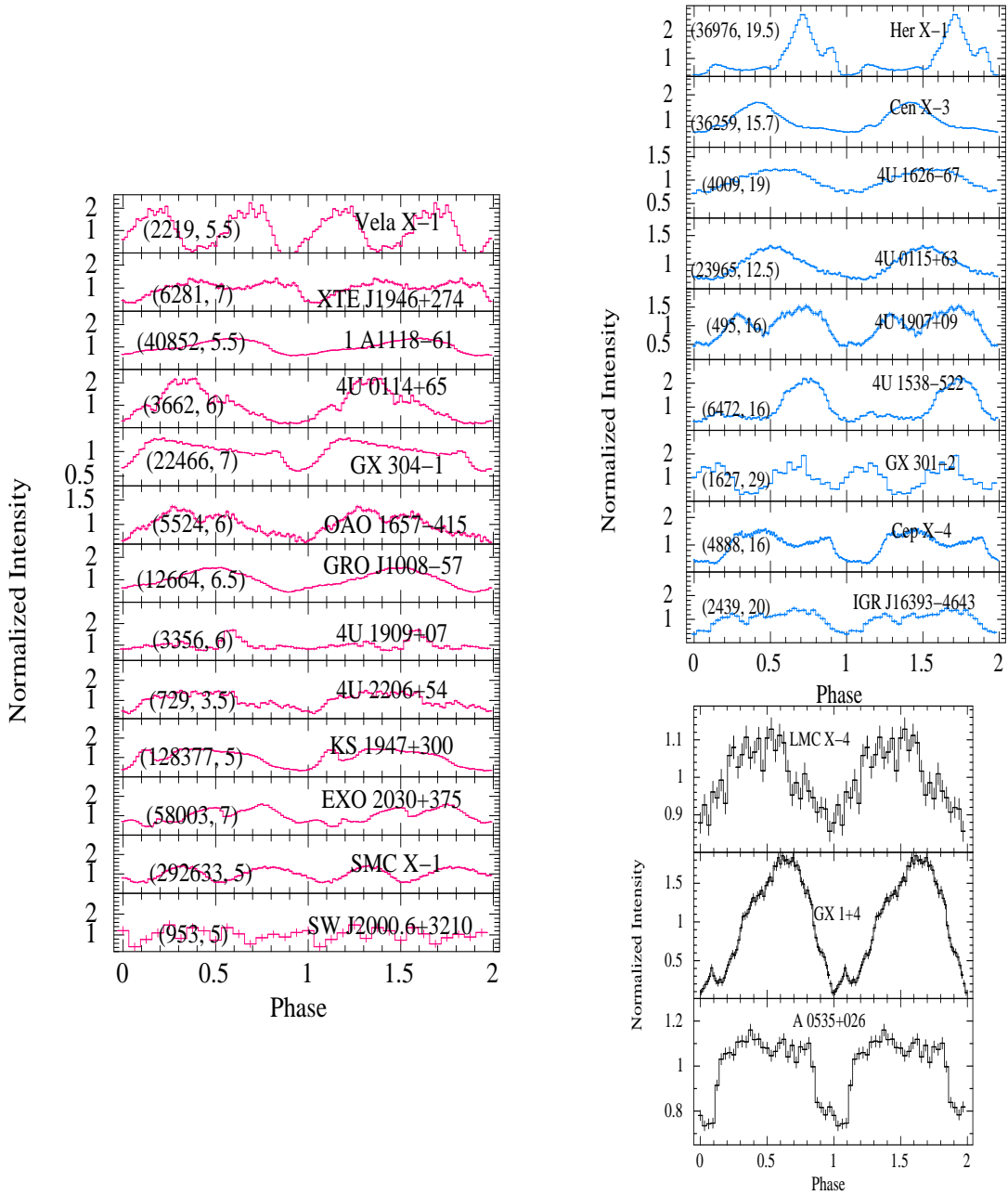

    \centering
\hspace{-5cm}
    \begin{minipage}{.45\textwidth}
        \centering
        \includegraphics[height=8cm, width=12cm, angle=-90]{pp_magenta1_2020_v3.ps}
   %     \caption{$dt=0.1$}
   %     \label{fig:prob1_6_2}
    \end{minipage}%
    \begin{minipage}{0.55\textwidth}
        \centering
        \includegraphics[height=6.5cm, width=10cm,angle=-90]{pp_blue1_2020_v3.ps}
        \includegraphics[height=5.5cm, width=7cm, angle=-90]{pp_lmc_gx_a05.ps}
  %      \caption{$dt =$}
  %      \label{fig:prob1_6_1}
    \end{minipage}
    \vspace{2.6cm}
    \caption{\rmfamily{Left: HXD-PIN (15-70 \rm{keV}) pulse profiles of pulsars in the magenta branch. Right (top): HXD-PIN (15-70 \rm{keV}) pulse profiles of pulsars in the blue branch. The first number in parenthesis in both the figures is the X-ray luminosity in units of 10$^{33}$ erg/s and the second number is the cutoff energy in keV. Right (bottom): Pulse profiles for three sources, LMC X-4, GX 1+4 and A 0535+026 which could not be fit with the HIGHECUT model. See text for details. }}
    \label{pp}
\end{figure}

Next, we extracted the PIN pulse profiles (15.0-70.0 \rm{keV}) for the pulsating sources in each branch to check if a different beaming 
mechanism (fan or pencil; \citealt[]{davidson1973, burnard1991}) could be at the origin of the observed bi-modality\footnote{We did not extract 
the pulse profiles of the two sources 4U 1822-37, due to the very low pulse fraction \citep[see][]{4u1822_pp}, and V 0332+53, due to the limited 
statistics of the available \emph{Suzaku} data.}. As shown in Fig.~\ref{pp}, the single and double peaked pulse profiles indicating pencil and fan beam respectively are spread out in both groups at random. We therefore also conclude that the current bi-modality in the distribution between \ecut\ and \lumin\ is not based on the beaming pattern of X-rays. As the beaming pattern depends on the accretion rate, $\dot{M}$, the bi-modality is independent of beaming pattern and the accretion rate as well. That the bi-modality in \ecut~ is independent of the X-ray luminosity (and hence the accretion rate), can also be see from Fig.~\ref{lumin_ecut}. We maintained the same colour scheme for the different sources in all other figures that we describe in the following sections to investigate alternative possibilities that could give rise to the observed behavior.

%{\bf We also made a tentative check on the inclination angles of the pulsars in both the groups which we have plotted in Fig.~\ref{inclination}. While it appears that the blue sources have a narrower span of inclination from 50-81 degrees (with the exception of 4U 1626-67 that have an inclination angle of 23 degrees) while the magenta sources have larger span of inclination angles from 14-84 degrees, we need to caution here that the distribution may only be apparent since there are less number of sources in the blue group compared to the magenta group. }

We inspected the ratio of the source spectra to Crab to see if the \ecut~ values measured are also evident in the Crab spectral ratio. We however found that since CRSF is rather strong for many sources, the spectral curvature is affected by this absorption feature in the X-ray spectrum for many sources and bi-modality therefore is not straightforward to interpret from this exercise.

Furthermore, in order to check the robustness of our results, with another model, we also tried fitting the 29 sources (that required a cutoff energy) with another phenomological model ‘FDCUT’ in \texttt{XSPEC} . We could reasonably fit 26 sources with FDCUT, albeit with much higher reduced chi-square than HIGHECUT for each source. We find that even when fitting with FDCUT, there are two branches distinguished in the \lumin\ versus \ecut\ plane – similar to what we see with HIGHECUT. Note, however that since the cutoff energies for both these models are mathematically different from each other, their numerical values do not match.

All these factors indicate that the bi-modality in the X-ray spectral shape of these groups are possibly real and further investigations with physical models will allow us to further understand the cause for this. Such detailed studies is however beyond the scope of this current work.

%\begin{figure*}
%\centering
%\hspace{-3cm}
%\includegraphics[height=6.5cm, width=9cm,angle=-90]{pp_blue1_2020_v2.ps}
%\hspace{-0.1cm}
%\includegraphics[height=6.5cm, width=9cm, angle=-90]{pp_magenta1_2020_v2.ps}
%\hspace{-0.1cm}
%\includegraphics[height=5.5cm, width=7cm, angle=-90]{pp_lmc_gx_a05.ps}
%\vspace{1.6cm}
%\caption{\rmfamily{To the left and middle are HXD-PIN (15-70 \rm{keV}) pulse profiles of pulsars in the blue and magenta branch. The first number in parenthesis is the X-ray luminosity in units of 10$^{33}$ erg/s and the second number is the cutoff energy in keV. To the right are the pulse profiles for three sources, LMC X-4, GX 1+4 and A 0535+026 which could not be fit with the HIGHECUT model. See text for details. }}
%\label{pp}
%\end{figure*}

%\begin{figure*}
%\includegraphics[height=9cm,width=12cm,angle=0]{inclination.png}
%\caption{Histogram of inclination of the pulsars in both the blue and magenta group. The blue sources have a narrower span of inclination from 50-81 degrees with the exception of 4U 1626-67 while the magenta sources have larger span of inclination angles from 14-84 degrees. We need to caution here that the distribution may only be apparent since there are less number of sources in the blue group compared to the magenta group. }
%\label{inclination}
%\end{figure*}

\subsubsection{\wgamma versus \lumin} 
We report the dependence of \wgamma as a function of \lumin\ in the case of neutron star HMXBs  (including Be XRBs, SGXBs, SFXTs and a few LMXBs), accreting high magnetic field NS in LMXBs spanning over five orders of magnitude in X-ray luminosity. As we show in top of Fig.~\ref{lumin_pho_efold}, we find that for all these sources \wgamma and \lumin\ are marginally anticorrelated with the best-fit to all our data as: $\Gamma$ = a log L$_{X33}$ + b, where a = -0.25 $\pm$ 0.04, b = 1.93 $\pm$ 0.15. The dependence of \wgamma as a function of \lumin\ in the case of Be XRBs have been previously investigated by \citet{bexrbs_reig} who found that \wgamma and \lumin\ anticorrelate with the X-ray flux during the low X-ray intensity (sub-critical) states of these sources, while a positive correlation is measured during the high intensity (super-critical) states. Their study spanned two orders of magnitude variation in \lumin.\ 

It should be noted that most of the sources in our sample are in the sub-critical luminosity regime (see section \ref{sec:disc}) and our findings are therefore consistent with the anti-correlation seen by \citet{bexrbs_reig} for individual sources. 

The luminosity span in {the current work} is however much larger than the RXTE observation of \citet{bexrbs_reig} and while the RXTE observations focused on the evolution of spectra for individual sources (using multiple observations of same Be XRB pulsars at different luminosities), we focus on the overall `class' behaviour of the sources as a function of luminosity. Another improvement with the previous paper is that for the RXTE observations, the authors froze the N$_H$ values for most of the sources while in our analysis - given the better low energy coverage of XIS - we were able to constrain both the photon index and the N$_H$ independently (see contour plots, Fig.~\ref{contour1} and Fig.~\ref{contour2} in Appendix) in most cases\footnote{except some systems where the line of sight absorption is very low, like LMC X-4, Her X-1, SMC X-1 etc; see Table \ref{continuum}}. 
%We note however that we were unable to fully verify the findings of \citet{bexrbs_reig}, as in our \emph{Suzaku} sample we do not have available multiple observations of Be XRBs at different luminosities. }

\subsubsection{\efold\ versus \lumin} 
\begin{figure}
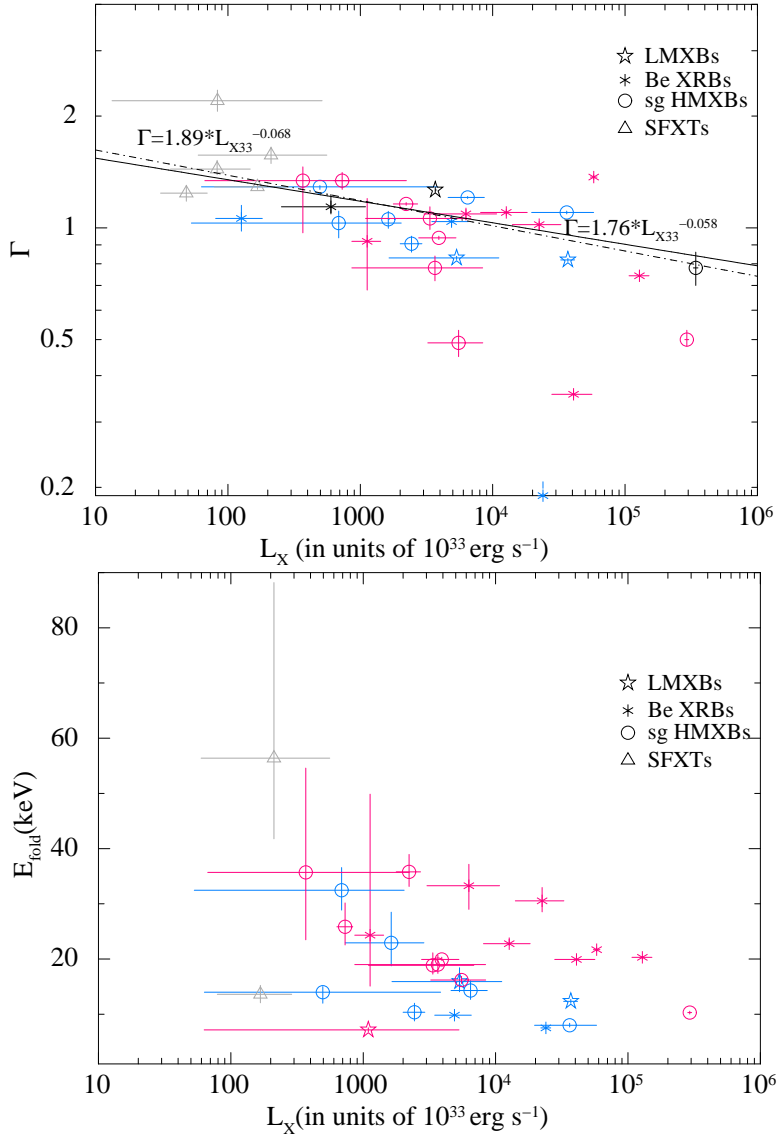

%\hspace{-5.cm}
\includegraphics[scale=0.55,angle=-90]{lumin_pho_2020_v6.ps} \\
\includegraphics[scale=0.55,angle=-90]{lumin_efold_2020_v5.ps}
\vspace{3cm}
\caption{\small{Top:~Variation of  photon index with luminosity. The black solid line represents the best-fit to the data ($\wgamma \propto \lumin ^ {-{\alpha}}$ with $\alpha$ = 0.058 $\pm$ 0.006 when sources fit with only XIS spectrum fits are excluded (see section \ref{sect:spectral analysis}, third-last paragraph). The best fit to all the data points (marked in dotted line) is $\alpha$ = 0.068 $\pm$ 0.006 ). The sources A 0535+26, GX 1+4 and LMC X-4 have been marked in black. Bottom: The variation of \efold~energy with X-ray luminosity. Both the figures are marked with the same colour coding as in Figure \ref{lumin_ecut}. }}
\label{lumin_pho_efold}
\end{figure}

In models of the X-ray spectrum, \efold~is a measure of the electron temperature of the infalling plasma. An anti-correlation between luminosity and \efold~ has been reported earlier for individual sources like A 0535+26, RX J0440.9+4431, for example \citep{muller2013,ferr2013}. It is for the first time that we present here a comprehensive behaviour of the \efold~dependence on \lumin~using a wider sample. In the bottom panel of Fig.~\ref{lumin_pho_efold}, we see that with increasing luminosity, the \efold~value show a weak anti-correlation with the Pearson co-efficient\footnote{\url{https://www.socscistatistics.com/tests/pearson/}} for these two quantities being -0.19.

\begin{figure} 
%\hspace{-1cm}
\includegraphics[scale=0.55,angle=-90]{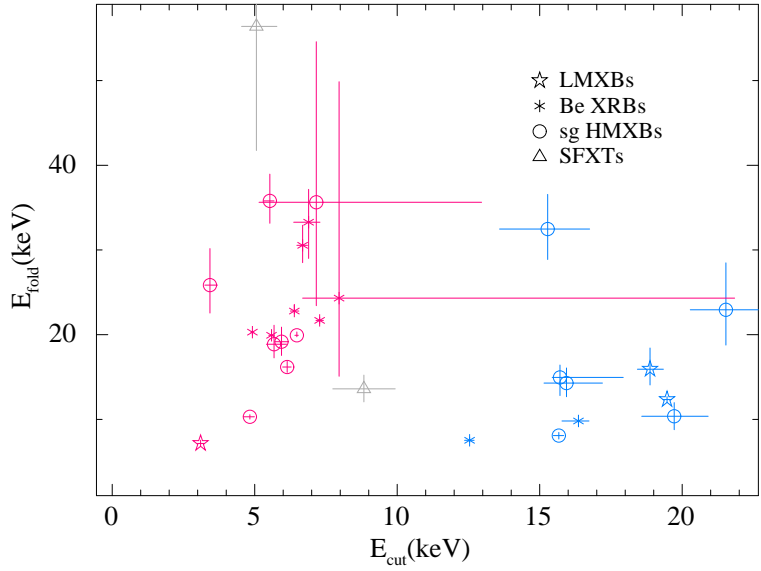}
\vspace{2.5cm}
\caption{Variation of \efold~with \ecut~marked with the same colour coding as in Figure \ref{lumin_ecut}. }
\label{ecut-efold}
\end{figure}

%\vspace{-1.cm}

\subsubsection{\efold\ versus \ecut}
We find that the sources in  Branch 2 (magenta) have a somewhat larger median of \efold\ than those in Branch 1 (blue). A plot of their variation is shown in Fig.~\ref{ecut-efold}. We remark here that we could independently constrain both the \ecut~and \efold~as seen in the confidence plots in Appendix (Fig.~\ref{ecut-efold-cont1} and Fig.~\ref{ecut-efold-cont2}) and there is no degeneracy between these two quantities for both groups.

A similar correlation between these two quantities has also been reported earlier by \citealt{makishima1999} where the cutoff energies seem to be divided into two groups (their Fig. 12 b) less than and above 10\,keV. The authors however do not comment about this in their paper.

\begin{figure}
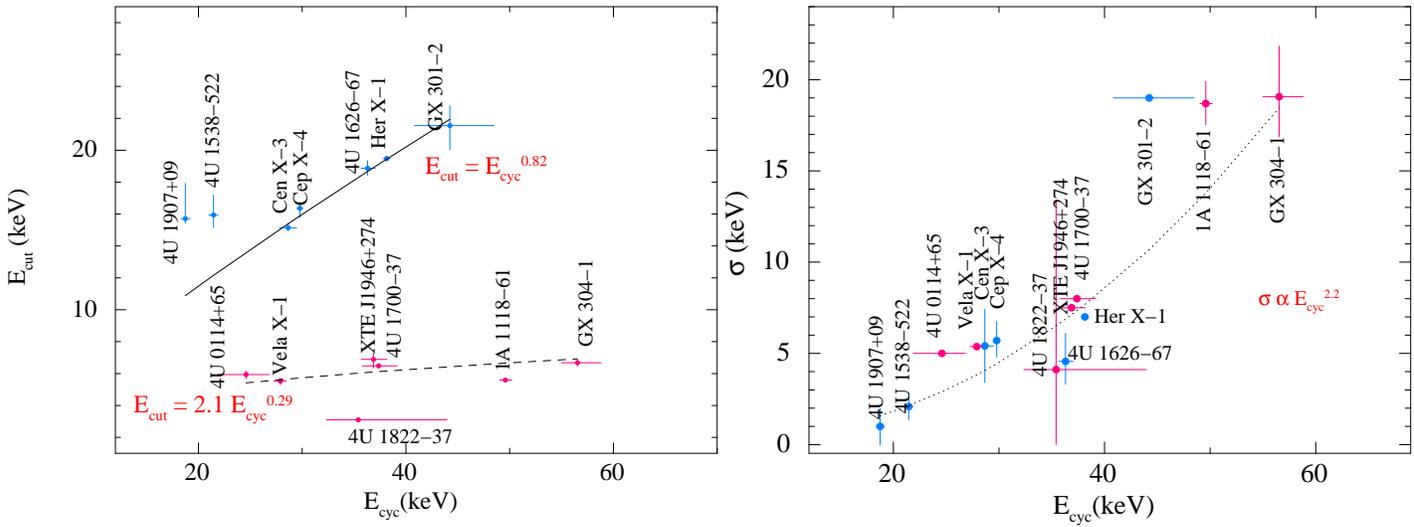

%[8]{r}{0.5\textwidth}
%\vspace{-1.cm}
%\includegraphics[height=8cm,width=5.4cm,angle=0]{lumin_ecyc_2020_v3.pdf}
%\hspace{-0.12cm}
%\centering
\hspace{-3.cm}
\includegraphics[scale=0.5,angle=-90]{ecyc_ecut_2020_v7-tofit.ps}
\hspace{-0.8cm}
\includegraphics[scale=0.5,angle=-90]{crsf_width_2020_v7.ps}
\vspace{2.5cm}
\caption{\small{Left: \ecut\ versus E$_{cyc}$ showing two scaling laws for two groups. Right: Positive correlation between CRSF width $W_{\rm c}$ versus E$_{cyc}$. All the figures are marked with the same colour coding as in Figure \ref{lumin_ecut}.}}
%\vspace{-1.cm}
\label{lumin_ecyc_ecut_width}
\end{figure}

\begin{figure}
\hspace{-3.cm}
\centering 
\includegraphics[scale=0.6,angle=-90]{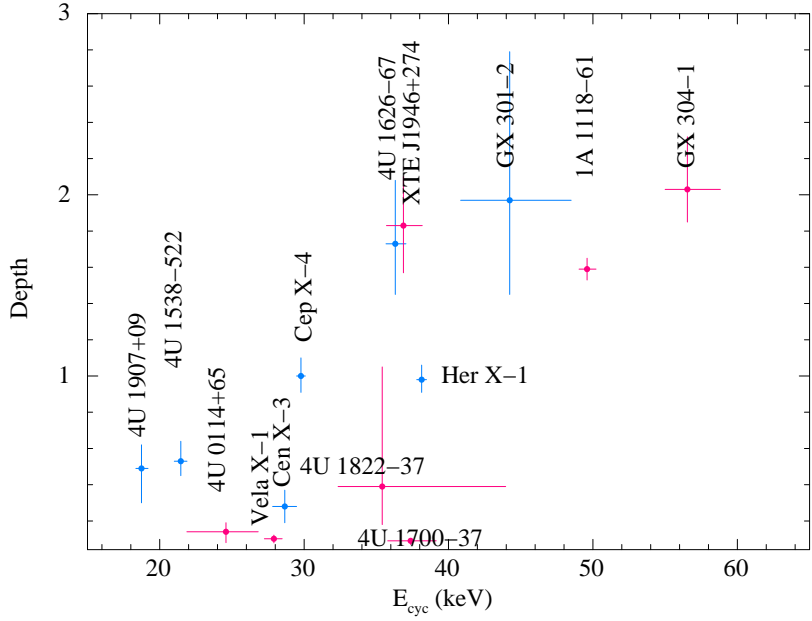}
\vspace{2.5cm}
\caption{\small{Plot of CRSF depth versus CRSF energy. All the figures are marked with the same colour coding as in Figure \ref{lumin_ecut}.}}
\label{ecyc_depth}
\end{figure}

%\subsubsection{E$_{C}$ versus \lumin}
%The cyclotron line energy variation with luminosity as shown by the left of Fig.~\ref{lumin_ecyc_ecut_width} show a marginal positive correlation. While there have been numerous studies on the variation of CRSF with luminosity for individual neutron stars, notable ones being Her X-1, Vela X-1 both of which show a positive correlation (see., \citealt{staubert2019} for review), this is the first time that a class variation of CRSF energy (of high magnetic field neutron stars) with \lumin~is reported.

\subsubsection{\ecut\ versus E$_{cyc}$}
The variation of \ecut~with E$_{cyc}$ for HMXBs has been investigated in earlier works. 
\citet{coburn2002} and \citet{makishima1999} used \emph{RXTE} and \emph{Ginga} data respectively to obtain 
the relationship E$_{cut}$ $\propto$ E$_{cyc}^{0.7}$ for 
E$_{cyc}$ below 35 \rm{keV}. 
%In \citet{staubert2003}, the predicted relationship between the two quantities was found to be E$_{cut}$ = 0.5$\times$E$_{C}$. 
Thanks to the broadband spectral capability of \emph{Suzaku}, in our analysis, we show that the correlation between these two quantities (contour plots in Fig.~\ref{ecut-ecyc-cont} in Appendix) is not unique 
but rather have a complex behaviour (left of Fig.~\ref{lumin_ecyc_ecut_width}).  The sources in blue branch can be fitted to a functional form of \ecut~$\propto$~E$_{cyc}^{0.82 \pm 0.08}$ (marked by the black line). The functional form for sources in magenta is \ecut~$\propto$~E$_{cyc}^{0.29 \pm 0.06}$ (black dashed line). 
In order to make a comparison with the latest result, we digitized the graph\footnote{\url{https://www.digitizeit.de}} in \citealt{staubert2003} to fit their Fig.~8 and find that in their work, they report: \ecut~$\propto$~E$_{cyc}^{0.47(-0.35, 0.39)}$ which is consistent with our results for both groups within errors.

It should however be noted that if we allow for an offset, we get a different best fit result of \ecut = 8 $+$ E$_{cyc}^{0.67}$ and \ecut = 5 $\times$ E$_{cyc}^{0.36}$. Therefore, while various forms work in the correlation between these quantities and the correlation is not mathematically unique, it is evident visually that the blue branch does have a steeper slope in this graph. We also remark here that the values of E$_{cyc}$ we obtain from this work are close to values obtained in \citealt{coburn2002}. The difference between the Figure 9 of \citealt{coburn2002} and Figure \ref{lumin_ecyc_ecut_width} of this work is a result of the determination of the values of \ecut~which is possibly not surprising since we do not expect a {\it one-to-one} correspondence on the values of cutoff energies from RXTE and \suzaku~as they cover different energy ranges and also because for some sources, the X-ray spectrum evolve with flux (see \citealt{bexrbs_reig}).

%while for the sources is green branch, \ecut~is almost constant around 6\,keV for more than a factor of two variation in E$_{cyc}$.

\subsubsection{CRSF width $W_{\rm c}$ versus E$_{cyc}$}
The variation of the CRSF width versus E$_{cyc}$ is shown on the right side of Fig.~\ref{lumin_ecyc_ecut_width}. In agreement with the 
results of \citet{coburn2002}, we observe a linear correlation between the two quantities. We are aware that using a HIGHECUT model can sometimes lead to artificial widening of the CRSF energy if E$_{cyc}$~and \ecut~are close to each other. To mitigate this, we took great care in constraining the width of the CRSF properly and cross-checked the values with those available in literature. 
For example, the three sources (1A 1118-61, GX 301-2 and GX 304-1), where we obtain large widths are consistent with their literature values \citep[]{suchy2011,suchy2011_1a,jaiswal2016}. From our fits of the broadband data, we find $W_c$ $\propto$ E$_{cyc}^{~(2.2 \pm 0.5)}$. Overall, as we see from Table \ref{continuum} and in Fig.~\ref{ecyc_depth} that the higher the value of cyclotron line energy, the deeper the line is.

\begin{figure}
\includegraphics[scale=0.6,angle=-90]{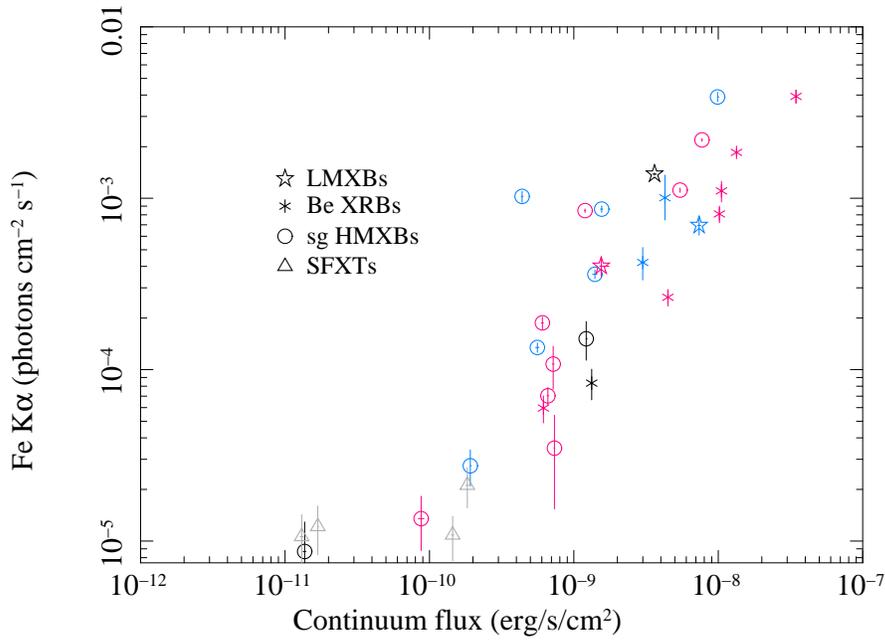}
\vspace{2.5cm}
\caption{\rmfamily{Plot of the iron line flux versus continuum flux marked with the same colour coding as in Figure \ref{lumin_ecut}. }}
\label{iron1}
\end{figure}

\subsubsection{Fe K$\alpha$ flux versus continuum flux}
From Fig.~\ref{iron1}, we see that the \emph{Suzaku} data of the considered sources suggest a strong correlation between the continuum flux and 
the Fe K$\alpha$ flux. Such a correlation\footnote{We remark here that all spectra used for the current analysis were cumulated outside X-ray 
eclipses.} is expected and was also previously reported by \citet{garcia} and \citet{torrejan} but over a smaller range of continuum flux.

\subsubsection{Equivalent Width (EW) of K$\alpha$ iron line versus \lumin}
 On the right side of Fig.~\ref{iron1}, we show the lack of any clear correlation between the equivalent width of the iron K$\alpha$ line and the source X-ray 
luminosity. An anticorrelation between these two parameters have been reported earlier in the literature by \citet{garcia}, \citet{Vasylenko2015}, and \citet{torrejan}. This anticorrelation is usually interpreted in terms of the so called Baldwin effect \citep{baldwin}. As previous studies were carried out in 
a limited energy range, we checked that our results did not change when only the XIS data are used to fit the spectra (with a powerlaw corrected for photoelectric line-of-sight absorption and Gaussian) of all the sources (0.8-10 \rm{}keV). 
We also performed the same study by using the restricted energy range 7.1-12.0 \rm{keV} to estimate the X-ray luminosity of each source (only photons above 7.1 keV 
contribute to the K$\alpha$ emission and the photoelectric cross section 
drops rapidly so that photons above 12 keV do not contribute much in this process) and the energy range 4.95-7.75 \rm{keV} as done in \citet{torrejan}. In none of these cases we could find a clear indication 
of the anticorrelation reported previously. In addition, we have also explored similar correlation using {\it XMM} data in a different work where we do not find any such relation between EW and luminosity (see Fig.~2 of \citealt{pradhan2018}).

The existence of this X-ray Baldwin effect has been a matter of much debate for AGNs and possibly depend on different class of AGNs.  A systematic analysis of many AGNs using XMM and INTEGRAL data showed a very weak correlation between the EW of Fe K$\alpha$ line and luminosity (see \citealt{Vasylenko2015} and references therein). We will return to this discussion in section \ref{sec:disc}.

\begin{figure}
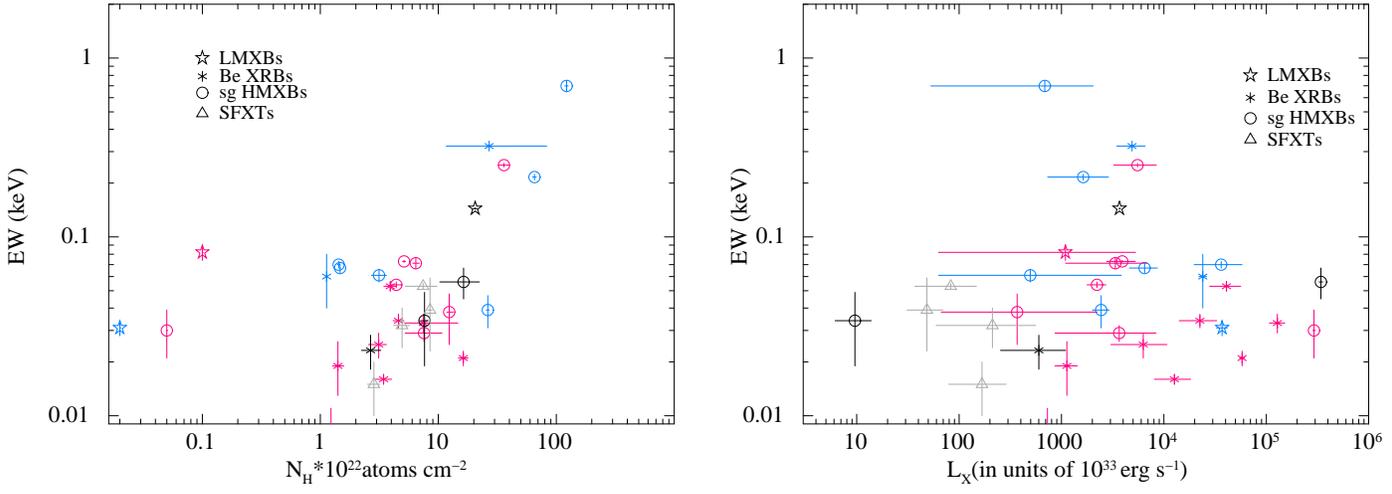

\hspace{-3.cm}
\centering 
\includegraphics[scale=0.47,angle=-90]{nh_eqw_2020_v4.ps}
\includegraphics[scale=0.47,angle=-90]{lumin_eqw_2020_v5.ps}
\vspace{2.5cm}
\caption{\small{Left: Plot of the equivalent width versus luminosity. Right: Plot of absorption column density versus the EW of iron line. We use for all plots the same color coding of Fig.~\ref{lumin_ecut}.}}
\label{iron2}
\end{figure}

\subsubsection{Equivalent Width (EW) of K$\alpha$ versus ${N_{\rm H}}$}

As visible from Fig.~\ref{iron2}, we found a clear linear correlation between the hydrogen column density of the different sources 
and the equivalent width of their K$\alpha$ lines.
Here ${N_{\rm H}}$ = ${N_{\rm H1}}$+${N_{\rm H2}}$*${C_{V}}$, where ${N_{\rm H1}}$ is the hydrogen column density along our line of sight to the source, ${N_{\rm H2}}$ accounts for local absorption and ${C_{V}}$ is the covering fraction, the latter having a wide range of 0.1-0.96 (see Table \ref{continuum}). \\

%\subsubsection{\wgamma versus ${N_{\rm H}}$}

Generally, it is well known that 
when a relatively limited energy range is used for the X-ray spectral fitting on data with limited statistical quality (eg 0.5-10.0 \rm{keV}), the derived values of the absorption column density and the power-law spectral index can be positively correlated (a larger absorption column density can be used in such restricted energy range to fit the 
spectrum equally well with a steeper power-law). We also checked {\it apriori} the variation of \wgamma as a function of ${N_{\rm H}}$ to cross-check some of our results and find that this is not the case for our spectral analysis, mainly thanks to the wide energy coverage of the instruments onboard \emph{Suzaku} (see contour plots, Fig.~\ref{contour1} and \ref{contour2} in the Appendix).

\section{Discussion}
\label{sec:disc}
In this paper, we took advantage of the broad energy coverage of the instruments on-board \emph{Suzaku} to perform a systematic study of all observed 
neutron star HMXBs, including the high magnetic field LMXB pulsars displaying a cyclotron absorption line in their spectra. The aim of our analysis was to explore a number of correlations between the spectral 
parameters of these sources reported previously in the literature. We used for the analysis of all sources a similar spectral model, in order to carry out a 
consistent investigation of all their spectral parameters. While detailed individual studies of most of these sources are already reported in literature, our aim with this paper is to outline a comprehensive behaviour of HMXBs as a class. Such a study was last carried out nearly two decades back in early 2000s \citep{coburn2002}.

We summarize the main results of the paper below:
\begin{enumerate}
 \item The broadband X-ray spectra of all HMXBs that we considered in this paper can be described by a powerlaw component modified by a cutoff energy (where PIN data is usable). This spectral shape is usually interpreted in terms of Comptonization of the seed photons produced in the thermal mound of the neutron star 
  accretion column by free electrons in the accreting material.  
 It is therefore expected that the spectral cutoff energies are proportional to the electron temperature. \\
 A lack of correlation between \ecut\ and \lumin\ has been reported earlier by \citet{white1983}. They point out, however, that the low luminosity 
 systems in their sample do exhibit a somewhat less sharp cutoffs at higher energies. \\ 
 In this study, we found all analyzed systems distributing on two different branches in the \ecut\ and \lumin\ plot. 
  For sources in Branch 1  (represented in blue in all plots), the cut-off energy,
 that is a measure of the electron temperature, vary from 12-24\,keV and increases with the mass accretion 
 rate (in turns regulating the X-ray luminosity).
 For sources following the Branch 2 (magenta), \ecut\ remains in a narrower range from 3-10 keV, even when \lumin\ changes by a factor of $\sim$ 1000. 
 This result indicates that 
 the cutoff energies of the sources do not depend \emph{only} on the luminosity. 
  As we showed in Section \ref{sect:spectral analysis}, this different behaviour cannot either be simply ascribed to switches between 
 a pencil and a fan beam emission geometry due to the lack of clear systematic variations in the pulse profiles of the analysed sources. 
 To the best of our knowledge, so far, no theoretical explanation has been proposed in the literature to interpret this behaviour.
 
 \item An anticorrelation between \wgamma and \lumin\ is known to exist in neutron star LMXBs and black hole systems (for luminosities lower than `critical' luminosity) where accretion takes place through a disc \citep[]{lmxb_gammalx, bhb_gammalx}. \citet{wijnands2014} suggested that in case of strongly magnetized neutron stars the presence of a magnetic field could significantly alter the spectra, such that a direct comparison between HMXBs and LMXBs is not possible. For the present study, we make use of systems containing in all cases strongly magnetized neutron stars (B $\gtrsim$ $10 ^{12}$ G) and accreting both from disks and from stellar winds. We found that the anticorrelation between \wgamma and \lumin\ still holds, indicating that in HMXBs, the Comptonization process makes the spectrum harder at increasing \lumin~at least in the sub-critical regime (even though it might break down at luminosities higher than $10^{37}$ ergs $s^{-1}$).

For black hole binaries, since this anti-correlation is valid for luminosities below a critical value, we also cross-checked the X-ray luminosity of each pulsar against the critical luminosity of each pulsar using Eqn.~\ref{eqn:lcrit} \citep{becker2012} below.  %In accreting X-ray pulsars, this `critical luminosity' $L_{\mathrm{crit}}$ divides the regimes of accretion into the super and sub-critical regime \citep{becker2012}. 
%The critical luminosity, $L_{\rm crit}$, according to \citet{becker2012}, depends on the magnetic field.  
%For typical neutron star parameters \citep[see][for details]{becker2012}, the critical luminosity can be estimated as
%\begin{equation}
%L_{\rm crit} \approx 1.28 \times 10^{37} (E_{\rm cyc}/10{\rm keV})^{16/15} \, \,  {\rm erg
%\, \,s^{-1}} 
%\end{equation}
\begin{equation}
    	L_{\rm crit} = 1.49 \times 10^{37}{\rm erg\,s}^{-1} \times \left( \frac{\Lambda}{0.1} \right)^{-7/5}  \times w^{-28/15} \times \left(\frac{M}{1.4{\rm\,M}_{\odot}} \right)^{29/30} \left( \frac{R}{10{\rm\,km}} \right)^{1/10} \left( \frac{B_{\rm surf}}{10^{12}{\rm\,G}} \right)^{16/15} 
	\label{eqn:lcrit}
\end{equation}
where $R$, $M$, and $B$ are, the radius, mass, and surface
magnetic field strength (B in terms of $\sim 10^{12}$ G) of the neutron star, $\Lambda$ $=$ 1, and $w = 1$ for wind-accreting systems. For all those sources with measured B (through CRSF line energy), we find that besides Her X-1, 4U 0115+63, Cen X-3 and 1A 1118-61, all other sources are indeed in the sub-critical regime. 
% Source B Lcrit Lx33
% Her X-1, 38, 2.9e37, 3.7e37 G
% 4U 0115, 10, 6.9e36, 2.4e37 G
% CEN X-3, 29, 2.1e37, 3.6e37 G
% 4U1626 36, 2.7e37, 0.4e37 L
% XTE1946 37, 2.8e37, 0.6e37 L
%VELA 27.9, 2e37, 0.22e37 L
% 1907 18.74, 1.4e37, 0.0494e37 L
%4U 1538 21.46, 1.6e37, 0.64e37 L
% GX301-2 44.20, 3.4e37, 0.16e37 L
% 1A118 50, 3.8e37, 4.0e37 G
% 4U0114 24, 1.7e37, 0.36e37 L
% GX304 56, 4.3e37, 2.2e37 L
%CEP 30, 2.2e37, 0.48e37 L
%1700 37, 2.8e37, 0.39e37 L
%A0535 42, 3.2e37, 0.06e37 L
%4U1822 35, 2.6e37, 0.11e37 L
%(1.49*10^(37))*((1/0.1)^(-7/5))*(1.^(-28/15)))*((Ekev)^(16/15))

 \item An anticorrelation of \efold~with \lumin~indicates that with increasing luminosity, \efold~(a proxy for electron temperature of the `infalling' plasma) decreases. 
As the luminosity increases, the radiation field from the neutron star begin to affect the accretion flow and Compton cooling becomes more efficient (since scattering rate of photons off electrons increase). This leads to the shift of electron temperature toward lower energies with increasing luminosity giving rise to this anticorrelation of \efold~with \lumin.

\item From the variation of folding energy with cutoff energies, we see that smaller cutoff energy imply larger folding energies. It has also been reported by earlier authors that the non cyclotron-line pulsars should exhibit larger values of folding energy, (less steep spectral breaks) due to their lack of the spectral trough caused by an absorption like feature in CRSF \citep{makishima1999}. However, when we compared the folding energy of cyclotron line sources with the folding energy of non-cyclotron line sources, we see no such trend.

 \item The variation of \ecut\ with E$_{cyc}$  has been studied in a number of earlier works \citep[]{makishima1990, makishima1999, coburn2002}. \citet{makishima1990} obtained a linear correlation between \ecut\ and E$_{cyc}$ (E$_{cyc}$ = 1.4-1.8 \ecut ). From the measurements of \ecut,\ they thus inferred that the magnetic field of HMXBs 
 was spanning a relatively narrow range (1-4 $\times 10^{12}$ G). Nine years later, these results were updated by \citet{makishima1999}. These authors used a larger sample of sources to prove that there exists a saturation of \ecut\ at high values of E$_{cyc}$ ( \ecut\ $\propto$ E$_{cyc}^{0.7}$). They interpreted this finding by suggesting that the presence of a thermal emission component can affect the position of the spectral energy break. \citet{coburn2002} reconfirmed the same relationship, 
adding also new constrains on the saturation of \ecut\ . The latter seems to be detected only below 35 \rm{keV}. 
They also suggested the possible presence of two linear relations between \ecut\ and E$_{cyc}$, one below 35 \rm{keV} and one above. 
%\citet{staubert2003} suggested instead that \ecut\ = 0.5 \ecyc\ . \\ 
Our current study has an added advantage of the broadband energy coverage compared to these previous works in literature. We find that for higher values of magnetic field, the cutoff energies tend to saturate for both these groups. This is possibly because there could be some other relativistic effects in the creation of the continuum that become important at at higher values of magnetic field.

%cutoff energy as proxy for CRSF is not as straightforward and is very different for the two branches. 

%A single relation of proportionality between \ecut\ and E$_{cyc}$ is not able to provide a satisfactory description of all sources (left panel of Fig.~\ref{lumin_ecyc_ecut_width}) and instead the sources follow at least two different scaling laws.

%A number of systems follows the linear relationship \ecut\ = \ecyc\ , while for others the relationship \ecut\ = 0.5 \ecyc\ $^{0.7}$ provides a better description of the data. Other systems placed in the middle of the \ecyc\\ - \ecut\ plot seem to suggest the existence of a third new scaling law.  \\ 
%Based on previous suggestions in the literature, it is likely that for sources following the  \ecut\ $\propto$ \ecyc\ $^{0.7}$ relation the thermal spectral component 
%could be more predominantly contributing to the spectral cutoff, leading to a faster saturation of \ecut\ . 
%For sources following the relation \ecut\ = \ecyc\ , this contribution is less significant. 
%However, the sources located in the middle of the left of Fig.~\ref{ecyc_ecut_pho} cannot be easily explained in this scenario. 

\item The right panel of Fig.~\ref{lumin_ecyc_ecut_width} shows that in the present study we found a relatively well marked linear correlation  between the energy of the cyclotron line and it's width. \citet{coburn2002} mentioned that the FWHM of a CRSF changes with the viewing angle $\theta$ of the 
 observer with respect to the magnetic field as follows: 
 \begin{equation}\label{eq:width}
   \eqfwhmpropto
\end{equation}

This implies that two quantities, the characteristic electron temperature, \kte\, and $\theta$ (angle between the observer line of sight and the neutron star magnetic field) for all sources considered in this paper are not dramatically different. The same linear correlation was also found by \citet{coburn2002} and are indicative of small values of \kte\ . The reason for this is that the matter being accreted onto the hot spots form an accretion column and in the steady state, the amount of matter falling in balances the matter spreading out at the base where the mound parameters (area, density etc) are similar for most sources irrespective of accretion rate. Our results support their argument since despite five orders of magnitude change in \lumin~in the current work, our findings are indicative of the same linear correlation. Alternatively, this correlation can also be explained if the temperature is tied to the magnetic field strength as seen in some simulations of gamma-ray bursts \citep{lamb1990}.

Another implication of this relation is that, any effect of $\theta$ is also very small. Since the bulk motion of the electrons (which in turn depend on the accretion rate) also play a role in line broadening, we should have expected this correlation to smear out.  This brings us to a very interesting point as has been discussed in \citet{coburn2002} that accretion could have an influence on the relative orientation of spin and magnetic axes, with the spin axis presumably being influenced by the age of the system and the accretion timescales making $\theta$ very small. This correlation can be further tested for individual sources by detailed pulse profile modelling of each source and measuring the width of the CRSF line with varying $\theta$.

It is interesting to mention here that in \citealt{staubert2019-all}, there are two groups in width of CRSF versus energy (their Fig. 12) where the authors argue that the second slope are mostly consist of `outlier" sources. We find similar scatter in the depth of CRSF versus CRSF energy (Fig.~\ref{ecyc_depth}) although the sources do not hold a on-to-one correspondence in both the works\footnote{Note that there is often a variation in CRSF parameters over different time and/or spin phases}.

%\item In Fig.~\ref{ecyc_width_ratio}, we also analyze a possible correlation between the width of the CRSF divided by the centroid energy and it's depth.  \citet{araya1999} showed that such two quantities should be anticorrelated because the cyclotron scattering cross-section of the resonant photons is larger (hence deeper CRSF) when the latter move perpendicularly (hence smaller value of cos $\theta$) with respect to the magnetic field orientation. As a consequence, we should expect deeper CRSF features at smaller values of cos $\theta$. In the present study, we could not detect any clear correlation or anticorrelation between the  width of the CRSF scaled to its centroid energy and the CRSF depth. \\
\item Our analysis also confirmed the presence of a strong correlation between the X-ray continuum of all the sources 
where a K$\alpha$ iron emission line is found and the line flux. 
This is expected, since the Fe K$\alpha$ line is thought to be produced in these systems due to the reprocessing of the X-ray continuum by iron atoms in the matter 
surrounding the neutron star (being either the material from the stellar winds in HMXBs or the disk in LMXBs). 
As was also discussed by \citet{torrejan} and \citet{garcia}, we thus expect the Fe K$\alpha$ line flux to increase with the flux of the continuum. \\

Historically, an inverse correlation was obtained between the EW of C IV lines and the UV luminosity in AGNs (Baldwin effect; \citealt{baldwin}). 
Analogously, correlation studies between the EW of K$\alpha$ lines and \lumin\ have been carried out for AGNs and XRBs providing indications for an inverse correlation
\citep[]{garcia, Vasylenko2015, torrejan}. This anticorrelation between the EW of K$\alpha$ lines and \lumin\ 
is called the X-ray Baldwin Effect. At odds with the results of \citet{garcia} and \citet{torrejan}, we did not detect 
any significant correlation between the EW of the Fe K$\alpha$ lines and the source X-ray luminosity. As discussed by \citet{torrejan}, we noticed in our case that a 
marginal correlation (not statistically significant) is visible when the EW of the K$\alpha$ line is plotted against the X-ray flux but the correlation disappears when the luminosity is used in place of the flux. 

The non-confirmation of the X-ray Baldwin effect for the Fe K$_\alpha$ in X-ray binaries is probably not surprising since there is no physical reason for the effect to hold true for XRBs with the neutral iron line. In case of AGNs the physical reason for the anti-correlation of EW of C$_{IV}$ lines with luminosity is that for higher luminosity the atoms are highly ionised and do not produce C$_{IV}$ lines in same proportion. However, in the case of the neutral 6.4\,keV iron line in X-rays, if the increase in luminosity cause the ionisation to be so high as to remove the K-shell electrons, we should be seeing the Hydrogen like and Helium like iron atoms and the iron line would instead be seen at 6.7 and 6.9\,keV instead of the neutral K$_\alpha$ line at 6.4\,keV. On the contrary, very few HMXBs like Cen X-3 \citep{cenx3_time}, OAO 1657-415 \citep{pradhan2019-oao},for example, show these emission lines. 

\item Finally, we also reported on the positive correlation between the EWs of the iron K$\alpha$ line and the absorption column density local to the HMXB sources (${N_{\rm H}}$). This correlation is expected because a larger ${N_{\rm H}}$ implies that more stellar wind material around the neutron star is involved in the formation of a stronger iron line (see e.g.,  \citealt{inoue1985,pradhan2014_oao,pradhan2015_0114,garcia,Pradhan_2019-igrj}) except in some sources with accretion disk like SMC X-1, where a relatively low column density is seen \citep{Pradhan_2020}. In others like the persistent Be XRB, SW J2000.6+3210, the iron line is almost undetectable since there is not enough matter surrounding the neutron star to facilitate fluorescence \citep{pradhan2013_swj}.

Although we have very few SFXTs compared to classical sg HMXBs in the current work, we note that the absorption column density (and EW) for these SFXTs are smaller when compared to classical sgXBs (see Table \ref{continuum}). This motivated us to look into our findings with more details by including a larger sample of SFXTs. We also therefore extended this study by including {\it XMM} data and have discussed the findings in detail in \citealt{pradhan2018} which we refer to the reader for further details. We find that the lower N$_H$ and lower EW should be expected for SFXTs. Such a difference is either due to faster (or rarer) stellar winds in SFXTs or due to the inhibition of accretion in SFXTs most of the time by other mechanisms like magnetic gating \citep{bozzo2008} leading to inefficient photoionization of the stellar wind.  

We reported in the plots of Fig.~\ref{iron1} and Fig.~\ref{iron2} also the LMXBs considered in this paper for completeness, but we do not expect them to follow a similar correlation 
as for the HMXBs. For LMXBs, the iron line is indeed supposed to have a completely different origin than in HMXBs (see, e.g. \citealt{cackett2013}).

\end{enumerate} 

The residuals from all the best fits to the data of the different sources  
are shown in Fig.~\ref{residuals1} and \ref{residuals2} and various contour plots are shown in the Appendix. 

\section{Summary}
In this section, we summarize the main results of our work discussed above. This class analysis of the broad band X-ray spectrum of 39 accreting neutron star X-ray binaries indicate some interesting findings. (i) We find that the relationship between the cut-off energy and X-ray luminosity follow a bi-modal behaviour. The interpretation of such a dichotomy is not straightforward and is not a result of different companion stars (LMXBs, Be XRBs, sgXBs) or beaming patterns (explored through pulse profiles). We encourage further studies using physical continuum and line models in order to investigate this behaviour. (ii) We also find that the dependence of cut-off energies on the CRSF energy is not unique. We also confirm the previous findings of \citealt{coburn2002} that the width (and depth) of the CRSF is linearly correlated to the CRSF energy. (iii) We confirm the correlation between the iron K$\alpha$ emission line and the X-ray continuum flux, between the EW of the iron K$\alpha$ line and absorption. This is expected since such lines are formed by the fluorescence of the X-ray continuum photons in interstellar matter around the NS in case of HMXBs. We also note that the EW and absorption are different between SFXTs and classical HMXBs and interpret that as being caused by difference in stellar wind properties between these two systems or a result of inhibited accretion in SFXTs during most times. Finally, (iv) we note that the photon index and \efold~ is negatively correlated with luminosity, thus suggesting that Compton cooling becomes more efficient at higher luminosities which makes the spectrum harder and also lowers the electron temperature of the plasma. 

Overall, we have updated the correlation between spectral parameters of accreting neutron stars taking advantage of the broadband energy coverage of \emph{Suzaku}. Such a study was long overdue in literature with the last such study  made almost two decades back.

\begin{landscape}
\begin{figure*}
\centering
\caption{Spectra of all considered sources together with residuals from the best fits. All best fit results are reported in Table \ref{continuum}}
\includegraphics[height=4.3cm,width=4.7cm,angle=0]{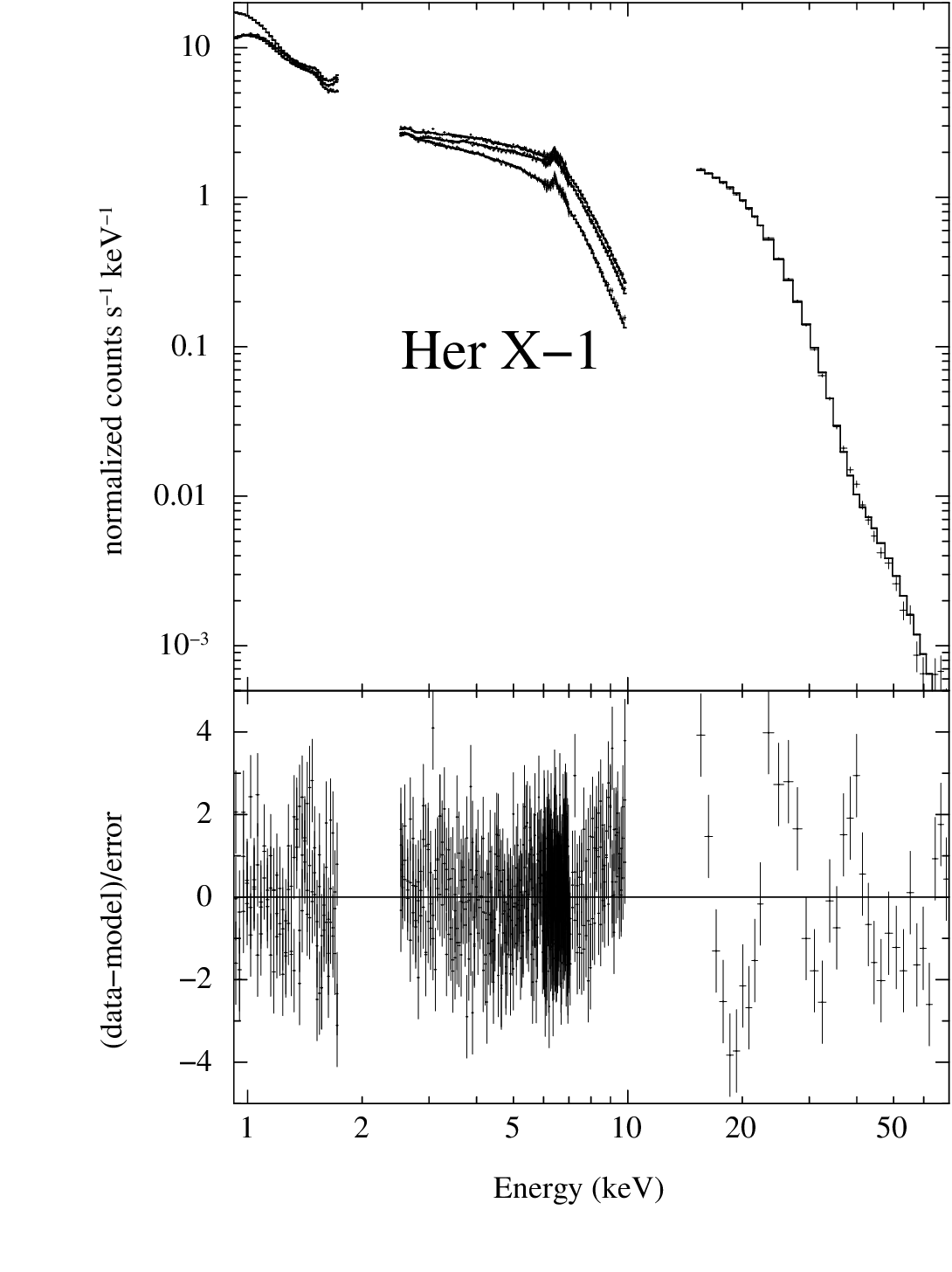}
\includegraphics[height=4.3cm,width=4.7cm,angle=0]{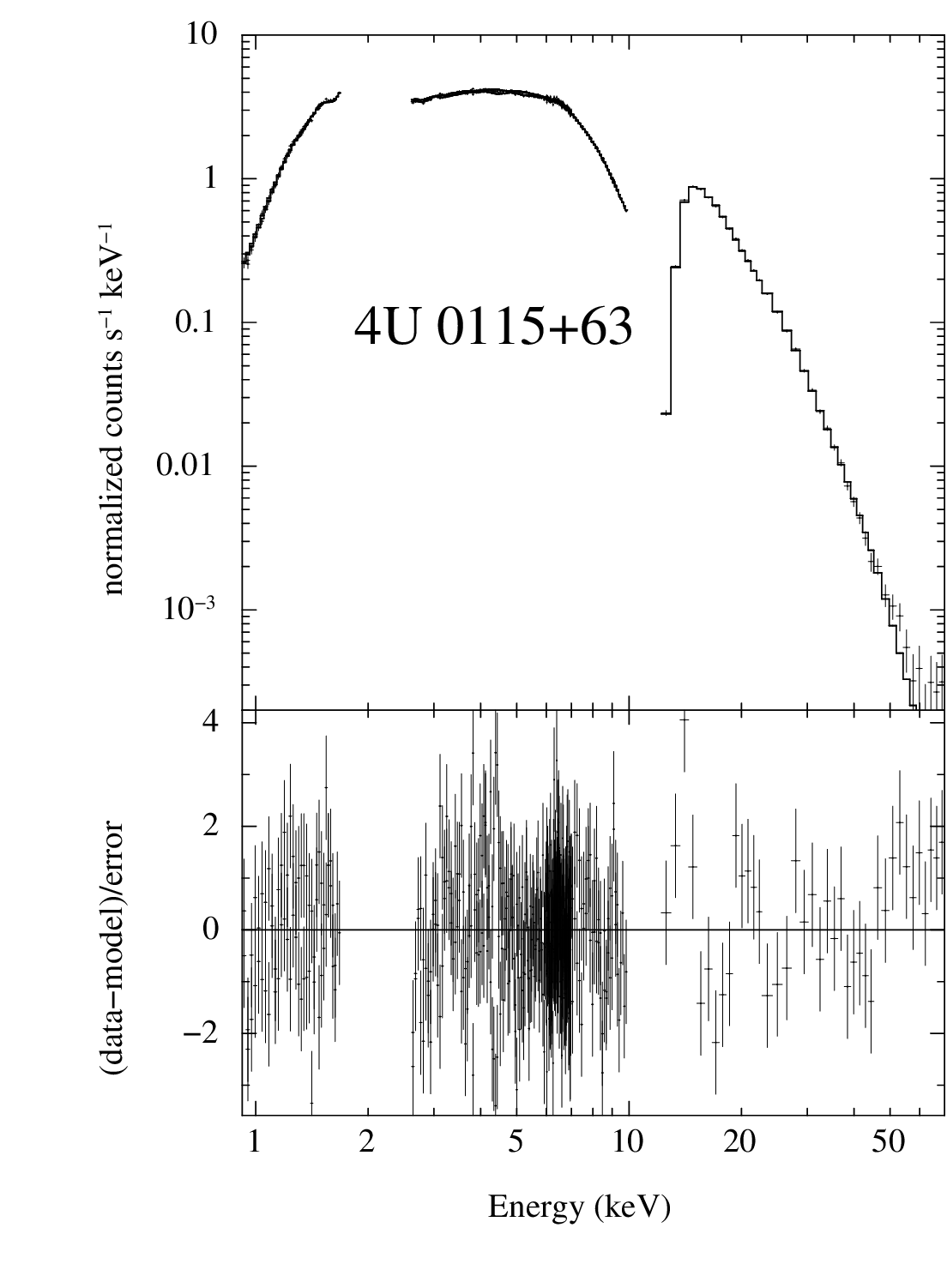}
\includegraphics[height=4.3cm,width=4.7cm,angle=0]{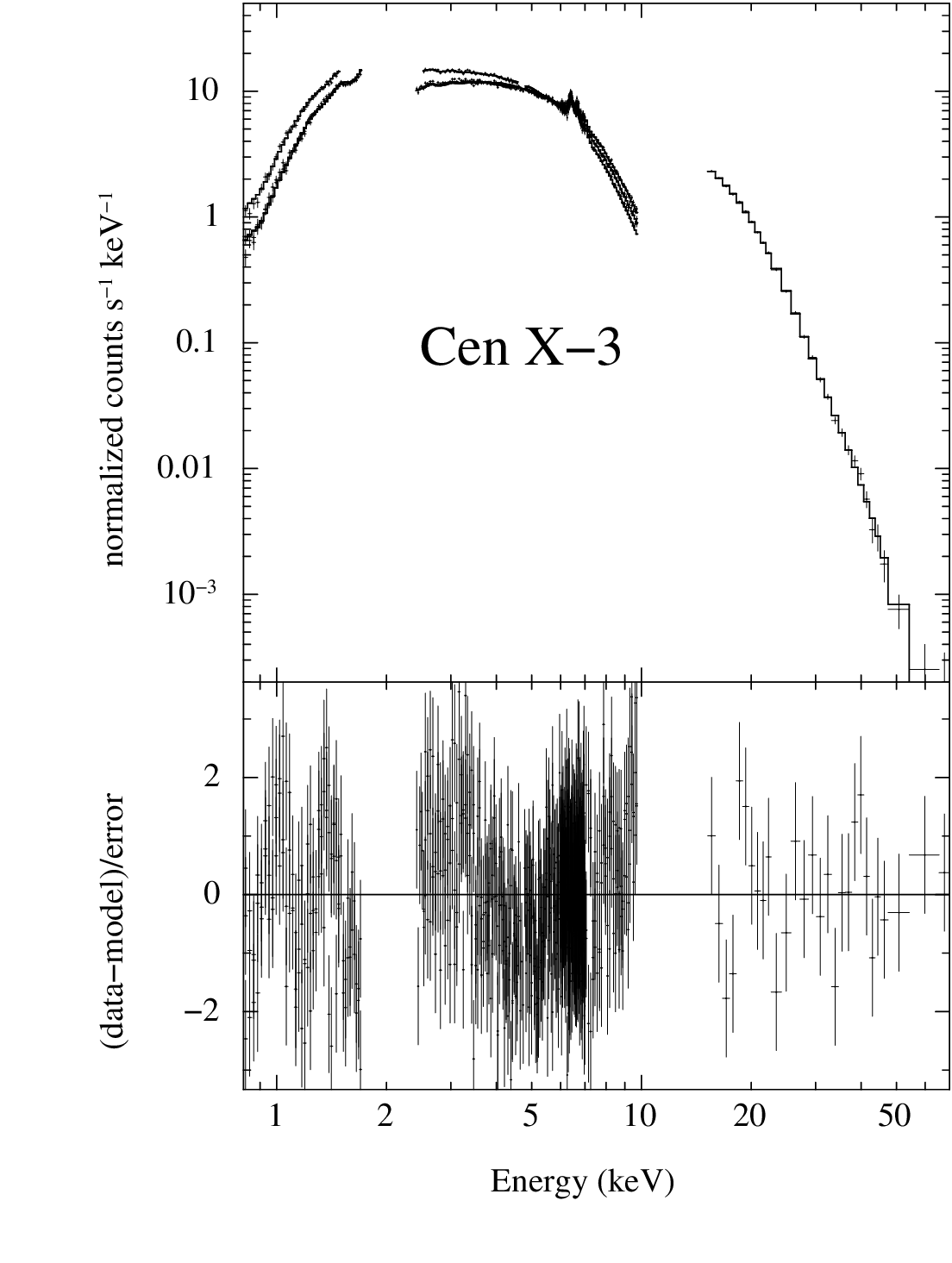}
\includegraphics[height=4.3cm,width=4.7cm,angle=0]{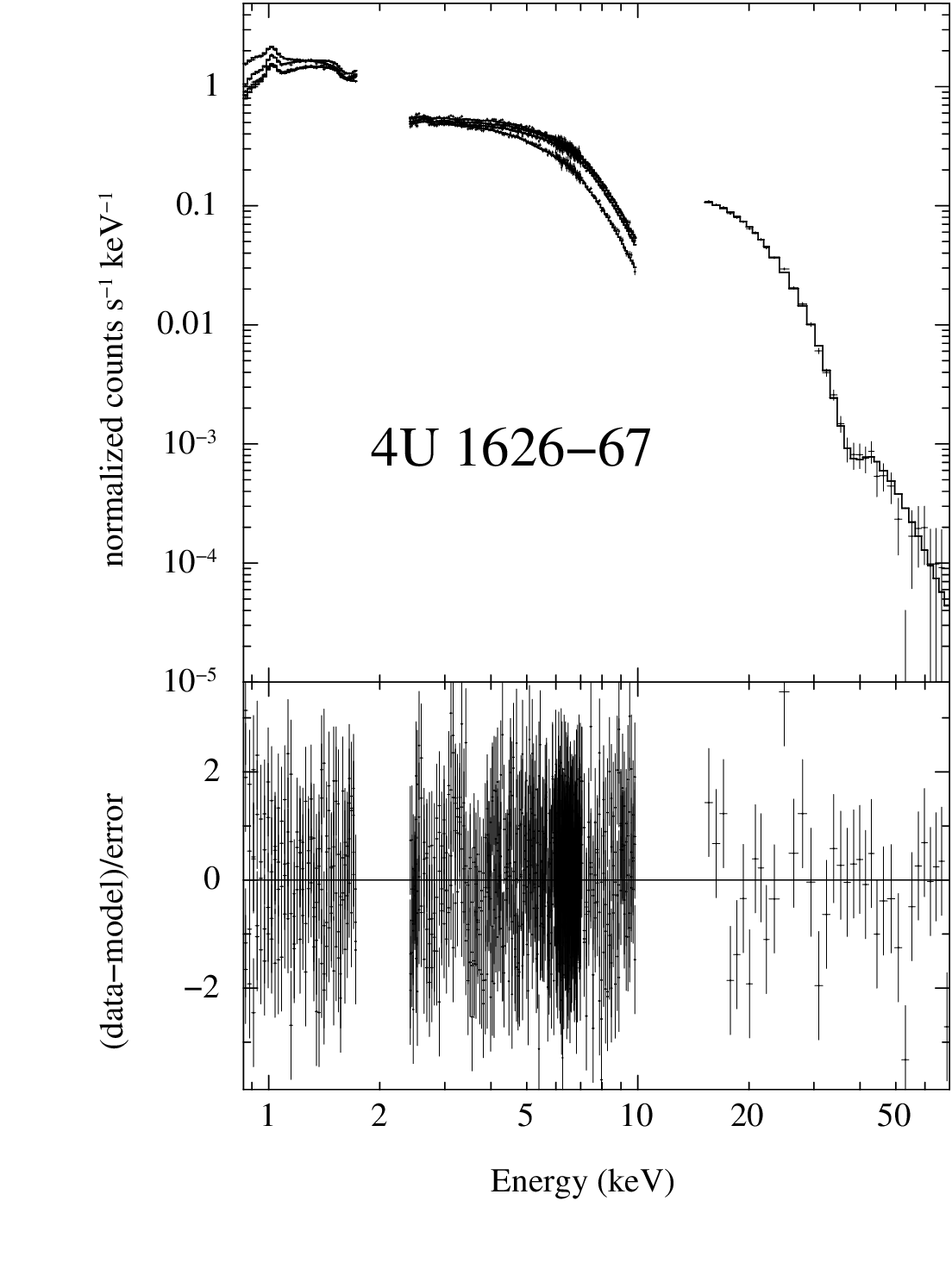}
\includegraphics[height=4.3cm,width=4.7cm,angle=0]{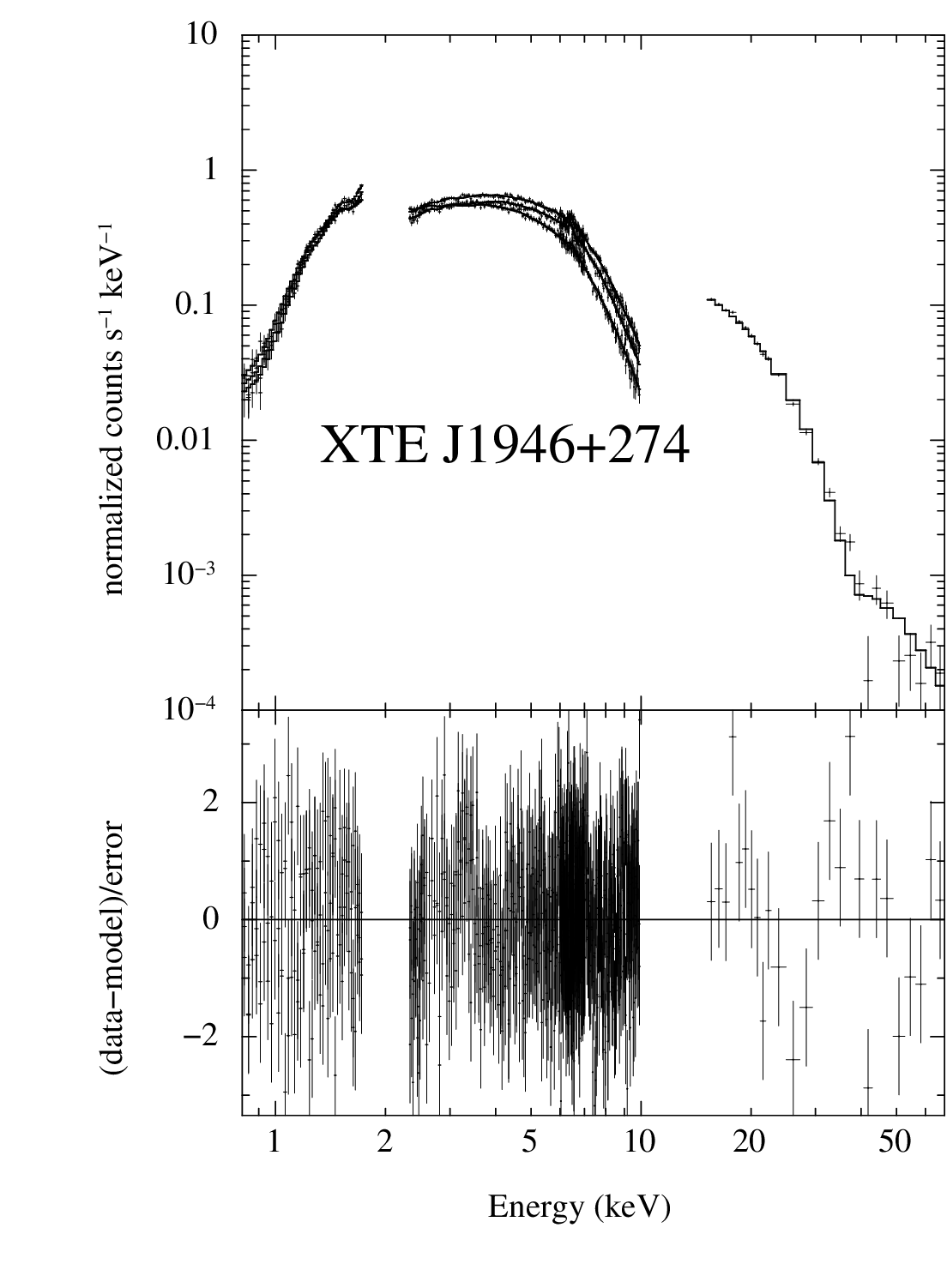}
\includegraphics[height=4.3cm,width=4.7cm,angle=0]{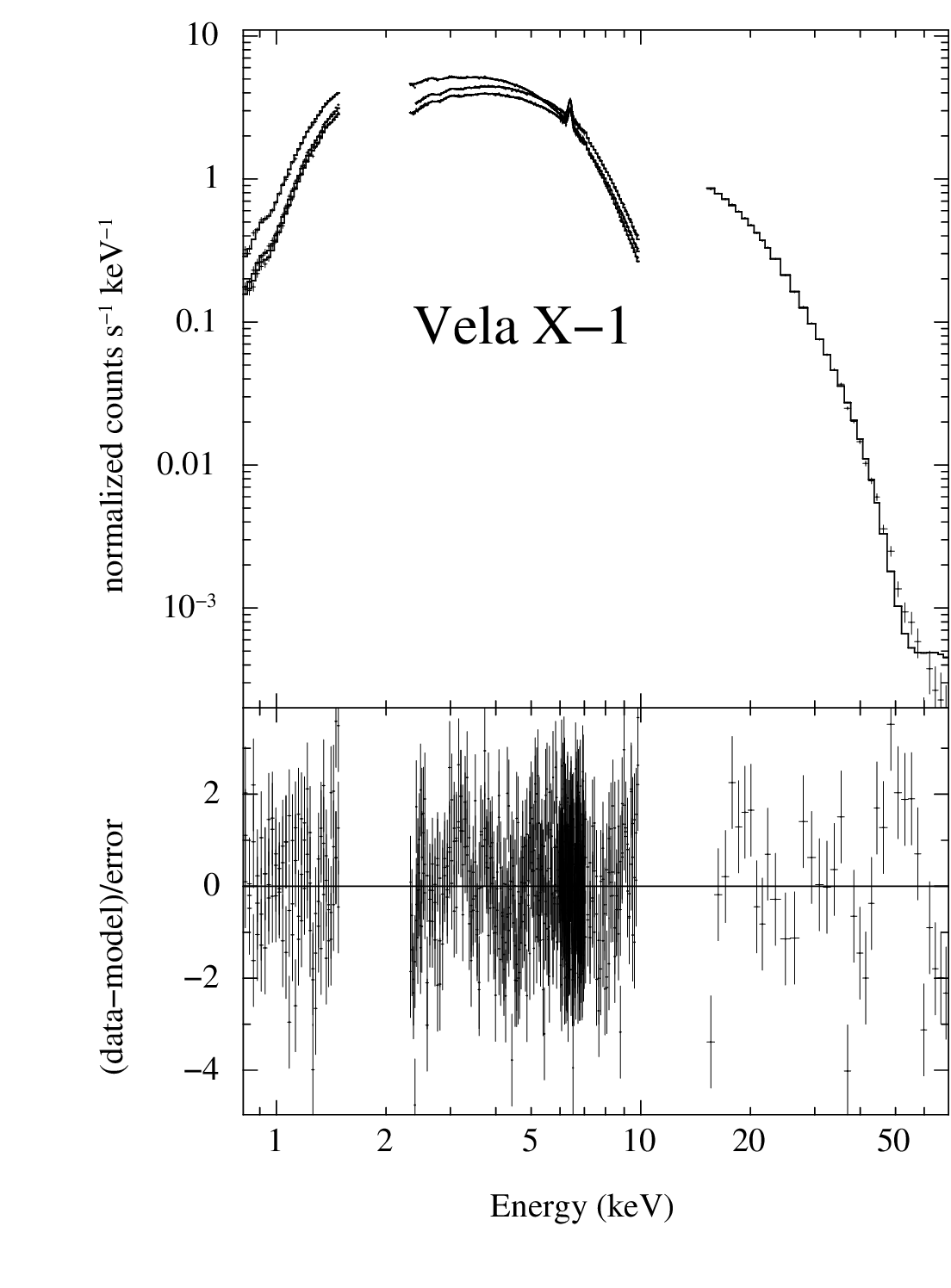}
\includegraphics[height=4.3cm,width=4.7cm,angle=0]{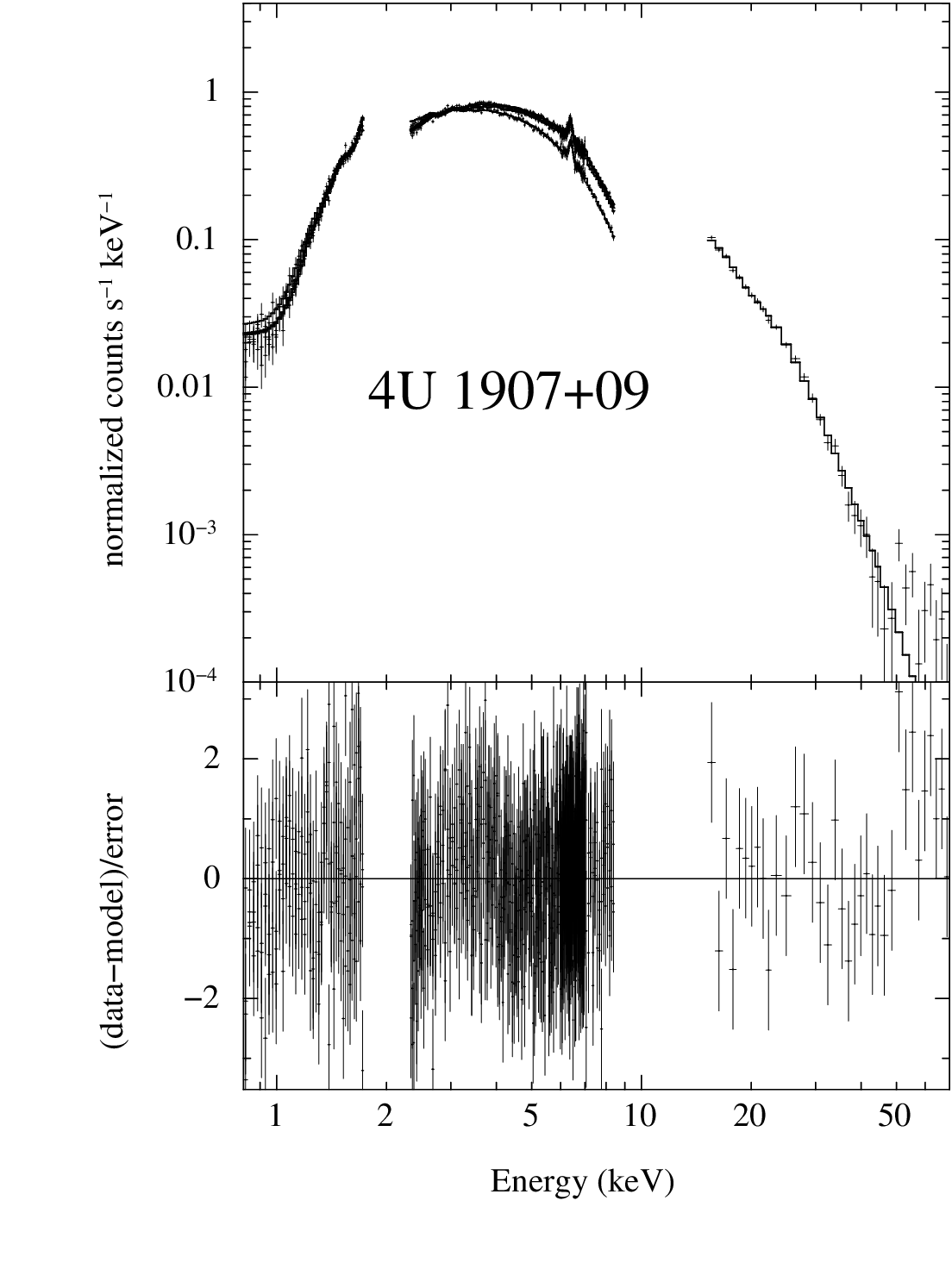}
\includegraphics[height=4.3cm,width=4.7cm,angle=0]{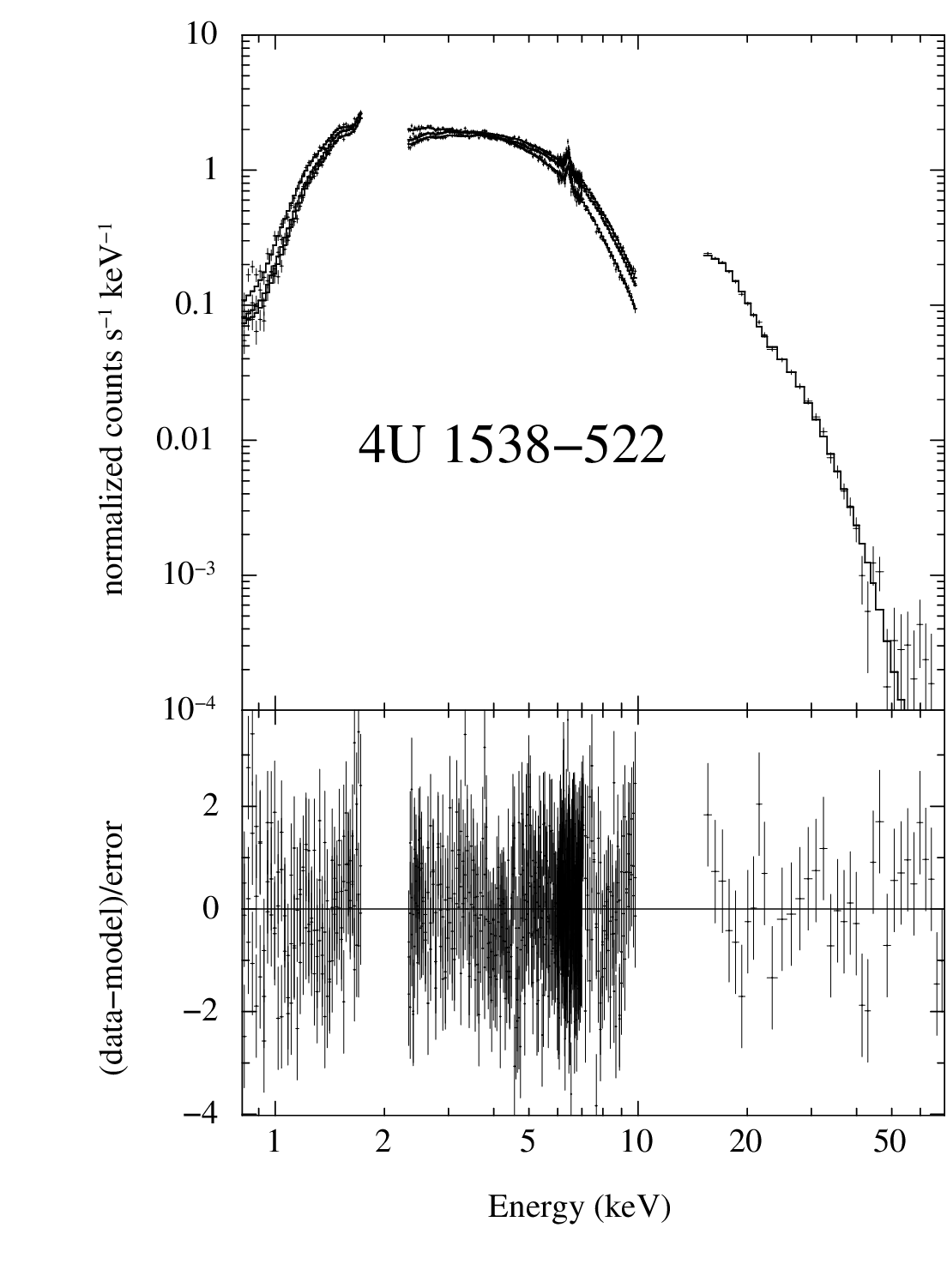}
\includegraphics[height=4.3cm,width=4.7cm,angle=0]{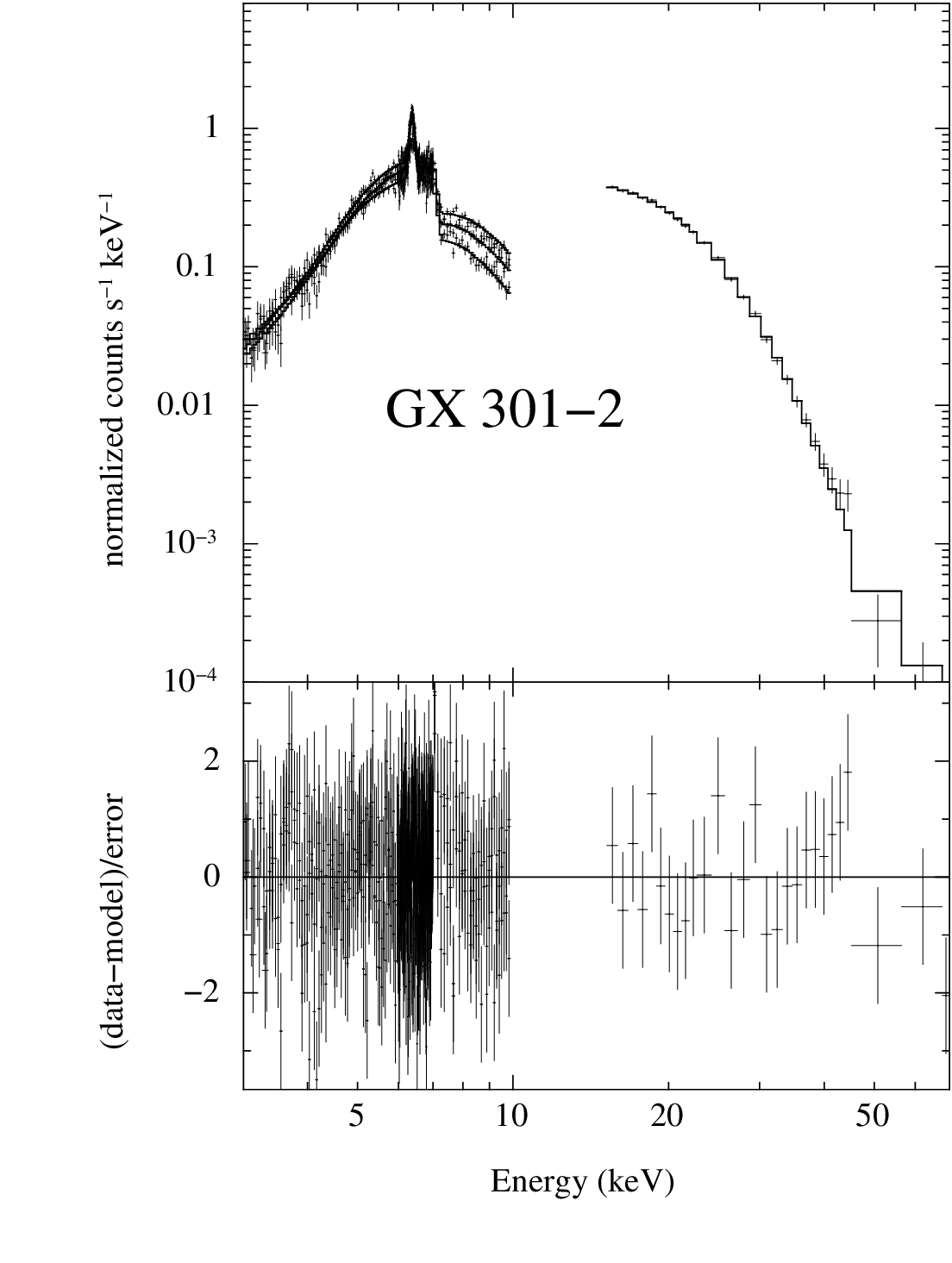}
\includegraphics[height=4.3cm,width=4.7cm,angle=0]{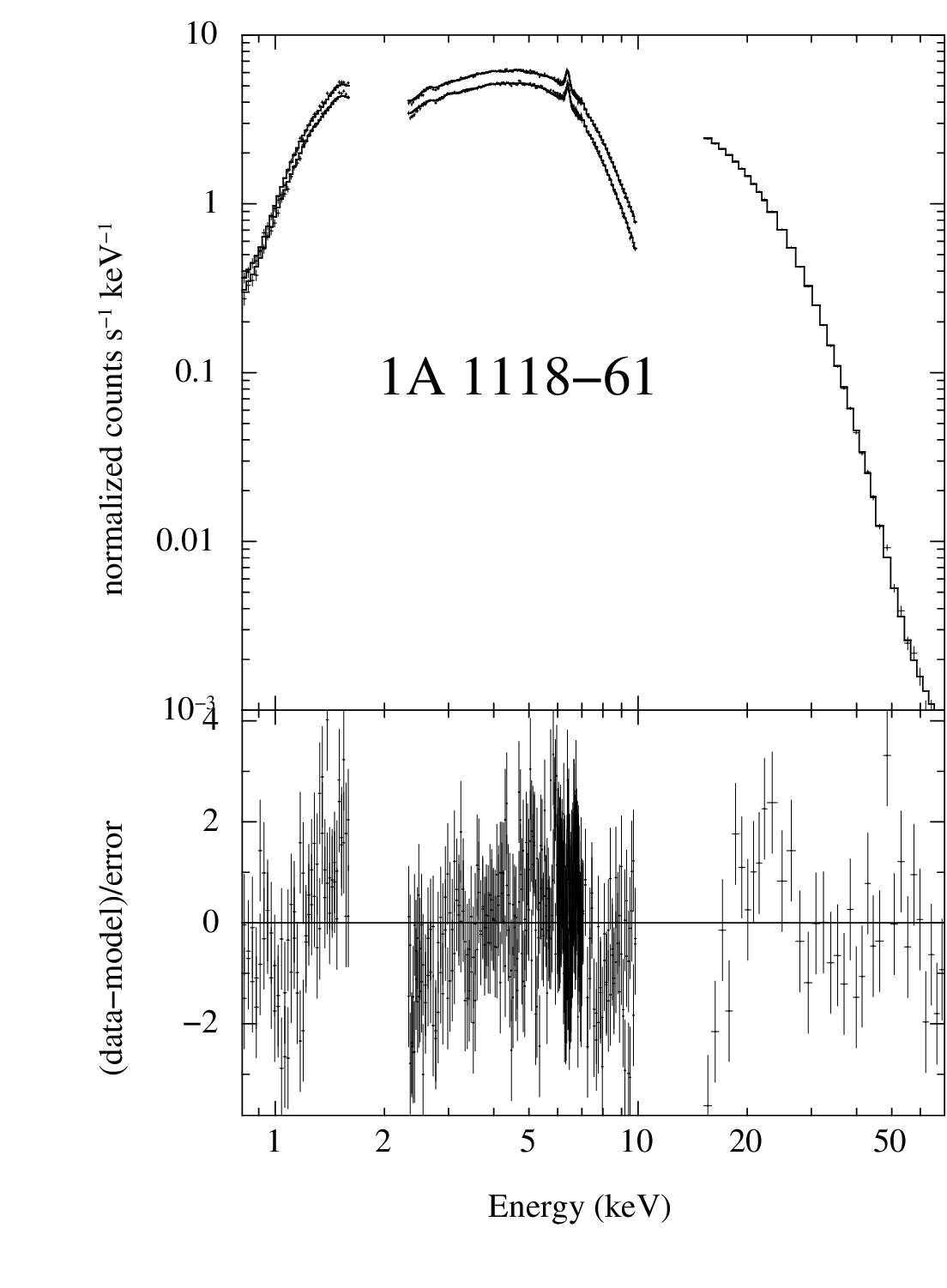}
\includegraphics[height=4.3cm,width=4.7cm,angle=0]{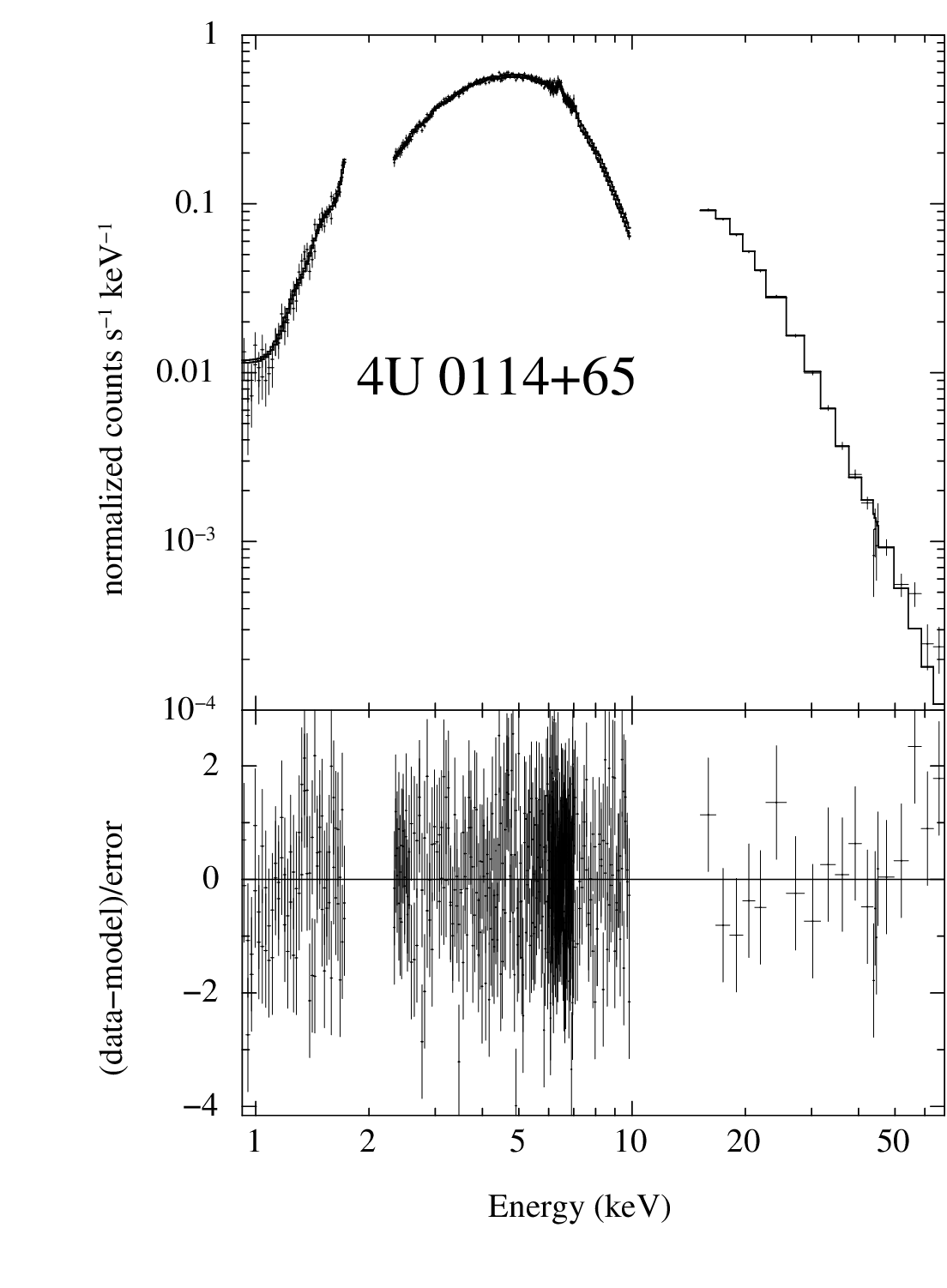}
\includegraphics[height=4.3cm,width=4.7cm,angle=0]{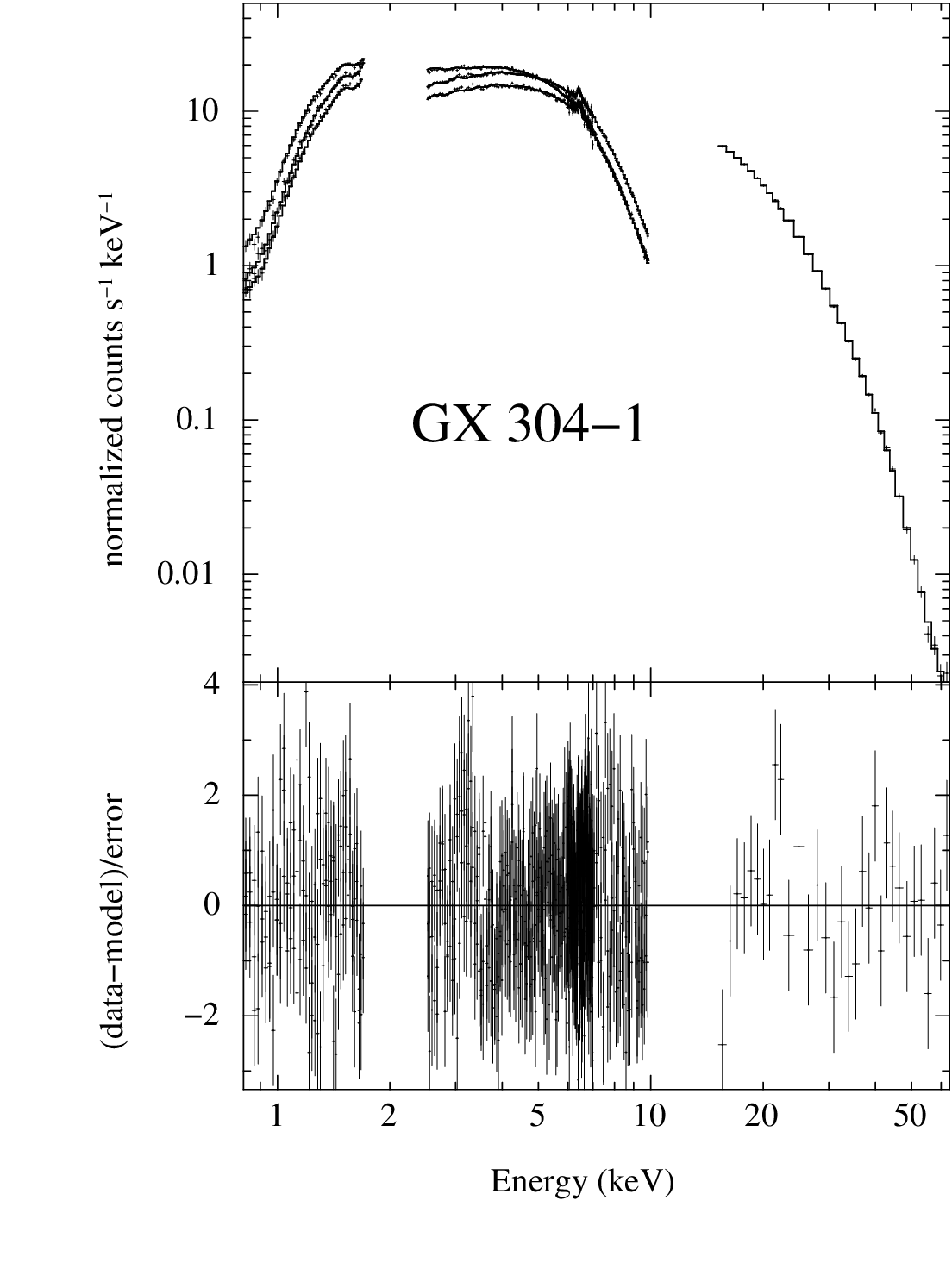}
\includegraphics[height=4.3cm,width=4.7cm,angle=0]{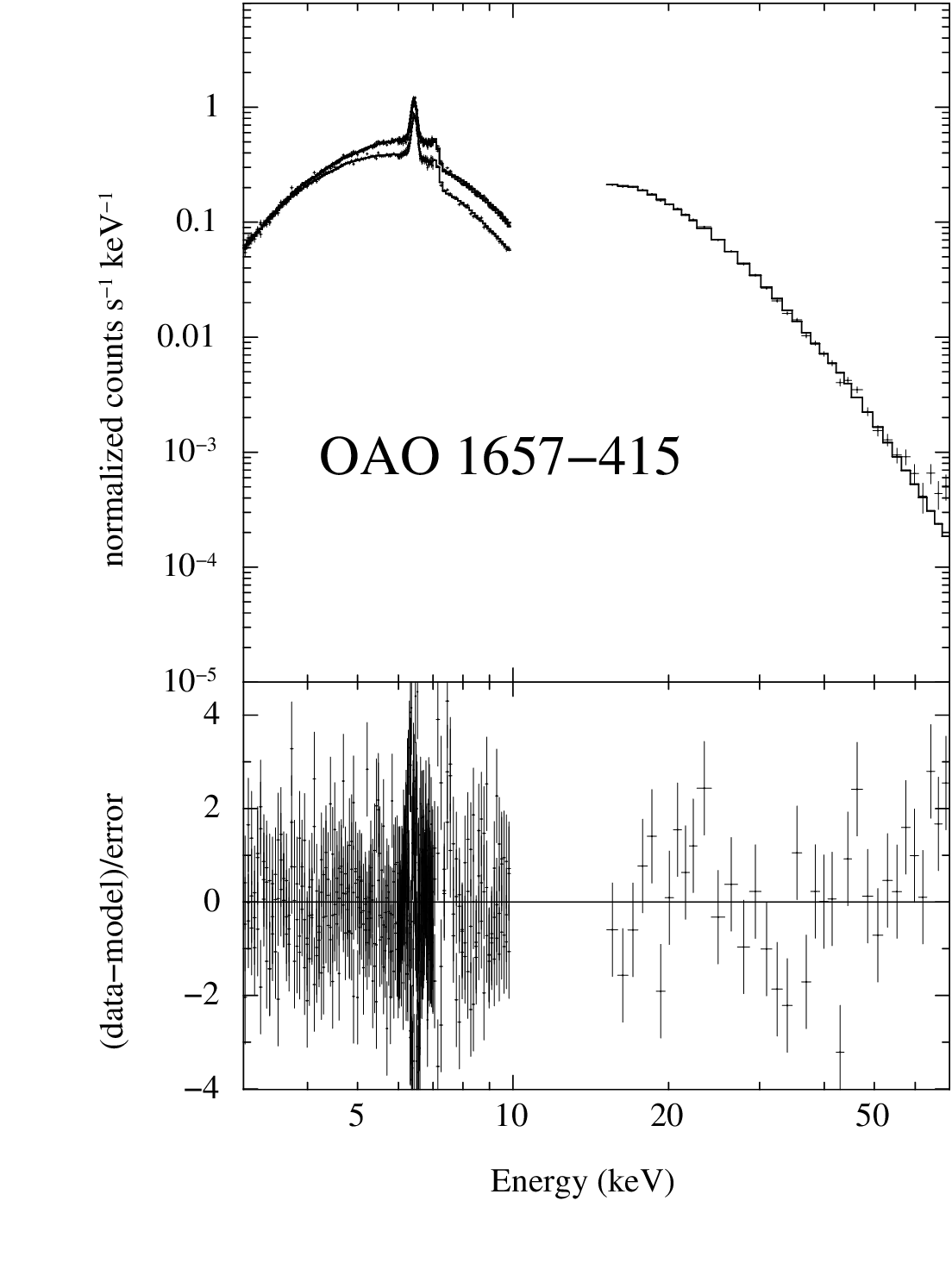}
\includegraphics[height=4.3cm,width=4.7cm,angle=0]{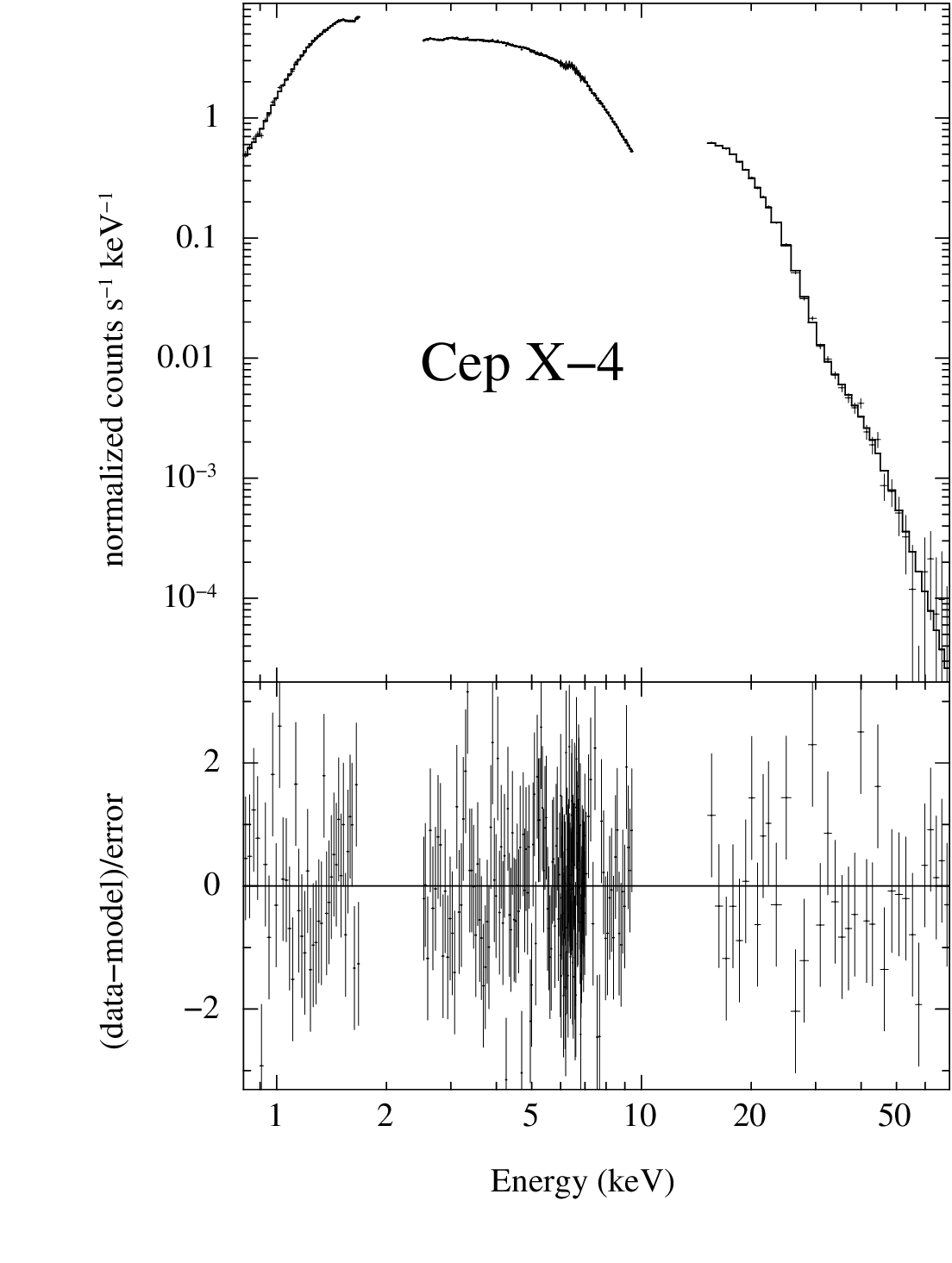}
\includegraphics[height=4.3cm,width=4.7cm,angle=0]{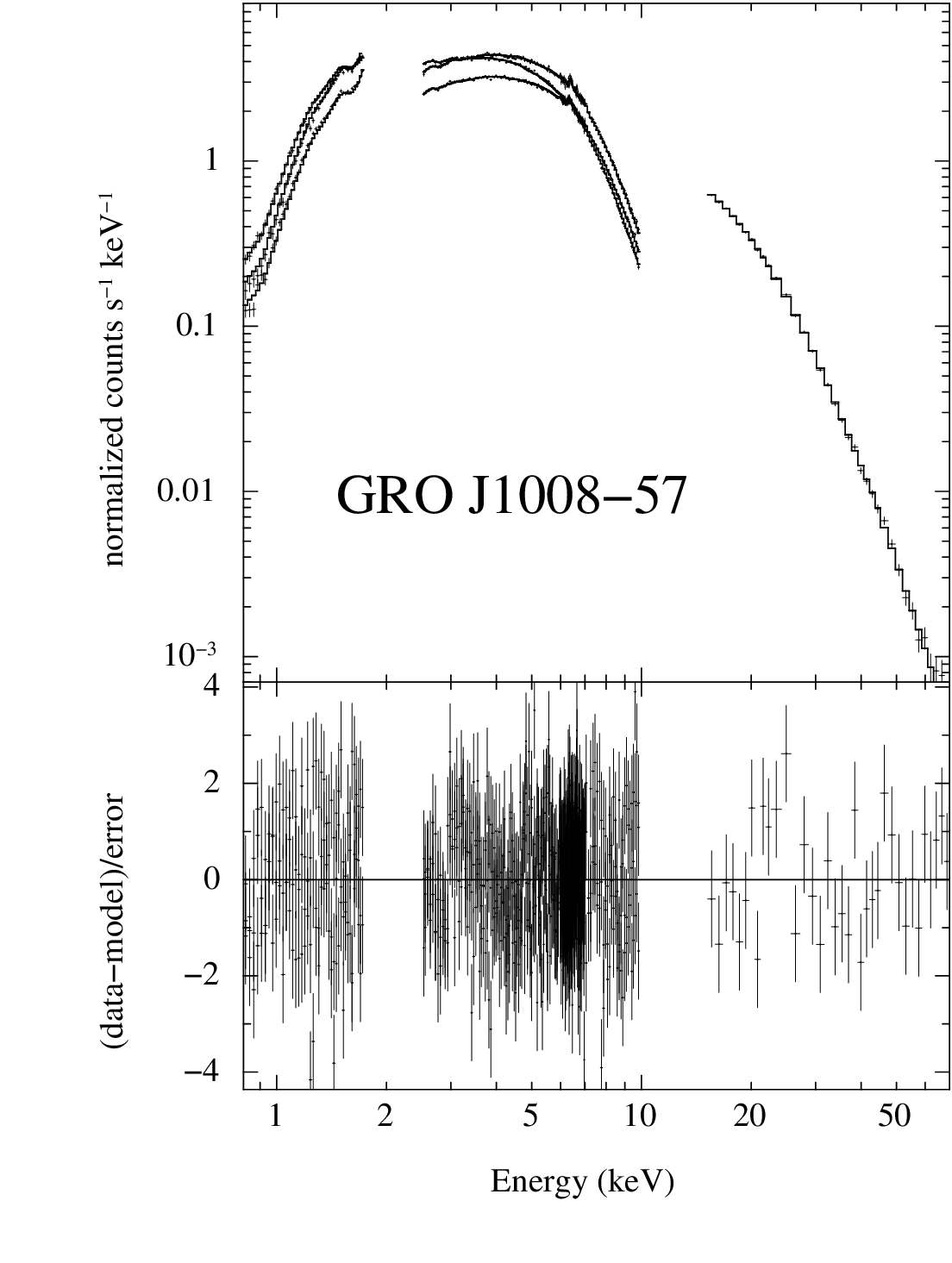}
\includegraphics[height=4.3cm,width=4.7cm,angle=0]{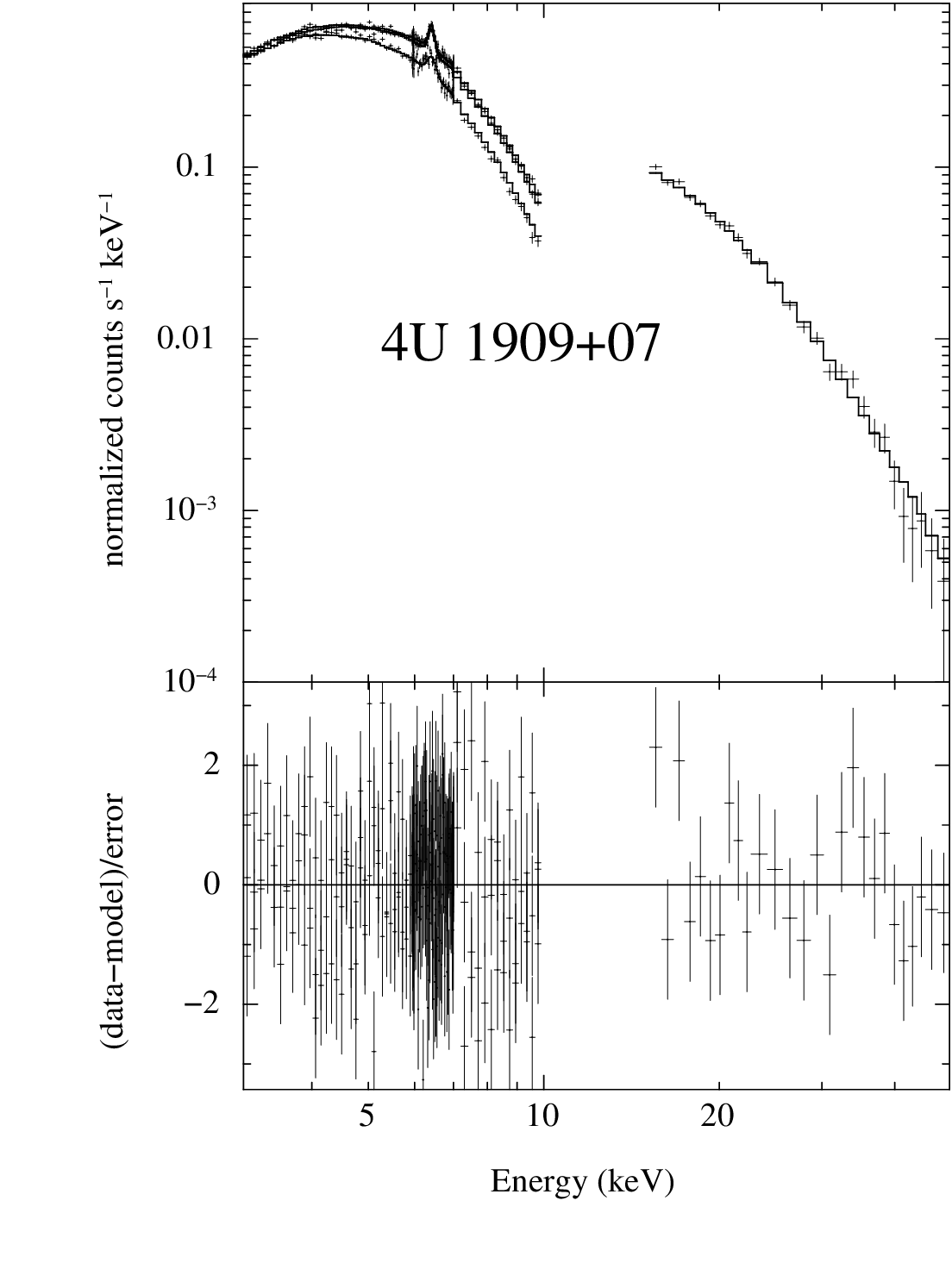}
\includegraphics[height=4.3cm,width=4.7cm,angle=0]{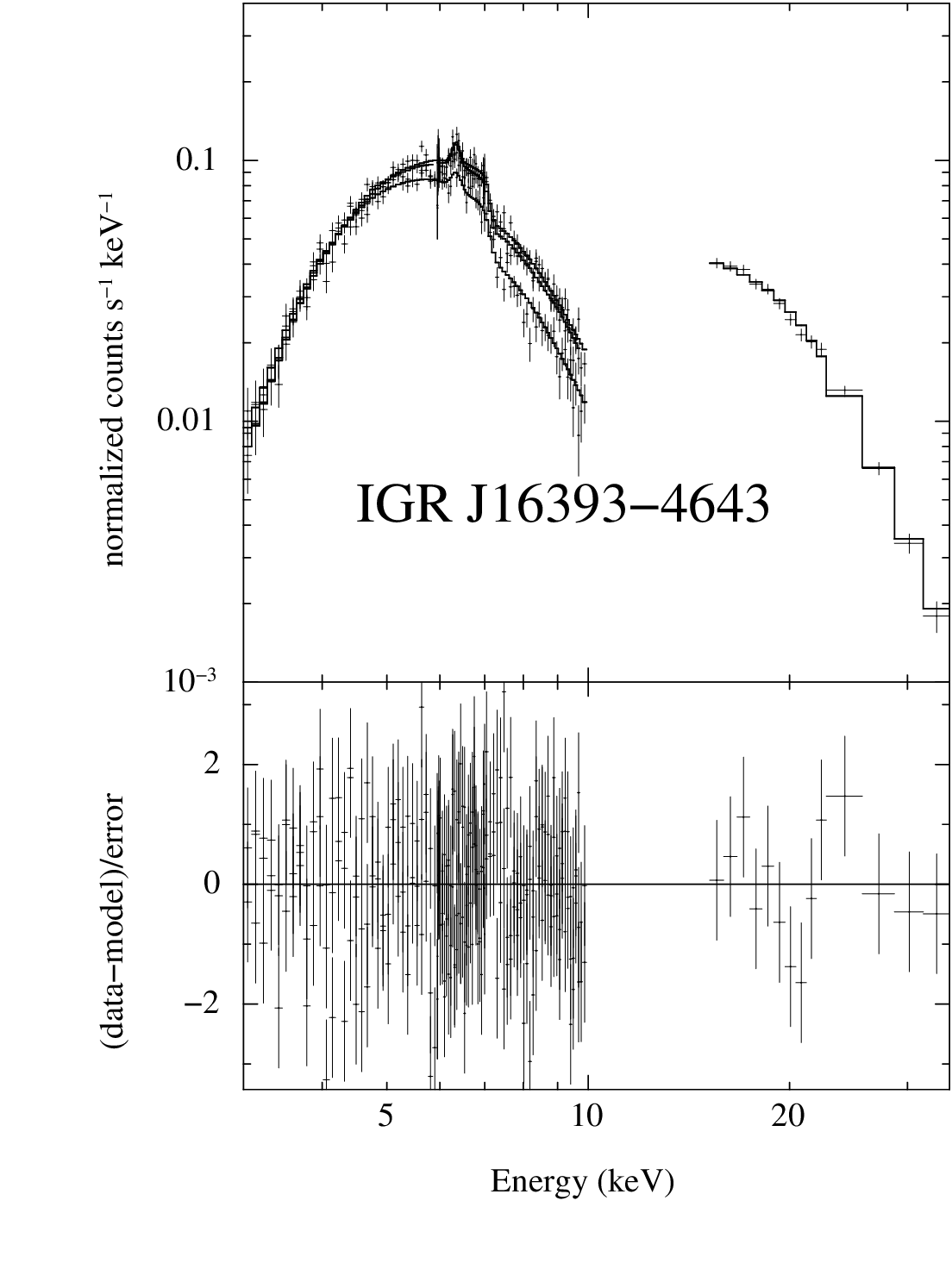}
\includegraphics[height=4.3cm,width=4.7cm,angle=0]{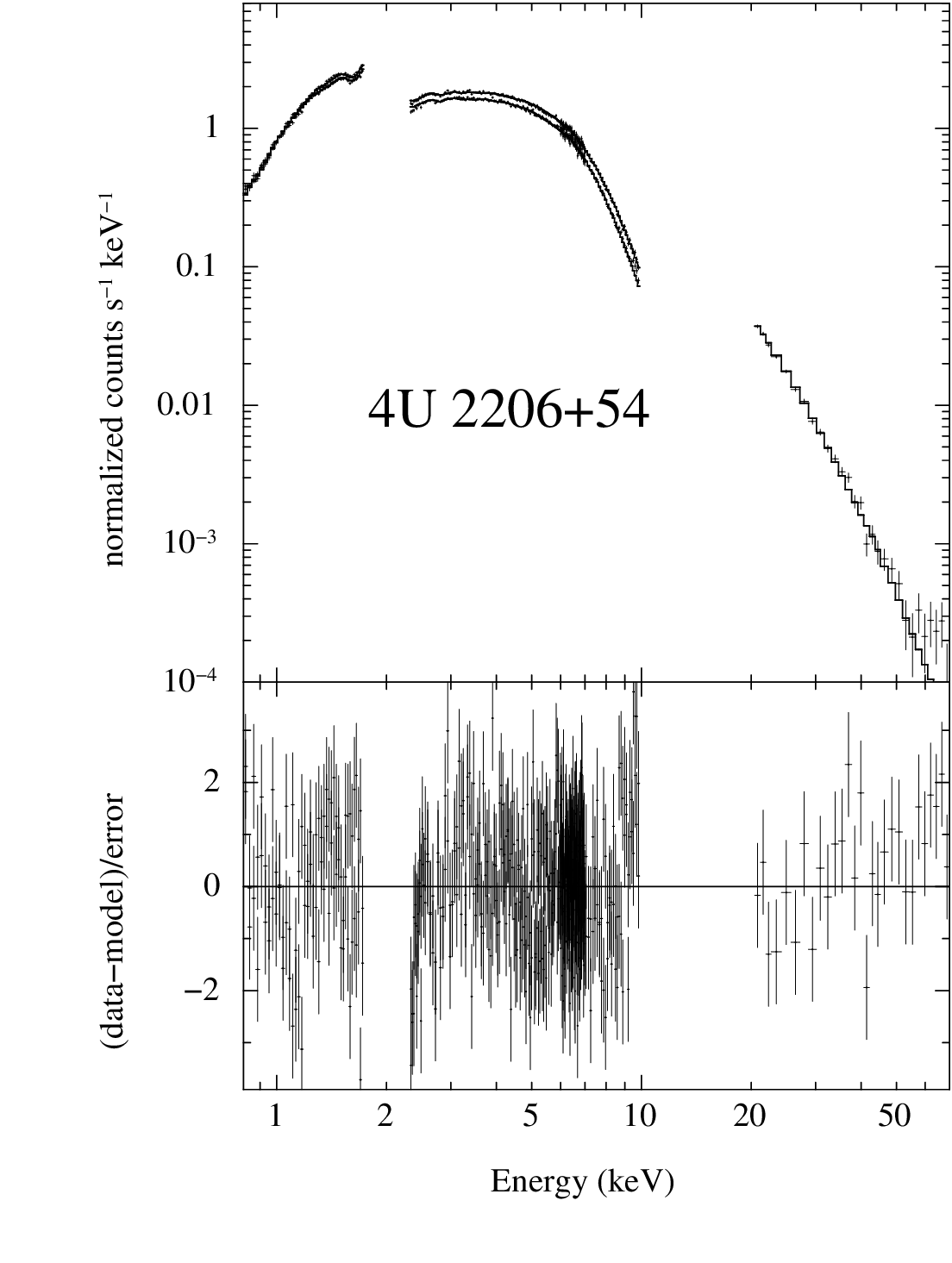}
\includegraphics[height=4.3cm,width=4.7cm,angle=0]{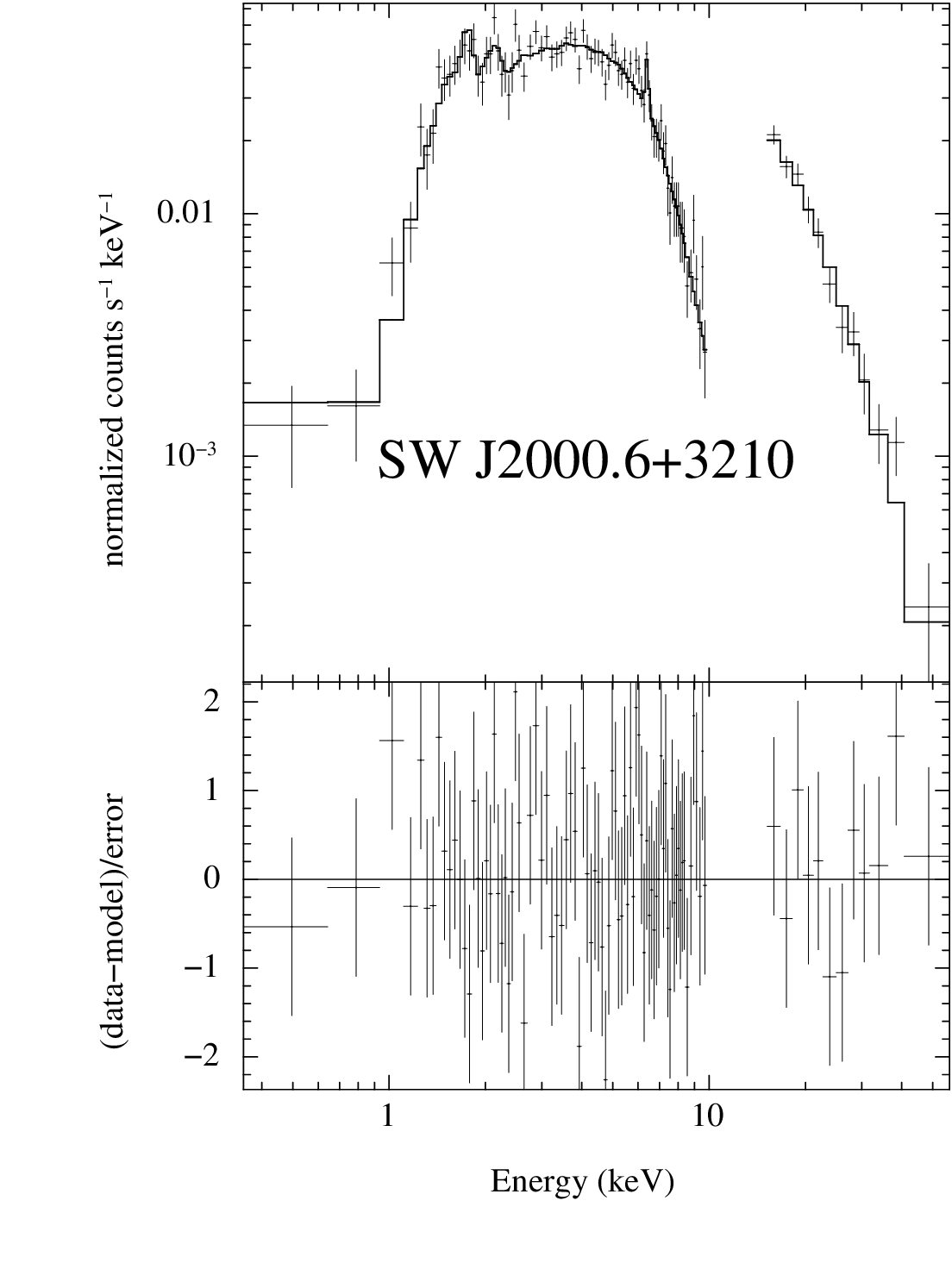}

\label{residuals1}
\end{figure*}
\begin{figure*}
\caption{Spectra of all considered sources together with residuals from the best fits. All best fit results are reported in Table \ref{continuum}..contd}
\includegraphics[height=4.3cm,width=4.7cm,angle=0]{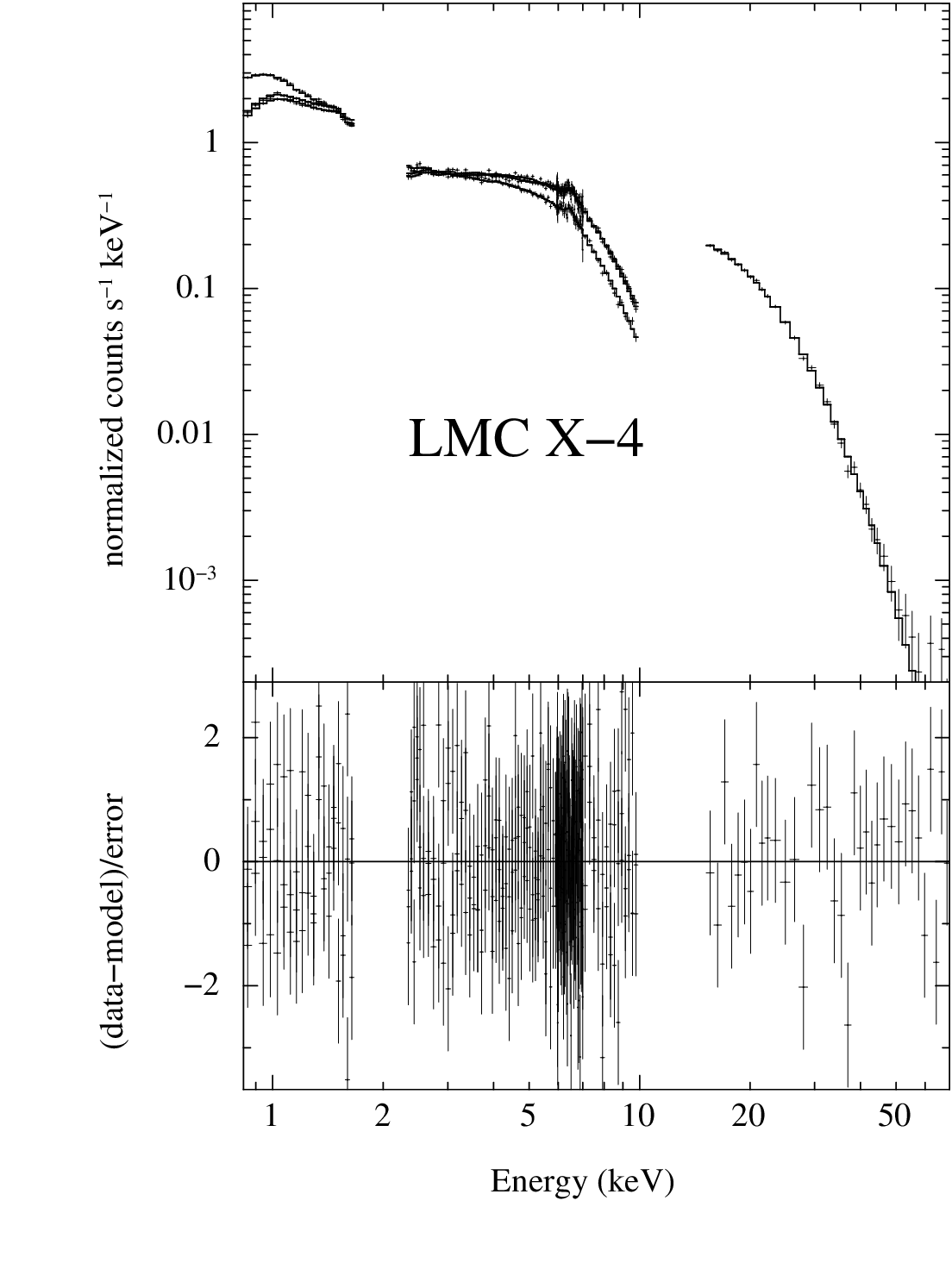}
\includegraphics[height=4.3cm,width=4.7cm,angle=0]{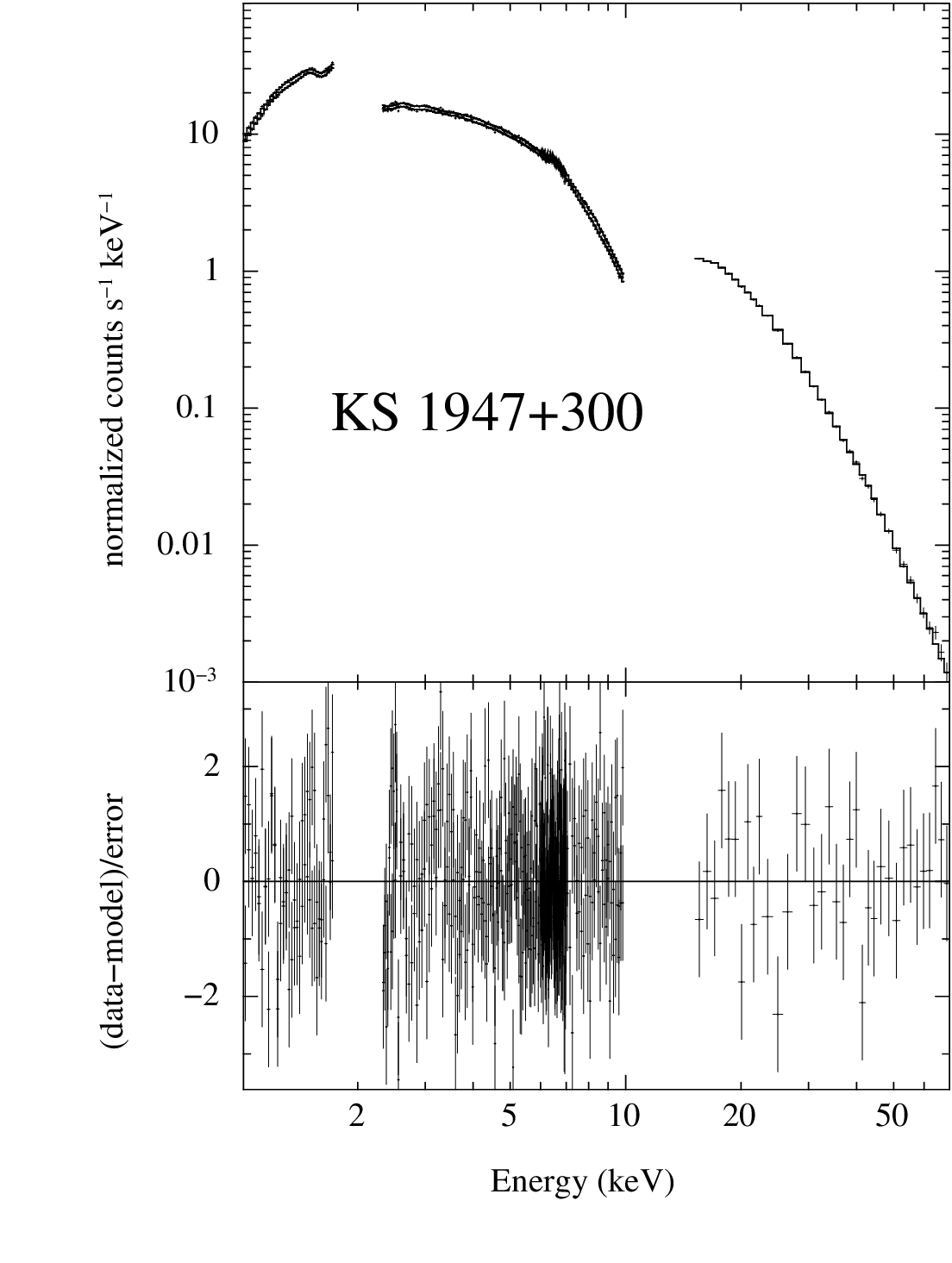}
\includegraphics[height=4.3cm,width=4.7cm,angle=0]{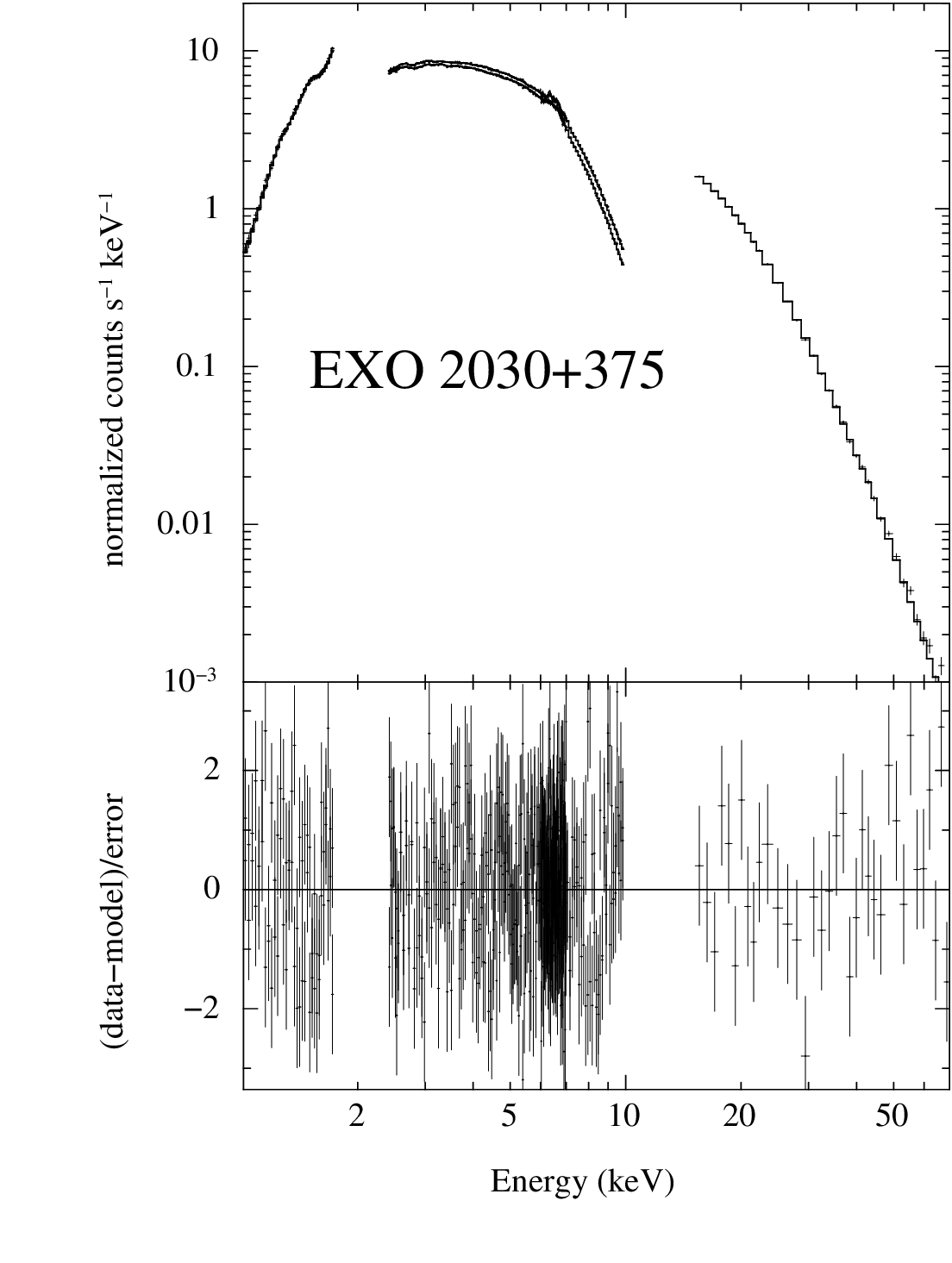}
\includegraphics[height=4.3cm,width=4.7cm,angle=0]{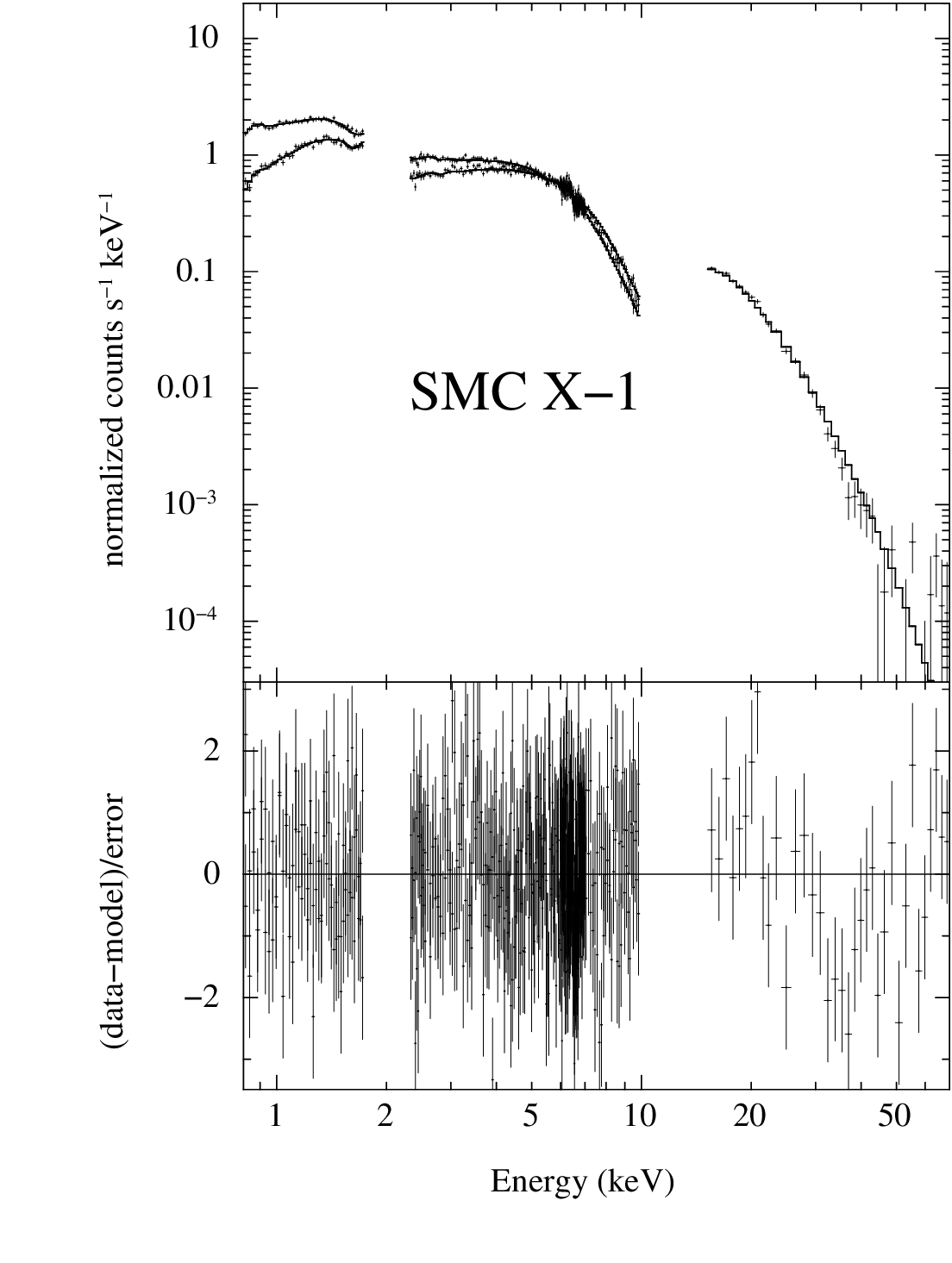}
\includegraphics[height=4.3cm,width=4.7cm,angle=0]{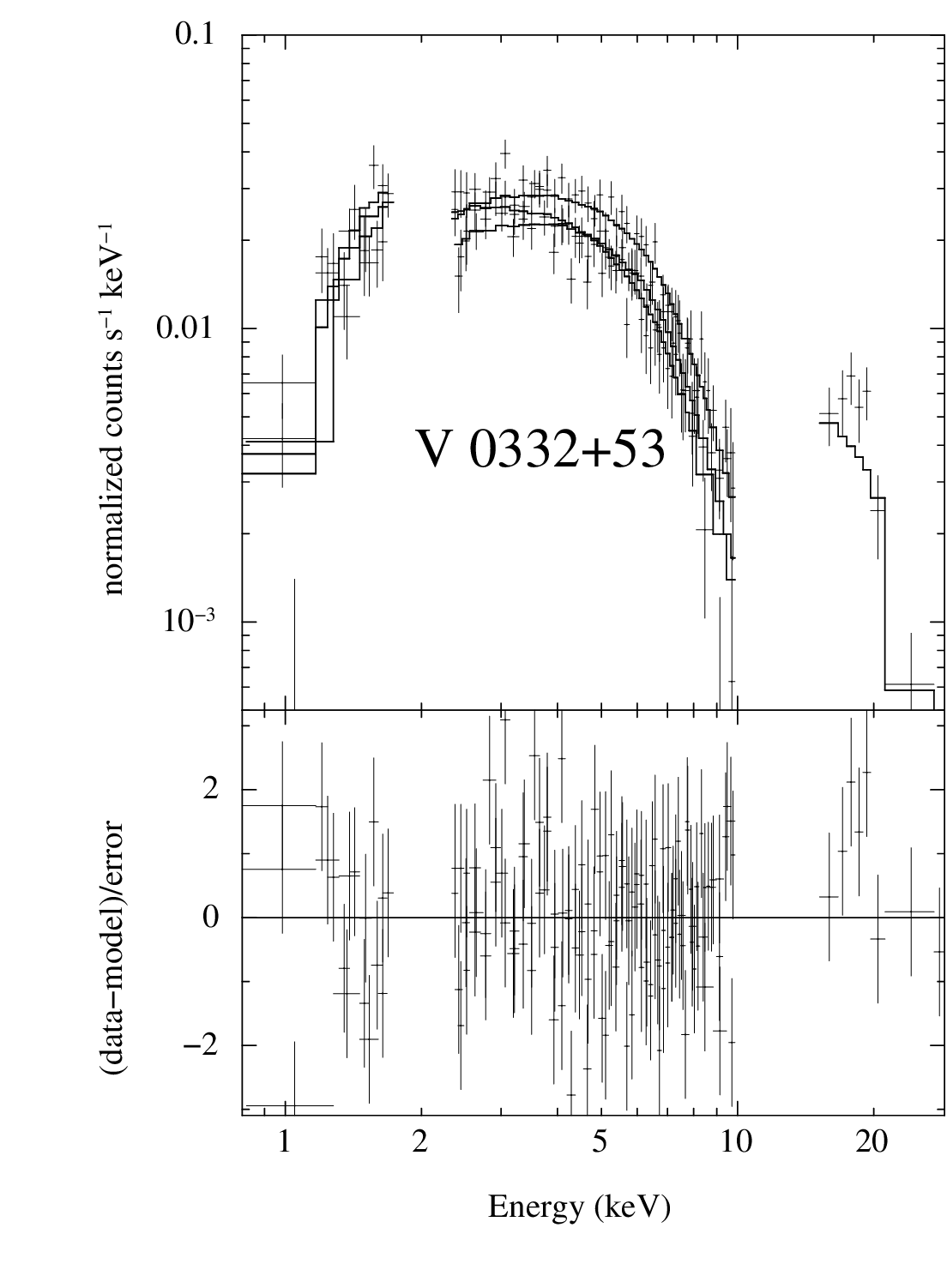}
\includegraphics[height=4.3cm,width=4.7cm,angle=0]{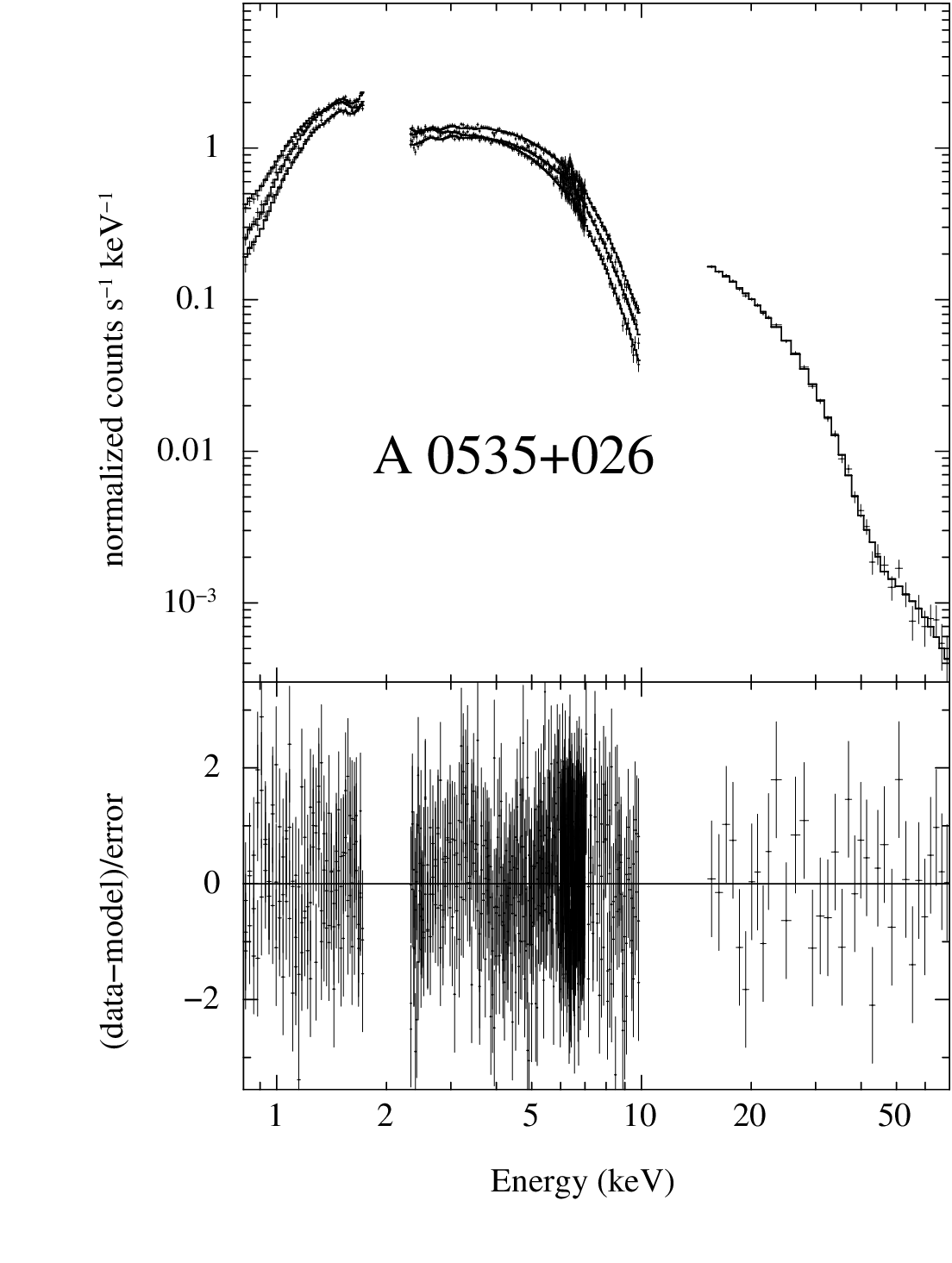}
\includegraphics[height=4.3cm,width=4.7cm,angle=0]{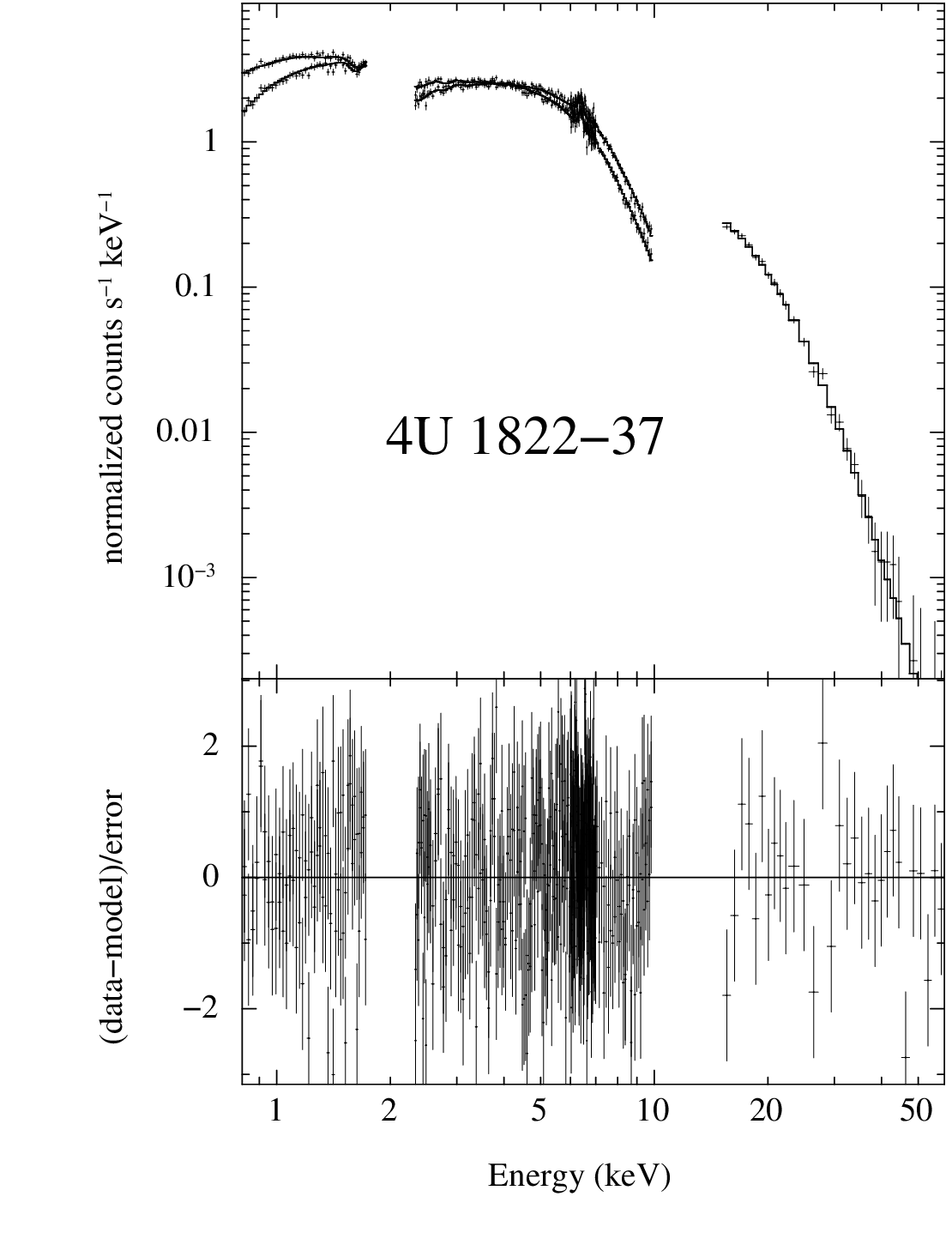}
\includegraphics[height=4.3cm,width=4.7cm,angle=0]{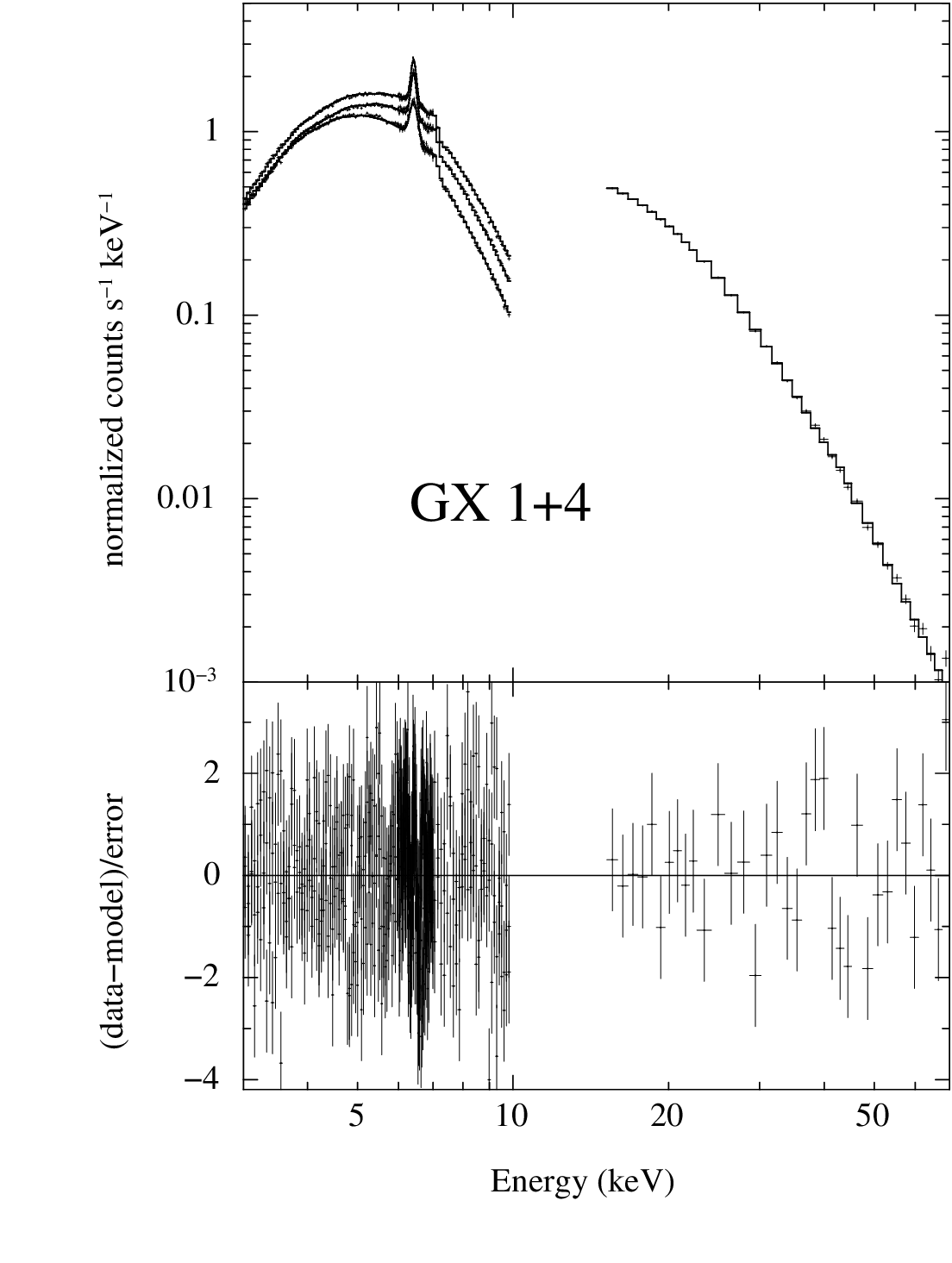}
\includegraphics[height=4.3cm,width=4.7cm,angle=0]{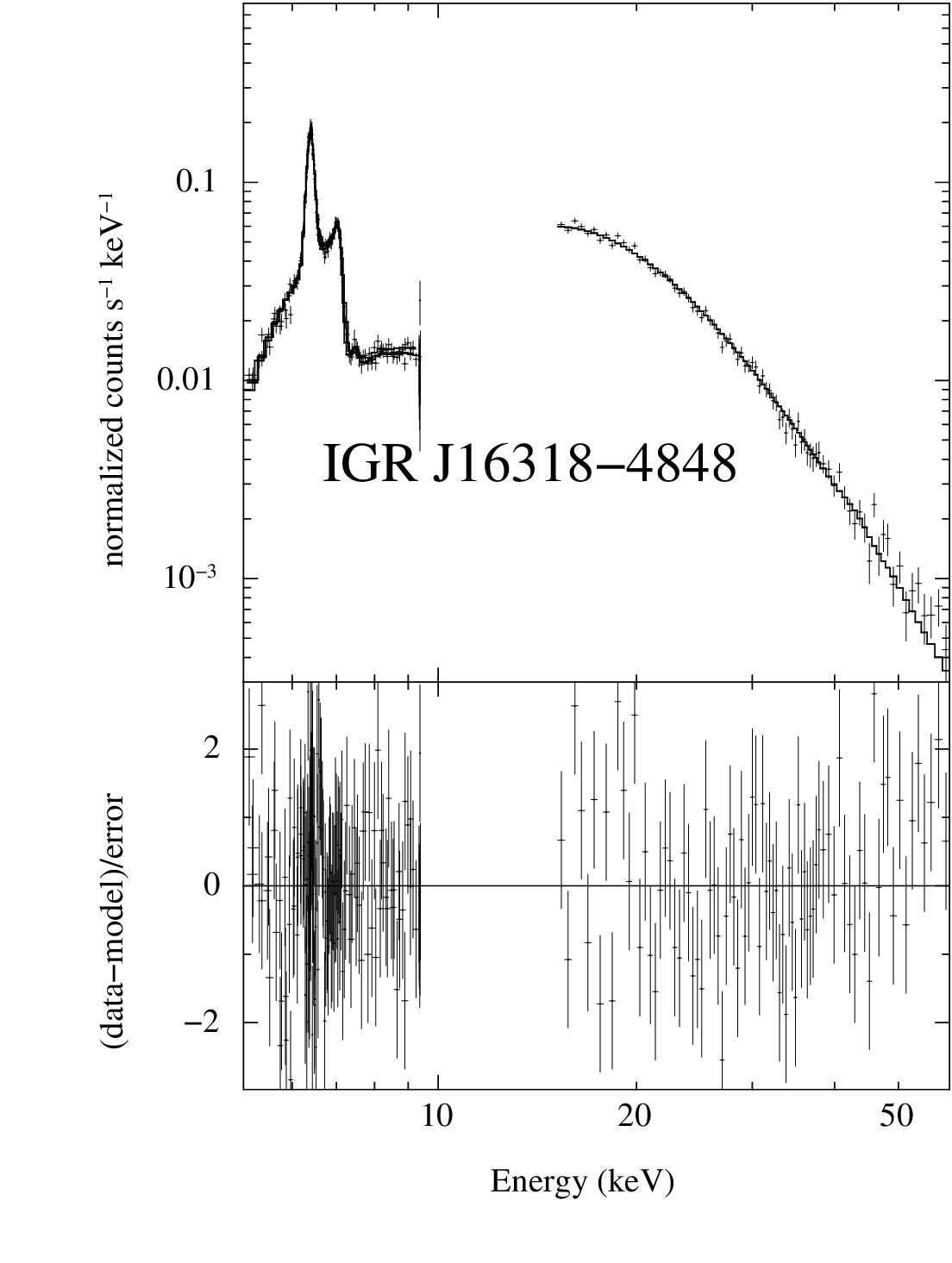}
\includegraphics[height=4.3cm,width=4.7cm,angle=0]{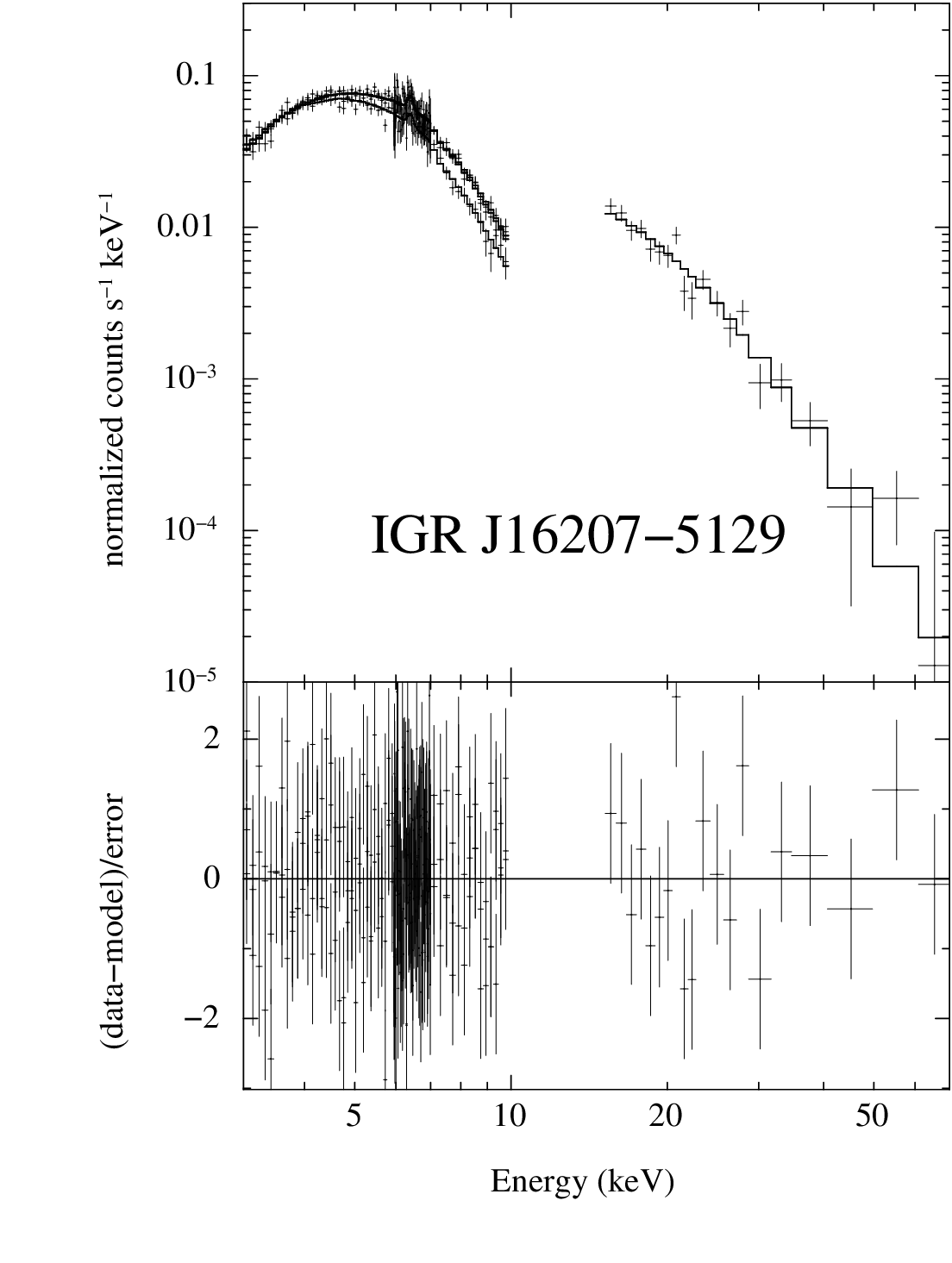}
\includegraphics[height=4.3cm,width=4.7cm,angle=0]{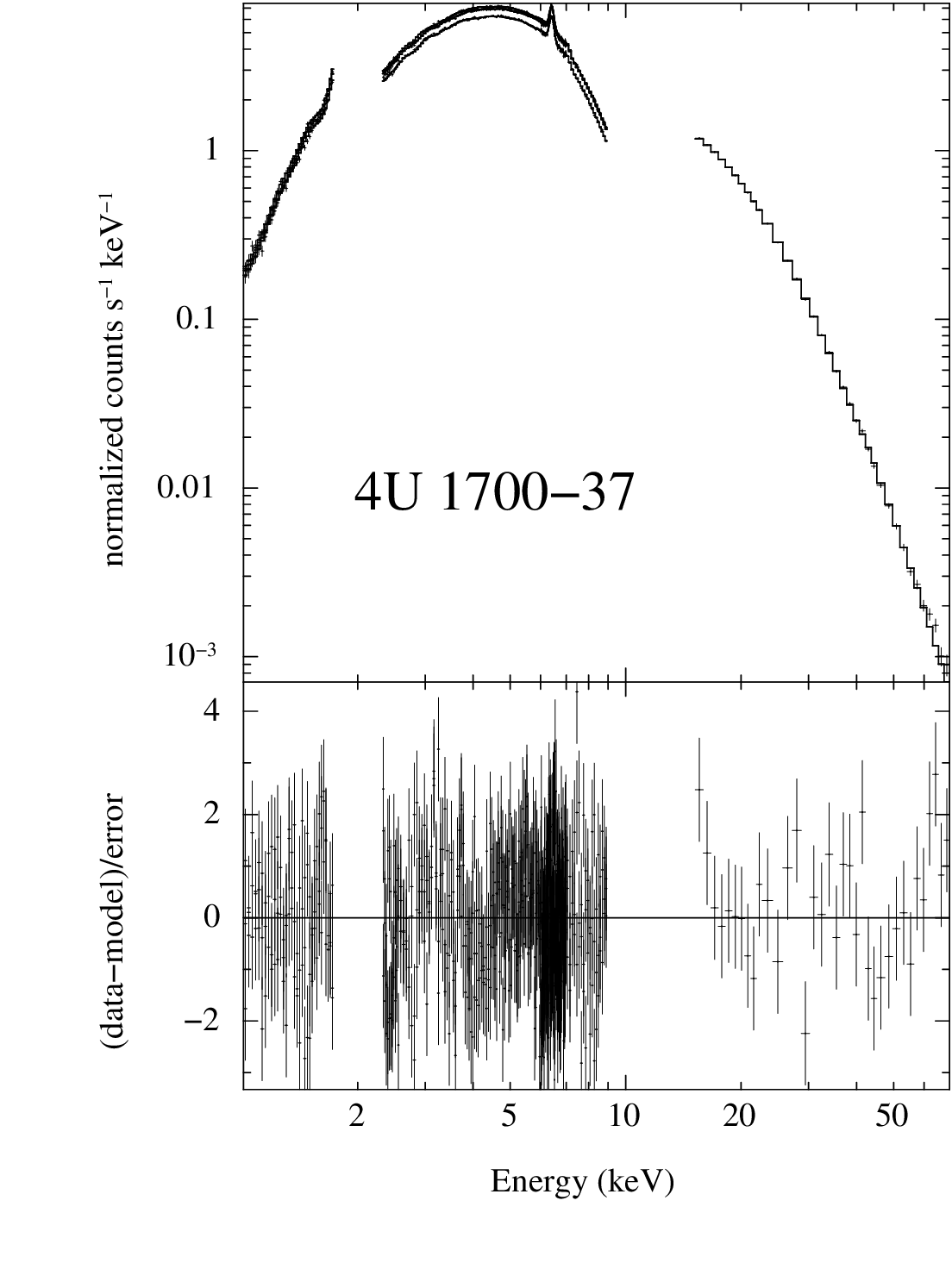}
\includegraphics[height=4.3cm,width=4.7cm,angle=0]{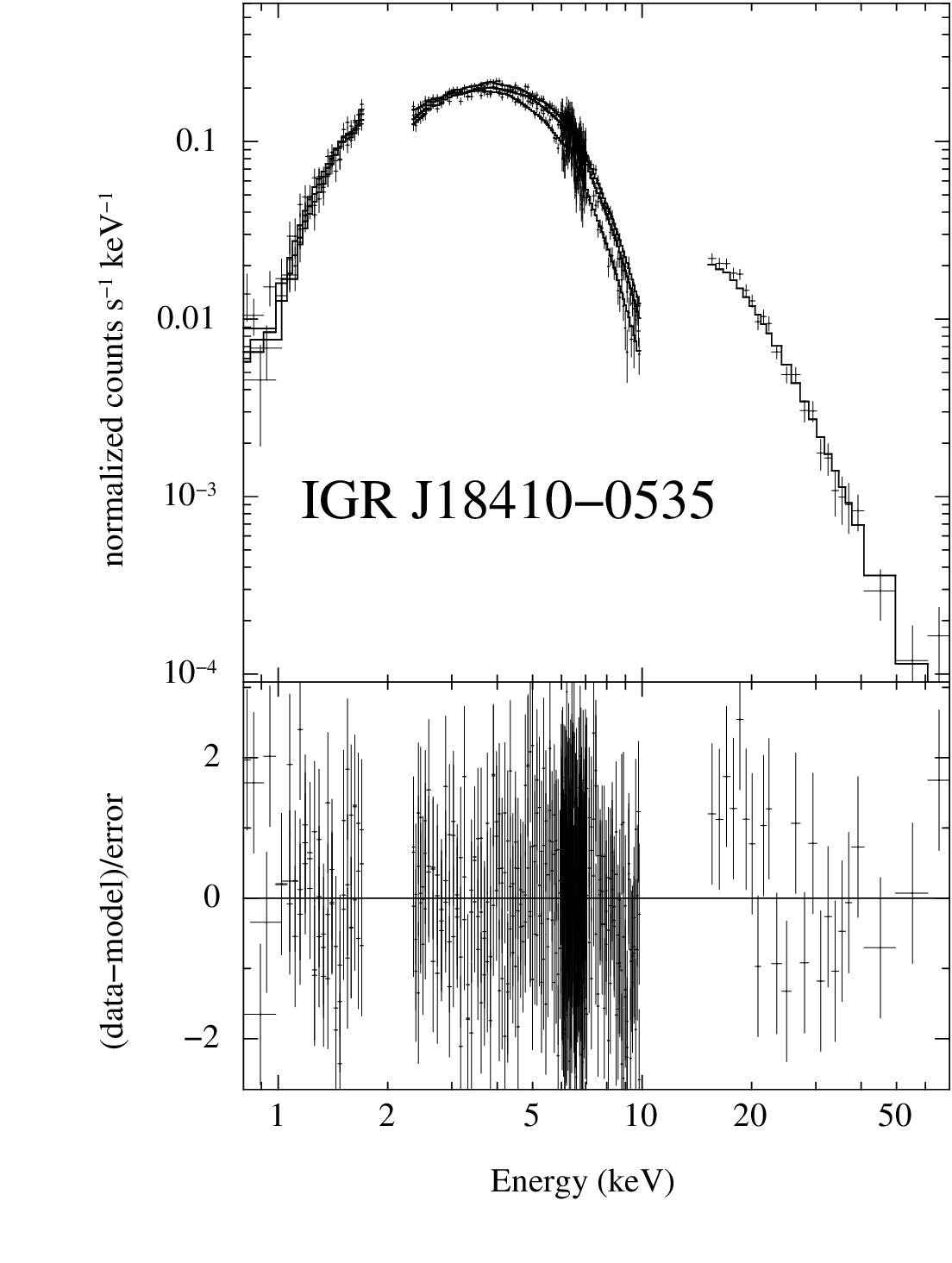}
\includegraphics[height=4.3cm,width=4.7cm,angle=0]{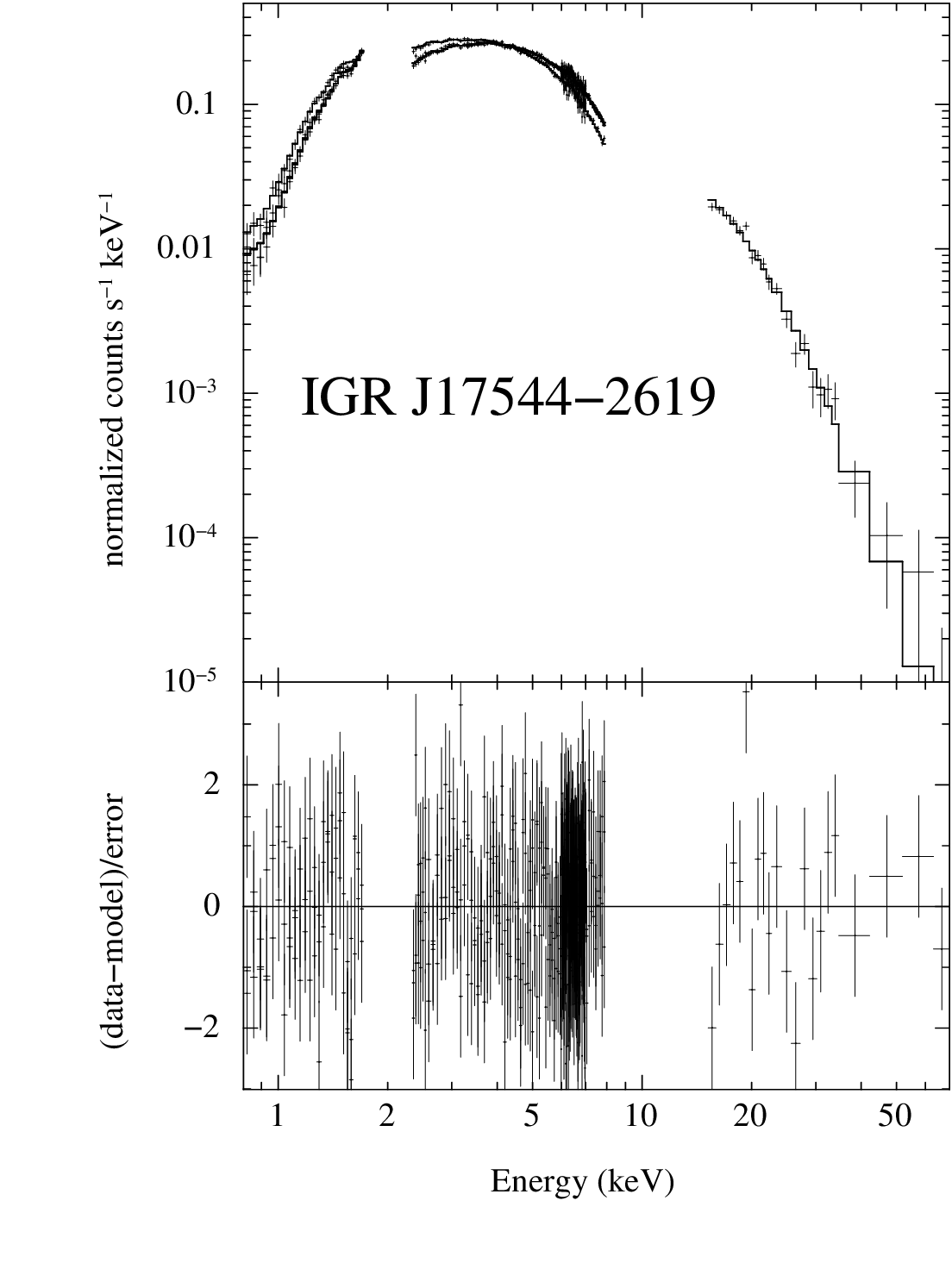}
\includegraphics[height=4.3cm,width=4.7cm,angle=0]{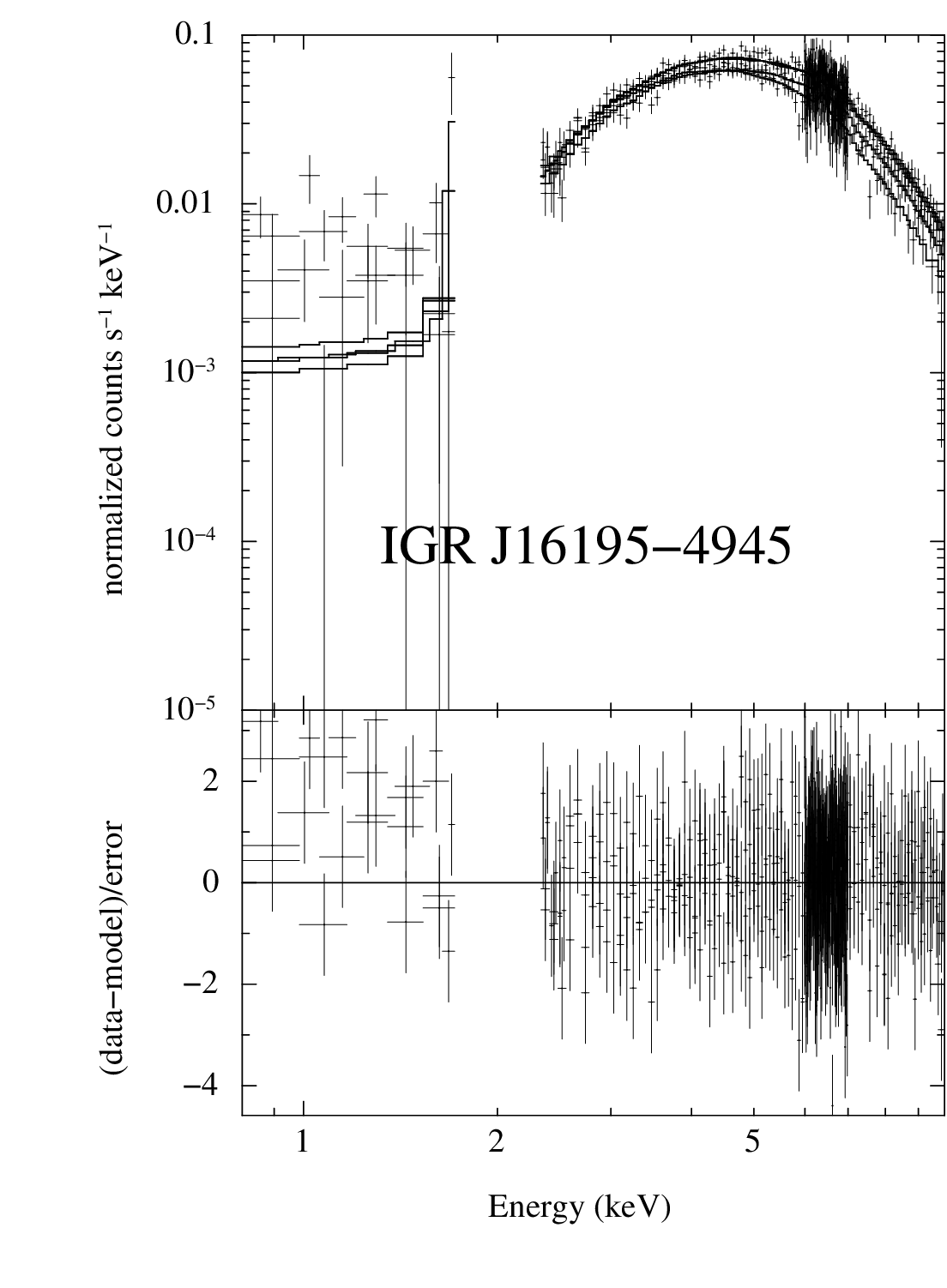} 
\includegraphics[height=4.3cm,width=4.7cm,angle=0]{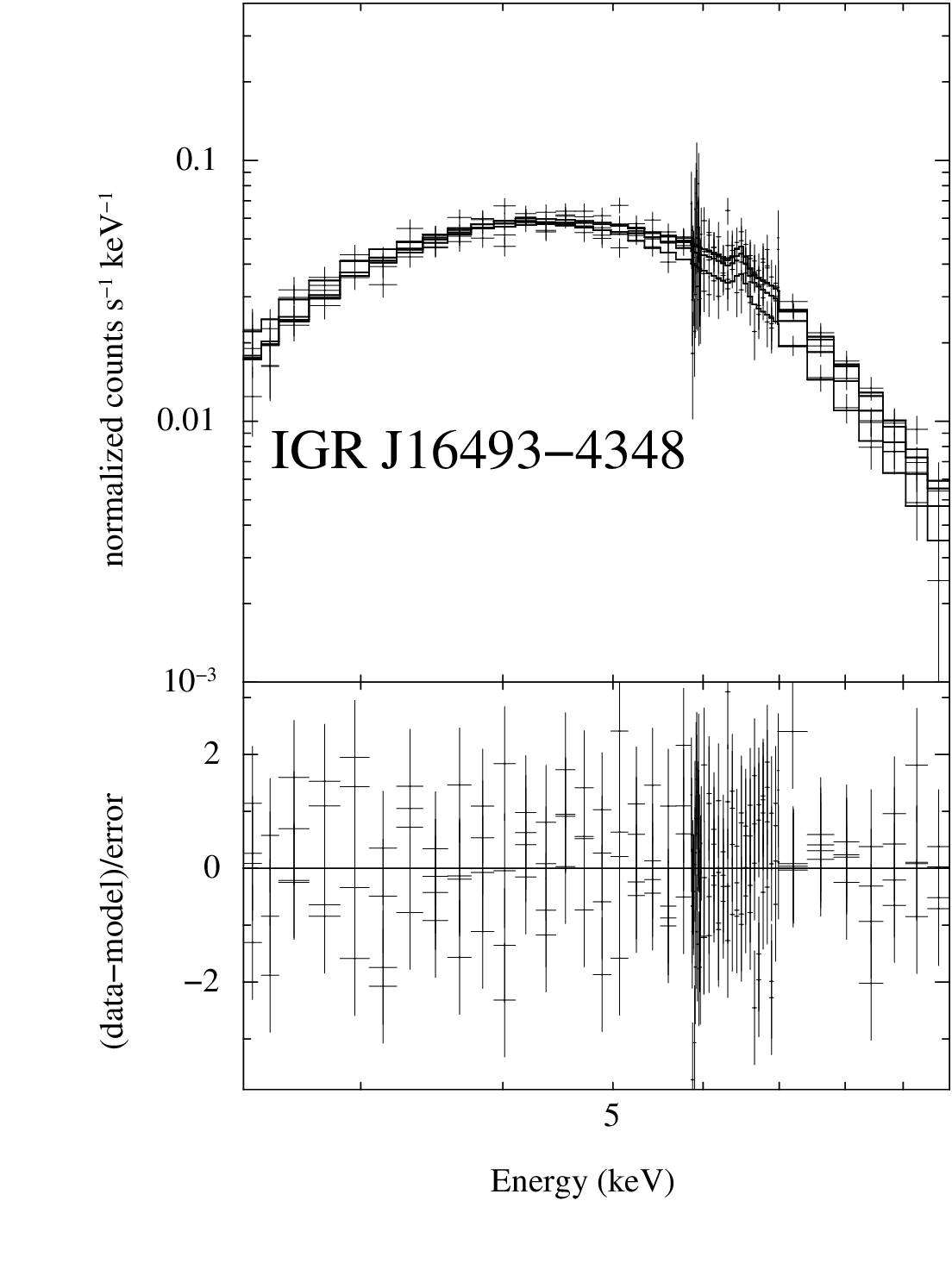}
\includegraphics[height=4.3cm,width=4.7cm,angle=0]{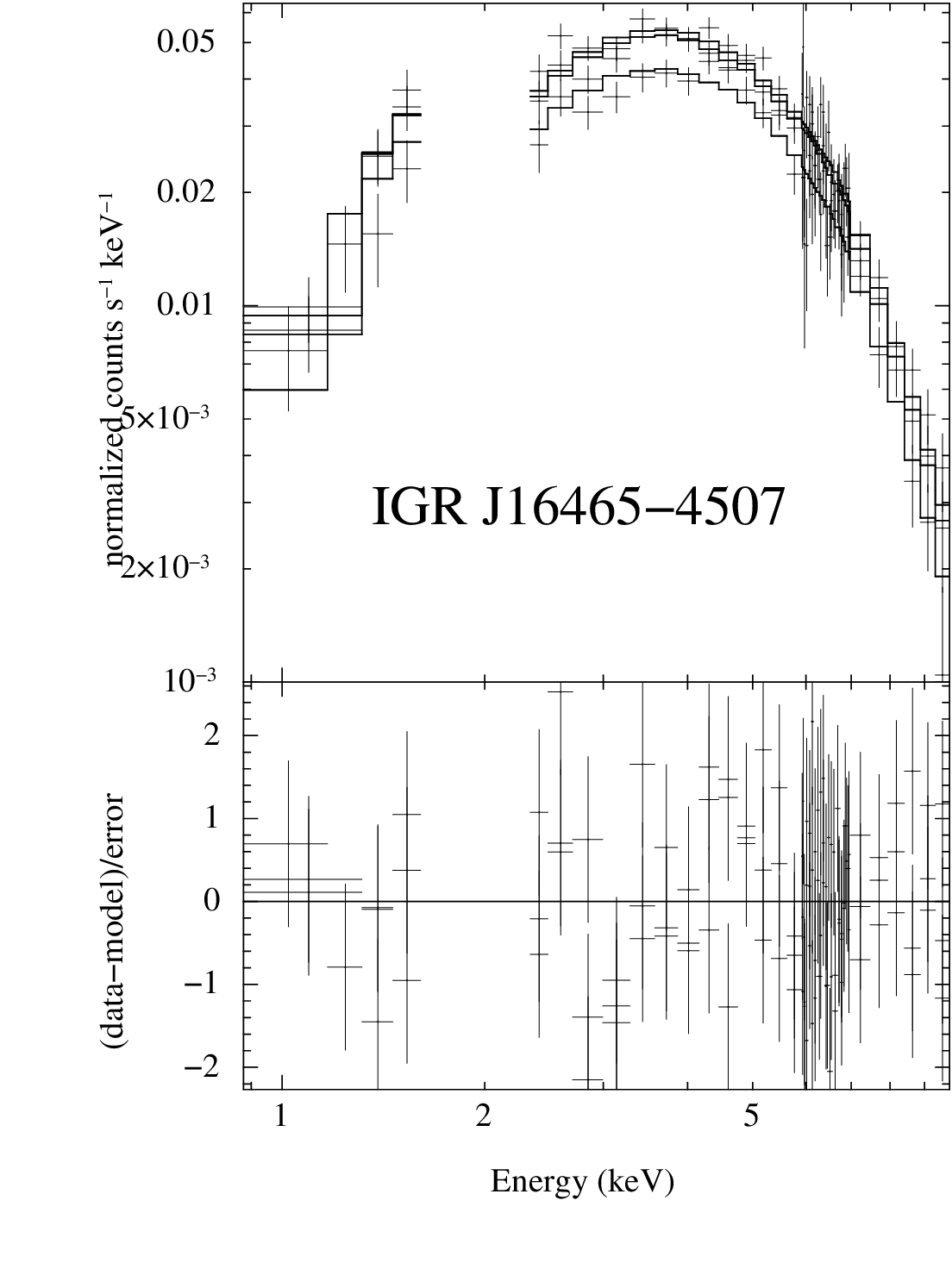}
\includegraphics[height=4.3cm,width=4.7cm,angle=0]{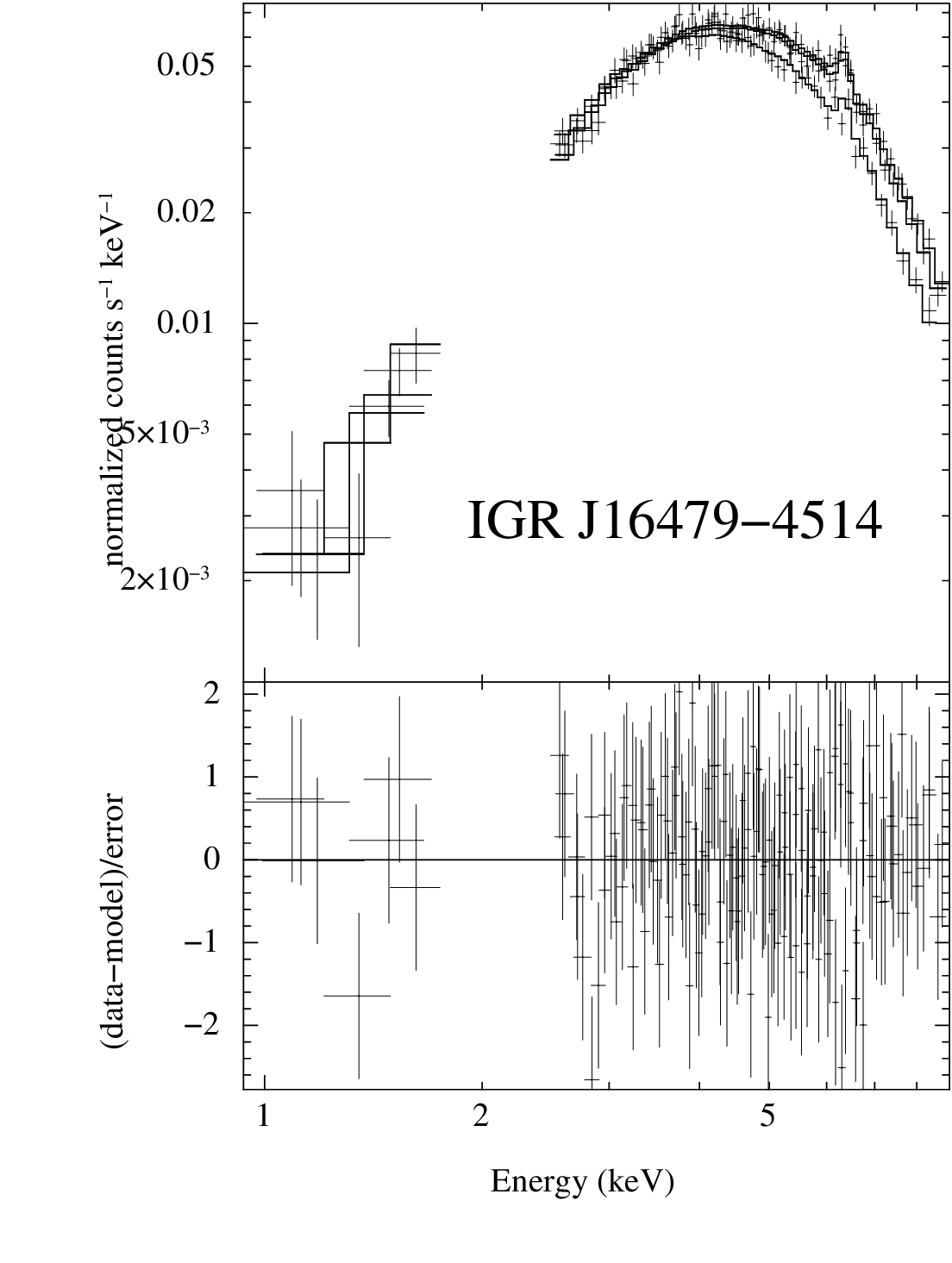}
\includegraphics[height=4.3cm,width=4.7cm,angle=0]{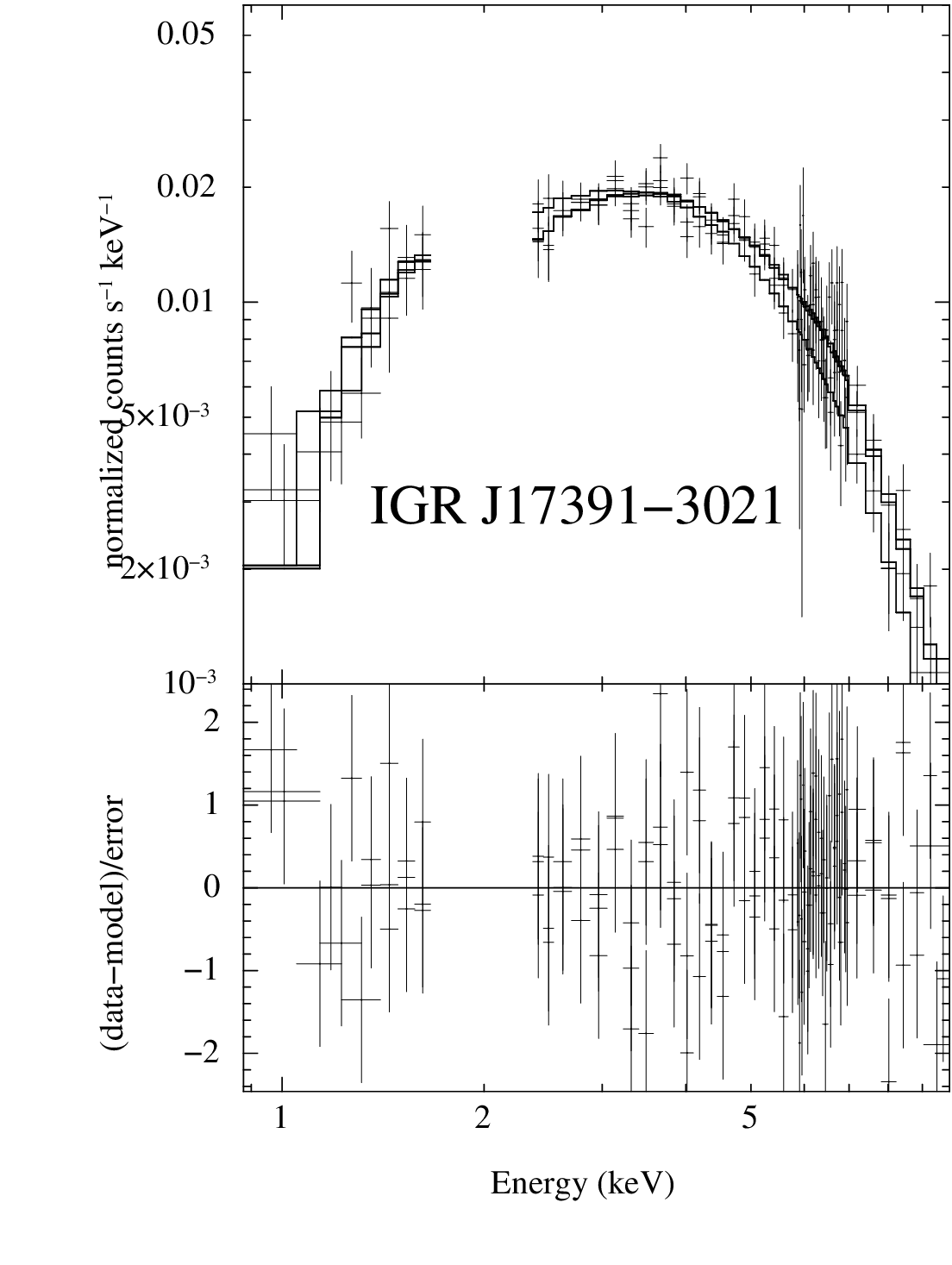}
\includegraphics[height=4.3cm,width=4.7cm,angle=0]{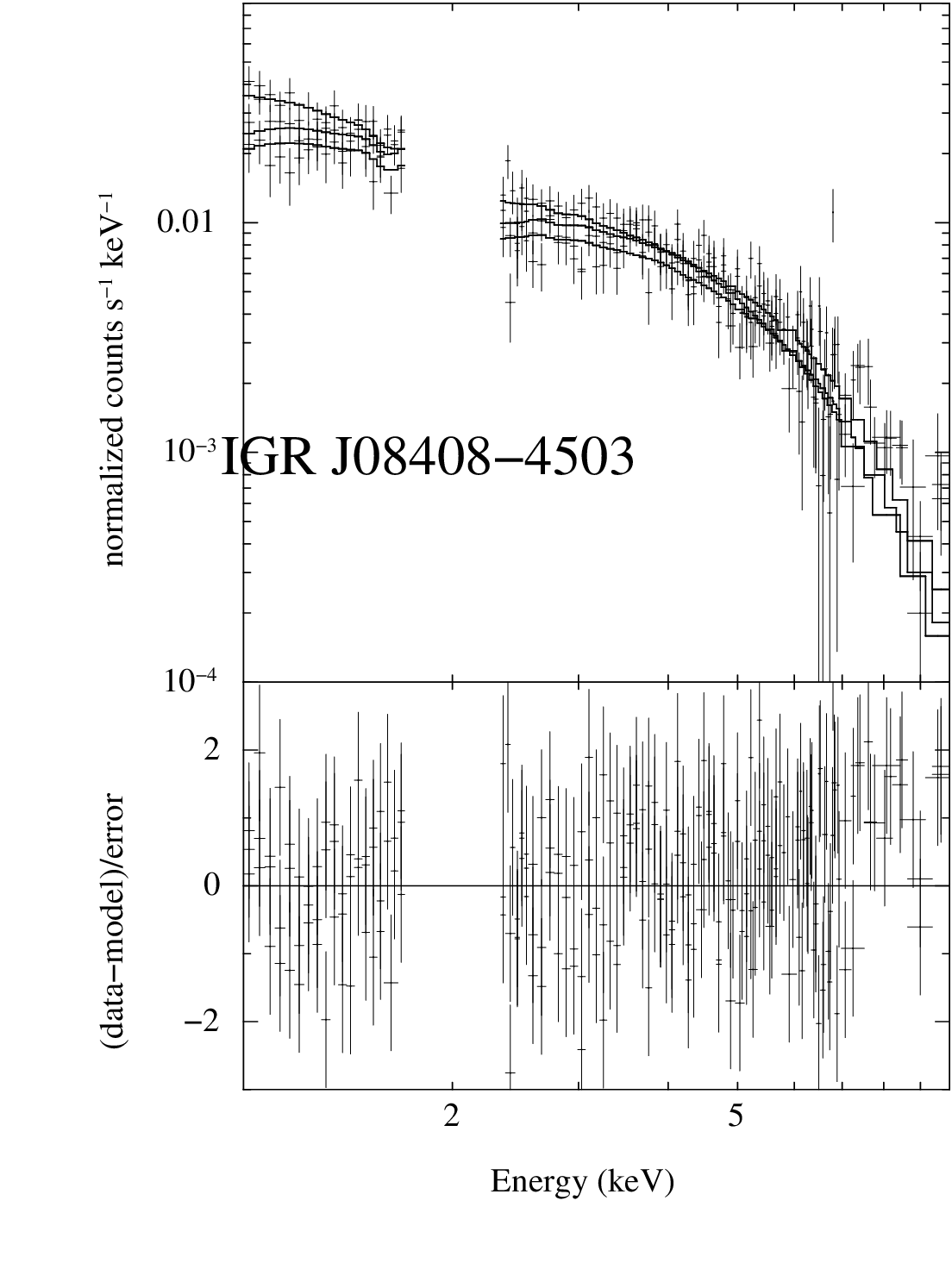}
\includegraphics[height=4.3cm,width=4.7cm,angle=0]{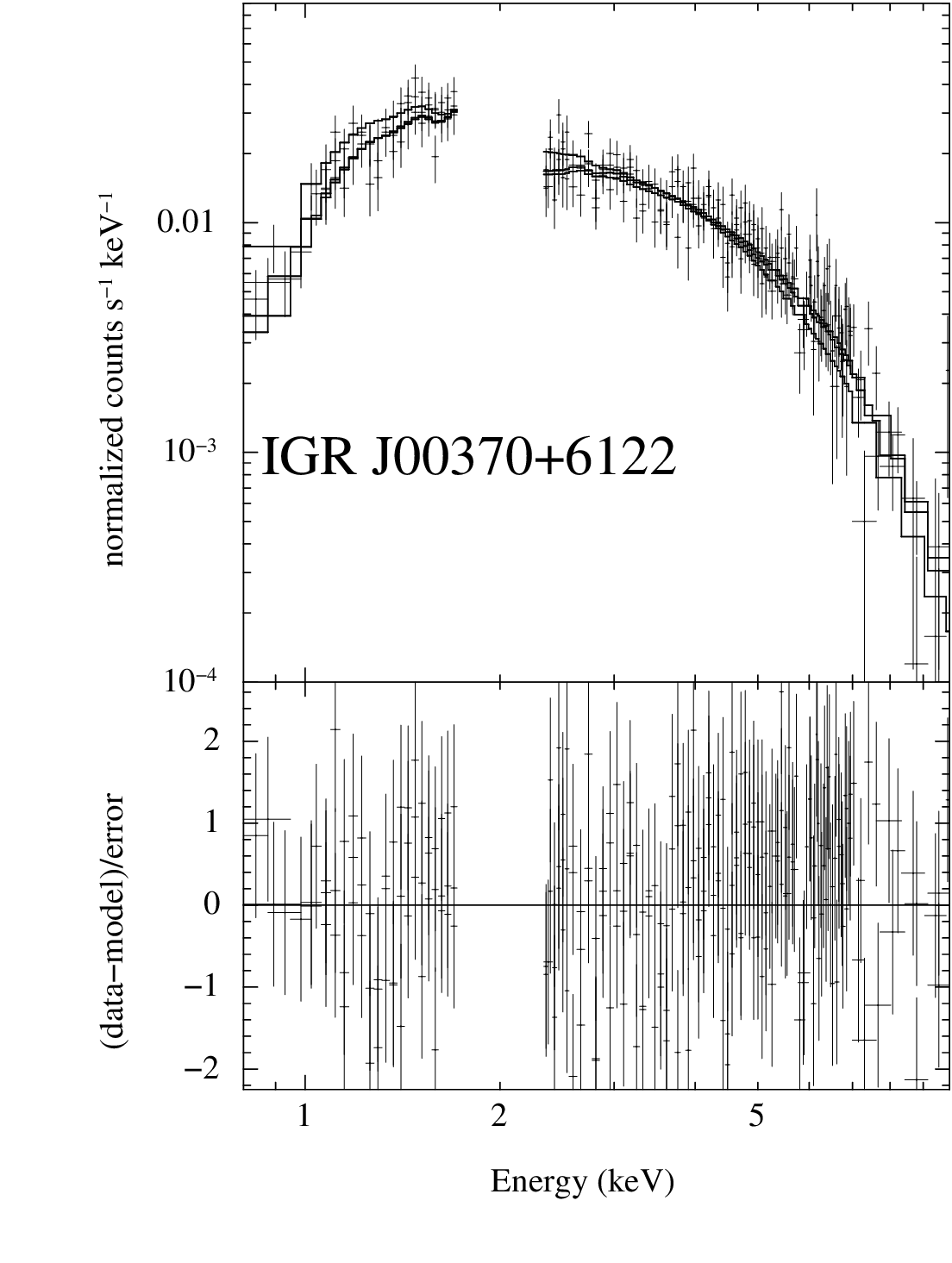}
\label{residuals2}
\end{figure*}
\end{landscape}

\begin{landscape}
\noindent
\pagestyle{plain}
\small\addtolength{\tabcolsep}{-1.0pt}

\begin{longtable}{cccccccccccccc}
\caption{Observation Log and distances to the different sources considered in the present paper.} \\
\toprule
Source & OBSID  & WINDOW MODE & DISTANCE (kpc) & Reference \\
Her X-1 & 100035010 & 1/8 & 6.6 $\pm$ 0.4 & \citealt{dist_herx1} & \\
4U 0115+63 & 406049010 & 1/4 & 7.0 $\pm$ 0.3 & \citealt{dist_more1} \\
Cen X-3 & 403046010 & 1/4 & 5.7 $\pm$ 1.5 & \citealt{dist_cenx3} \\
4U 1626-67 & 400015010 & 1/8 & 5-13 & \citealt{4u1626_dist_deepto} \\ 

%7.8 & \citealt{dist_4u1626} \\
XTEJ1946+274 & 405041010 & 1/4 & 9.5 $\pm$ 2.9 & \citealt{dist_xte1946} \\
Vela X-1 & 403045010 & 1/4 & 1.9 $\pm$ 0.2 & \citealt{dist_velax1} \\
4U 1907+09 & 401057010 & 1/4 & $2.8_{-1.8}^{+5.0}$ & \citealt{dist_many2}\\
4U 1538-522 &407068010& 1/4 & 6.4 $\pm$ 1 & \citealt{dist_4u1538} \\
GX 301-2 & 403044010 & 1/4 & 3.04$^{*}$ & \citealt{dist_gx301-21} \\
1 A1118-61 & 403049010 & 1/4 & 5.2 $\pm$ 0.9 & \citealt{dist_more1} \\
4U 0114+65 & 406017010 & 1/4 & 7.0 $\pm$ 3.6 & \citealt{dist_4u0114} \\
GX 304-1 & 905002010 & 1/4 & 2.4 $\pm$ 0.5 & \citealt{dist_gx304-1} \\
OAO 1657-415 & 406011010 & 1/4 & 6.4 $\pm$ 1.5 & \citealt{dist_oao} \\
Cep X-4 & 409037010 & 1/4 & 3.8 $\pm$ 0.6 & \citealt{dist_cepx4} \\
4U 1700-37 & 401058010 & 1/4 & 2.12 $\pm$ 0.34 & \citealt{dist_4u17001} \\
A 0535+026 & 404054010 & 1/4 & 2.00 $\pm$ 0.7 & \citealt{dist_a0535} \\
4U 1822-37 & 401051010 & 1/4 & $2.5_{-1.9}^{+3.0}$ & \citealt{dist_4u1822} \\
GX 1+4 & 405077010 & 1/4 & 3-15 & \citealt{dist_gx1+4} \\
GRO J1008-57 & 902003010 & 1/4 & 5$^{*}$ & \citealt{dist_groj} \\
4U1909+07 & 405073010 & 1/4 & 7.00 $\pm$ 3.0 & \citealt{dist_4u1909} \\
IGR J16393-4643 & 404056010 & Off & 10.6$^{*}$ & \citealt{dist_igrj16479-4514} \\
4U 2206+54 & 402069010 & 1/4 & 2.90 $\pm$ 0.20 & \citealt{dist_more1} \\
SW J2000.6+3210 & 401053020 & Off & 8$^{*}$ & \citealt{dist_swj200} \\
LMC X-4 & 702036020 & 1/8 & 50$^{*}$ & \citealt{dist_lmcx4} \\
KS1947+300 & 908001020 & 1/4 & 10.4 $\pm$ 0.9 & \citealt{dist_more1} \\
EXO 2030+375 & 402068010 & 1/4 & 7.1 $\pm$ 0.2 & \citealt{dist_exo} \\
SMC X-1 & 706030100 & Off & 60.0$^{*}$ & \citealt{dist_smcx1} \\
V 0332+53 & 904004010 & 1/4 & 7.5 $\pm$ 1.50 & \citealt{dist_v0332} \\
IGR J16493-4348 & 401054010 & Off & 2-3 & \citealt{dist_igrj16493}  \\
IGRJ16318-4848 & 401094010 & Off & 3.60 $\pm$ 2.60 & \citealt{dist_igrj16318-4848} \\
IGRJ16207-5129 & 402065020 & 1/4 & $6.10_{-3.50}^{+8.90}$ & \citealt{dist_many2}\\
IGR J18410-0535 & 505090010 & Off &  $3.2_{-1.5}^{+2.0}$ & \citealt{dist_many2}\\
IGR J17544-2619 & 402061010 & 1/4 & 3.20 $\pm$ 1.00 & \citealt{dist_igrj17544-2619} \\
IGR J16195-4945 & 401056010 & Off & 5$^{*}$ & \citealt{dist_igrj16195-4945} \\
IGR J16465-4507 & 401052010 & Off & $9.5_{-5.7}^{+14.1}$ & \citealt{dist_igrj16465-4507} \\
IGR J16479-4514 & 406078010 & Off & 7.50 $\pm$ 2.50 & \citealt{dist_igrj16479-4514} \\
IGR J17391-3021 & 402066010 & 1/4 & 2.7$^{*}$ & \citealt{dist_igrj17391-3021} \\
IGR J08408-4503 & 404070010 & Off & 3.0$^{*}$ & \citealt{dist_igrj08408-4503} \\
IGR J00370+6122 & 402064010 & 1/4 & 3.3$^{*}$ & \citealt{dist_igrj00370} \\
\midrule
\bottomrule
\label{obslog} 
$^*$ Distance error is not known so error is assumed as 1\,kpc. 
\end{longtable}

\scriptsize
\begin{longtable}{ccccccccccccccc}
\endfirsthead
\caption{Parameters of the continuum and absorption features. Errors quoted are for 90 per cent confidence range} \\
\toprule
Source & ${N_{\rm H1}^{_a}}$ & ${N_{\rm H2}^{_a}}$ & ${C_{V}}$ & $\wgamma$ & $\wgamma_{norm}^{b}$ & E$_{cut}$ & E$_{fold}$  &
E$_{C1}$  & D$_1$ & W$_1$ & bb (\rm{kT}) & $bb_{norm}^{c}$  & $\chi^{2}_{\rm red}$/dof & \\

& & & & & & (\rm{keV}) & (\rm{keV}) & (\rm{keV}) &  & (\rm{keV}) & & & \\

Her X-1 &0.02 &- &- & $0.81_{-0.01}^{+0.02}$&$0.084_{-0.002}^{+0.004}$ & $19.47_{-0.14}^{+0.16}$ & $12.34_{-0.34}^{+0.35}$ & $38.14_{-0.34}^{+0.33}$ & $0.98_{-0.07}^{+0.08}$ & 7 & $0.55_{-0.03}^{+0.02}$ &$0.00077_{-0.00001}^{+0.00001}$&1.55/661 &  \\
\\
4U 0115+63 & $1.13_{-0.02}^{+0.02}$ & - & - & $0.19_{-0.02}^{+0.03}$ & $0.025_{-0.002}^{+0.001}$ & $12.54_{-0.18}^{+0.16}$ & $7.53_{-0.25}^{+0.39}$ & 9.57 & 0.67 & 1.0 & $0.59_{0.01}^{0.01}$ $0.0012_{-0.0001}^{+0.0001}$ & 1.44/436 &  \\
\\
Cen X-3 & $1.43_{-0.02}^{+0.03}$ & - & - & $1.09_{-0.01}^{+0.01}$ & $0.392_{-0.079}^{+0.089}$ & $15.14_{-0.15}^{+0.17}$ & 
$8.07_{-0.31}^{+0.37}$ & $28.66_{-0.84}^{+0.82}$ & $0.28_{-0.09}^{+0.09}$ & $5.41_{-2.03}^{+2.42}$ & $0.107_{-0.006}^{+0.006}$ & $0.049_{-0.011}^{+0.015}$ & 1.34/658 \\
\\
4U 1626-67 & $0.09_{-0.02}^{+0.02}$ & - & - & $0.830_{-0.008}^{+0.008}$ & $0.0092_{-0.0001}^{+0.0001}$ & $18.87_{-0.44}^{+0.47}$ & $15.92_{-1.85}^{+2.51}$ & 
$36.31_{-0.64}^{+0.74}$ & $1.73_{-0.28}^{+0.35}$ & $4.56_{-1.25}^{+1.54}$ & $0.28_{-0.01}^{+0.01}$ & $0.00014_{-0.00002}^{+0.00002}$ & 1.29/910 \\
\\
XTE 1946+274 & $1.37_{-0.03}^{+0.03}$ & $7.00_{-1.01}^{+1.11}$ & $0.25_{-0.04}^{+0.03}$ & $1.09_{-0.04}^{+0.03}$ & $0.018_{-0.001}^{+0.001}$ & $6.89_{-0.52}^{+0.40}$ & 
$33.28_{-4.28}^{+3.87}$ & $36.88_{-1.18}^{+1.30}$ & $1.83_{-0.26}^{+0.27}$ & 7.5 & - & - & 1.12/816 \\
\\
Vela X-1 & $1.29_{-0.04}^{+0.03}$ & $3.02_{-0.21}^{+0.20}$ & $0.43_{-0.03}^{+0.02}$ & $1.16_{-0.02}^{+0.02}$ & $0.19_{-0.01}^{+0.01}$ & $5.53_{-0.23}^{+0.15}$ & $35.79_{-2.64}^{+3.15}$
& $27.90_{-0.66}^{+0.57}$ & $0.102_{-0.023}^{+0.025}$ & $5.37$ & - & - & 1.64/664 \\
\\
4U 1907+09 & $2.47_{-0.08}^{+0.06}$ & $2.81_{-0.78}^{+0.86}$ & $0.24_{-0.06}^{+0.08}$ & $1.29_{-0.02}^{+0.01}$ & $0.0240_{-0.0008}^{+0.0004}$ & $15.71_{-0.27}^{+2.22}$ & 
$14.96_{-2.15}^{+1.46}$ & $18.74_{-0.39}^{+0.45}$ & $0.49_{-0.19}^{+0.13}$ & $1.00_{-1.00}^{+0.99}$ & - & - & 1.07/882 \\
\\
4U 1538-522 & $1.47_{-0.01}^{+0.01}$ & - & - & $1.208_{-0.008}^{+0.007}$ & $0.0506_{-0.0006}^{+0.0005}$ & $15.94_{-0.79}^{+1.26}$ & 
$14.29_{-1.02}^{+0.86}$ & $21.46_{-0.45}^{+0.43}$ & $0.53_{-0.08}^{+0.11}$ & $2.09_{-0.74}^{+0.76}$ & - & - & 1.22/710 \\
\\
GX 301-2 & $14.98_{-0.73}^{+0.72}$ & $53.02_{-1.97}^{+1.91}$ & 0.95 & $1.055_{-0.058}^{+0.053}$ & $0.068_{-0.009}^{+0.010}$ & $21.54_{-1.25}^{+1.45}$ & 
$22.93_{-4.16}^{+5.55}$ & $44.20_{-3.40}^{+4.25}$ & $1.97_{-0.52}^{+0.82}$ & 19 & - & - & 1.11/529 \\
\\

1A 1118-61 & $0.74_{-0.01}^{+0.01}$ &  &  & $0.36_{-0.01}^{+0.01}$ & $0.069_{-0.001}^{+0.001}$ & $5.58_{-0.09}^{+0.10}$ & 
$19.92_{-0.57}^{+0.61}$ & $49.94_{-0.51}^{+0.56}$ & $1.59_{-0.06}^{+0.06}$ & $18.69_{-1.16}^{+1.22}$ & - & - & 1.61/468 \\
\\
4U 0114+65 & $4.79_{-0.22}^{+0.16}$ & $23.15_{-9.85}^{+15.61}$ & $0.12_{-0.04}^{+0.05}$ & $0.78_{-0.06}^{+0.06}$ & $0.011_{-0.001}^{+0.001}$ & $5.94_{-0.24}^{+0.22}$ & 
$19.17_{-1.62}^{+1.72}$ & $24.53_{-2.71}^{+2.22}$ & $0.14_{-0.06}^{+0.05}$ & 5 & $0.138_{-0.006}^{+0.007}$ & $0.0180_{-0.0070}^{+0.0093}$ & 1.17/453 \\
\\
GX 304-1 & $1.19_{-0.01}^{+0.02}$ & $12.19_{-0.78}^{+0.88}$ & $0.28_{-0.02}^{+0.02}$ & $1.02_{-0.03}^{+0.02}$ & $0.816_{-0.04}^{+0.04}$ & $6.68_{-0.20}^{+0.17}$ & 
$30.55_{-2.04}^{+2.37}$ & $56.54_{-1.54}^{+2.28}$ & $2.03_{-0.18}^{+0.29}$ & $19.06_{-2.18}^{+2.78}$  & - & - & 1.39/673 \\
\\
OAO 1657-415 & $16.67_{-0.58}^{+0.54}$ & $43.06_{-4.56}^{+5.00}$ & $0.45_{-0.03}^{+0.03}$ & $0.49_{-0.04}^{+0.04}$ & $0.011_{-0.001}^{+0.001}$ & $6.14_{-0.16}^{+0.14}$ & 
$16.18_{-0.62}^{+0.67}$ & - & - & -  & - & - & 1.31/420 \\
\\
Cep X-4 & $0.79_{-0.02}^{+0.02}$ & $90.15_{-30.79}^{+59.39}$ & $0.29_{-0.07}^{+0.43}$ & $1.04_{-0.03}^{+0.03}$ & $0.11_{-0.01}^{+0.08}$ & $16.36_{-0.58}^{+0.37}$ & 
$9.82_{-0.58}^{+0.66}$ & $29.78_{-0.30}^{+0.30}$ & $1.03_{-0.09}^{+0.10}$ & $5.70_{-0.90}^{+1.04}$  & $1.04_{-0.06}^{+0.08}$ & $0.0011_{-0.0003}^{+0.0003}$ & 1.28/234 \\
\\
4U 1700-37 & $2.03_{-0.04}^{+0.04}$ & $4.43_{-0.10}^{+0.11}$ & $0.70_{-0.01}^{+0.01}$ & $0.94_{-0.01}^{+0.01}$ & $0.184_{-0.004}^{+0.005}$ & $6.48_{-0.07}^{+0.05}$ & 
$19.93_{-0.17}^{+0.39}$ & $37.38_{-1.59}^{+1.77}$ & $0.09_{-0.02}^{+0.01}$ & 8  & - & - & 1.45/654 \\
\\
A0535+026$^n$ & $0.74_{-0.04}^{+0.07}$ & $5.50_{-0.64}^{+0.86}$  & $0.35_{-0.03}^{+0.04}$  & $1.14_{-0.05}^{+0.09}$ & $0.048_{-0.003}^{+0.005}$ & $13.15_{-1.42}^{+4.06}$ &- & 
$41.91_{-0.75}^{+1.02}$  & $10.14_{-1.99}^{+3.83}$  & $1.43_{-0.20}^{+0.37}$ & $0.089_{-0.015}^{+0.013}$ & $0.0029_{-0.0012}^{+0.0044}$ & 1.09/706 \\
\\
4U 1822-37 & 0.1 & - & - & $0.058_{-0.037}^{+0.040}$ & $0.0131_{-0.0004}^{+0.0005}$ & $3.10_{-0.11}^{+0.14}$ & $7.16_{-0.31}^{+0.39}$ & $35.41_{-3.06}^{+8.55}$ & 
$0.39_{-0.21}^{+0.66}$ & $4.11_{-4.10}^{+9.27}$ & $0.189_{-0.008}^{+0.007}$ & $0.0003_{-1.06427e-05}^{+1.08845e-05}$ & 1.24/477 \\
\\
\emph{GX 1+4}$^{nhe}$ &  $14.52_{-0.59}^{+0.43}$ & $25.06_{-6.26}^{+7.21}$ & $0.24_{-0.04}^{+0.06}$ & $1.265_{-0.018}^{+0.021}$ & $0.110_{-0.005}^{+0.007}$ & $23.94_{-1.01}^{+1.23}$
& $41.63_{-2.71}^{+1.04}$ & - & - & - & - & - & 1.65/519 \\
\\
%\caption{Parameters of the continuum and absorption features..contd} \\
GRO J1008-57$^{**}$ & $1.35_{-0.08}^{+0.09}$ & $6.15_{-0.70}^{+0.94}$ & $0.34_{-0.03}^{+0.03}$ & $1.10_{-0.02}^{+0.02}$ & $0.129_{-0.005}^{+0.006}$ & $6.39_{-0.14}^{+0.16}$ & 
$22.79_{-0.32}^{+0.79}$ & - & -  & - & $0.127_{-0.03}^{+0.02}$ & $0.0014_{-0.0010}^{+0.0010}$ & 1.44/680 \\
\\
4U 1909+07 & $6.45_{-0.38}^{+0.45}$ & - & - & $1.06_{-0.07}^{+0.08}$ & $0.0196_{-0.002}^{+0.003}$ & $5.68_{-0.27}^{+0.44}$ & 
$18.87_{-1.61}^{+2.25}$ & - & -  & - & - & - & 1.27/288 \\
\\
IGR J16393-4643 & $26.28_{-0.74}^{+0.75}$ & - & - & $0.906_{-0.046}^{+0.044}$ & $0.0036_{-0.0004}^{+0.0004}$ & $19.72_{-1.14}^{+1.19}$ & 
$10.37_{-1.58}^{+1.62}$ & - & -  & - & - & - & 1.22/282 \\
\\
4U 2206+54 & $0.40_{-0.04}^{+0.04}$ & $1.41_{-0.18}^{+0.19}$ & $0.59_{-0.03}^{+0.04}$ & $1.34_{-0.07}^{+0.07}$ & $0.041_{-0.004}^{+0.004}$ & $3.43_{-0.18}^{+0.26}$ & 
$25.85_{-3.29}^{+4.32}$ & - & -  & - &  $1.98_{-0.04}^{+0.05}$ & $0.0024_{-0.0001}^{+0.0001}$ & 1.39/665 \\
\\
SW J2000.6+3210 & $1.41_{-0.14}^{+0.17}$ & - & - & $0.92_{-0.24}^{+0.28}$ & $0.0010_{-0.0010}^{+0.0013}$ & $5.18_{-0.50}^{+0.38}$ & 
$15.78_{-5.08}^{+7.82}$ & - & -  & - &  $1.34_{-0.22}^{+0.29}$ & $0.00022_{-0.00009}^{+0.00007}$ & 1.16/780 \\ 
\\
LMC X-4 & 0.1 & $52.49_{-7.76}^{+8.78}$ & $0.31_{-0.07}^{+0.06}$ & $0.78_{-0.08}^{+0.08}$ & $0.016_{-0.003}^{+0.003}$ & $25.83_{-2.78}^{f+1.13}$ & 
$8.79_{-0.78}^{+0.83}$ & - & -  & - &  $0.27_{-0.09}^{+0.07}$ & $0.00015_{-0.00003}^{+0.00021}$ & 1.23/382 \\
\\
KS 1947+300 & $0.367_{-0.003}^{+0.003}$ & $7.39_{-0.67}^{+0.10}$ & $0.96_{-0.21}^{+0.96}$ & $0.743_{-0.001}^{+0.024}$ & $0.151_{-0.004}^{+0.009}$ & $4.92_{-0.06}^{+0.06}$ & 
$20.30_{-0.35}^{+0.08}$ & - & -  & - &  $0.695_{-0.012}^{+0.001}$ & $0.017_{-0.00004}^{+0.00062}$ & 1.19/461 \\ 
\\
EXO 2030+375 & $1.99_{-0.01}^{+0.01}$ & $65.08_{-3.89}^{+4.35}$ & $0.22_{-0.01}^{+0.01}$ & $1.37_{-0.02}^{+0.02}$ & $0.641_{-0.023}^{+0.024}$ & $7.28_{-0.13}^{+0.13}$ & 
$21.69_{-0.37}^{+0.38}$ & - & -  & - &  - & - & 1.35/452 \\ 
\\
SMC X-1 & 0.05 & - & - & $0.50_{-0.03}^{+0.02}$ & $0.0102_{-0.0004}^{+0.0003}$ & $4.80_{-0.23}^{+0.16}$ & 
$10.31_{-0.27}^{+0.25}$ & - & -  & - &  $0.178_{-0.041}^{+0.028}$ & $0.00012_{-0.0004}^{+0.0010}$ & 1.17/477 \\ 
\\
V 0332+53 &  $1.46_{-0.15}^{+0.16}$ & - & - & $1.06_{-0.08}^{+0.09}$ & $0.0006_{-0.0001}^{+0.0001}$ & $22.12_{-1.32}^{+1.53}$ & 
0.06 & - & -  & - &  - & - & 1.13/193 \\ 
\\
IGR J16493-4348 & $7.64_{-0.52}^{+0.54}$ & - &- &  $1.43_{-0.08}^{+0.09}$  & $0.0034_{-0.0005}^{+0.0006}$ & -  & - & 
- & - & - & - & - & 1.26/259 \\
\\
IGRJ16318-4848 & $122.43_{-4.37}^{+3.88}$ & - &- &  $1.03_{-0.09}^{+0.08}$  & $0.009_{-0.002}^{+0.003}$ & $15.28_{-1.69}^{+1.47}$  & $32.47_{-3.59}^{+4.09}$ & 
- & - & - & - & - & 1.26/223 \\
\\
IGRJ16207-5129 & $12.38_{-0.62}^{+1.01}$ & - &- &  $1.34_{-0.37}^{+0.12}$  & $0.004_{-0.002}^{+0.001}$ & $7.16_{-2.01}^{+5.84}$  & $35.63_{-12.21}^{+18.95}$ & 
- & - & - & - & - & 0.99/297 \\
\\
IGR J18410-0535 & $1.74_{-0.15}^{+0.15}$ & $4.69_{-0.55}^{+0.62}$ & $0.68_{-0.05}^{+0.04}$ &  $1.57_{-0.08}^{+0.08}$  & $0.0128_{-0.0007}^{+0.0020}$ 
& $5.06_{-0.52}^{+0.72}$  & $56.39_{-14.64}^{+31.77}$ & - & - & - & - & - & 1.07/554 \\
\\
IGR J17544-2619 & $1.39_{-0.08}^{+0.07}$ & $2.43_{-0.26}^{+0.28}$ & $0.60_{-0.04}^{+0.05}$ &  $1.29_{-0.02}^{+0.02}$  & $0.0081_{-0.0004}^{+0.0004}$ 
& $8.83_{-1.09}^{+1.22}$  & $13.61_{-1.54}^{+1.62}$ & - & - & - & - & - & 1.07/496 \\
\\
IGR J16195-4945 & $8.50_{-0.38}^{+0.39}$ & - &- &  $1.24_{-0.06}^{+0.06}$  & $0.0031_{-0.0003}^{+0.0003}$ & -  & - & 
- & - & - & - & - & 1.20/697 \\
\\
IGR J16465-4507 & $1.81_{-0.30}^{+0.32}$ & $5.83_{-1.06}^{+1.26}$ & $0.78_{-0.07}^{+0.06}$ &  $2.20_{-0.14}^{+0.15}$  & $0.0064_{-0.0015}^{+0.0022}$ & -  & - & 
- & - & - & - & - & 1.03/695 \\
\\
IGR J16479-4514 & $2.01_{-0.33}^{+1.72}$ & $5.77_{-0.86}^{+0.40}$ & $0.94_{-0.18}^{+0.04}$ &  $1.44_{-0.05}^{+0.08}$  & $0.0032_{-0.0003}^{+0.0005}$ & -  & - & 
- & - & - & - & - & 0.85/150 \\
\\
IGR J17391-3021 & $1.39_{-0.61}^{+0.52}$ & $4.26_{-1.16}^{+1.43}$ & $0.82_{-0.13}^{+0.09}$ &  $2.05_{-0.19}^{+0.19}$  & $0.0019_{-0.0006}^{+0.0008}$ & -  & - & 
- & - & - & - & - & 0.89/211 \\
\\
IGR J08408-4503 & 0.1 & $4.02_{-0.81}^{+0.88}$ & $0.80_{-0.06}^{+0.05}$ &  $2.48_{-0.18}^{+0.18}$  & $0.0012_{-0.0003}^{+0.0004}$ & -  & - & 
- & - & - & - & - & 1.22/504 \\
\\
IGR J00370+6122 & $0.93_{-0.21}^{+0.19}$ & $3.43_{-1.11}^{+1.61}$ & $0.75_{-0.07}^{+0.05}$ &  $2.79_{-0.21}^{+0.28}$  & $0.0029_{-0.0009}^{+0.0018}$ & -  & - & 
- & - & - & - & - & 0.96/521 \\
\midrule
\bottomrule
\label{continuum}
\end{longtable} 
\noindent
$^{_a}$In units of $10^{22}$ atoms $cm^{-2}$; \\
$^{_b}$ In units of photons $keV^{-1}$ $cm^{-2}$ $s^{-1}$ at 1 \rm {keV}; \\
$^{_c}$ In units of ${L_{39}}$/${D_{10}}$, where ${D_{10}}$ is the distance to the source in units of 10 kpc. \\
$^{*}$ The cyclotron line features are at 10 \rm{keV}, an energy range that is not covered by \emph{Suzaku} data. \\
$^{**}$ The cyclotron line feature is at 86 \rm{keV}, i.e., beyond the energy range covered by \emph{Suzaku} data in this paper. \\
$^f$ Here FDCUT is used, $^n$ NPEX is used, $^nhe$ NEWHCUT is used with width = 20 keV \\
%\end{landscape}

%\begin{landscape}
\small
\begin{longtable}{ccccccccccc}
\caption{Cyclotron line harmonics detected in a few sources among those used in the present paper.} \\
\toprule
Source & E$_{C1}$  & D$_1$ & W$_1$ & E$_{C2}$ & D$_2$ & W$_2$ 

%& E$_{C3}$ & D$_3$ & W$_3$ 
\\
 & (\rm{keV}) &  & (\rm{keV}) & (\rm{keV}) &  & (\rm{keV}) 
 %& (\rm{keV}) & (\rm{keV}) & (\rm{keV}) 
 \\
 \midrule
 4U 0115+63 & $17.89_{-0.13}^{+0.12}$ & 0.68 $\pm$ 0.01 & $6.03_{-0.28}^{+0.27}$ & $30.48_{-1.21}^{+1.27}$ & 0.29 $\pm$ 0.04 & 7  \\
\\
 Vela X-1 & $51.42_{-0.63}^{+0.71}$ & $2.13_{-0.18}^{+0.19}$ & 10.5 & - & - & - & 
 %- & - & - &   
 \\
\\
 4U 1538-522 & 52 & 10 & $1.18_{-0.82}^{+1.09}$ & - & - & - & 
 %- & - & - &   
 \\
 \\
 4U 0114+65 & $36.45_{-3.84}^{+3.20}$ & $0.25_{-0.13}^{+0.14}$ & 5 & - & - & - & \\
 \\
\endhead
\\
\endlastfoot
\midrule
\bottomrule
\label{harmonics} 
\end{longtable} 
%\end{landscape}
\noindent

\small
\begin{longtable}{ccccccccccccc}
\caption{Emission Lines detected in the sources present in the present paper} \\
\toprule
Source & $EL$ (\rm{keV}) & $EQ (\rm{keV})$ & $EL$ (\rm{keV}) &  $EQ (\rm{keV})$ & $EL$ (\rm{keV}) &  $EQ (\rm{keV})$
& $EL$ (\rm{keV}) &  $\sigma (\rm{keV})$ \\
%& $EL_{5}$ (\rm{keV}) &  $\sigma_{5} (\rm{keV})$ &$EL_{6}$ (\rm{keV}) &  $\sigma_{6} (\rm{keV})$ &
%$EL_{7}$ (\rm{keV}) &  $\sigma_{7} (\rm{keV})$ &\\
\midrule

Her X-1
%\footnote{An evident `hump' is seen in the energy range 4-9 keV in the Her X-1 spectrum \citep{asami2014}.} 
& $6.51_{-0.05}^{+0.04}$ & $0.23_{-0.005}^{+0.005}$ & $6.65_{-0.02}^{+0.02}$ & $0.056_{-0.005}^{+0.005}$ & $6.40_{-0.01}^{+0.18}$ & $0.031_{-0.003}^{+0.003}$ &  
$0.90_{-0.02}^{+0.02}$ & $0.198_{-0.016}^{+0.014}$ \\
& &  &  &  &  &  & 1.06 & $0.300_{-0.054}^{+0.054}$ \\
4U 0115+63 & $6.65_{-0.06}^{+0.08}$ & $0.06_{-0.02}^{+0.02}$ & - & -& -& -& -& - &  \\
Cen X-3 & $6.409_{-0.005}^{+0.006}$ & $0.070_{-0.003}^{+0.003}$ & $6.658_{-0.009}^{+0.009}$ & $0.047_{-0.003}^{+0.003}$ & 
$6.977_{-0.018}^{+0.018}$ & $0.038_{-0.002}^{+0.003}$
& - & - \\
4U 1626-67 & $6.70_{-0.09}^{+0.09}$ & $0.029_{-0.003}^{+0.003}$ & - & -& -& -& $0.913_{-0.004}^{+0.005}$ & $0.03_{-0.01}^{+0.01}$ \\
& - & - & - & - & - & - & $1.015_{-0.001}^{+0.002}$ & 0.009 \\
XTE 1946+274 & $6.43_{-0.02}^{+0.02}$ & $0.025_{-0.004}^{+0.004}$ & - & -& -& -& - & -  \\
Vela X-1 & $6.403_{-0.003}^{+0.001}$ & $0.054_{-0.002}^{+0.002}$ & $7.09_{-0.03}^{+0.02}$ & $0.0075_{-0.001}^{+0.001}$ &-&-& $0.904_{-0.008}^{+0.006}$ & 1.83E-05 & \\
%&  &  &  &  &  & & $1.61_{-0.01}^{+0.01}$ & $0.15_{-0.02}^{+0.03}$ \\
%& &  &  &  &  &  &  $2.09_{-0.36}^{+0.21}$ & $0.79_{-0.11}^{+0.12}$  \\
&  &  & &  &  &  & $1.361_{-0.007}^{+0.007}$ & $2.15E-05$  \\
4U 1907+09 & $6.393_{-0.004}^{+0.006}$ & $0.061_{-0.005}^{+0.005}$ & $7.01_{-0.03}^{+0.04}$ & $0.0091_{-0.0007}^{+0.0007}$ & -& -& - & -  \\
4U 1528-522 & $6.417_{-0.005}^{+0.007}$ & $0.067_{-0.004}^{+0.004}$ & - & - & -& -& - & -\\
GX 301-2 & $6.378_{-0.003}^{+0.005}$ & $0.216_{-0.008}^{+0.008}$ & -& - & - &- & $2.95_{-0.03}^{+0.03}$ & 1.67E-05  \\
1A 1118-61 & $6.399_{-0.004}^{+0.005}$ & $0.053_{-0.003}^{+0.003}$ & $7.07_{-0.03}^{+0.03}$ & $0.010_{-0.003}^{+0.003}$ & -& -& - & -  \\
4U 0114+65 & $6.42_{-0.01}^{+0.01}$ & $0.029_{-0.003}^{+0.003}$ & $7.09_{-0.04}^{+0.03}$ & $0.012_{-0.001}^{+0.001}$ & -& -& - & -  \\
GX 304-1 & $6.46_{-0.01}^{+0.01}$ & $0.034_{-0.003}^{+0.003}$ & - & - & -& -& - & - \\
OAO 1657-415 & $6.461_{-0.003}^{+0.001}$ & $0.252_{-0.006}^{+0.006}$ & $7.130_{-0.008}^{+0.009}$ & $0.094_{-0.002}^{+0.002}$ & -& -& - & -  \\
Cep X-4 & $6.46_{-0.03}^{+0.02}$ & $0.322_{-0.006}^{+0.006}$  & - & -  & - & - & - & - \\
4U 1700-37 & $6.419_{-0.003}^{+0.003}$ & $0.073_{-0.001}^{+0.001}$  & $7.09_{-0.01}^{+0.01}$ & $0.014_{-0.001}^{+0.001}$  & - & - & - & - \\
A0535+026 &  $6.40_{-0.02}^{+0.02}$ & $0.0232_{-0.005}^{+0.005}$ & - & - & -& -& - & - \\
4U 1822-37  & $6.41_{-0.01}^{+0.01}$ & $0.082_{-0.008}^{+0.008}$ 
& $6.97_{-0.03}^{+0.03}$ & $0.025_{-0.006}^{+0.006}$  & - & - &  \\
GX 1+4  & $6.42_{-0.003}^{+0.003}$ & $0.144_{-0.001}^{+0.001}$ & $7.09_{-0.01}^{+0.01}$ & $0.028_{-0.003}^{+0.003}$ & $7.46_{-0.04}^{+0.04}$ & $0.061_{-0.007}^{+0.007}$ & - & -  \\
GRO J1008-57 & $6.39_{-0.01}^{+0.01}$ & $0.016_{-0.001}^{+0.001}$ & $7.07_{-0.06}^{+0.06}$ & $0.0033_{-0.0003}^{+0.0003}$ & -& -& - & -  \\
4U 1909+07  & $6.416_{-0.009}^{+0.008}$ & $0.0713_{-0.005}^{+0.005}$ & - & - & -& -& - & - \\
IGR J16393-4643 & $6.36_{-0.02}^{+0.03}$ & $0.039_{-0.008}^{+0.008}$ & - & - & -& -& - & -\\
4U 2206+54 & 6.4 & $0.007_{-0.004}^{+0.004}$ & - & - & -& -& - & - \\
SW J2000.6+3210 & $6.29_{-0.06}^{+0.07}$ & $0.019_{-0.006}^{+0.007}$ & - & - & -& -& - & - \\
LMC X-4 & $6.49_{-0.03}^{+0.03}$ & $0.056_{-0.011}^{+0.011}$ & - & - & -& - & $0.69_{-0.12}^{+0.15}$ & $0.23_{-0.09}^{+0.05}$ \\
%\caption{Emission lines....contd} \\
KS 1947+300 & $6.51_{-0.03}^{+0.03}$ & $0.033_{-0.004}^{+0.004}$ & $7.30_{-0.07}^{+0.07}$ & $0.022_{-0.003}^{+0.003}$ & -& -& - & -  \\
EXO 2030+375 & $6.38_{-0.0}^{+0.0}$ & $0.021_{-0.002}^{+0.002}$ & $6.64_{-0.0}^{+0.0}$ & $0.019_{-0.001}^{+0.001}$ & -& -& 2.5 & $0.12_{-0.04}^{+0.04}$   \\
& & & & & & & $3.26_{-0.03}^{+0.03}$ & 4.8E-02 \\
SMC X-1 & $6.32_{-0.03}^{+0.03}$ & $0.030_{-0.009}^{+0.009}$ & $5.89_{-0.17}^{+0.16}$ & $0.025_{-0.010}^{+0.011}$ & -& - & $1.30_{-0.03}^{+0.05}$ & $0.08_{-0.08}^{+0.08}$ \\
& & & & & & & $0.88_{-0.01}^{+0.02}$ & $0.00018_{-0.00009}^{+0.00008}$ \\
V 0332+53 & - & - & - & - & - & - & - & - \\
IGR J16493-4348 &  $6.49_{-0.07}^{+0.05}$ & $0.034_{-0.015}^{+0.015}$ & - & - & -& -& - & - \\
IGRJ16318-4848 & $6.39_{-0.02}^{+0.02}$ & $0.698_{-0.050}^{+0.048}$  & $7.11_{-0.02}^{+0.01}$ & $0.153_{-0.011}^{+0.011}$  & $7.44_{-0.05}^{+0.05}$ & $0.079_{-0.006}^{+0.006}$ &
- & - \\
IGRJ16207-5129 &  $6.40_{-0.03}^{+0.03}$ & $0.038_{-0.013}^{+0.010}$ & - & - & -& -& - & - \\
IGR J18410-0535 & $6.38_{-0.03}^{+0.03}$ & $0.032_{-0.008}^{+0.008}$  & - & -  & - & - & - & - \\
IGR J17544-2619  & $6.38_{-0.02}^{+0.03}$ & $0.015_{-0.005}^{+0.005}$  & - & -  & - & - & - & - \\
IGR J16195-4945 & $6.39_{-0.03}^{+0.03}$ & $0.039_{-0.016}^{+0.020}$  & - & -  & - & - & - & - \\
%IGR J16493-4348 & $6.48_{-0.06}^{+0.06}$ & $0.036_{-0.018}^{+0.018}$  & - & -  & - & - & - & - \\
IGR J16465-4507 & 6.4 & 0.001  & - & -  & - & - & - & - \\
IGR J16479-4514 & $6.36_{-0.04}^{+0.04}$ & $0.053_{-0.002}^{+0.002}$  & - & -  & - & - & - & - \\
IGR J17391-3021 &- & - & - & -  & - & - & - & - \\
IGR J08408-4503 &- & - & - & -  & - & - & - & - \\
IGR J00370+6122 & - & - & - & - & - & - & - & - \\
%\caption{Parameters of the continuum, including observation ID, state of the source, model (see Table~\ref{tab:models}), artificial gain (offset or slope), reduced $\chi^2$, null hypothesis probability ($P_0$), luminosity between 1-10~keV ($L_{1-10\,keV}$), total $N_H$ (adding every absorption component), photon index ($\Gamma$), and temperature of thermal components ($kT_i$): bbody, diskbb, bremss, mekal or cemekl (depending on the model). Parameters frozen or unbounded are included without an error estimation. Therefore they are not used in the plots and the subsequent discussion. The possible states are: quiescence(Q), flare(F), eclipse ingress/egress(I/E), eclipse(E) and dip(D). } \\
%\endlastfoot
\midrule
\bottomrule
\label{emission} 
%\bottomrule
\end{longtable} 
\end{landscape}

\begin{table}
\centering
\caption{Literature values of cut-off energies from RXTE data. To maintain consistency with the cutoff energy measurements, we take these values from the same instrument.}
\begin{tabular}{llllll}

Source          &  & Ecut  & positive error & negative error & Ref \\
\hline \\
Her X-1         &  & 22    & 1.4     & 0.8     &  \citep{coburn2002} \\
4U 0115+6       &  & 10    & 0.5     & 0.4     &  \citep{coburn2002} \\
Cen X-3         &  & 21.3  & 0.2     & 0.4     &  \citep{coburn2002}\\
4U 1626-67    &  & 6.8   & 0.3     & 0.3     &  \citep{coburn2002}\\
XTE J1946+274   &  & 22    & 0.8     & 0.9     & \citep{coburn2002} \\
Vela X-1        &  & 17.9  & 0.3     & 0.4     &  \citep{coburn2002}\\
4U 1907+09      &  & 13.5  & 0.2     & 0.2     &  \citep{coburn2002}\\
4U 1538-52    &  & 13.57 & 0.04    & 0.05    &  \citep{coburn2002}\\
GX 301-2      &  & 17.3  & 0.1     & 0.2     &  \citep{coburn2002}\\
               
1A 1118–61      &  & 5.88  & 0.17    & 0.19    &  \citep{devasia2011_1a1118}\\
                &  &       &         &         &  \\
GX 304-1        &  & 6.08  & 0.16    & 0.19    & \citep{rothschild2017} \\
                &  &       &         &         &  \\
Cep X-4         &  & 16.3  & 0.1     & 0.1     &  \citep{koyama1991_cepx4}\\
                &  &       &         &         &  \\
4U 1909+07      &  & 7.8   & 0.5     & 0.5     &  \citep{furst2011_1909}\\
                &  &       &         &         &  \\
KS 1947+300     &  & 6.5   & 0.5     & 0.5     & \citep{Tsygankov2005_ks1947} \\
                &  & 15.8  & 0.5     & 0.5     & \citep{Tsygankov2005_ks1947} \\
                &  &       &         &         &  \\
4U 2206+54      &  & 7.3   & 0.1     & 0.1     &  \citep{Corbet_2001}\\
                &  &       &         &         &  \\
EXO 2030+375    &  & 7.7   & 0.2     & 0.2     &  \citep{Epili2017_exo2030}\\
                &  &       &         &         &  \\
SMC X-1         &  & 13.7  & 3.4     & 3.4     &  \citep{inam2010_smcx1}\\
                &  & 6.6   & 1.6     & 1.6     &  \citep{inam2010_smcx1} \\
%                &  &       &         &         &  \\
%V 0332+53       &  &       &         &         &  \\
%                &  &       &         &         &  \\
%IGR J16207-5129 &  &       &         &         &  \\
%                &  &       &         &         &  \\
%SFXTs           &  &       &         &         & \\
\hline 
\end{tabular}
\label{literature-cutoff}
\end{table}

\section*{Acknowledgment}
The authors would like to acknowledgment the referee for his/her thorough and constructive comments that greatly improved the quality of the paper. PP would like to thank Raman Research Institute, Bengaluru and St Joseph's College, Darjeeling for the infrastructure facilities provided during the preparation of this work. PP would also like to thank Dipankar Bhattacharya for useful discussions and also the members of XMAG for their useful comments. Finally, PP would like to acknowledge the grant received as a part of Minor Research Project from University Grants Commission, India that partly supported this work. 
This research has made use of data and/or software provided by the High Energy Astrophysics Science Archive Research Center (HEASARC), which is a service of the Astrophysics Science Division at NASA/GSFC.

\section*{Data Availability }
The data underlying this article will be shared on reasonable request to the corresponding author.

\normalsize

\bibliography{reference.bib}{}
\bibliographystyle{mn2e}

 \clearpage
\appendix
\label{appendix}
\section{}
 \clearpage

\begin{figure*}
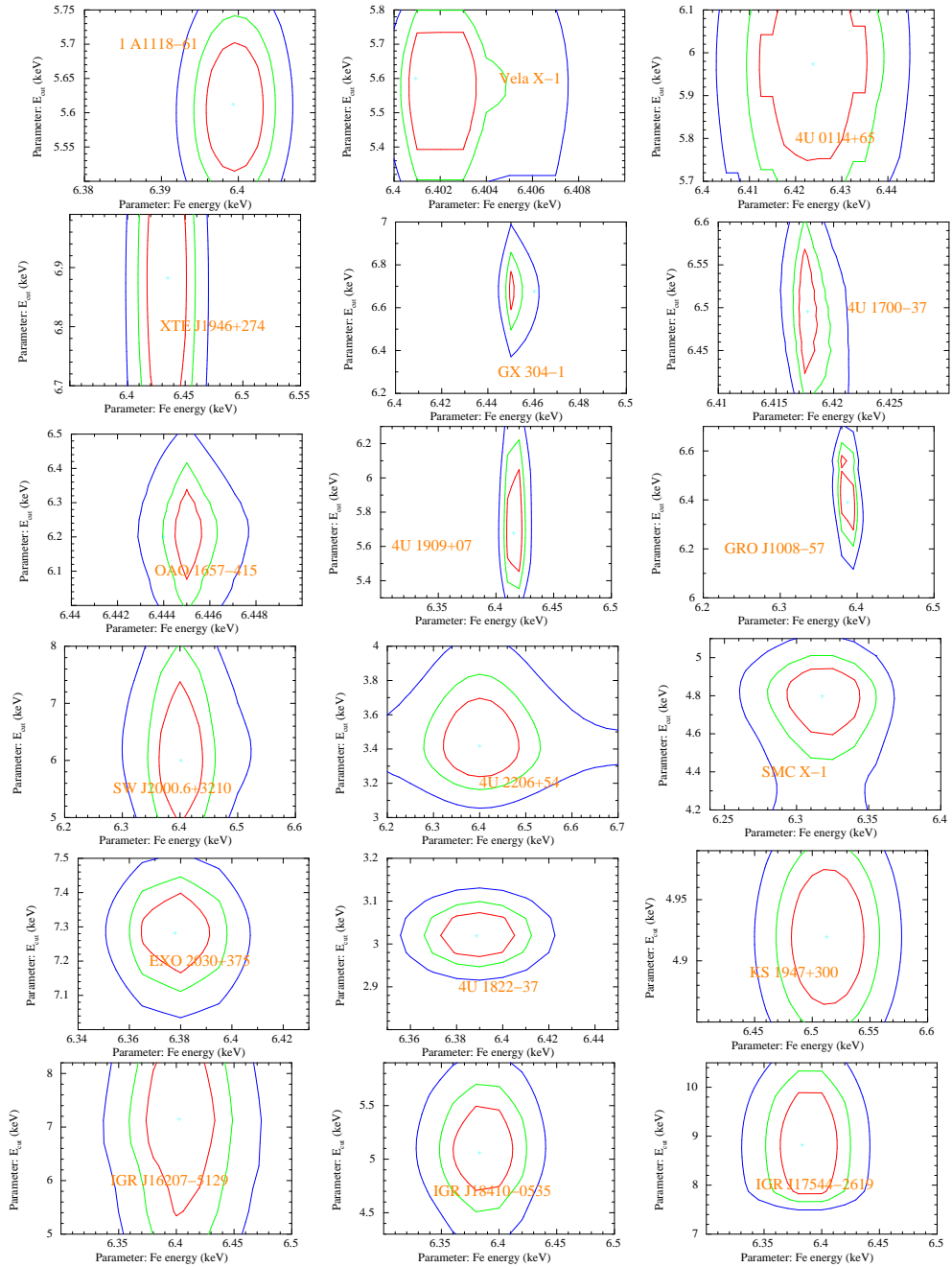

\includegraphics[scale=0.2,angle=-90]{irone-ecut-1a1118.ps}
\includegraphics[scale=0.2,angle=-90]{irone-ecut-vela.ps}
\includegraphics[scale=0.2,angle=-90]{irone-ecut-4u0114.ps}
\includegraphics[scale=0.2,angle=-90]{irone-ecut-xtej.ps}
\includegraphics[scale=0.2,angle=-90]{irone-ecut-gx304.ps}
\includegraphics[scale=0.2,angle=-90]{irone-ecut-4u1700.ps}
\includegraphics[scale=0.2,angle=-90]{irone-ecut-oao.ps}
\includegraphics[scale=0.2,angle=-90]{irone-ecut-4u1909+0.ps}
\includegraphics[scale=0.2,angle=-90]{irone-ecut-groj1008-57.ps}
\includegraphics[scale=0.2,angle=-90]{irone-ecut-swj200.ps}
\includegraphics[scale=0.2,angle=-90]{irone-ecut-4u2206.ps}
\includegraphics[scale=0.2,angle=-90]{irone-ecut-smcx1.ps}
\includegraphics[scale=0.2,angle=-90]{irone-ecut-exo2030.ps}
\includegraphics[scale=0.2,angle=-90]{irone-ecut-4u1822.ps}
\includegraphics[scale=0.2,angle=-90]{irone-ecut-ks1947.ps}
\includegraphics[scale=0.2,angle=-90]{irone-ecut-igrj16207-5129.ps}
\includegraphics[scale=0.2,angle=-90]{irone-ecut-igrj18410-0535.ps}
\includegraphics[scale=0.2,angle=-90]{irone-ecut-igrj17544-2619.ps}
\vspace{2.5cm}
\caption{Contour plots between cutoff energy and iron line energy for sources in group 2 where these two values are close to each other. Note that for most of the sources, the iron line energy is centered at the neutral K$\alpha$ line energy of 6.4\,keV.}
\label{cutoff-ironenergy}
\end{figure*}

\begin{figure*}
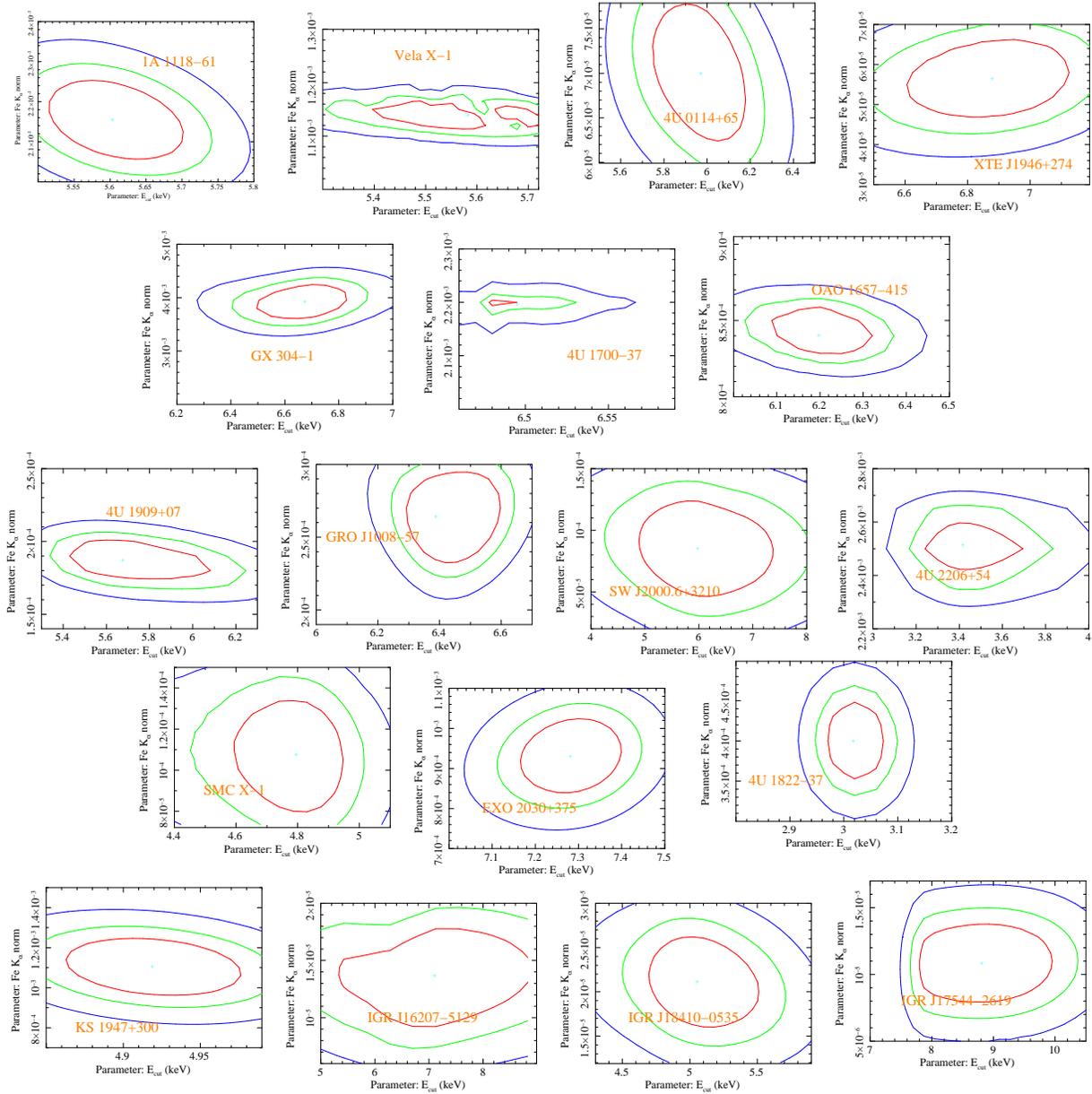

\includegraphics[scale=0.2,angle=-90]{1a1118-61_iron_ecut.ps}
\includegraphics[scale=0.2,angle=-90]{ironc-ecut-vela.ps}
\includegraphics[scale=0.2,angle=-90]{ironc-ecut-4u0114.ps}
\includegraphics[scale=0.2,angle=-90]{ironc-ecut-xtej.ps}
\includegraphics[scale=0.2,angle=-90]{ironc-ecut-gx304.ps}
\includegraphics[scale=0.2,angle=-90]{ironc-ecut-4u1700.ps}
\includegraphics[scale=0.2,angle=-90]{ironc-ecut-oao.ps}
\includegraphics[scale=0.2,angle=-90]{ironc-ecut-4u1909+07.ps}
\includegraphics[scale=0.2,angle=-90]{ironc-ecut-groj1008-57.ps}
\includegraphics[scale=0.2,angle=-90]{ironc-ecut-swj200.ps}
\includegraphics[scale=0.2,angle=-90]{ironc-ecut-4u2206.ps}
\includegraphics[scale=0.2,angle=-90]{ironc-ecut-smcx1.ps}
\includegraphics[scale=0.2,angle=-90]{ironc-ecut-exo2030.ps}
\includegraphics[scale=0.2,angle=-90]{ironc-ecut-4u1822.ps}
\includegraphics[scale=0.2,angle=-90]{ironc-ecut-ks1947.ps}
\includegraphics[scale=0.2,angle=-90]{ironc-ecut-igrj16207-5129.ps}
\includegraphics[scale=0.2,angle=-90]{ironc-ecut-igrj18410-0535.ps}
\includegraphics[scale=0.2,angle=-90]{ironc-ecut-igrj17544-2619.ps}
\vspace{2.5cm}
\caption{Contour plots between cutoff energy and iron line normalization for sources in group 2 (magenta) where these two values are close to each other}
\label{cutoff-ironnorm}
\end{figure*}

\begin{figure*}
%\hspace{-2cm}
\includegraphics[scale=0.2,angle=-90]{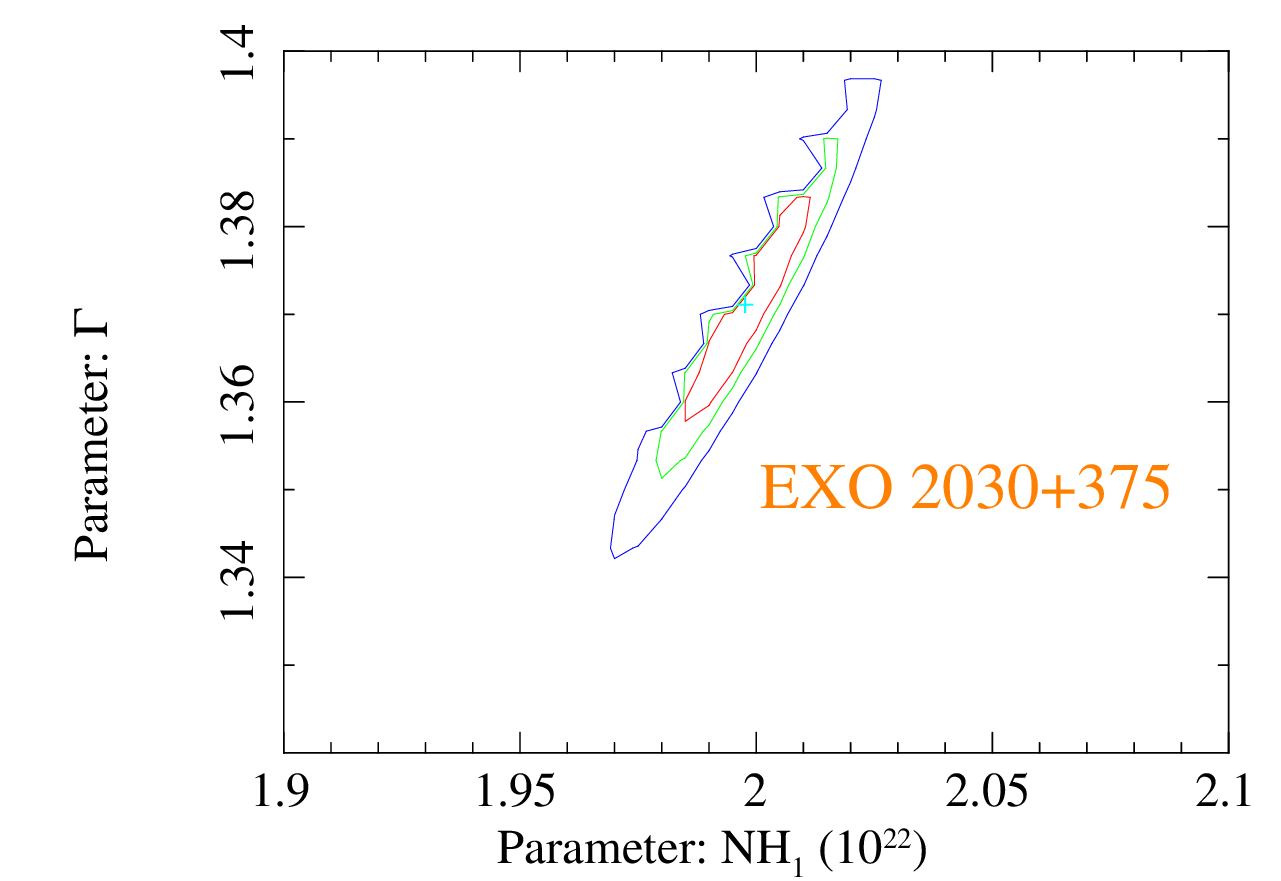}
\includegraphics[scale=0.2,angle=-90]{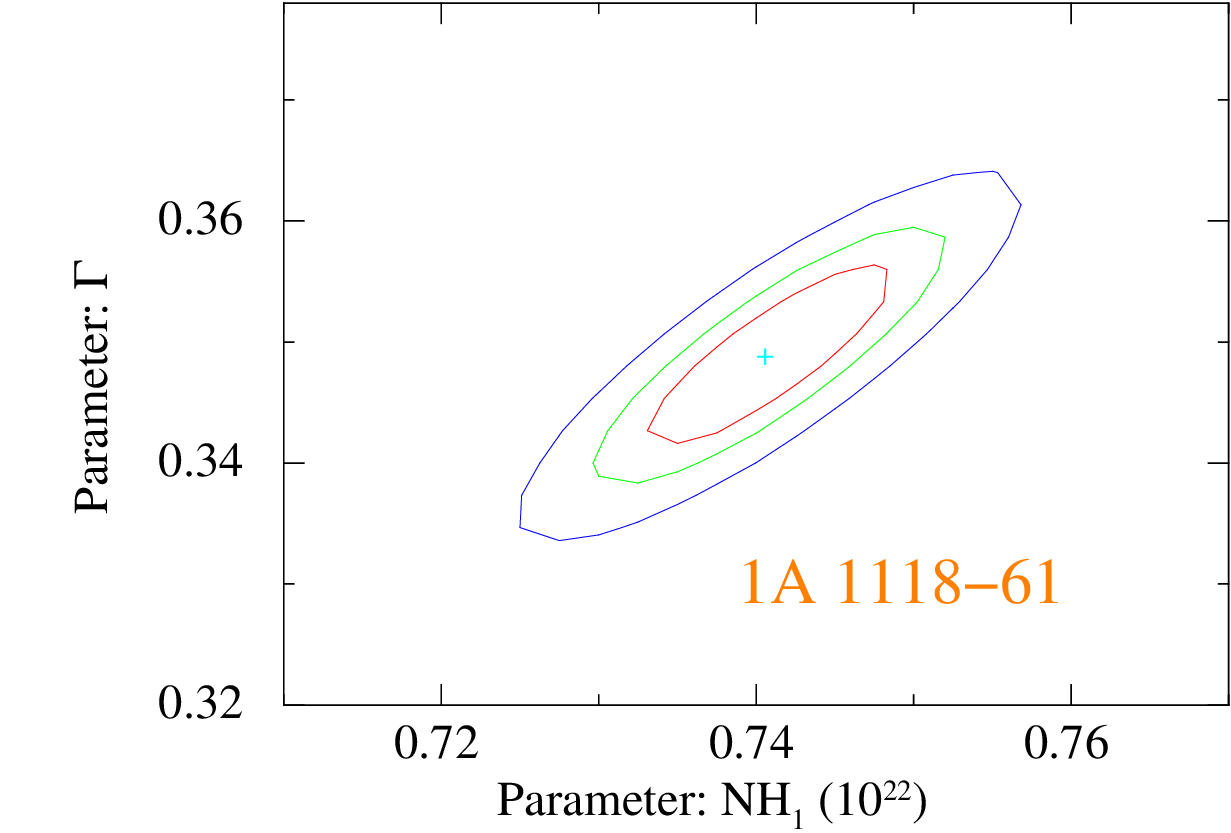}
\includegraphics[scale=0.2,angle=-90]{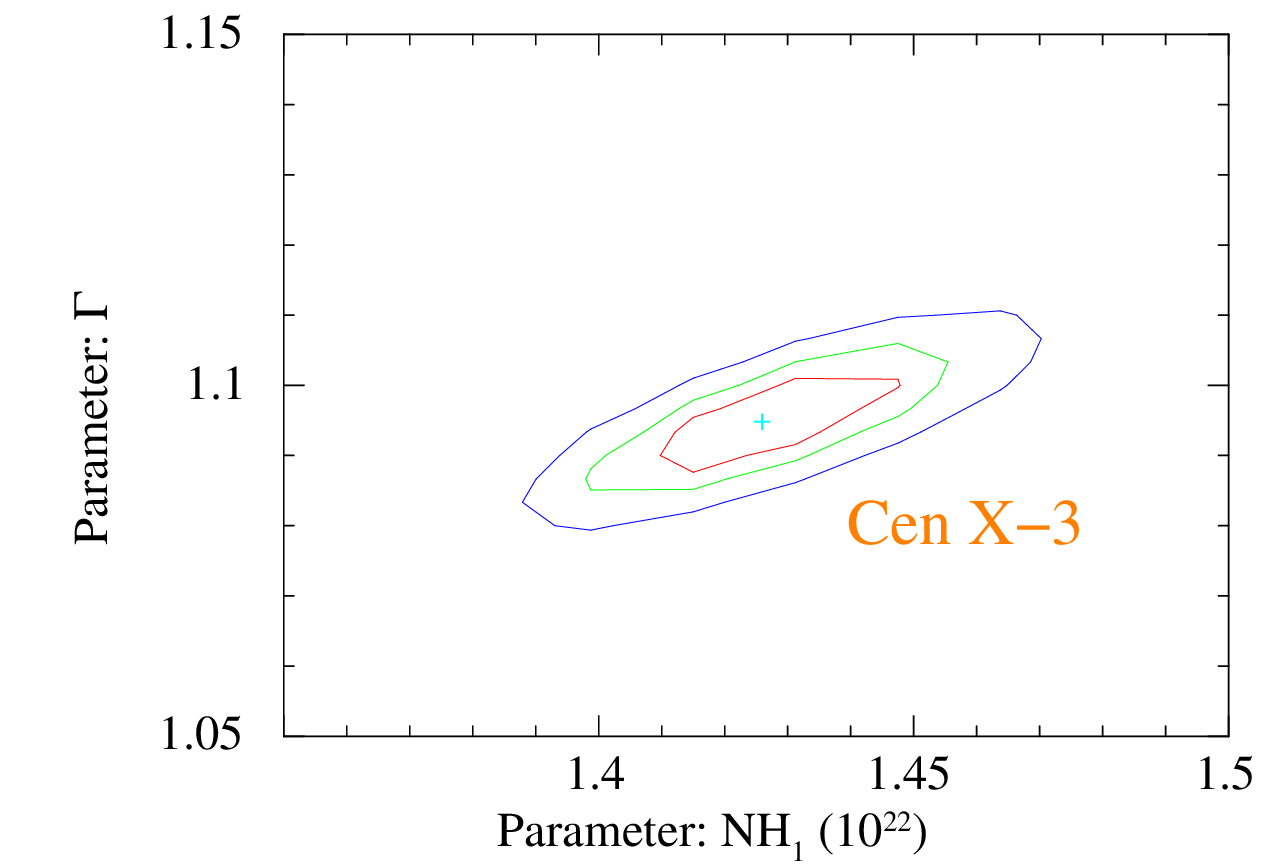}
\includegraphics[scale=0.2,angle=-90]{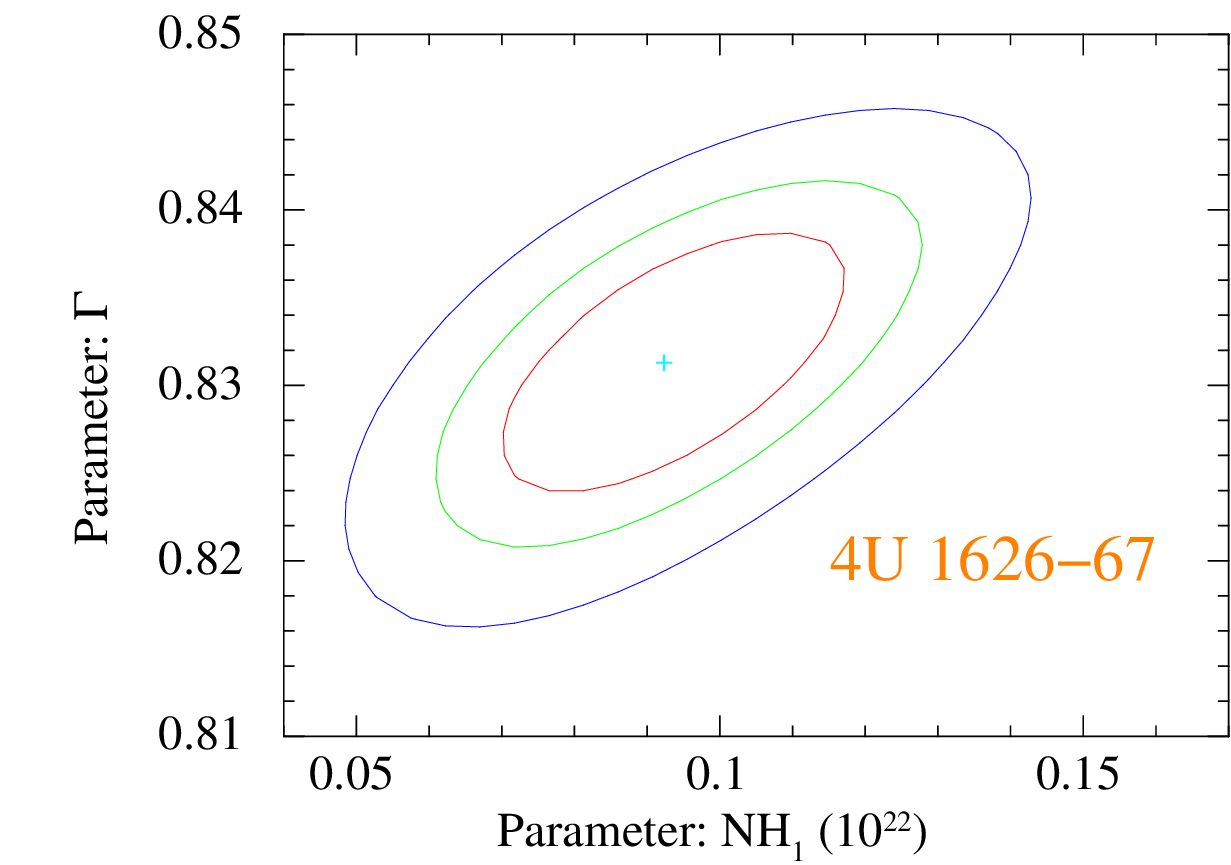}
\includegraphics[scale=0.2,angle=-90]{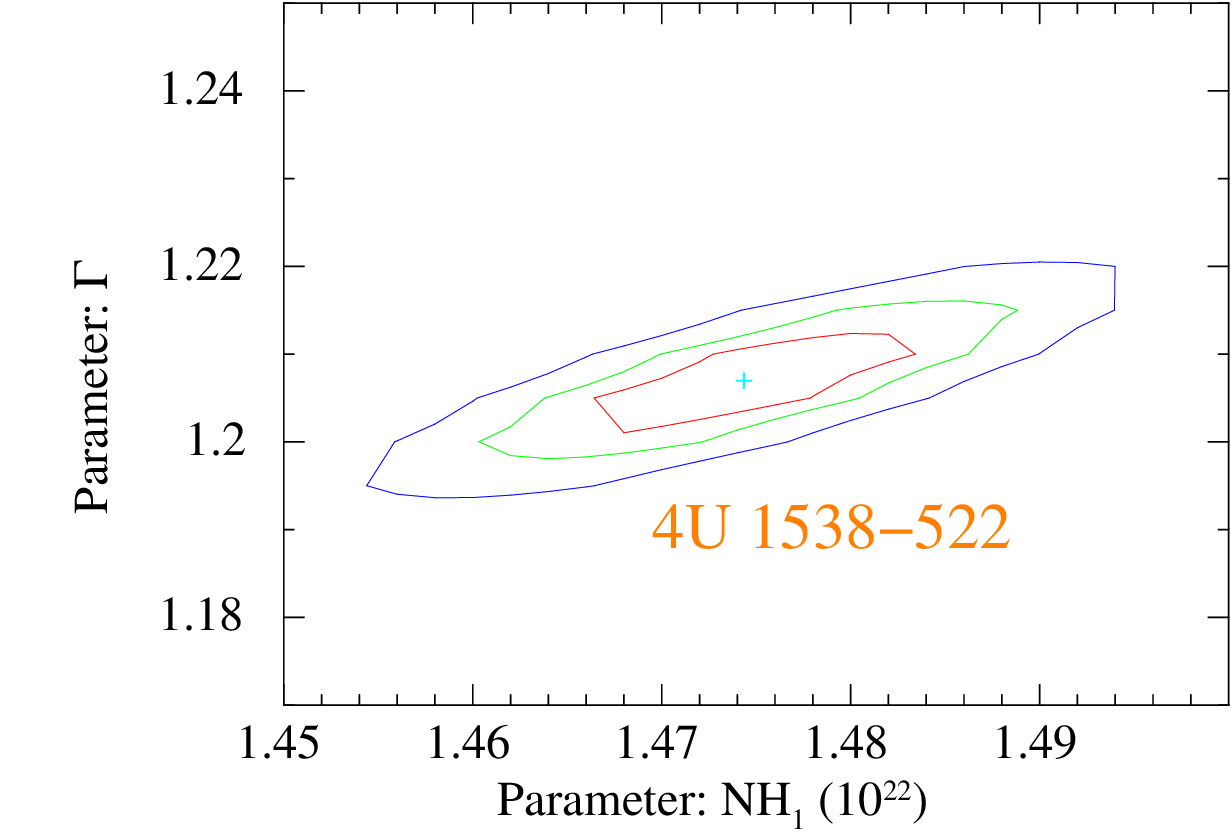}
\includegraphics[scale=0.2,angle=-90]{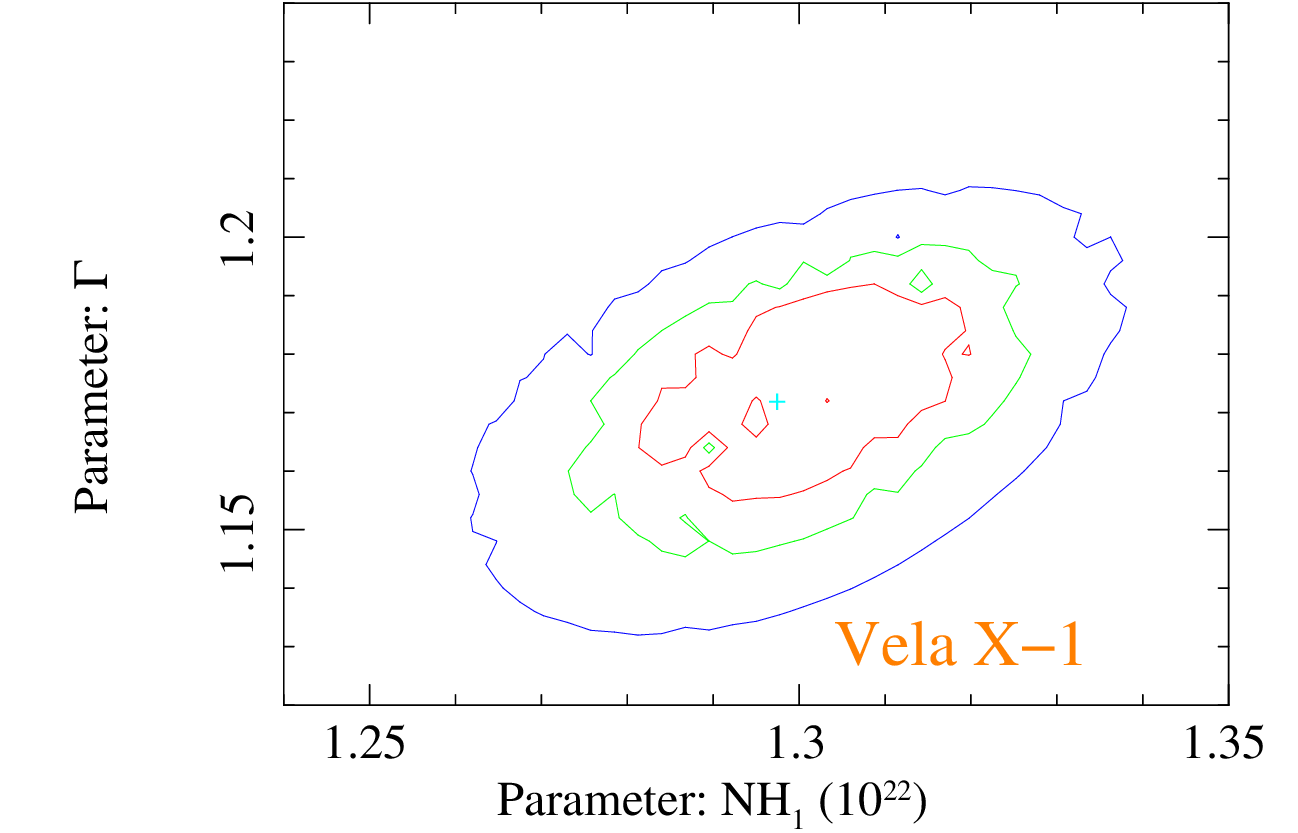}
\includegraphics[scale=0.2,angle=-90]{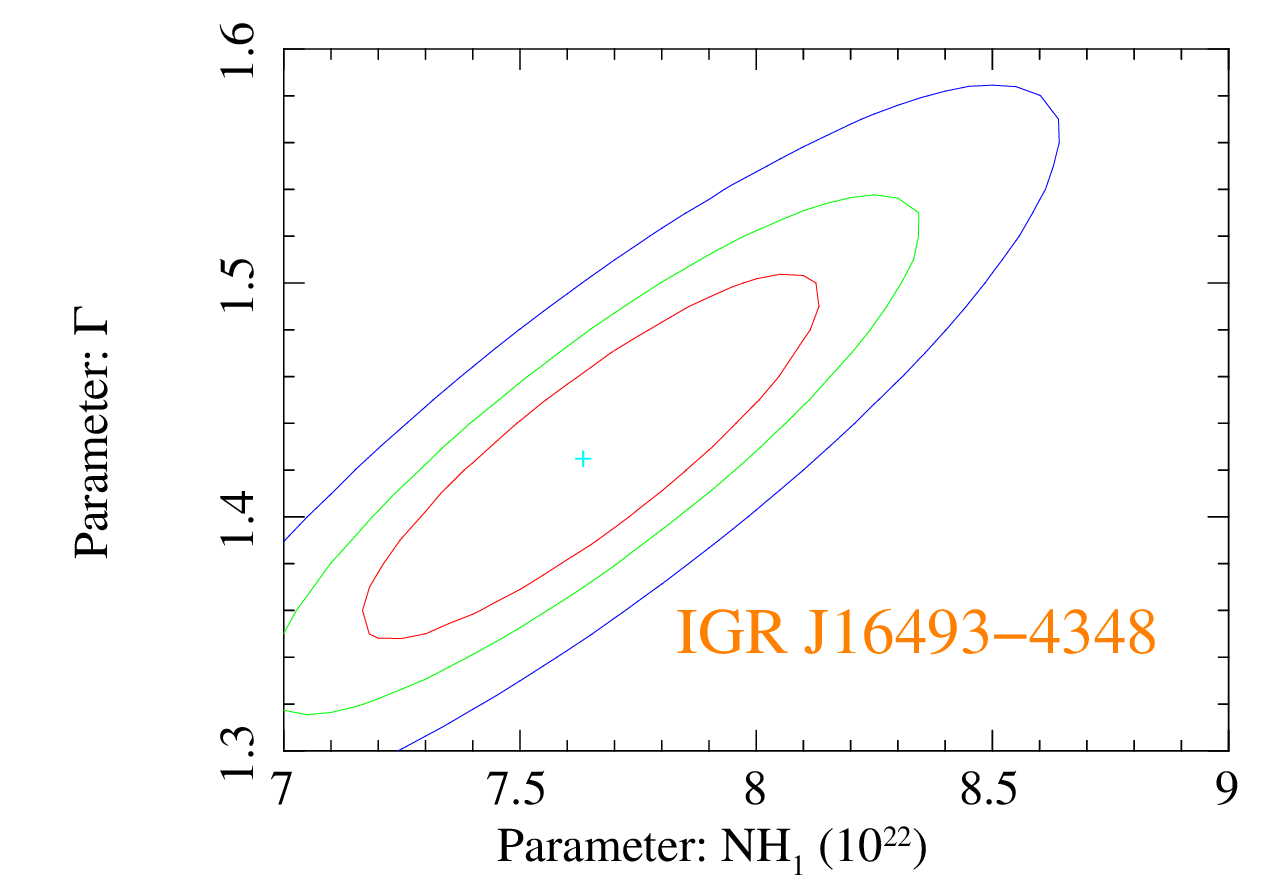}
\includegraphics[scale=0.2,angle=-90]{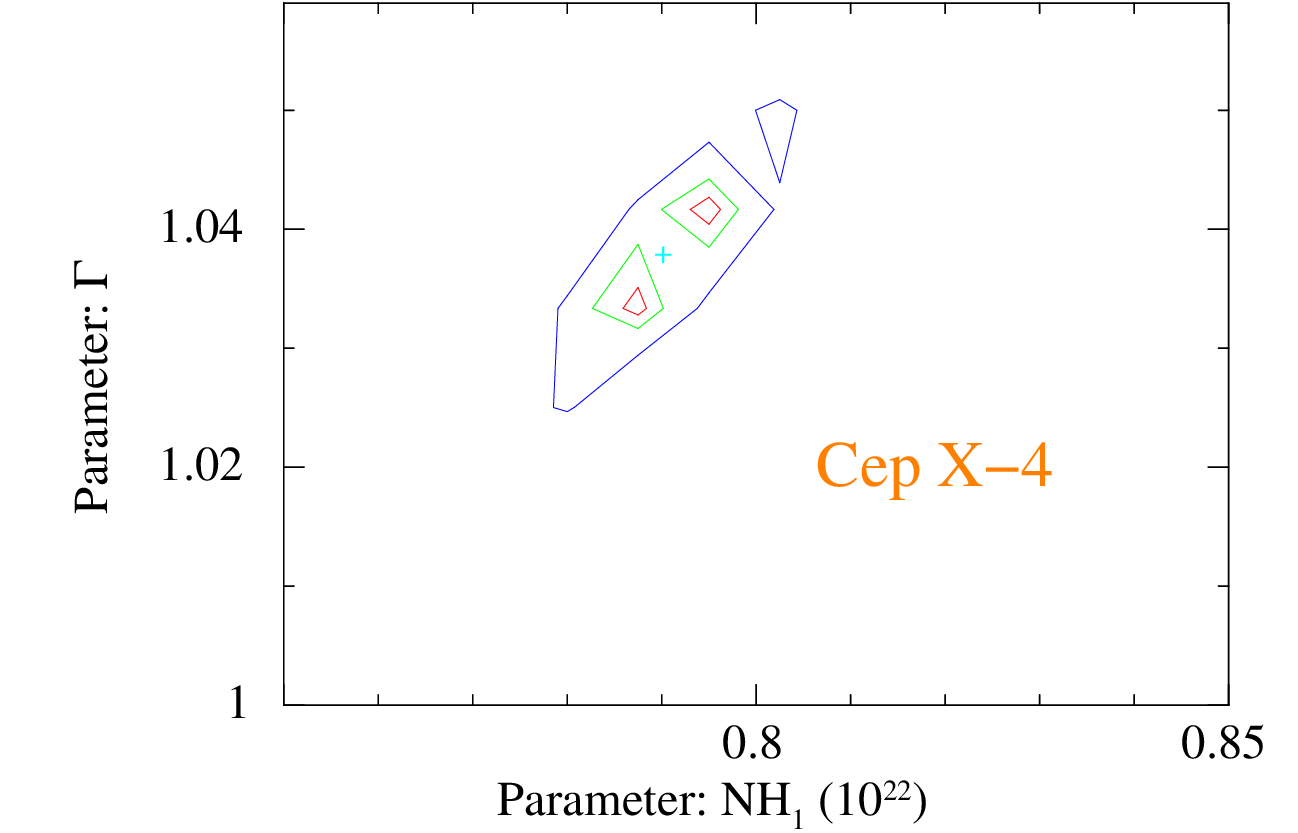}
\includegraphics[scale=0.2,angle=-90]{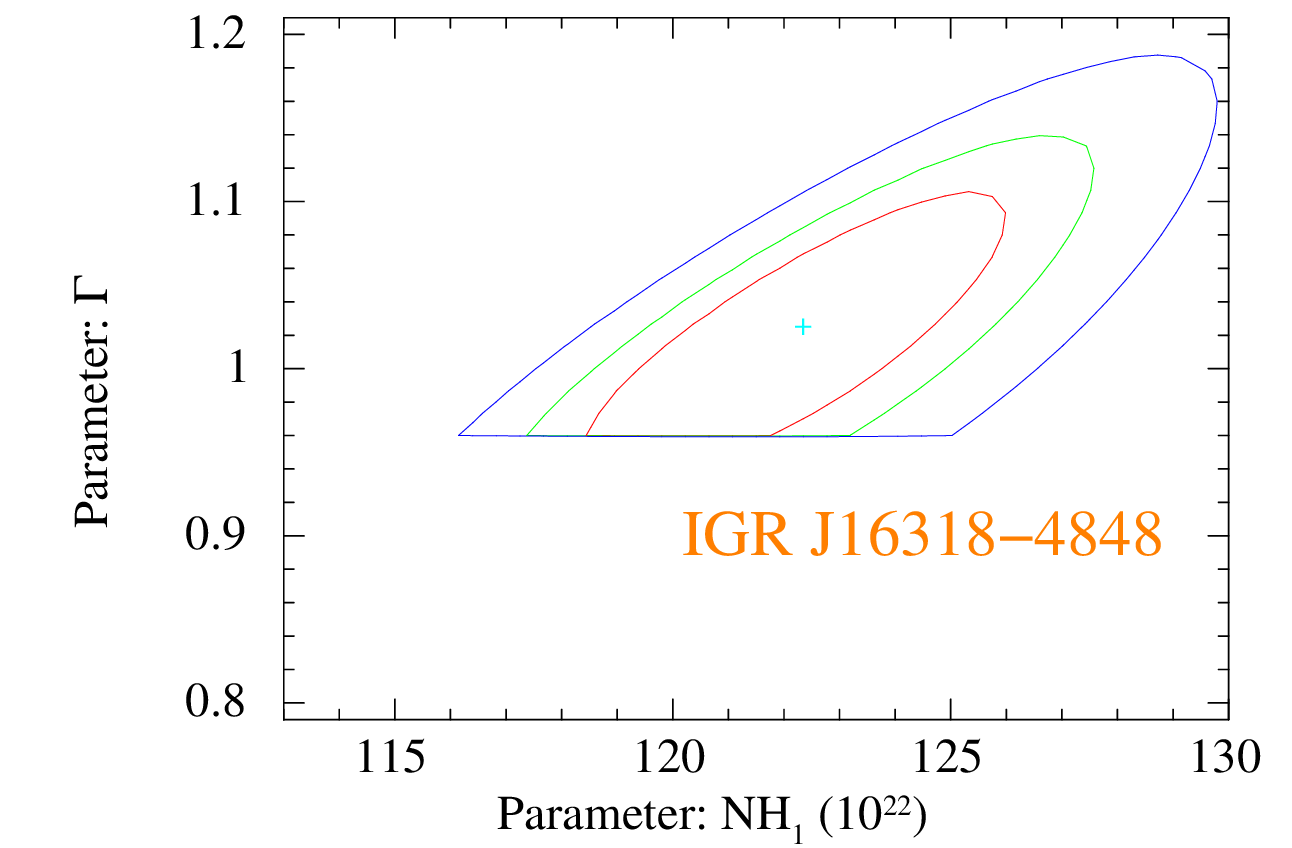}
\includegraphics[scale=0.2,angle=-90]{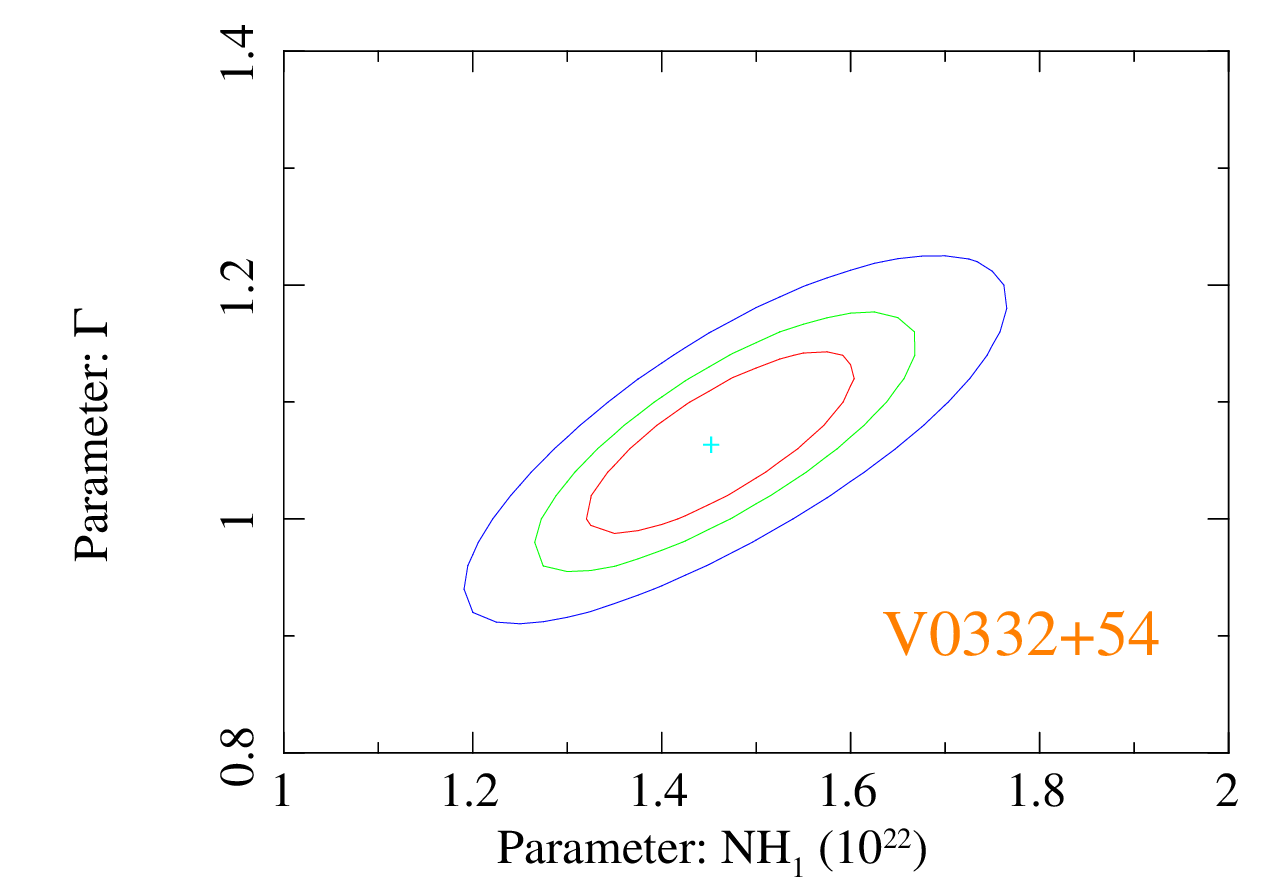}
\includegraphics[scale=0.2,angle=-90]{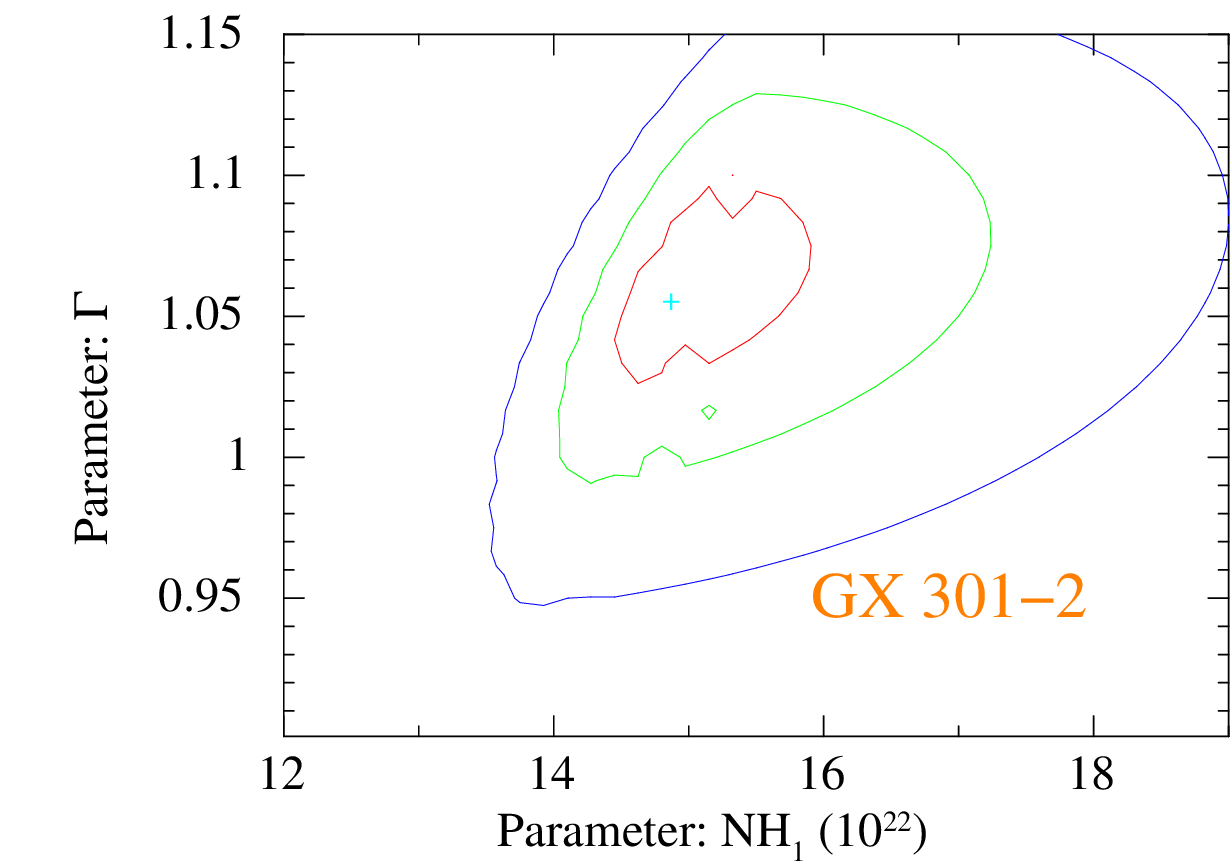}
\includegraphics[scale=0.2,angle=-90]{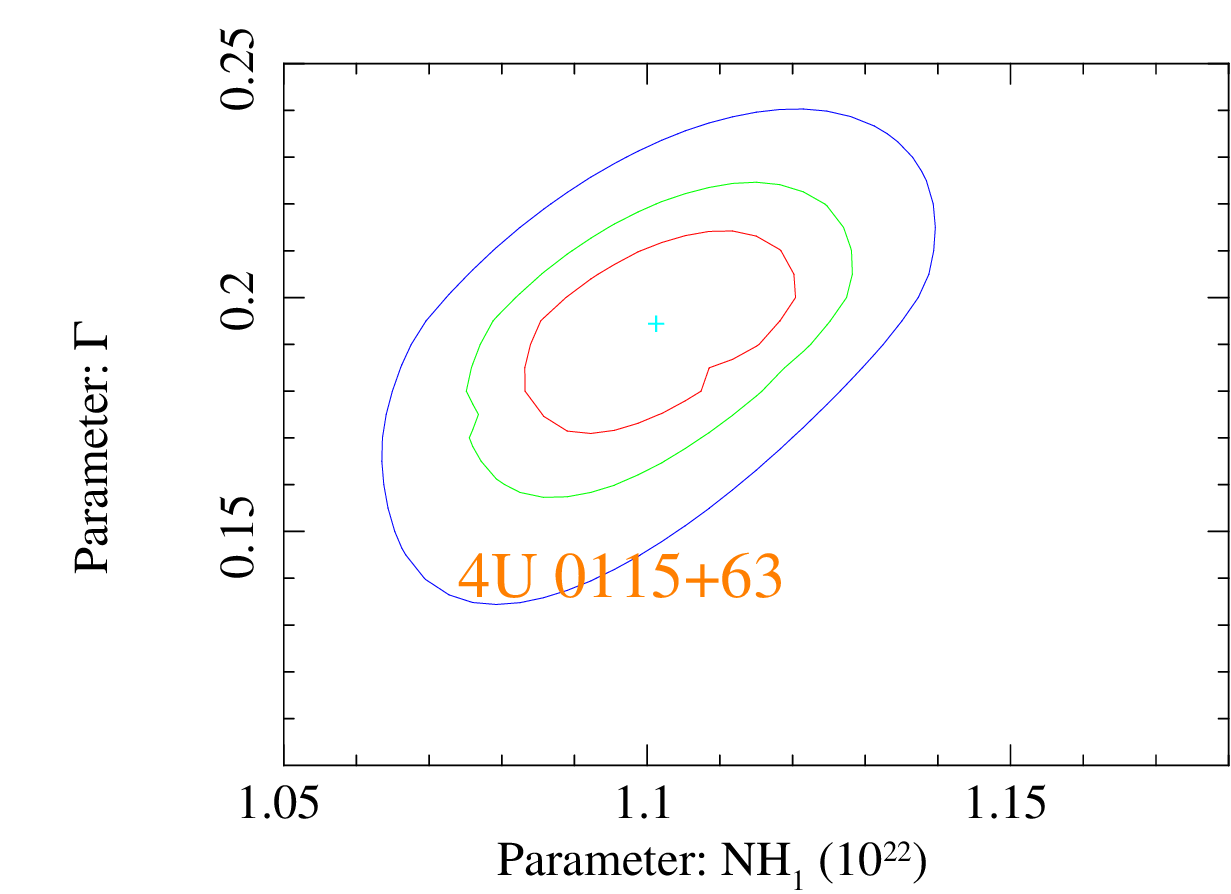}
\includegraphics[scale=0.2,angle=-90]{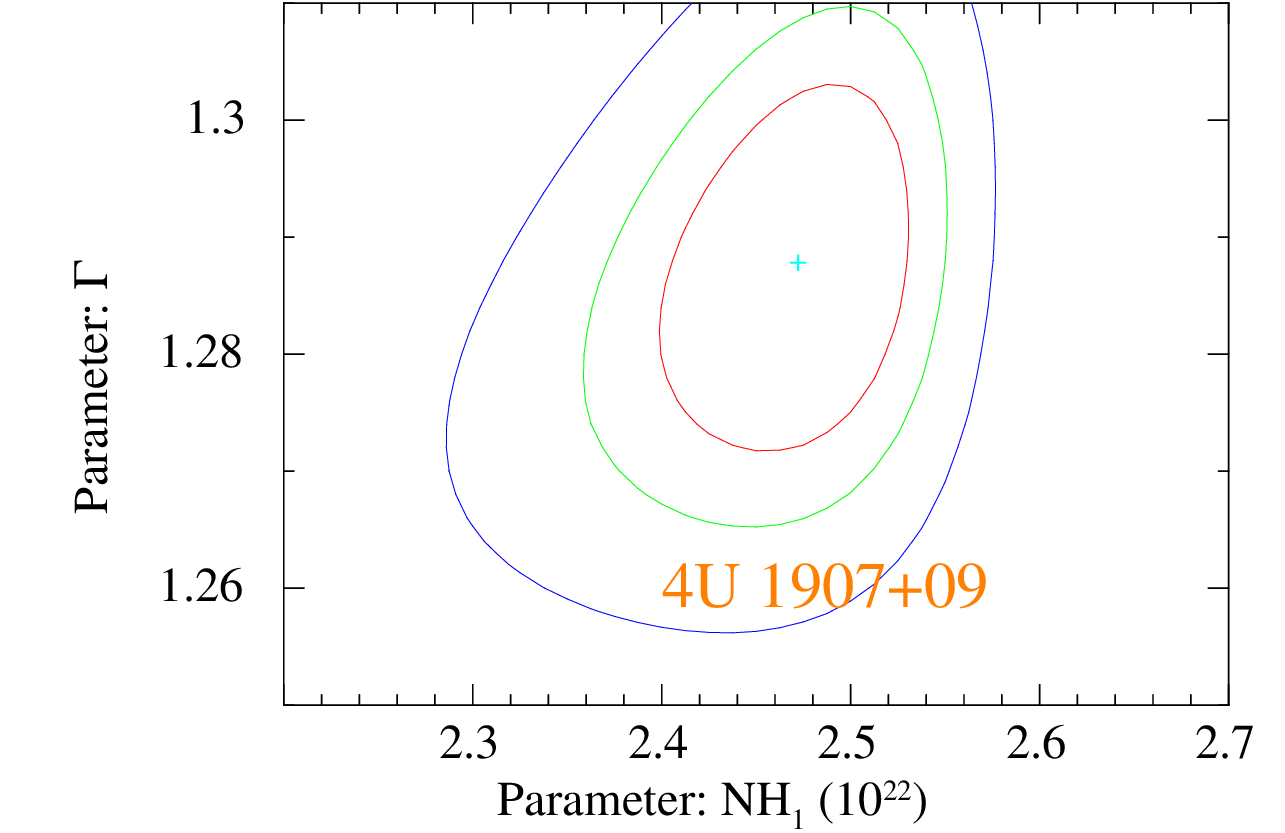}
\includegraphics[scale=0.2,angle=-90]{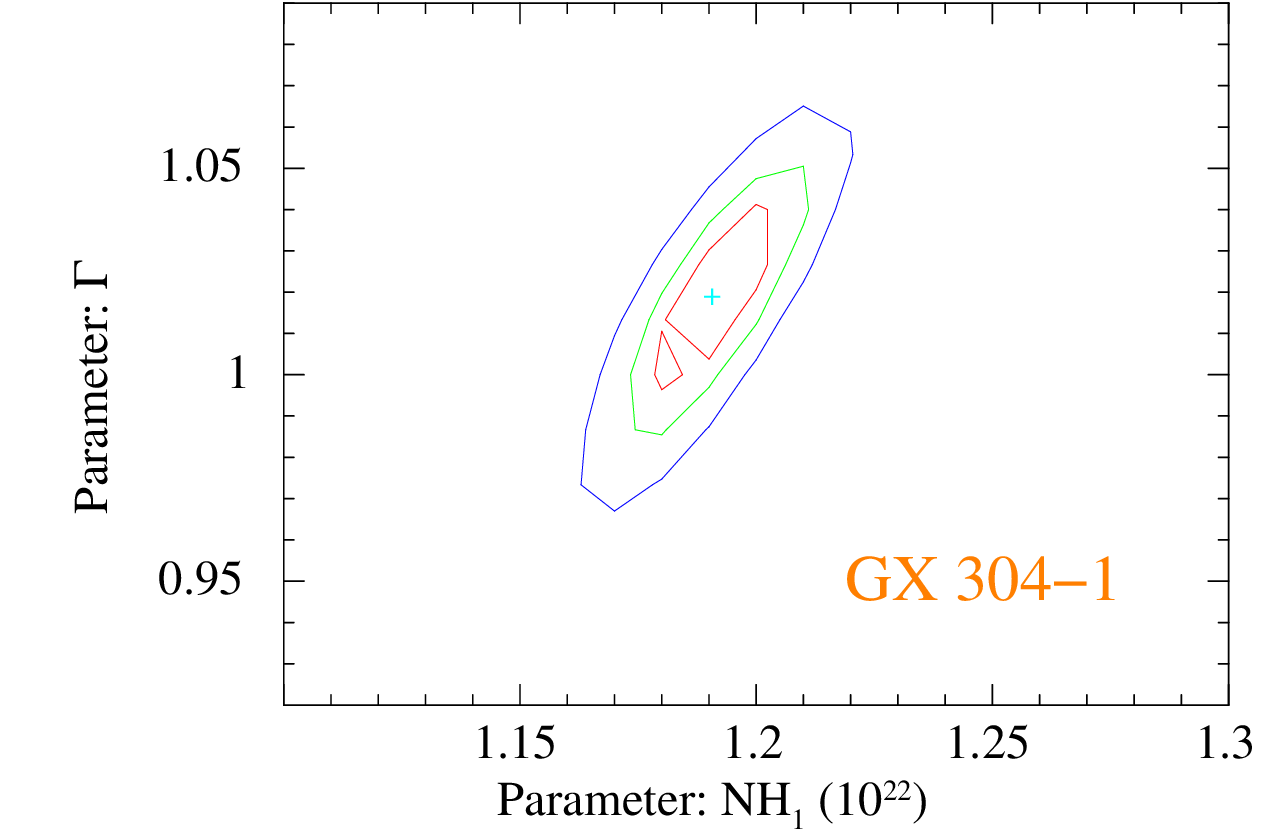}
\includegraphics[scale=0.2,angle=-90]{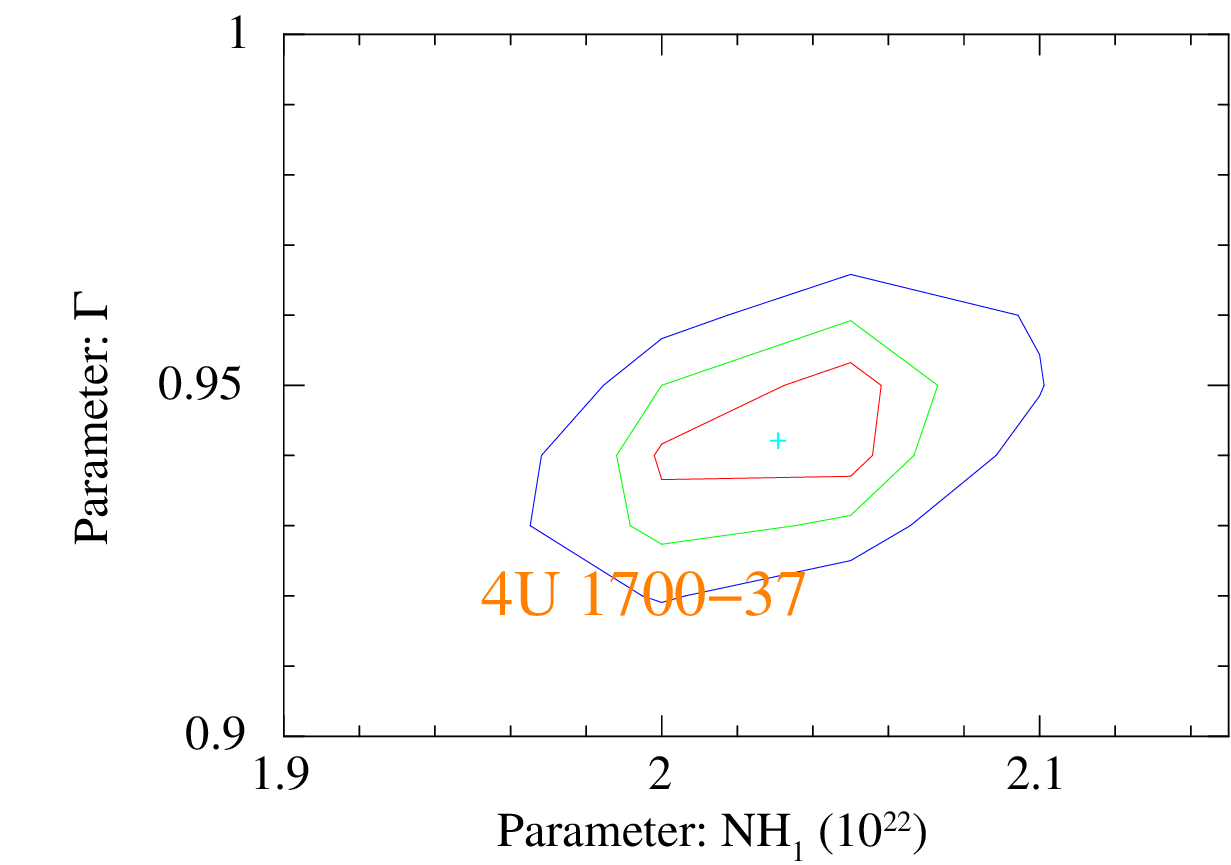}
\includegraphics[scale=0.2,angle=-90]{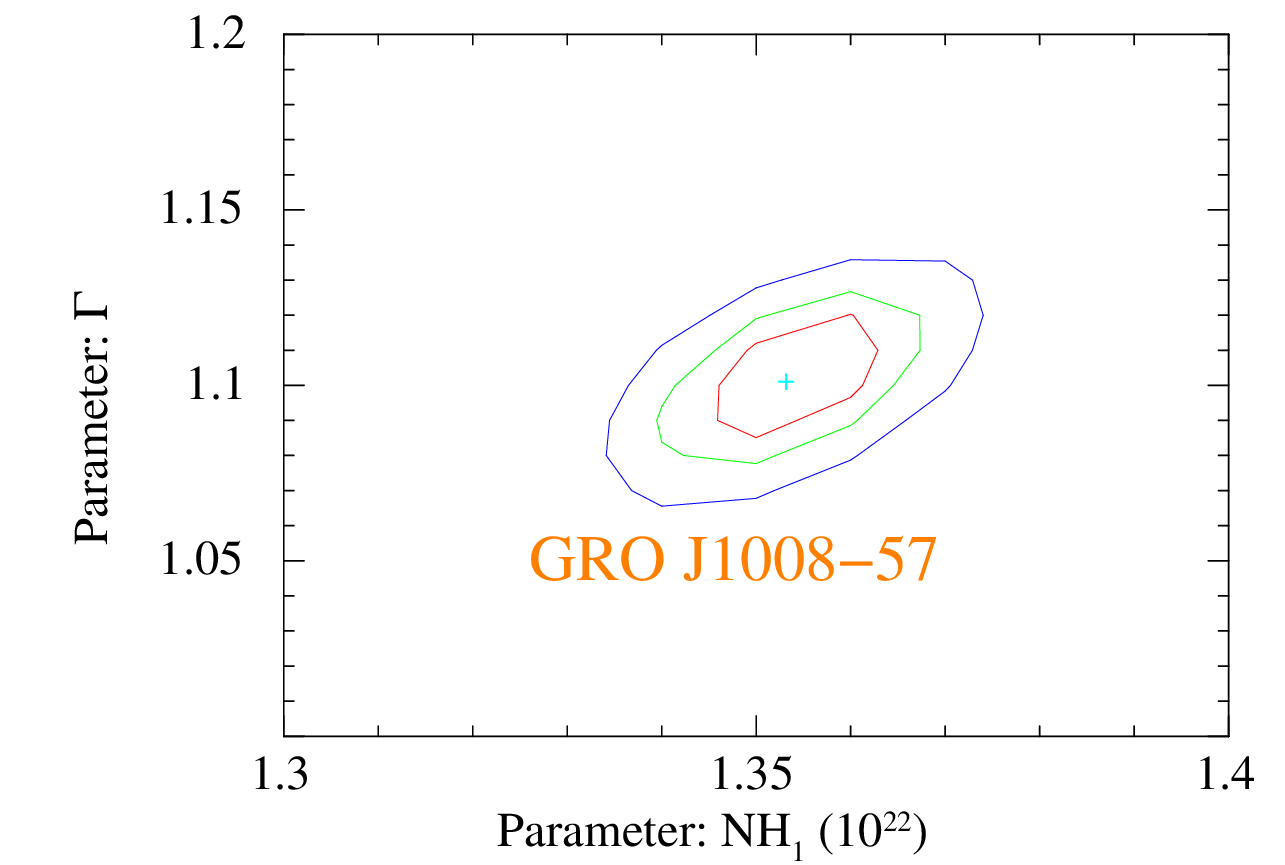}
\includegraphics[scale=0.2,angle=-90]{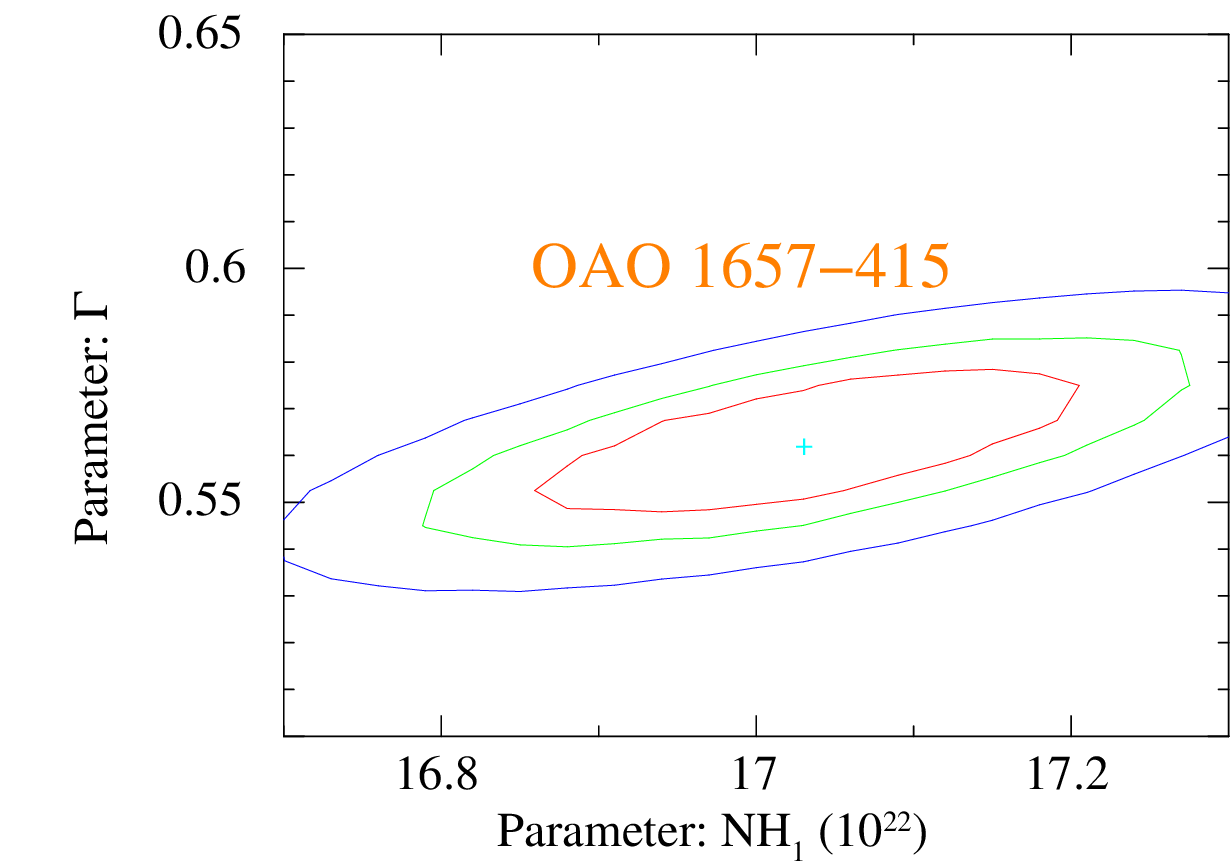}
\includegraphics[scale=0.2,angle=-90]{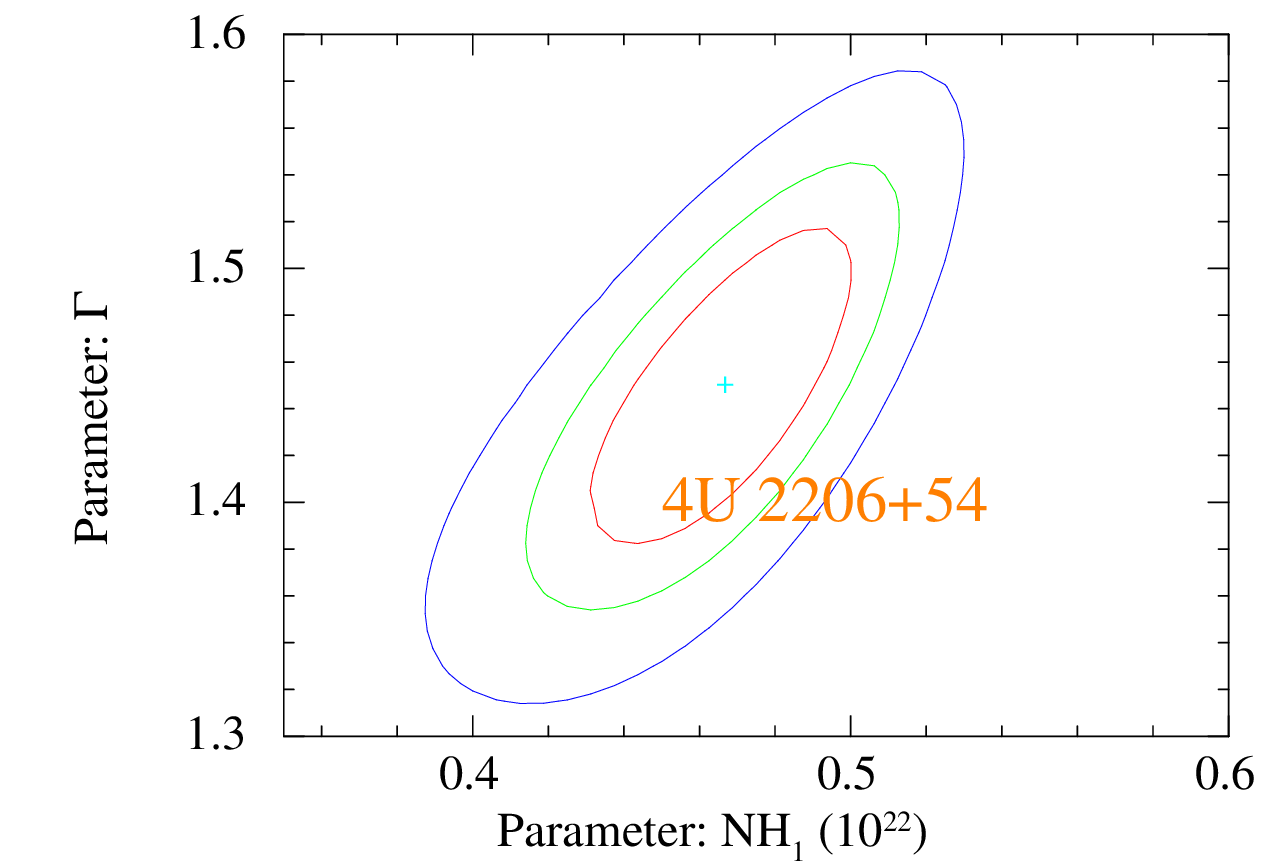}

\vspace{2.5cm}
\caption{Contour plots between NH1 and gamma}
\label{contour1}
\end{figure*}

\begin{figure*}
%\hspace{-2cm}
\includegraphics[scale=0.2,angle=-90]{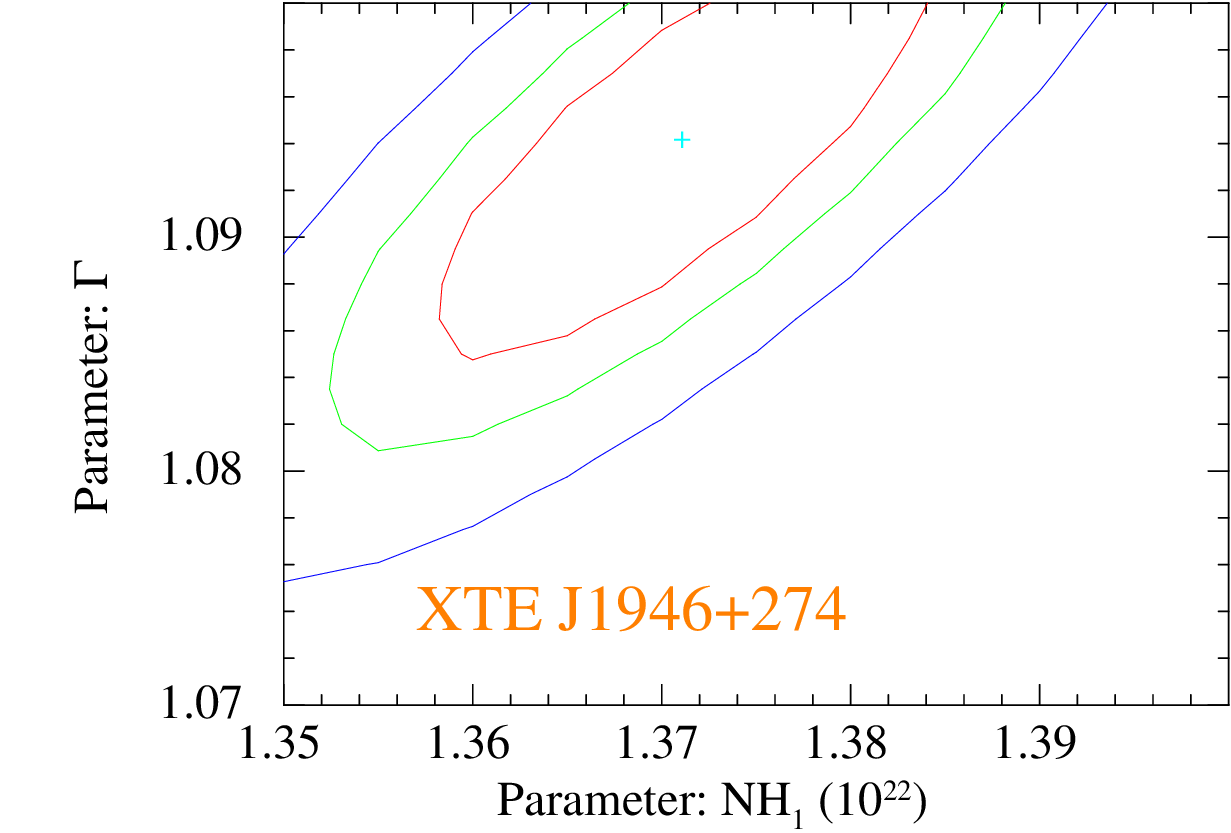}
\includegraphics[scale=0.2,angle=-90]{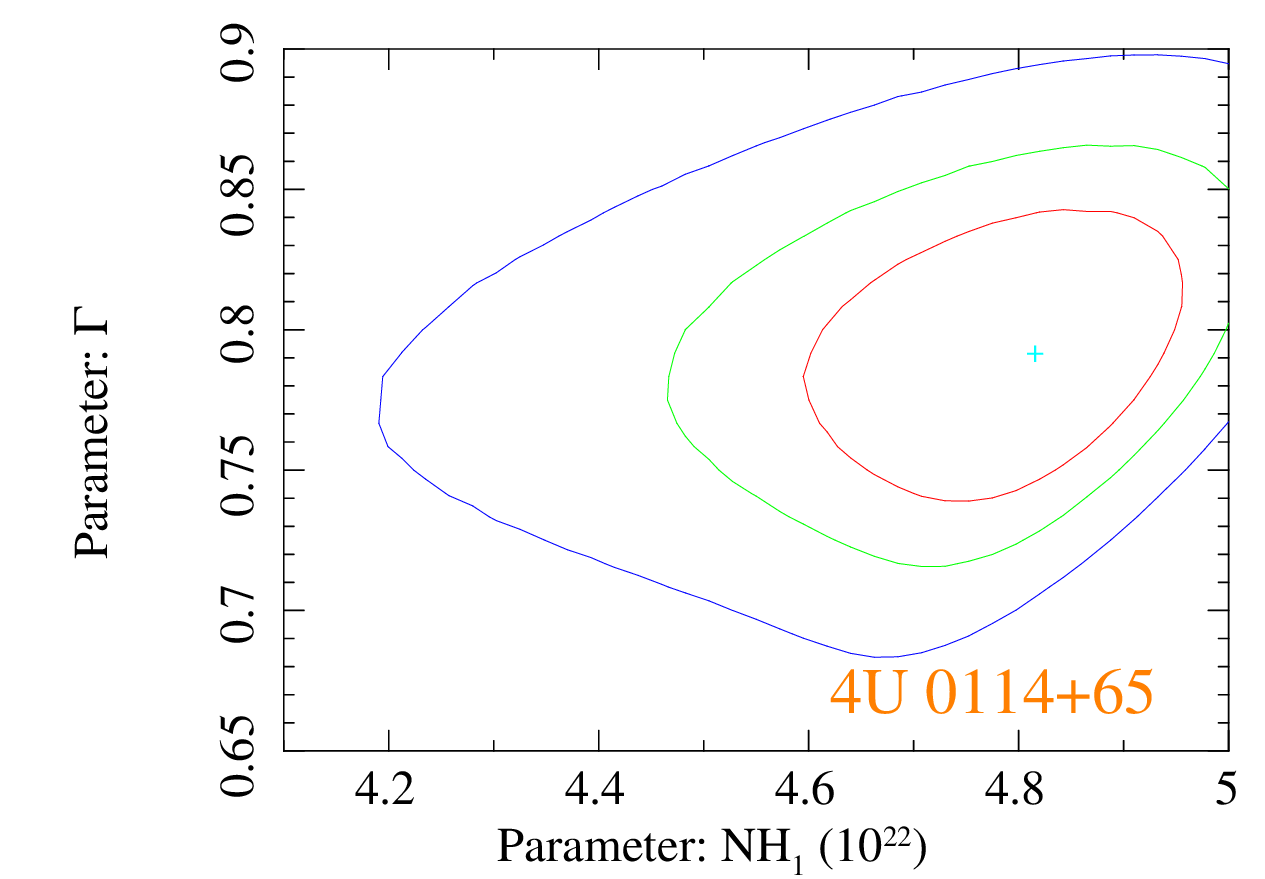}
\includegraphics[scale=0.2,angle=-90]{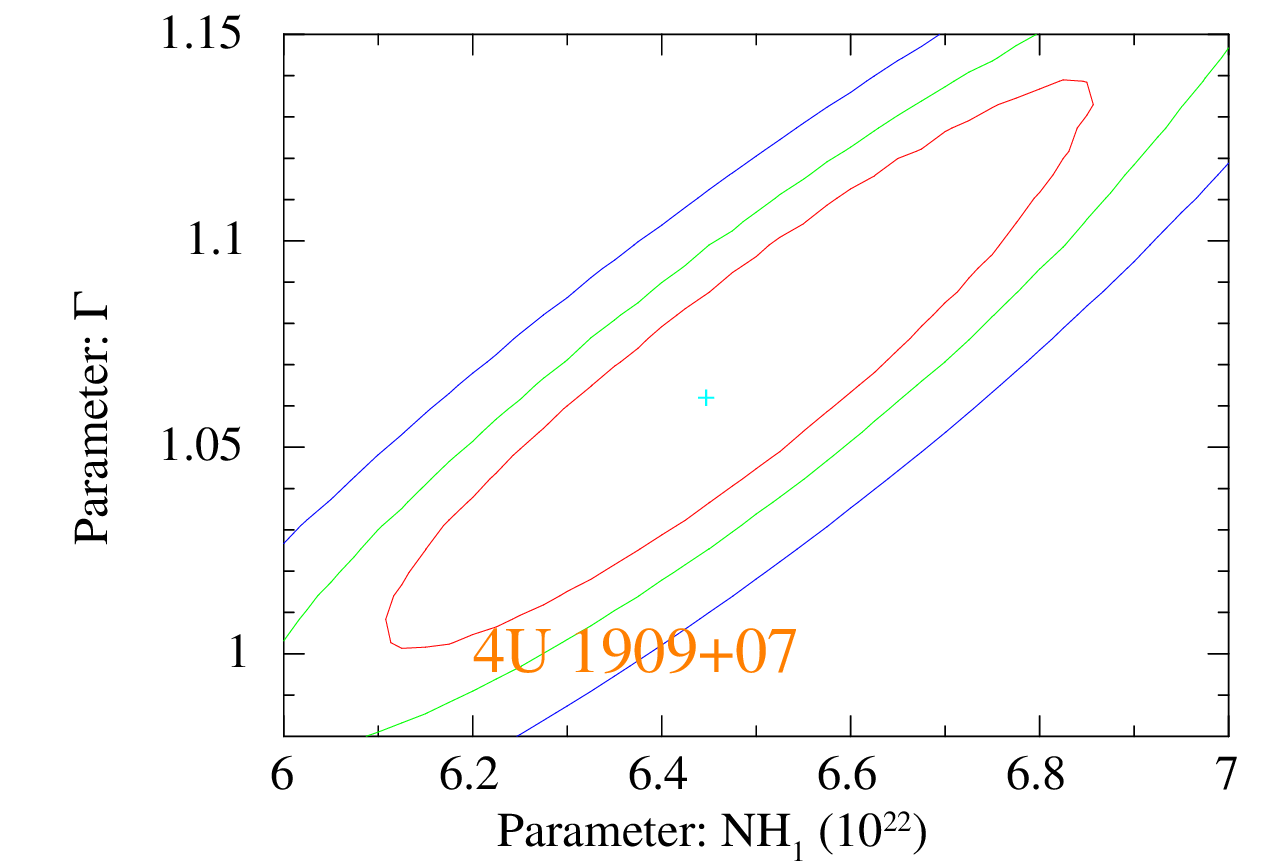}
\includegraphics[scale=0.2,angle=-90]{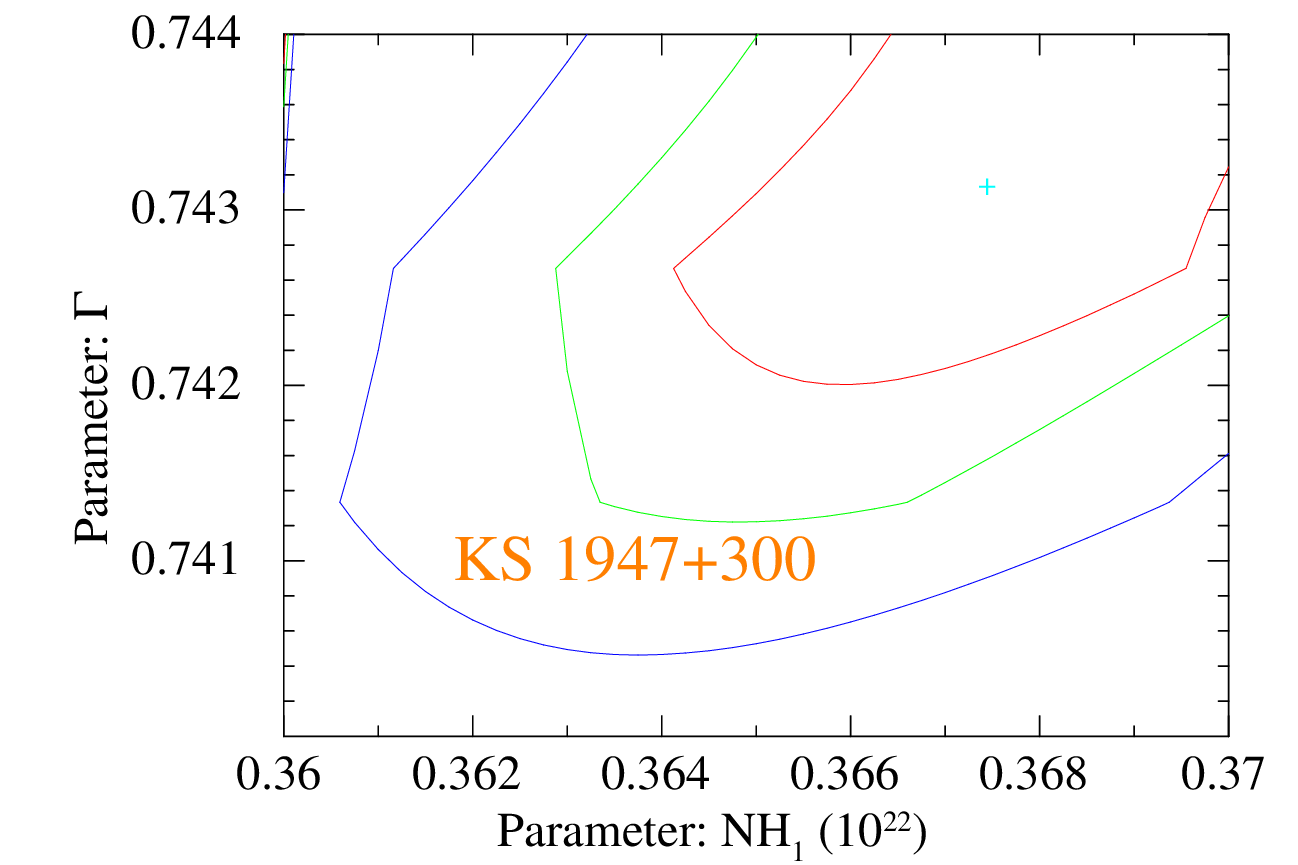}
\includegraphics[scale=0.2,angle=-90]{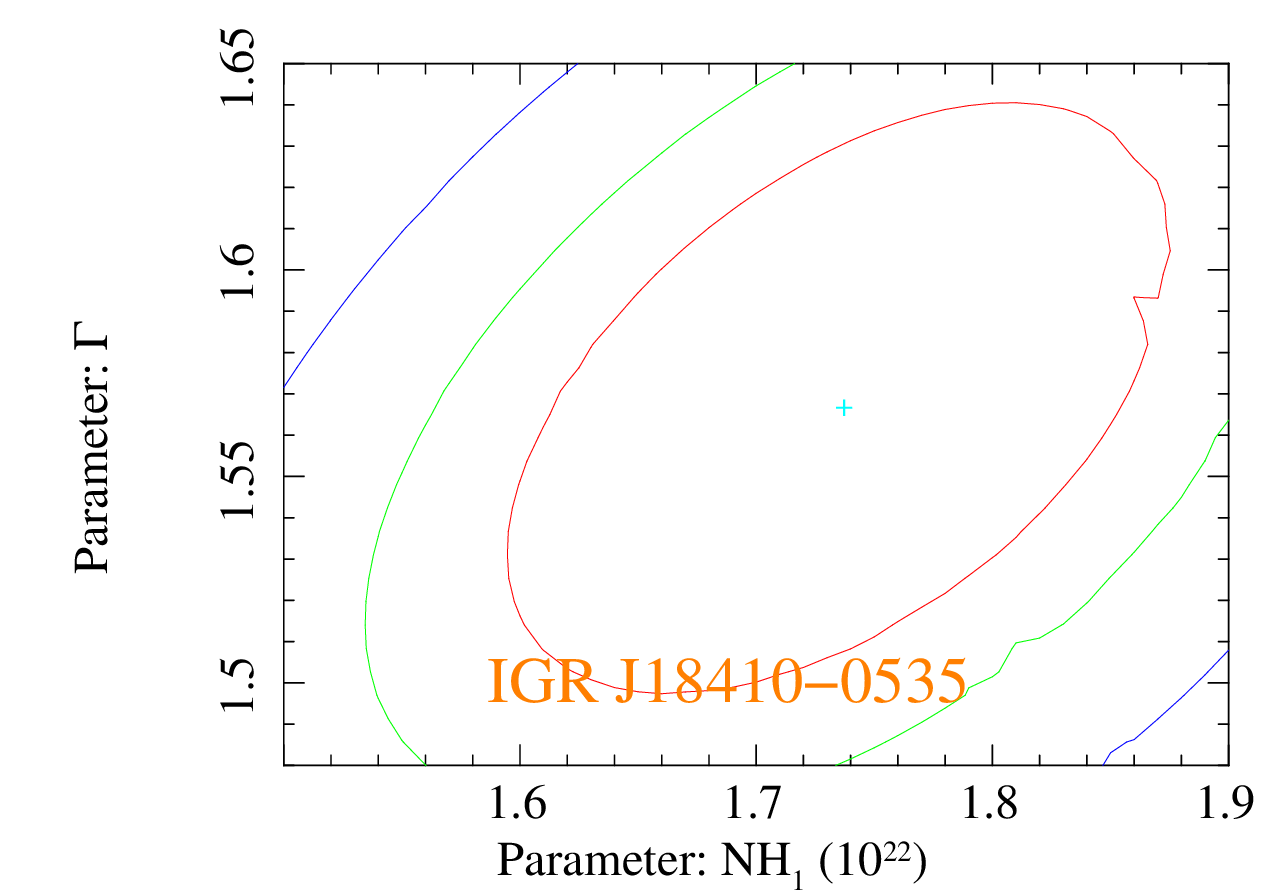}
\includegraphics[scale=0.2,angle=-90]{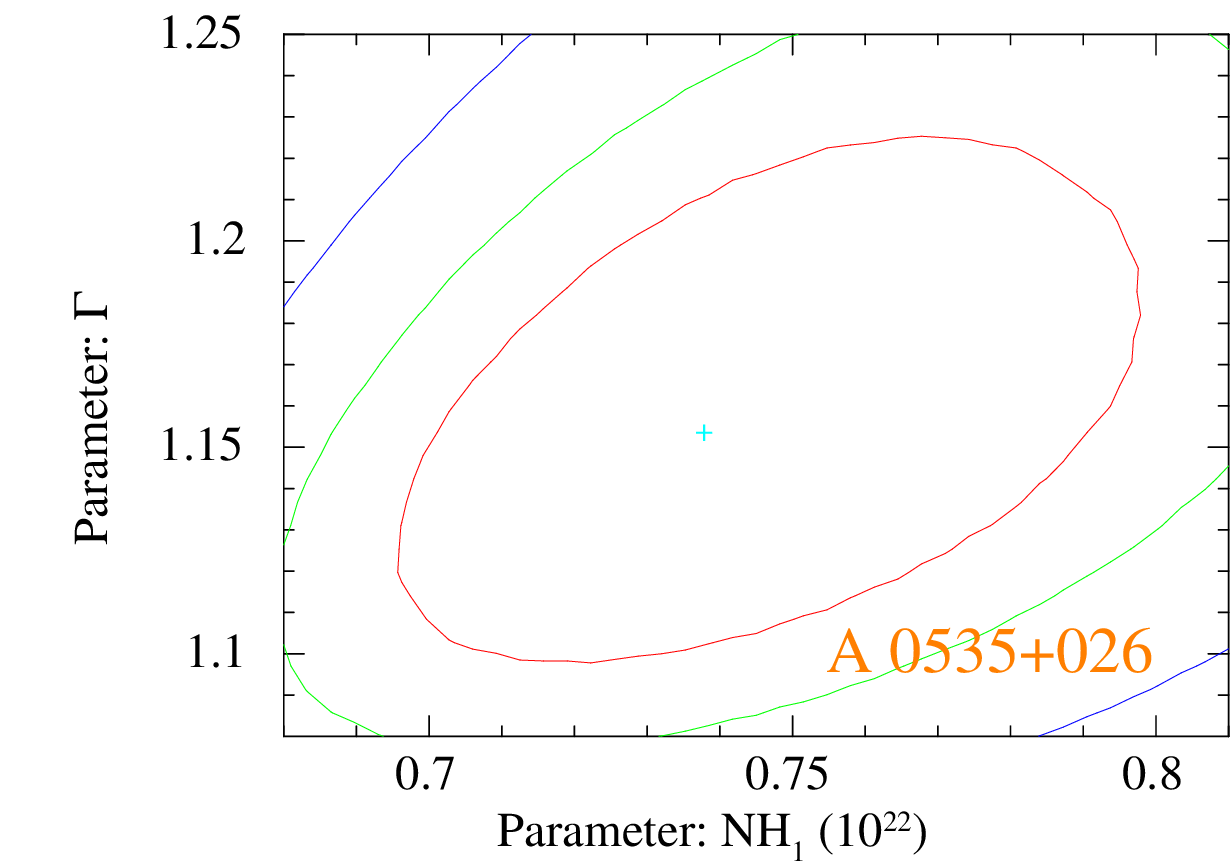}
\includegraphics[scale=0.2,angle=-90]{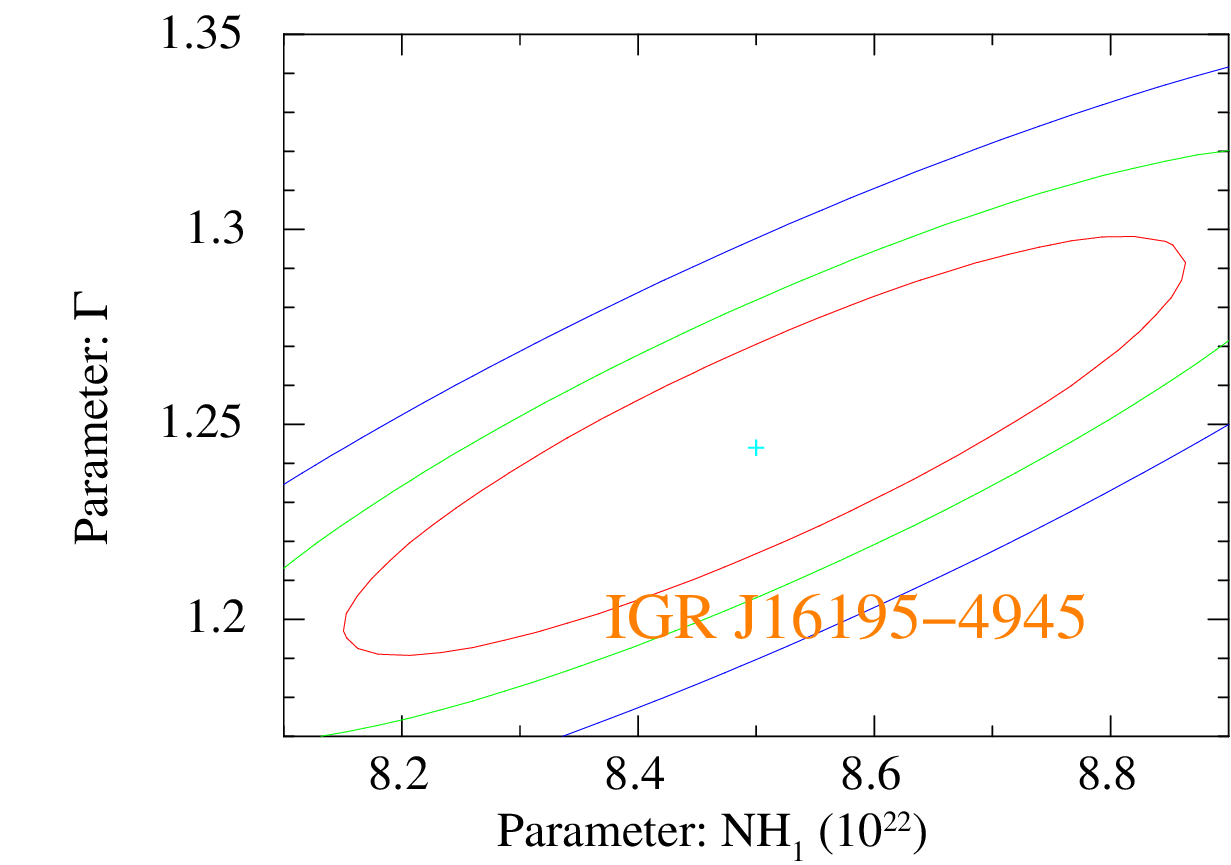}
\includegraphics[scale=0.2,angle=-90]{cont-igrj16493.ps}
\includegraphics[scale=0.2,angle=-90]{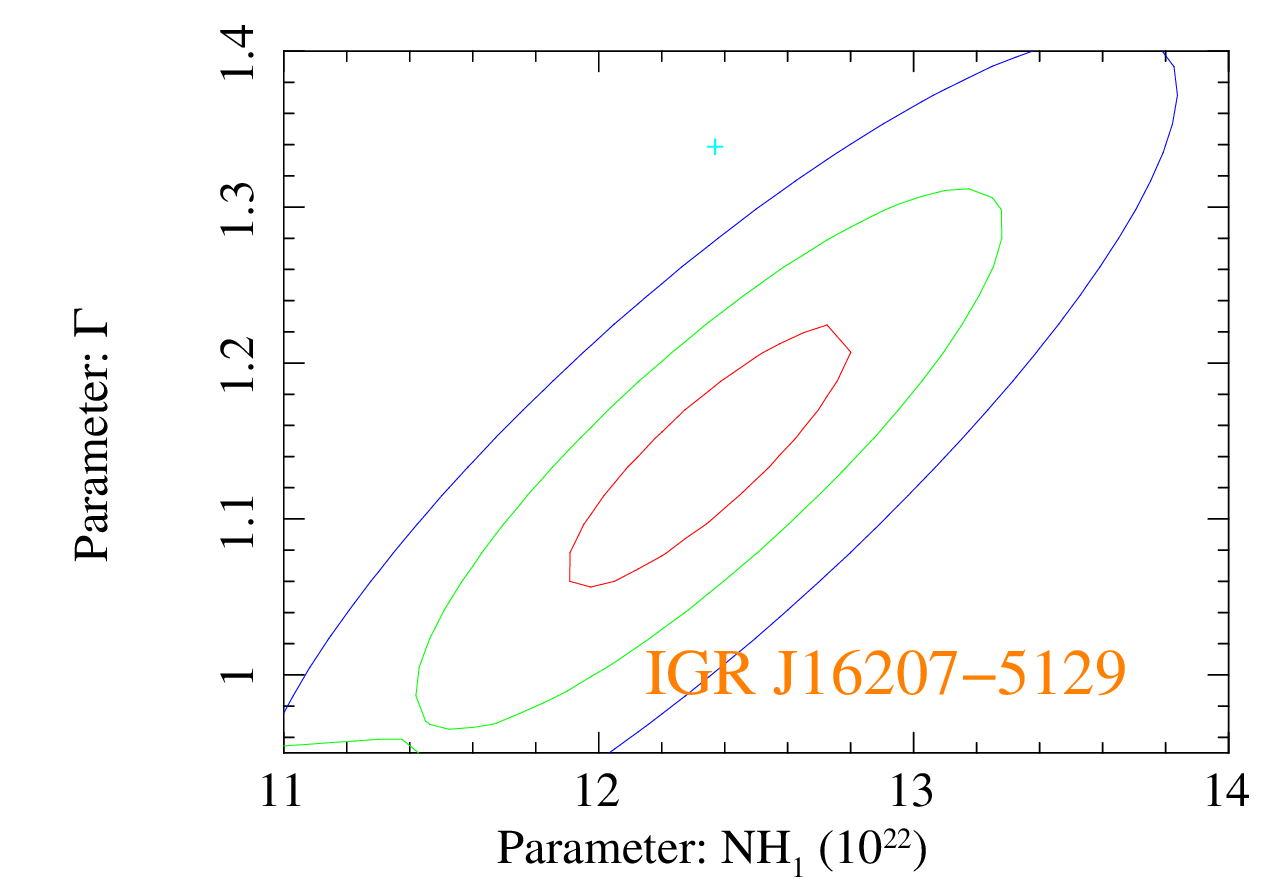}
\includegraphics[scale=0.2,angle=-90]{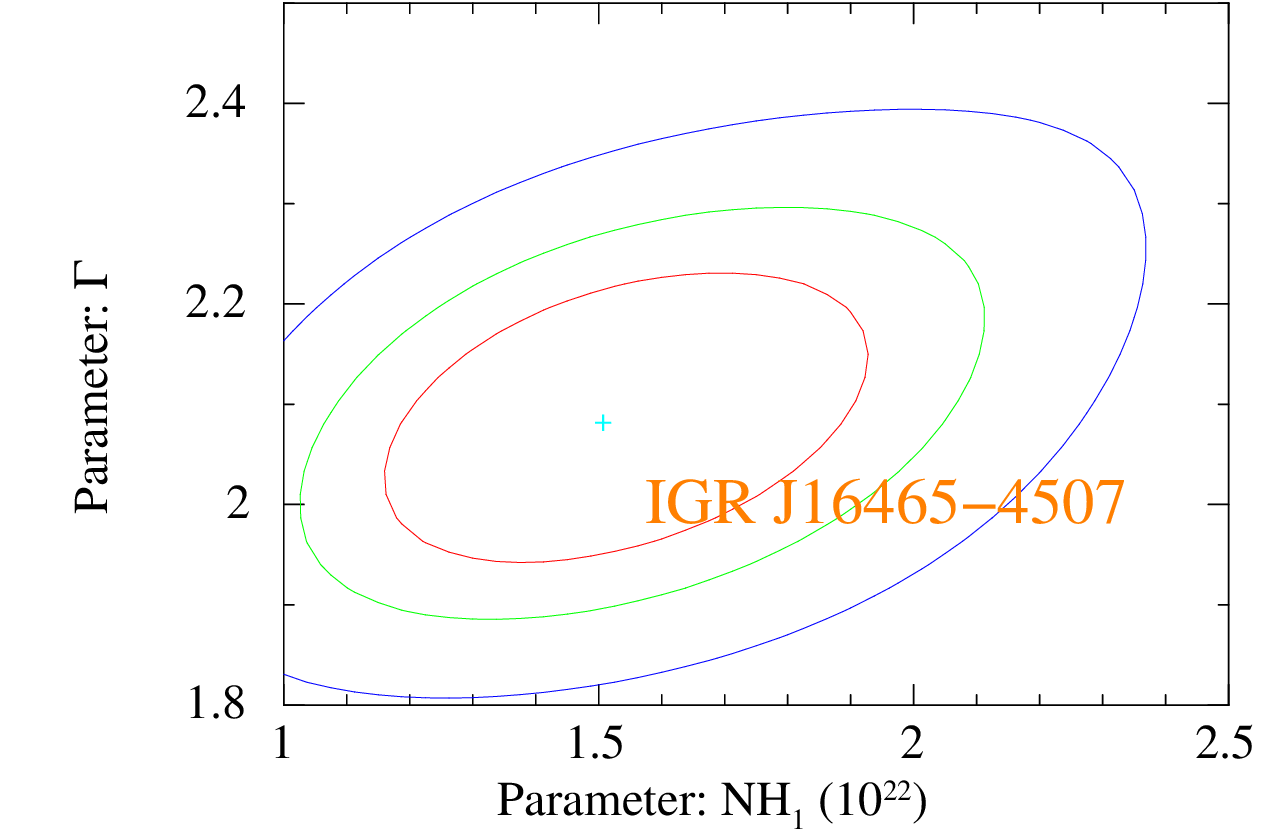}
\includegraphics[scale=0.2,angle=-90]{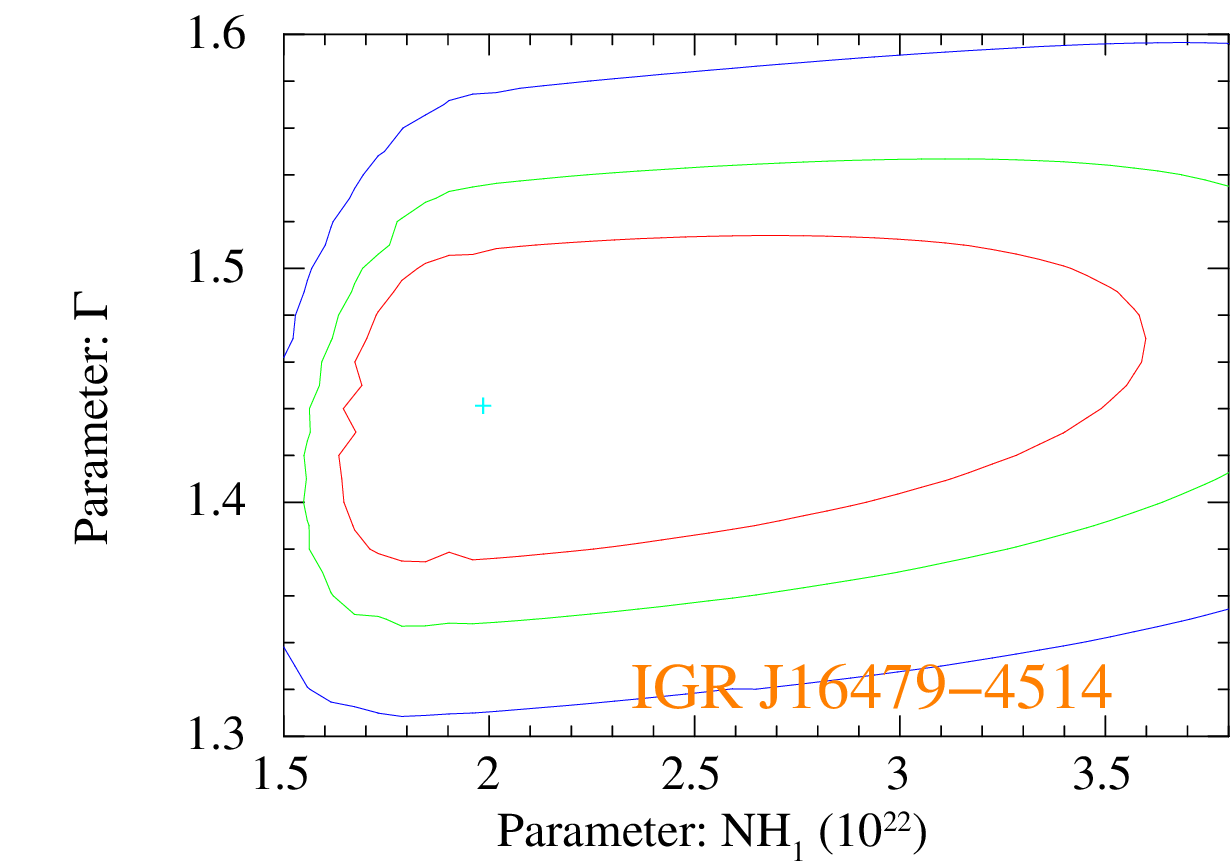}
\includegraphics[scale=0.2,angle=-90]{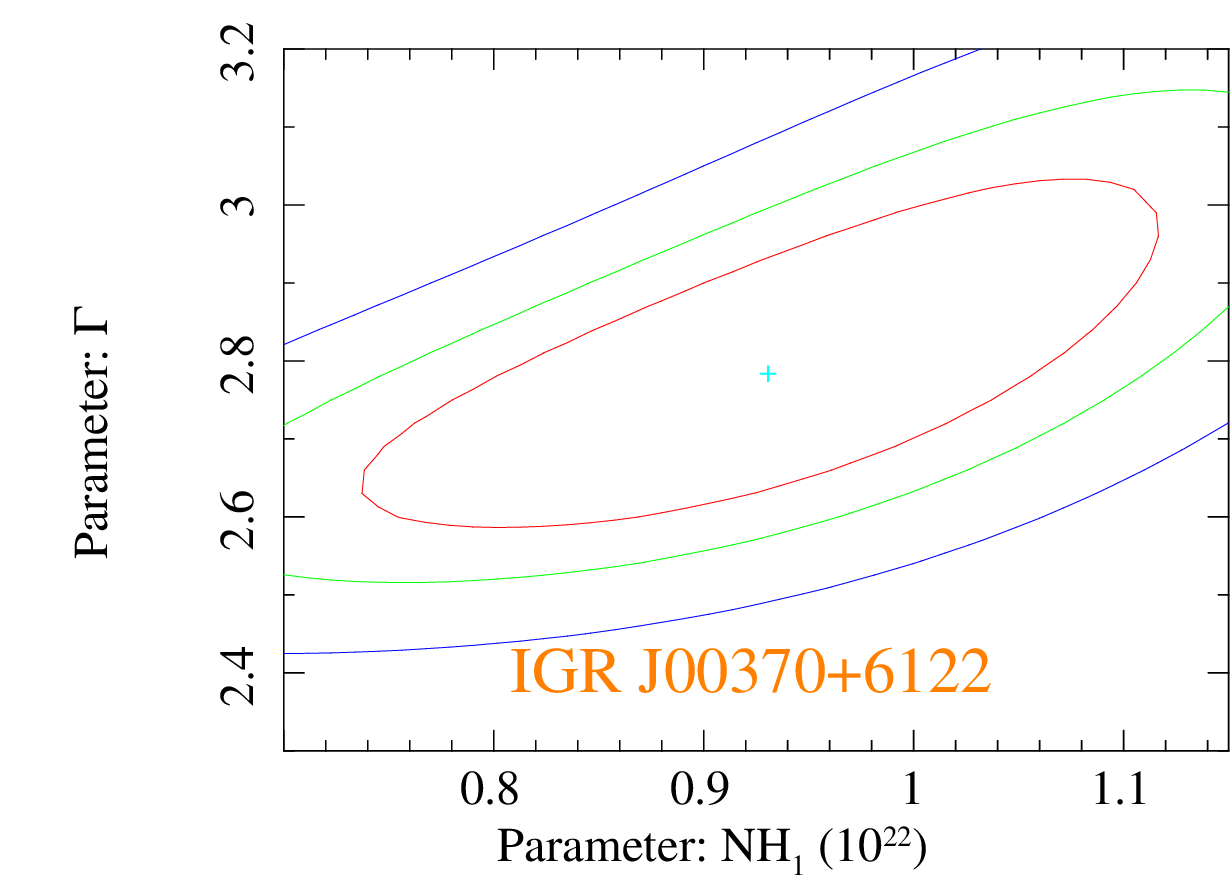}
\includegraphics[scale=0.2,angle=-90]{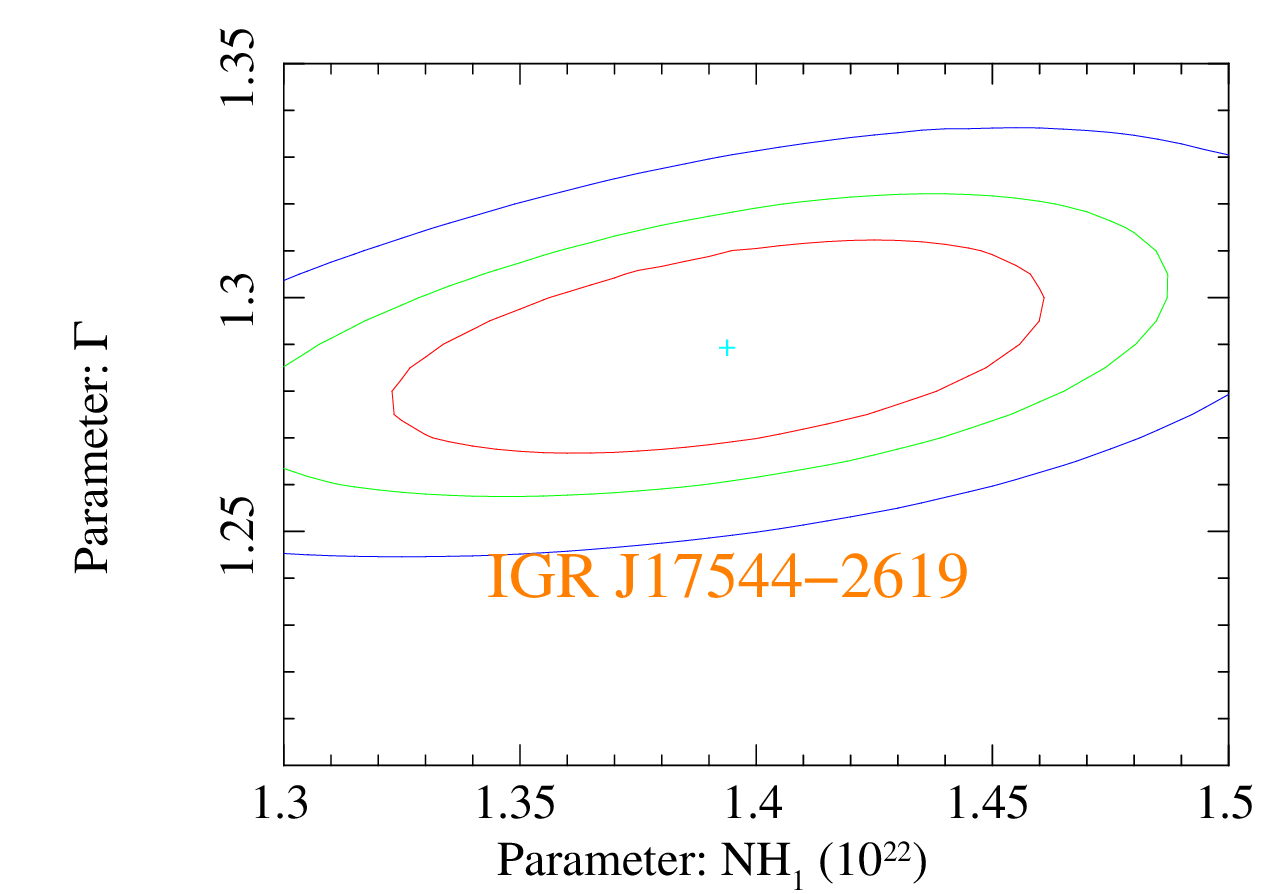}
\includegraphics[scale=0.2,angle=-90]{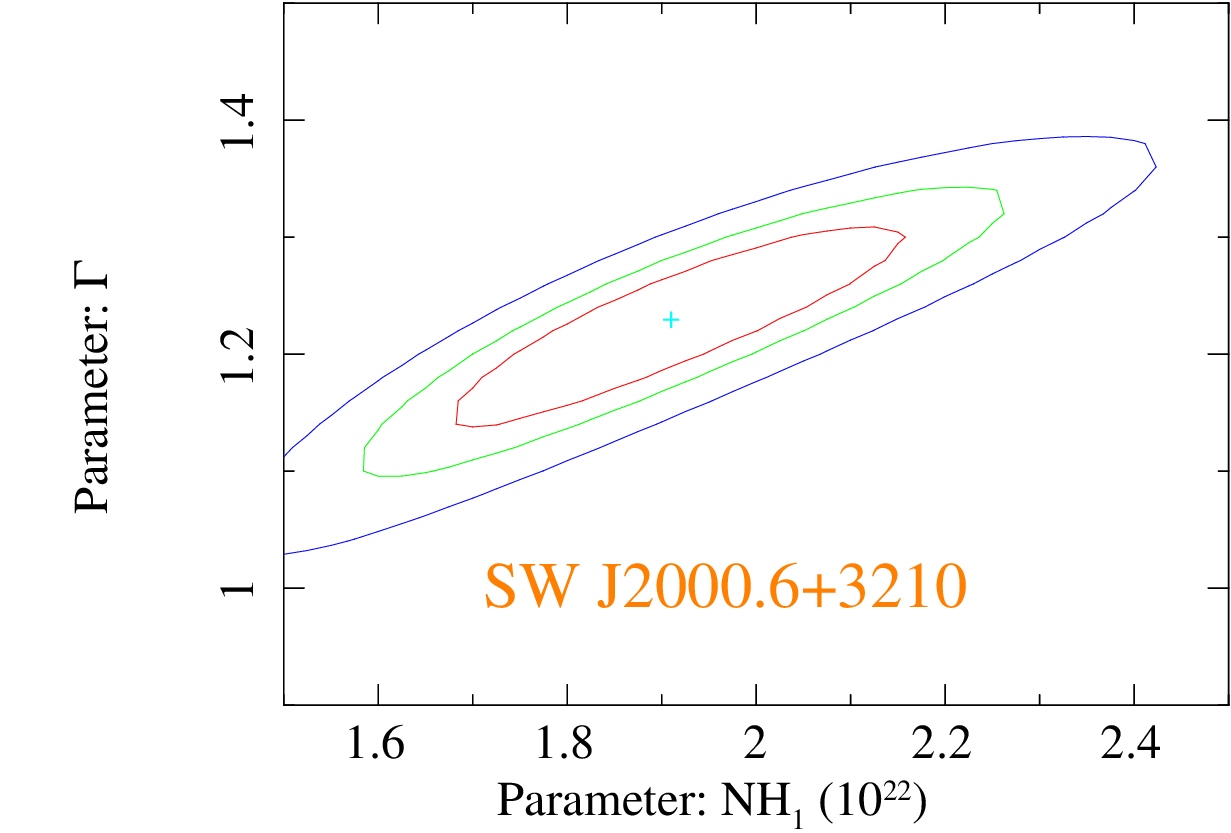}
\includegraphics[scale=0.2,angle=-90]{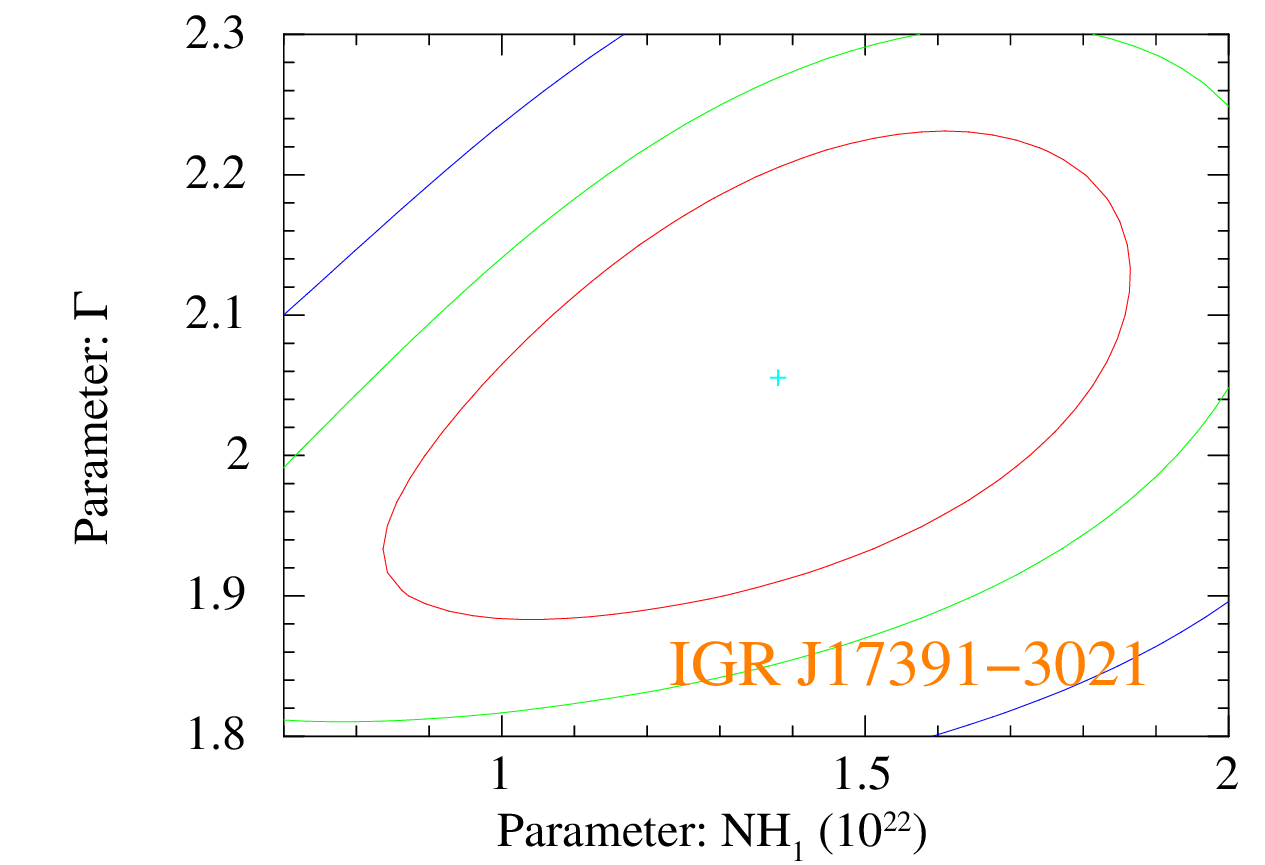}
\includegraphics[scale=0.2,angle=-90]{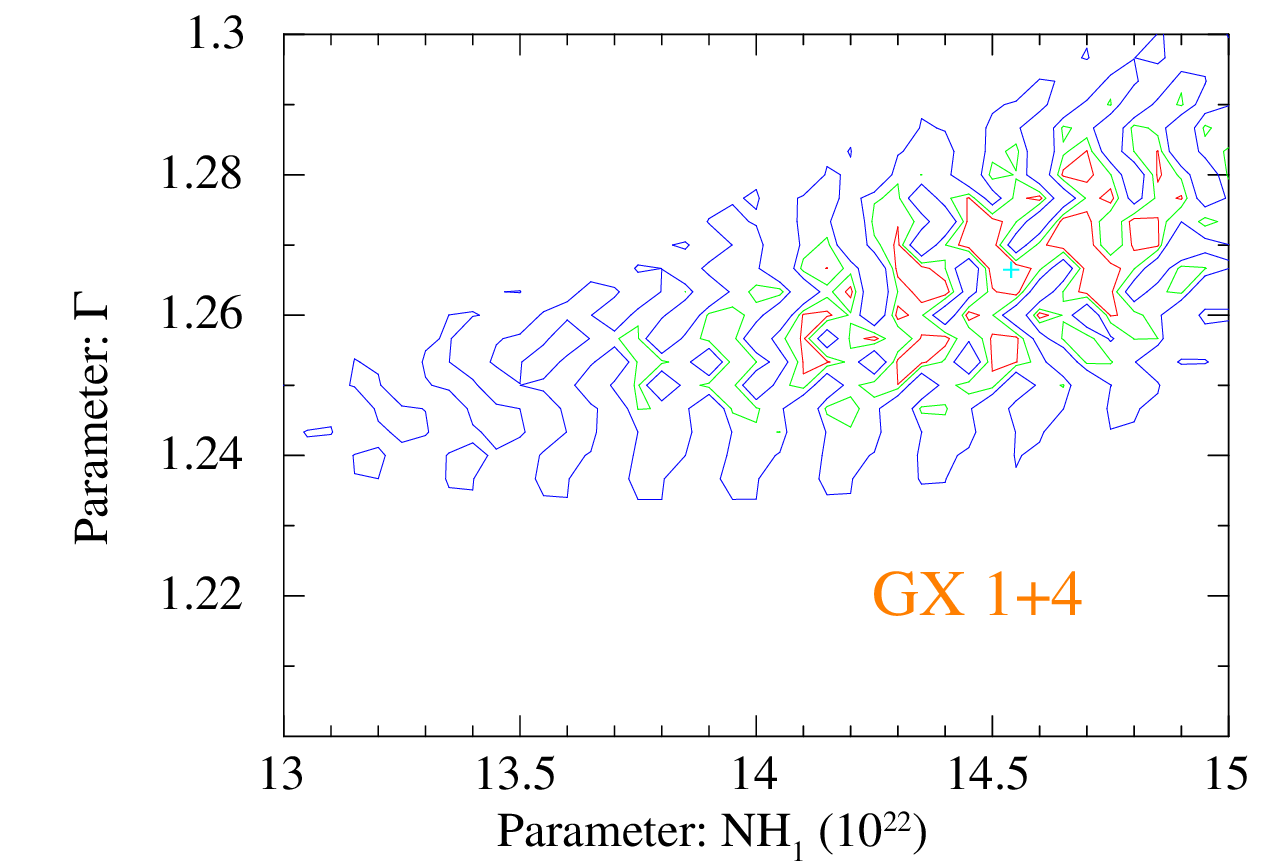}
\vspace{2.5cm}
\caption{Contour plots between NH1 and gamma}
\label{contour2}
\end{figure*}

\begin{figure*}
%\hspace{-2cm}
\includegraphics[scale=0.2,angle=-90]{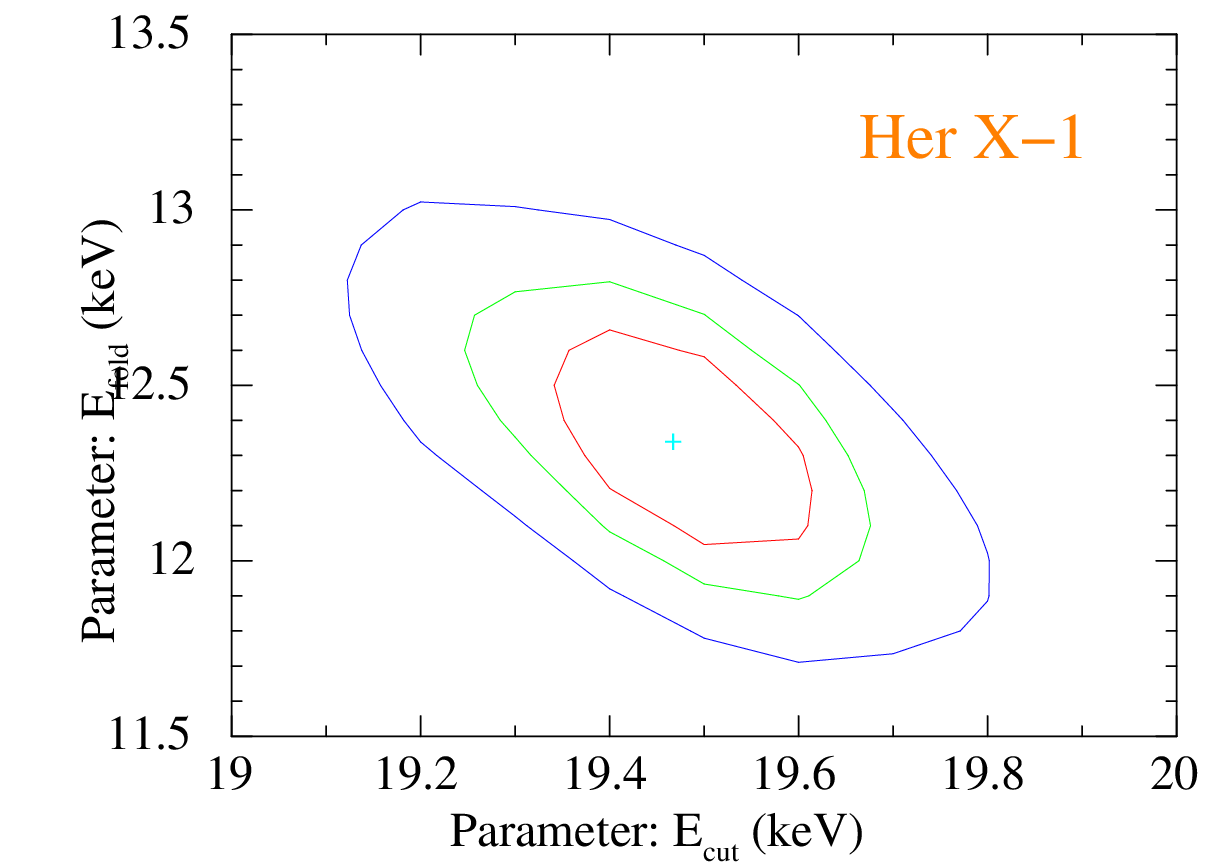}
\includegraphics[scale=0.2,angle=-90]{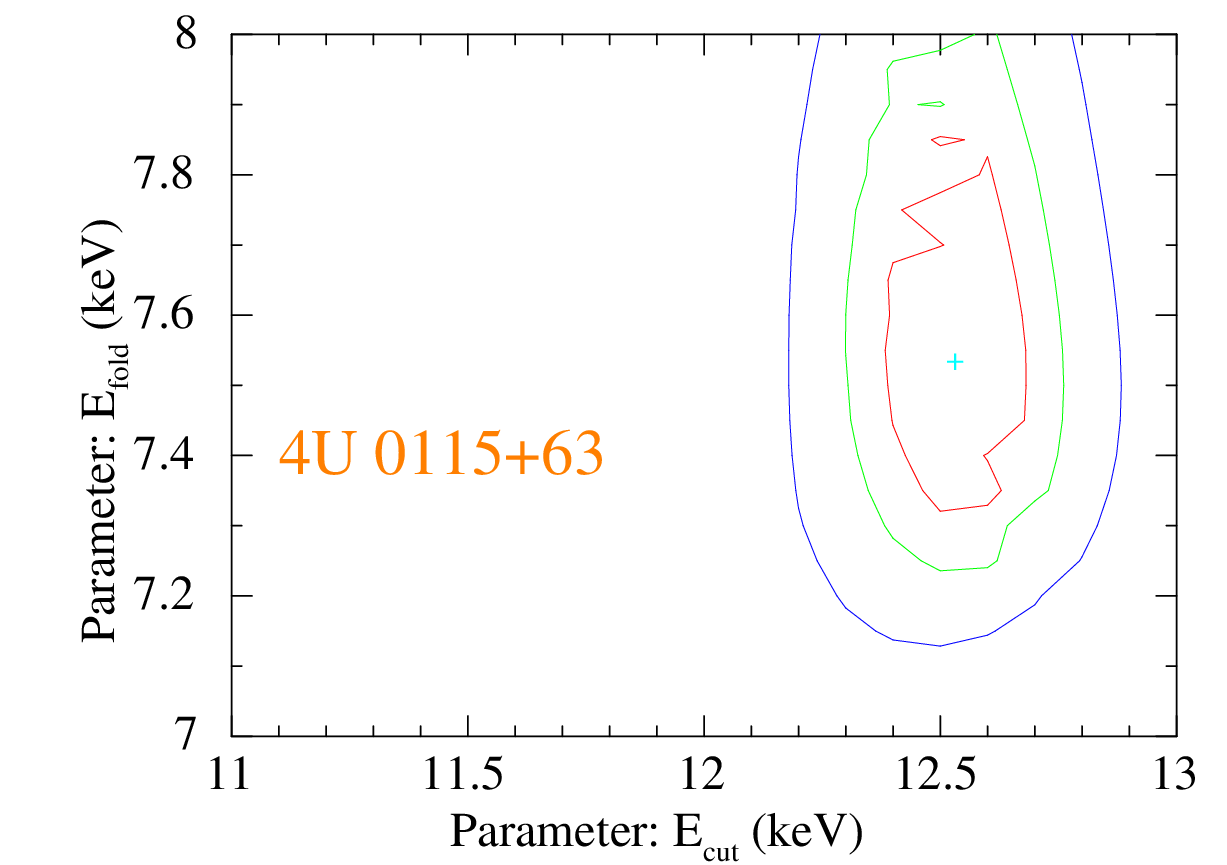}
\includegraphics[scale=0.2,angle=-90]{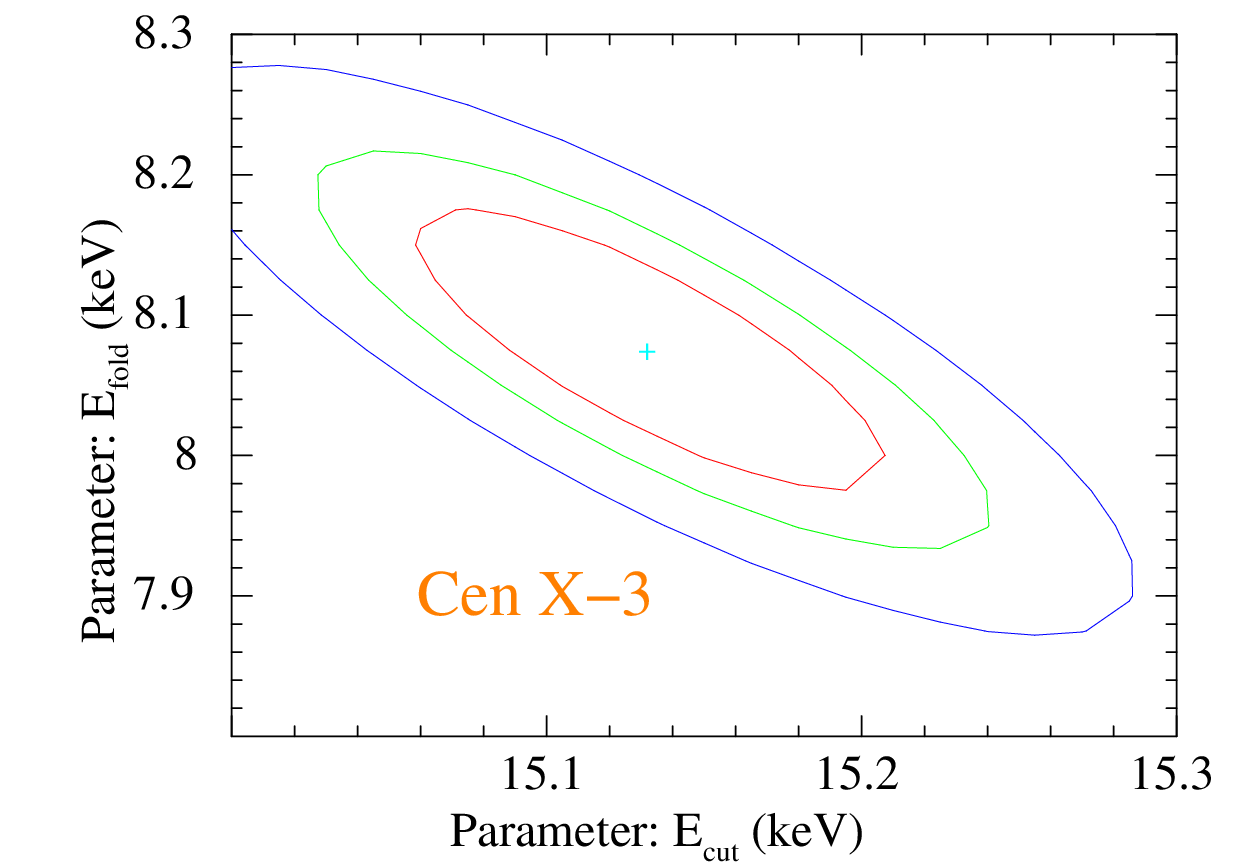}
\includegraphics[scale=0.2,angle=-90]{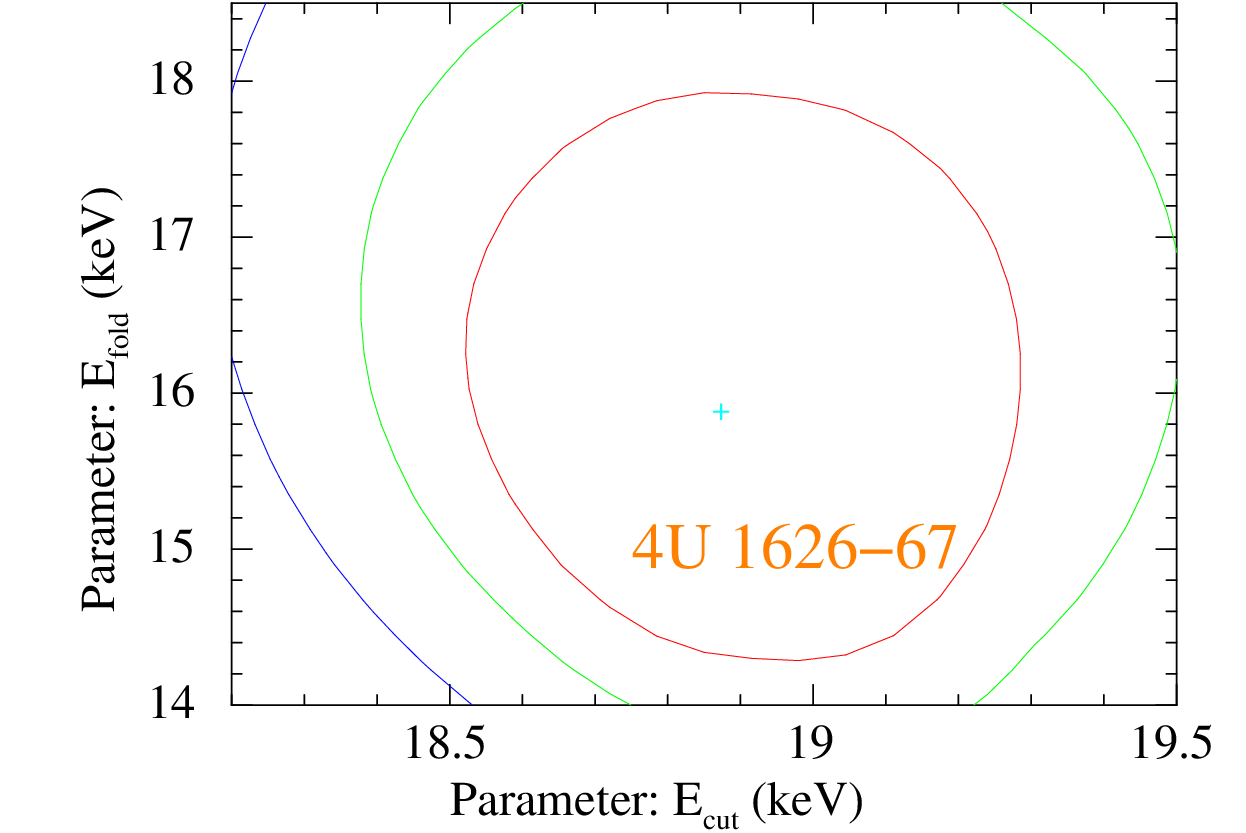}
\includegraphics[scale=0.2,angle=-90]{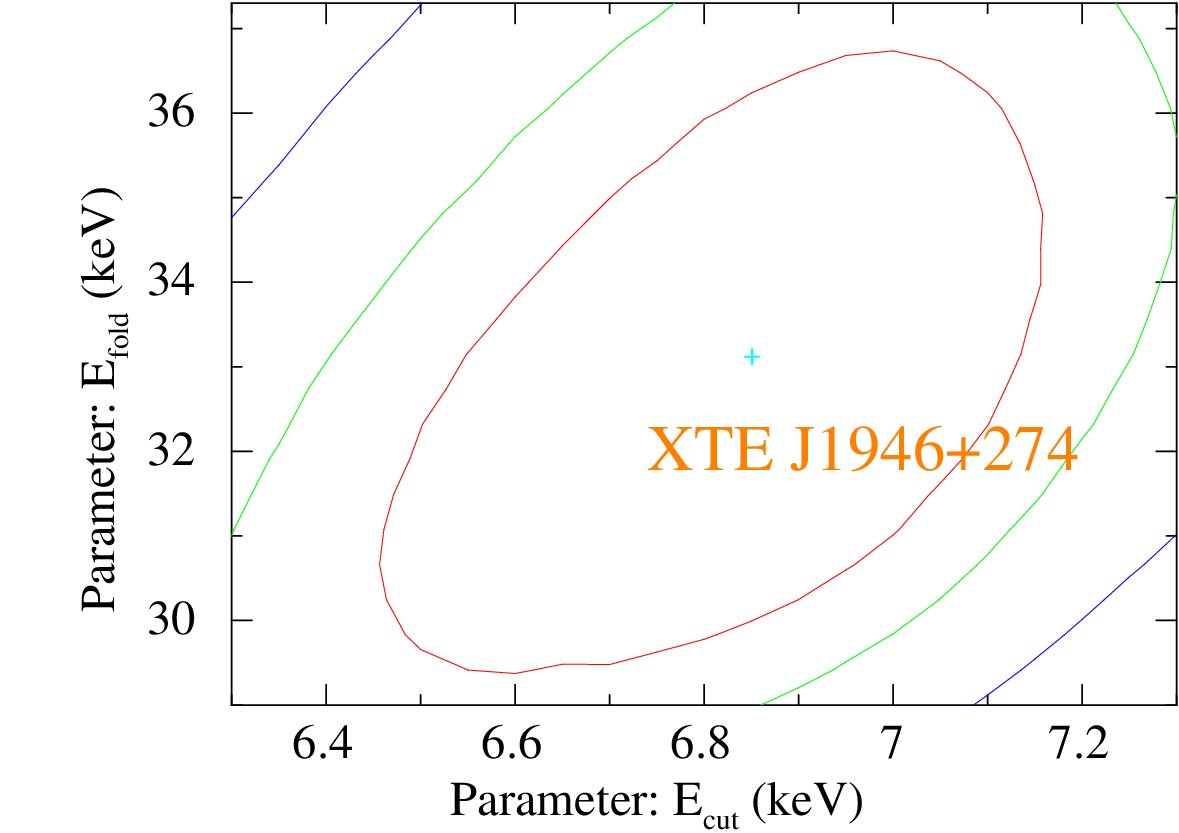}
\includegraphics[scale=0.2,angle=-90]{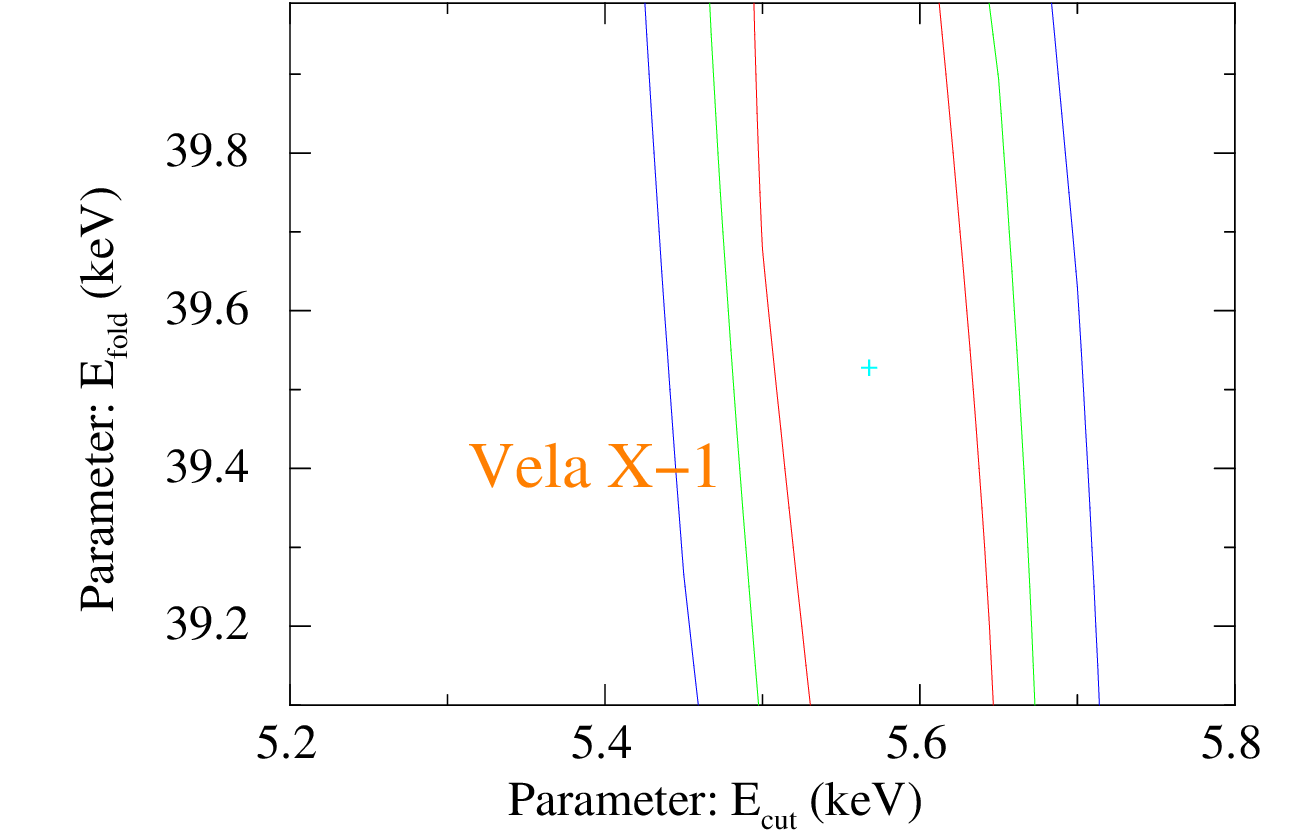}
\includegraphics[scale=0.2,angle=-90]{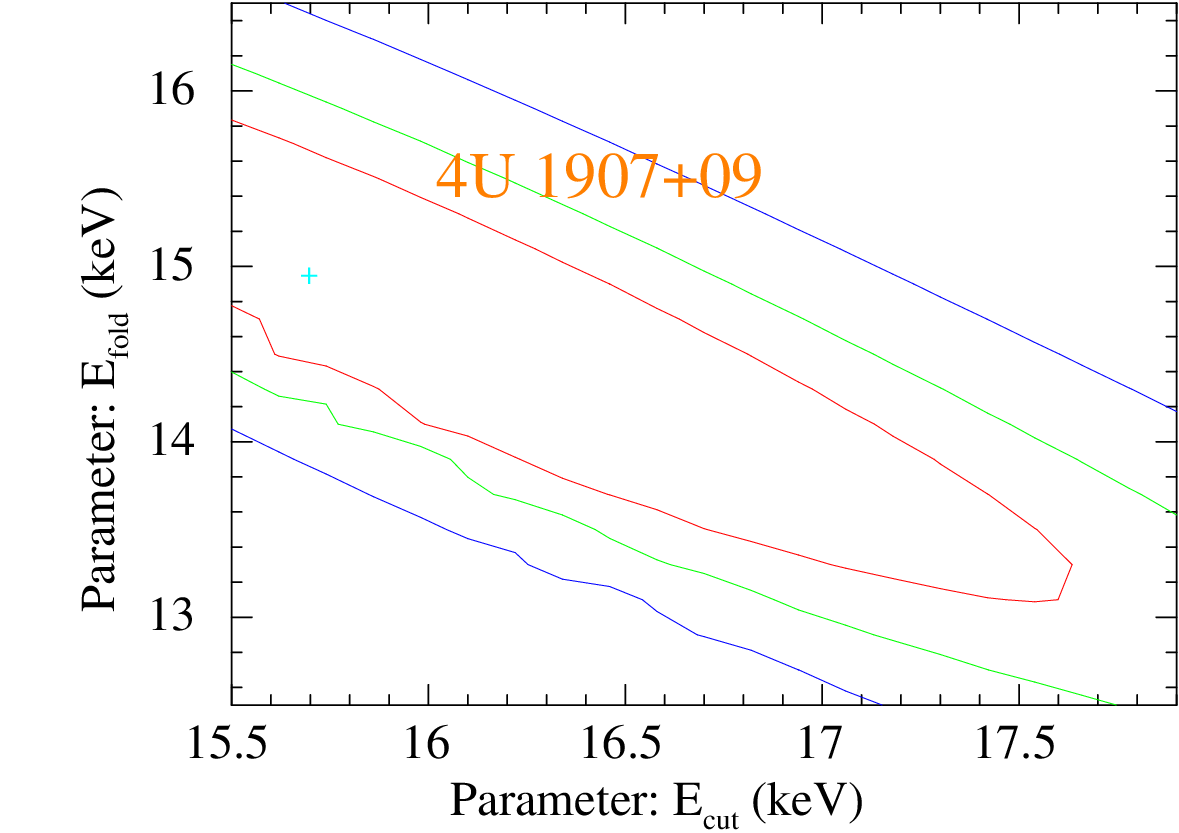}
\includegraphics[scale=0.2,angle=-90]{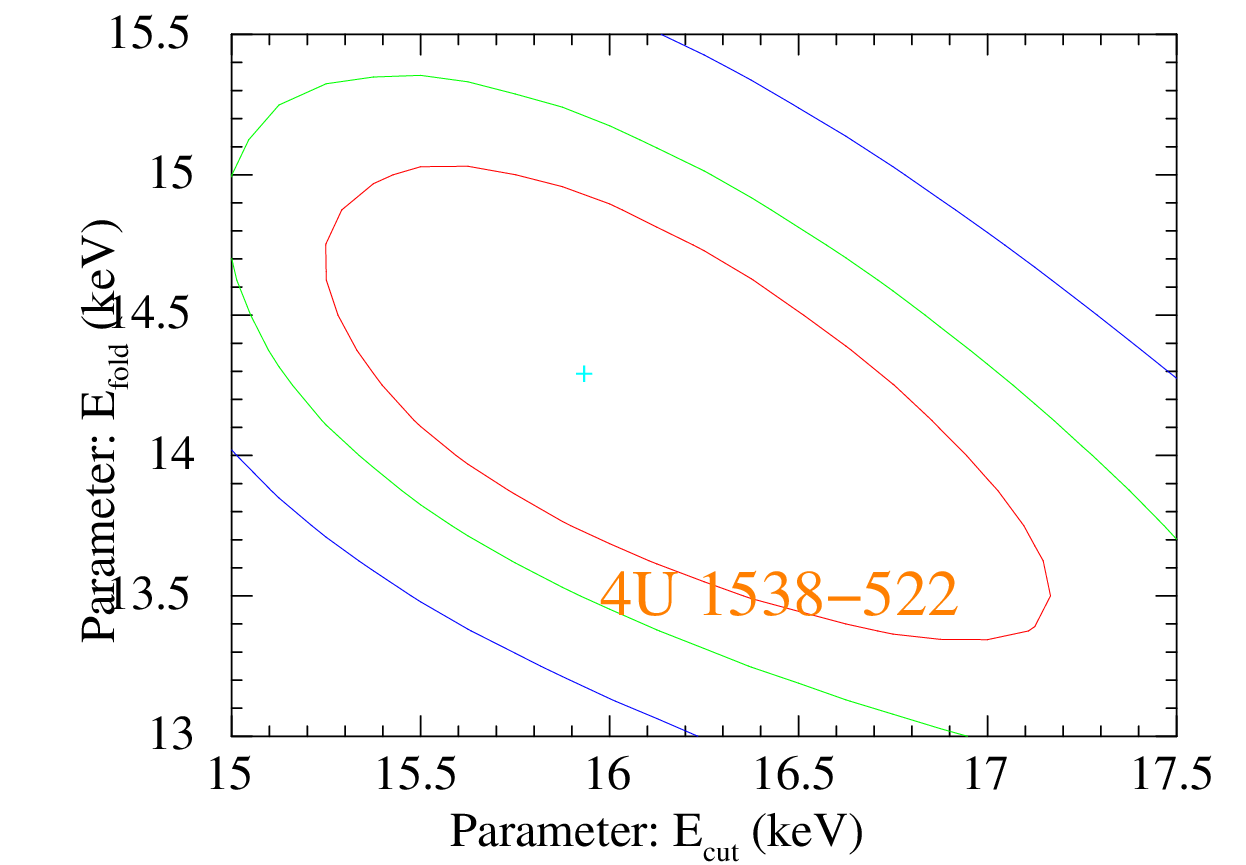}
\includegraphics[scale=0.2,angle=-90]{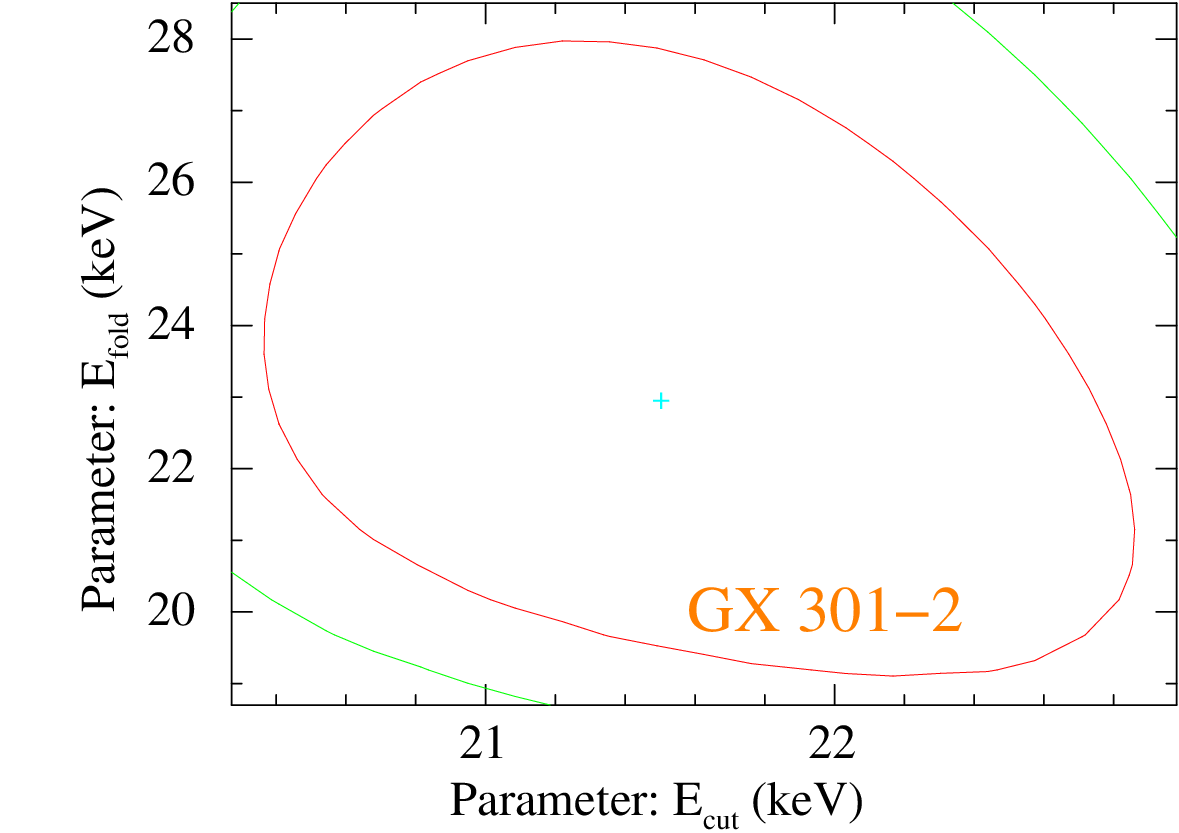}
\includegraphics[scale=0.2,angle=-90]{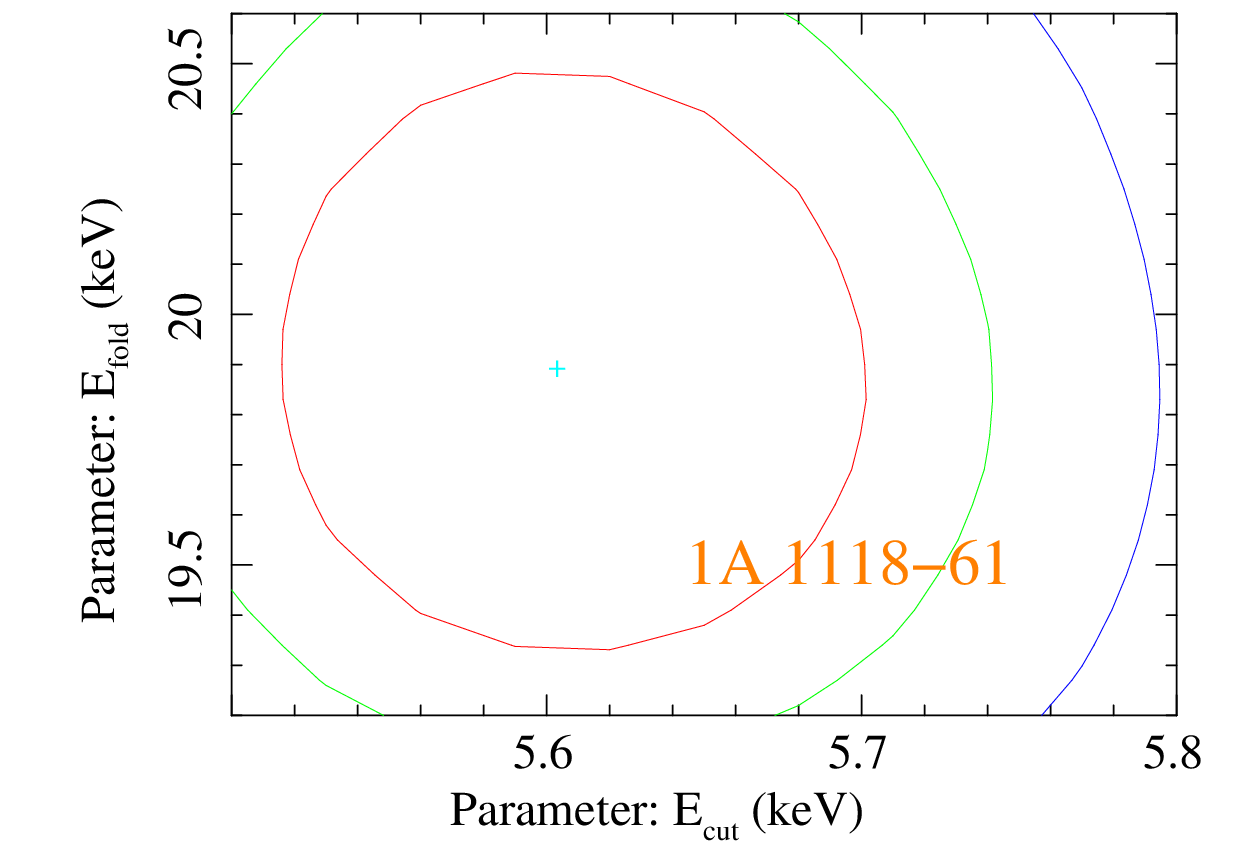}
\includegraphics[scale=0.2,angle=-90]{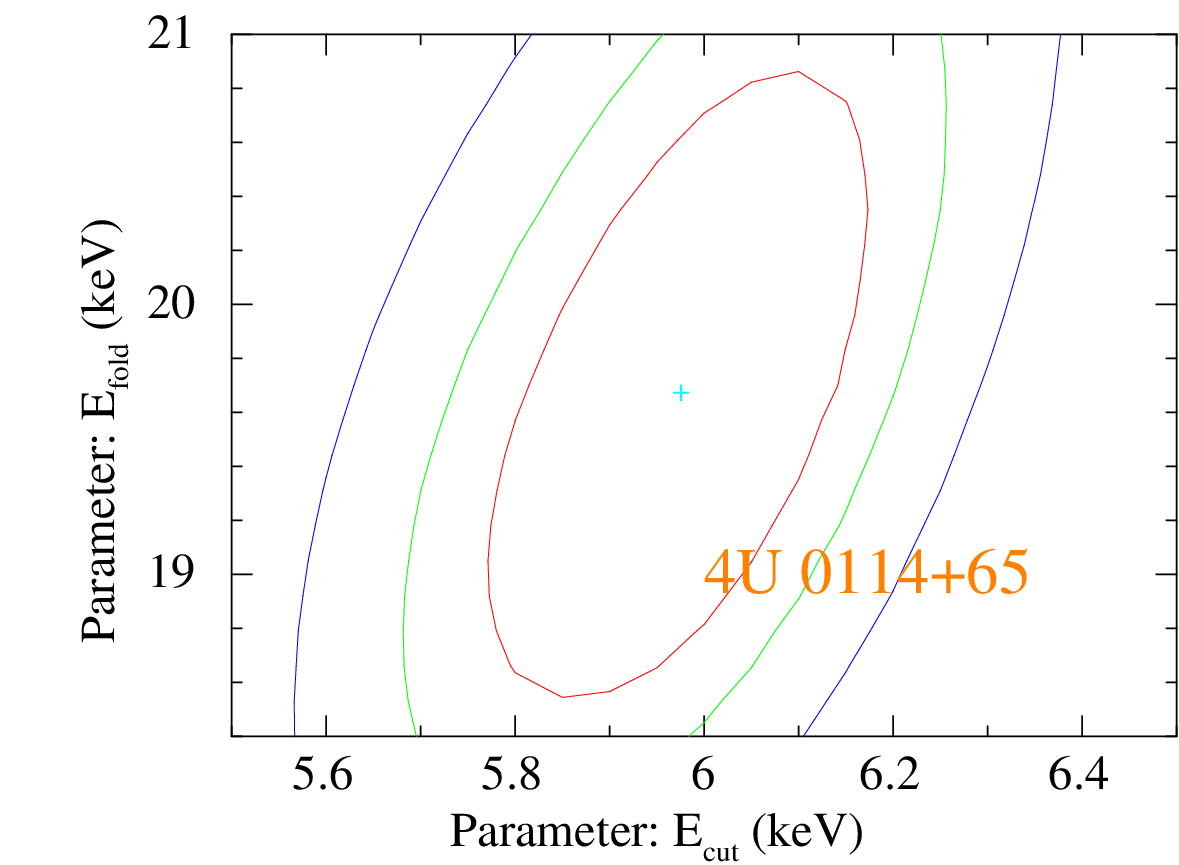}
\includegraphics[scale=0.2,angle=-90]{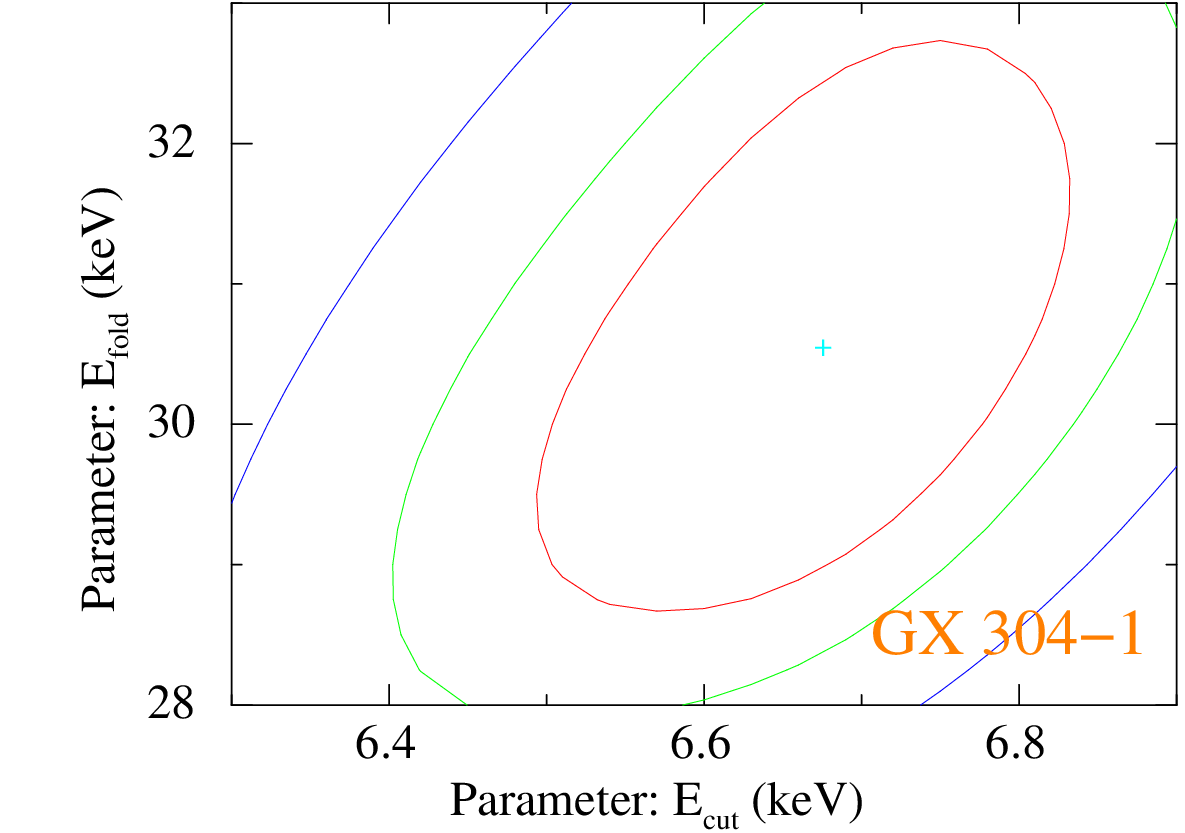}
\includegraphics[scale=0.2,angle=-90]{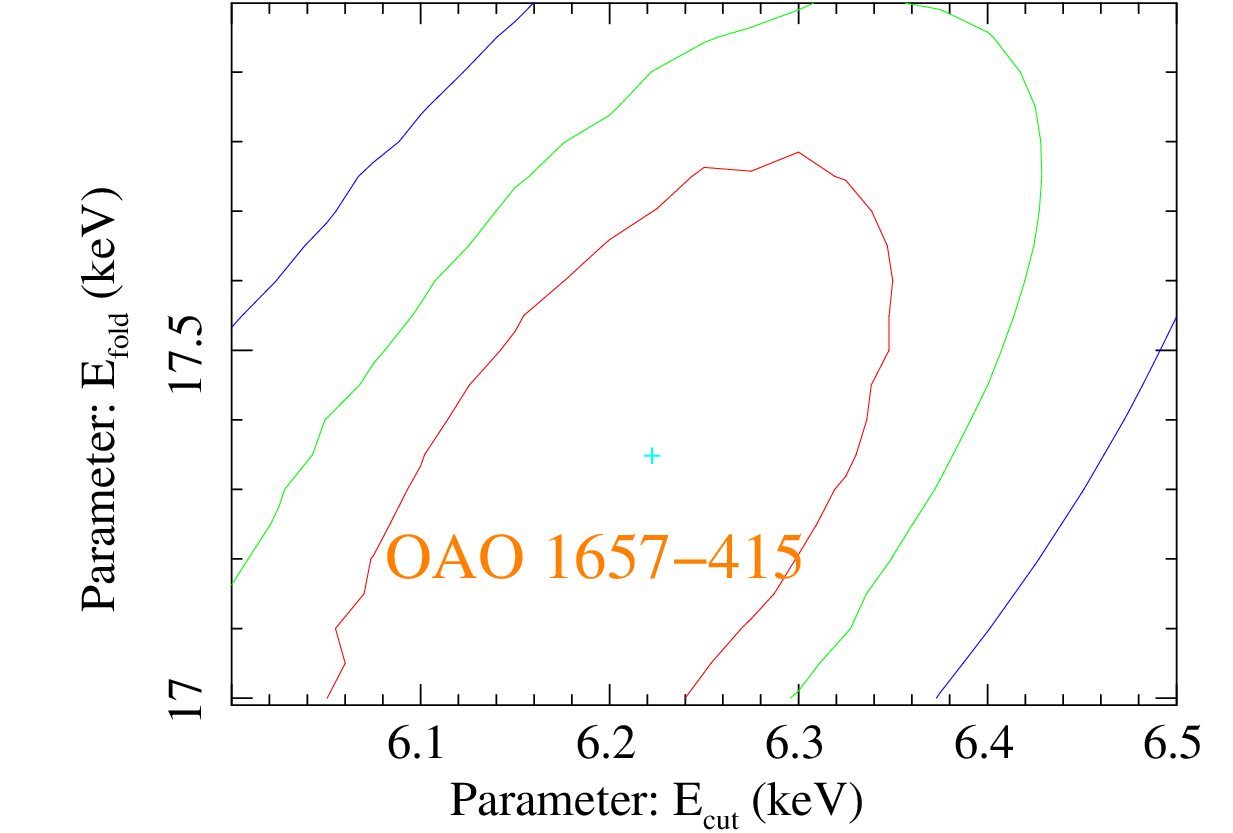}
\includegraphics[scale=0.2,angle=-90]{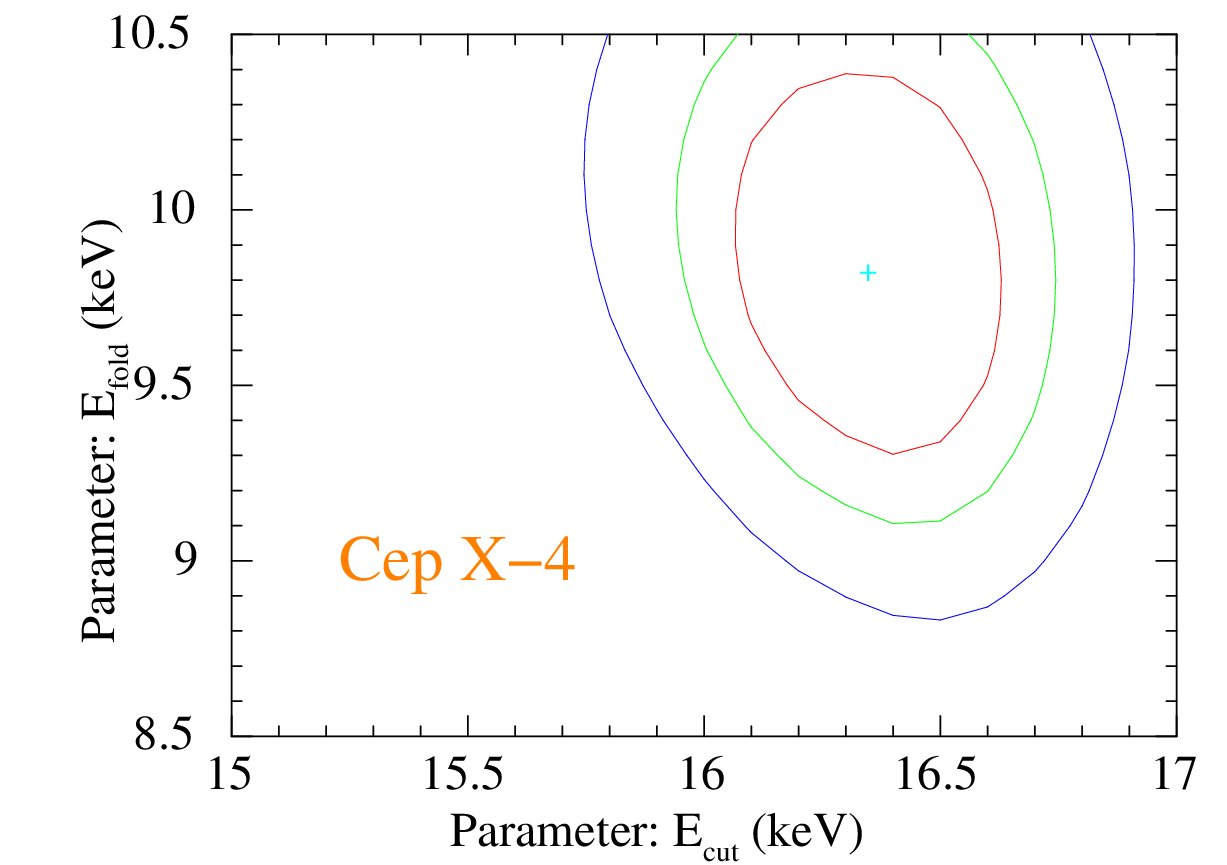}
\vspace{2.5cm}
\caption{Contour plots between cutoff energy and folding energy}
\label{ecut-efold-cont1}
\end{figure*}

\begin{figure*}
%\hspace{-2cm}
\includegraphics[scale=0.2,angle=-90]{ecut_efold_conf_4u1700.ps}
\includegraphics[scale=0.2,angle=-90]{ecut-efold_cont_4u1822.ps}
\includegraphics[scale=0.2,angle=-90]{ecut_efold_conf_groj1008.ps}
\includegraphics[scale=0.2,angle=-90]{ecut_efold_conf_4u1909.ps}
\includegraphics[scale=0.2,angle=-90]{ecut_efold_conf_igrj16393-4643.ps}
\includegraphics[scale=0.2,angle=-90]{ecut_efold_conf_4u2206.ps}
\includegraphics[scale=0.2,angle=-90]{ecut_efold_conf_swj200.ps}
\includegraphics[scale=0.2,angle=-90]{ecut_efold_conf_ks1947.ps}
\includegraphics[scale=0.2,angle=-90]{ecut_efold_conf_exo.ps}
\includegraphics[scale=0.2,angle=-90]{ecut_efold_conf_smcx1.ps}
\includegraphics[scale=0.2,angle=-90]{ecut_efold_conf_igrj16318-4848.ps}
\includegraphics[scale=0.2,angle=-90]{ecut_efold_conf_igrj16207-5129.ps}
\includegraphics[scale=0.2,angle=-90]{ecut_efold_conf_igrj18410-0535.ps}
\includegraphics[scale=0.2,angle=-90]{ecut-efold-17544.ps}
\vspace{2.5cm}
\caption{Contour plots between cutoff energy and folding energy}
\label{ecut-efold-cont2}
\end{figure*}

\begin{figure*}
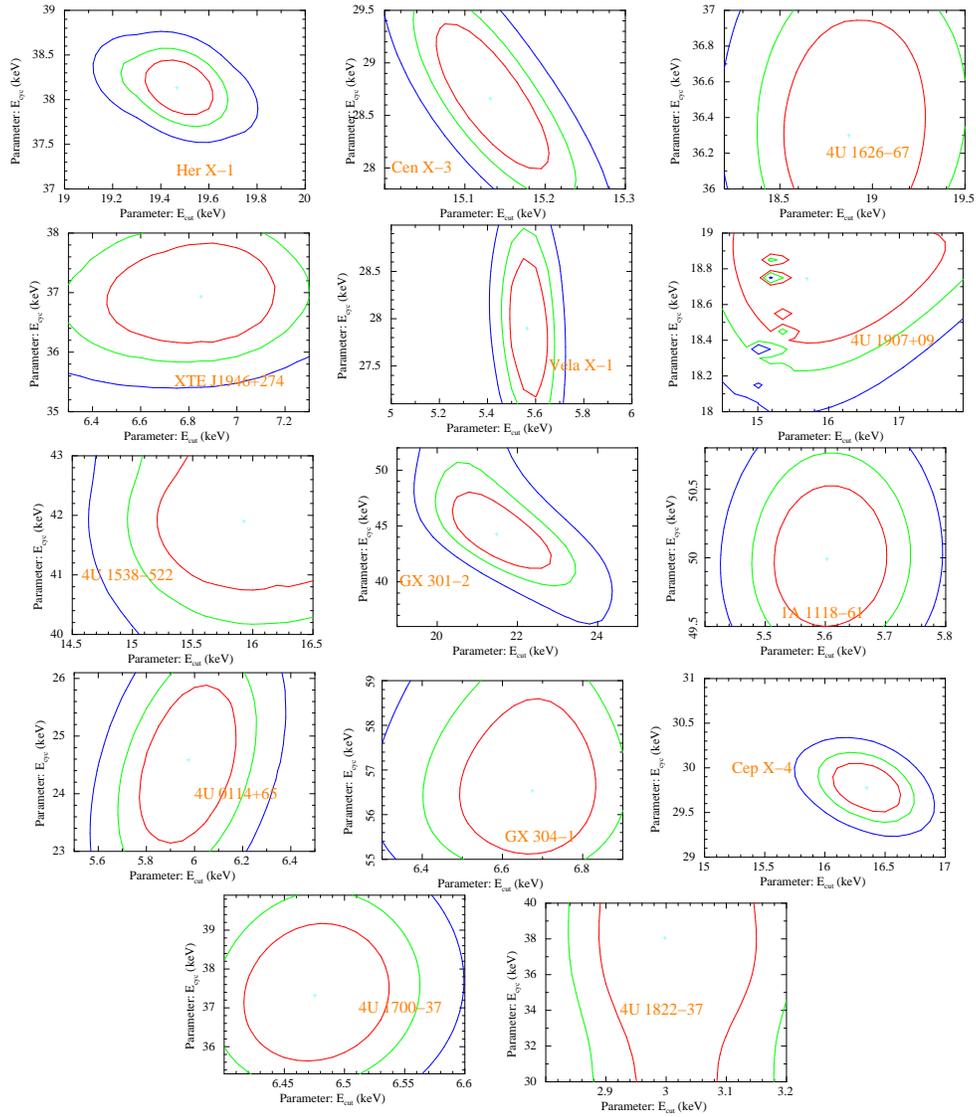

%\hspace{-2cm}
\includegraphics[scale=0.2,angle=-90]{ecut-ecyc-cont_herx1.ps}
\includegraphics[scale=0.2,angle=-90]{ecut_ecyc_conf_cenx3.ps}
\includegraphics[scale=0.2,angle=-90]{ecut_ecyc_cont_4u1626.ps}
\includegraphics[scale=0.2,angle=-90]{ecut_ecyc_conf_xte1946+274.ps}
\includegraphics[scale=0.2,angle=-90]{ecut_ecyc_cont_vela.ps}
\includegraphics[scale=0.2,angle=-90]{ecut_ecyc_conf_4u1907.ps}
\includegraphics[scale=0.2,angle=-90]{ecut_ecyc_conf_4u1538.ps}
\includegraphics[scale=0.2,angle=-90]{ecut-ecyc_gx301.ps}
\includegraphics[scale=0.2,angle=-90]{ecut-ecyc_1a1118.ps}
\includegraphics[scale=0.2,angle=-90]{ecut_ecyc_conf_4u0114.ps}
\includegraphics[scale=0.2,angle=-90]{ecut_ecyc_gx304.ps}
\includegraphics[scale=0.2,angle=-90]{ecut_ecyc_conf_cepx4.ps}
\includegraphics[scale=0.2,angle=-90]{ecut_ecyc_conf_4u1700.ps}
\includegraphics[scale=0.2,angle=-90]{ecut_ecyc_conf_4u1822.ps}
\vspace{2.5cm}
\caption{Contour plots between cutoff energy and cyclotron line energy}
\label{ecut-ecyc-cont}
\end{figure*}

\end{document}